\documentclass[12pt, a4paper]{article}
\usepackage[T2A]{fontenc}
\usepackage[utf8]{inputenc}
\usepackage{cite}
\usepackage{mathtext}
\usepackage{amssymb,amsthm,amsmath}
\usepackage[mathscr]{eucal}
\usepackage{mathrsfs}
\usepackage{colortbl}
\usepackage{bbm}
\usepackage[mathscr]{eucal}
\usepackage{array}
\usepackage[english]{babel}
\usepackage[title,titletoc]{appendix}
\usepackage{graphicx}
\usepackage{float}
\usepackage{upgreek}
\usepackage{xcolor}
\usepackage{color}
\usepackage{relsize}

\begin{document}
\author{\bf Yu.A.\,Markov$\!\,$\thanks{e-mail:markov@icc.ru},\,
M.A.\,Markova$\!\,$\thanks{e-mail:markova@icc.ru},\,
N.Yu.\,Markov$\!\:$\thanks{e-mail:NYumarkov@gmail.com}}

\title{Hamiltonian formalism for Bose excitations\\ in a plasma with a
non-Abelian interaction:\\ plasmon bremsstrahlung}
%
\date{\it\normalsize
\begin{itemize}
\item[]Matrosov Institute for System Dynamics and Control Theory, Siberian Branch, Russian Academy of Sciences, Irkutsk, 664033 Russia
\vspace{-0.3cm}
%
\end{itemize}}
%
\thispagestyle{empty}
\maketitle{}
%
%
\def\theequation{\arabic{section}.\arabic{equation}}
\vspace{-0.2cm}
{
\[
\mbox{\bf Abstract}
\]
The classical Hamiltonian wave theory describing plasmon bremsstrahlung radiation occurring upon collision of high-energy color-charged particles in 
a hot quark-gluon plasma is proposed. A generalization of the Lie-Poisson bracket to the case of a hot QCD medium, involving two independent components $a^{\phantom{\ast}\!\!(1)	\hspace{0.02cm}a}_{\hspace{0.02cm}{\bf k}}$ and $a^{\phantom{\ast}\!\!(2)\hspace{0.02cm}a}_{\hspace{0.02cm}{\bf k}}$ of the bosonic normal field variable $a^{\phantom{\ast}\!\!a}_{\hspace{0.03cm}{\bf k}}$ and two non-Abelian color charges $Q^{\,a}_1$ and $Q^{\,a}_2$ is suggested and the corresponding Hamilton equations are presented. The canonical transformations including both bosonic degrees of freedom of the soft collective excitations of the hot quark-gluon plasma and degrees of freedom of two hard test particles related to their color charges are written out. Two systems of the canonicity conditions for these transformations based on the Lie-Poisson bracket are derived. The most general structure of the canonical transformations in the form of integro-power series in the components $c^{\phantom{\ast}\!\!\!(\alpha)\hspace{0.03cm}a}_{\hspace{0.02cm}{\bf k}},\,\alpha = 1, 2$, of the new normal field variable $c^{\,a}_{\hspace{0.02cm}{\bf k}}$ and in the new color charges ${\mathcal Q}^{\hspace{0.03cm}a}_{\alpha}$ up to the terms of sixth order is presented. 
To construct kinetic equations the multiple-time-scale perturbation theory is used. The notion of the plasmon number density ${\mathcal N}^{\,(\alpha,\hspace{0.03cm}\alpha^{\prime})\hspace{0.03cm}a\hspace{0.03cm}a^{\prime}_{\phantom{1}}\!}_{\hspace{0.02cm}{\bf k}}\!(\tau)$, which is a nontrivial matrix both in the effective color space and in effective two-particle space, is introduced. An explicit form of the effective fifth-order Hamiltonian ${\mathcal H}^{(5)}$ describing the bremsstrahlung emission of longitudinal waves during a collision of two hard color-charged particles in leading tree-level order, is found. The self-consistent system of four Boltzmann-type kinetic equations taking into account the time evolution of the averaged color charges of the hard particles is obtained. The derivation of the equations of motion for the expected value of the color charges $Q^{\,a}_1(t)$ and $Q^{\,a}_2(t)$ is considered. In the approximation of the fixed color charges of two interacting particles the exact solution of the system of kinetic equations defining the dynamics of the colorless $N^{\hspace{0.03cm}(1)}_{\bf k}$, $W^{\hspace{0.03cm}(1)}_{\bf k}$ and color $N^{\hspace{0.03cm}(2)}_{\bf k}$, $W^{\hspace{0.03cm}(2)}_{\bf k}$ components of the plasmon number density, is obtained.
 
}


\newpage

\section{Introduction}
\setcounter{equation}{0}

In the previous work \cite{Markov:2024}, we have developed the Hamiltonian wave theory for collective longitudinally polarized gluon excitations (plasmons) interacting with a classical hard color-charged particle propagating through a hot quark-gluon plasma. We have obtained a generalization of the Lie-Poisson bracket involving bosonic normal field variable $a^{\phantom{\ast}\!\!a}_{\hspace{0.03cm}{\bf k}}$ and a non-Abelian color charge $Q^{\hspace{0.03cm}a}(t),\, a = 1,\,\ldots\,,
N^{\hspace{0.03cm}2}_{c} - 1$ and presented the corresponding Hamilton equations. We have written out the canonical transformations including simultaneously both bosonic degrees of freedom of the soft collective excitations and the degree of freedom of hard test particle connected with its color charge in the quark-gluon plasma, and derived a complete system of the canonicity conditions for these transformations. An important notion of the plasmon number density ${\mathcal N}^{\;a\hspace{0.03cm}a^{\prime}_{\phantom{1}}\!}_{\hspace{0.02cm}{\bf k}}$, which is a nontrivial matrix in an effective color space, has been introduced. We have found an explicit form of the effective fourth-order Hamiltonian describing the elastic scattering of a plasmon off the hard color particle and obtained a self-consistent system of Boltzmann-type kinetic equations taking into account the time evolution of the averaged value of the color charge $\bigl\langle\hspace{0.03cm}\mathcal{Q}^{\hspace{0.03cm}a}(t)
\hspace{0.03cm}\bigr\rangle$ of the hard particle.\\
\indent In the present work, we generalize the previous analysis to the case of the interaction of {\it two} classical hard particles with color charges $Q^{\,a}_1(t)$ and $Q^{\,a}_2(t)$ in the quark-gluon plasma, each of which obeys the Wong equation
\begin{equation}
	\frac{d\hspace{0.03cm}Q^{\hspace{0.03cm}a}_{\alpha}(t)}{d\hspace{0.03cm}t} = gf^{\hspace{0.03cm}a\hspace{0.03cm}b\hspace{0.03cm}c}\hspace{0.02cm}
	(v_{\alpha}\cdot A^{b}(x))\hspace{0.03cm}Q^{\hspace{0.03cm}c}_{\alpha}(t),
	\qquad
	Q^{\hspace{0.03cm}a}_{\alpha}(t)\hspace{0.02cm}
	|_{\hspace{0.02cm}t\hspace{0.02cm}=\hspace{0.02cm}t_{0}}
	=
	Q^{\hspace{0.03cm}a}_{\alpha\hspace{0.03cm}0},
\label{eq:1q}
\end{equation}
where $v^{\hspace{0.02cm}\mu}_{\alpha} = (1,{\bf v}_{\alpha})$ and ${\bf v}_{\alpha}$ is a velocity of the hard particle with the label $\alpha = 1,2$. To make the considerations as simple as possible, just like our previous work, \cite{Markov:2024}, we assume the hard massless particles to be moving along their own straight lines with constant velocities
\begin{equation}
{\bf x}_{\alpha} = {\bf x}_{\alpha 0} + {\bf v}_{\alpha}(t - t_{0}).
\label{eq:1w}
\end{equation} 
Here, ${\bf x}_{\alpha 0}$ is the initial positions of the hard particles.
This rather nontrivial generalization makes it possible to study a fundamentally new interaction process related to the bremsstrahlung of color plasmons in the QGP. To achieve the proposed aim, as the guiding principle, we again use a general Hamiltonian approach to the description of wave processes in nonlinear media with nondecay dispersion laws developed by Zakharov, Kuznetsov, Falkovich {\it et al.} \cite{Zakharov:1971, Zakharov:1974, Zakharov:1985, Zakharov:1992, Zakharov:1997}. We extend the Lie-Poisson bracket to a system with distributed parameters, including bosonic field variables and two non-Abelian charges of two hard particles, construct corresponding canonical transformations, and obtain an effective fifth-order interaction Hamiltonian. This made it possible, in particular, to determine the effective amplitude describing the process of bremsstrahlung of collective gluon excitations, to obtain an appropriate system of kinetic equations for the bremsstrahlung plasmon number density and a self-consistent system of equations of motion for the average values of two non-Abelian charges of colliding particles.\\
\indent The problem of the interaction of two or more color-charged particles in the framework of (quasi)classical theory (mote precisely, with the use of the
Wong equation) has already been considered earlier within various physical settings. So in the works \cite{Khriplovich_1978, Sikivie:1978, Ball:1982} equations and boundary conditions are obtained for the field produced by two point Yang-Mills charges at rest. In this paper, it was pointed out a possible connection between this model and the problem of quark confinement. Further in the papers \cite{Kosyakov:1993, Kosyakov:1994} it was discussed the problem of construction of an exact retarded solution to the classical Yang-Mills equations with the source composed of two arbitrarily mowing color point charges. It is shown that the confinement description turns out to be impossible with the scope of the classical Yang-Mills theory.\\
\indent Also, we point out that there is a fairly extensive number of works where the interaction of classical color-charged particles has been considered without direct references to the use of the Wong equation (see, for example, the papers devoted the problem of two slowly moving color charges \cite{Goloviznin:1993, Cassing:2013, Voronyuk:2015}, with the corresponding extension to the case of relativistic color charges \cite{Zadora:2016}). However, in fact, the Wong-type equations for point-like color charges in these papers inevitably arise as a result of fulfilling the condition of compatibility of the Yang-Mills equations with a source. The source takes the form of the covariant four-current with the delta function for point-like particles. These equations can be derived from (\ref{eq:1q}) by expressing the gauge potential $A^{a}_{\mu}(x)$ in the form of expansions in the color charges $Q^{\hspace{0.03cm}a}_{1}(t)$ and $Q^{\hspace{0.03cm}a}_{2}(t)$ with coefficients playing the role of unknown functions. In fact, we perform such an expansion in the section \ref{section_7}, Eq.\,(\ref{eq:7q}), just not for the potential itself, but for the corresponding correlation function.\\
\indent The above-mentioned works clearly point to the difficulty of the problem of the interaction of two color-charged particles, even while remaining within the framework of (quasi)classical approximations. We generalize this problem to the case of the interaction of two hard color particles moving in an external medium, namely, in the hot quark-gluon plasma. It enables us to calculate an effective amplitude of the plasmon bremsstrahlung in the collision of two ultrarelativistic color particles in the  quark-gluon plasma and to derive the corresponding kinetic equation (more precisely, a self-consistent system of kinetic equations) for the bremsstrahlung plasmon number density.\\
\indent The problems of the theory of bremsstrahlung of longitudinal (Langmuir) and transverse (electromagnetic) waves has been extensively studied in the context of ordinary electron-ion plasma in the semi-classical limit both isotropic as well as anisotropic nonequilibrium systems (see, e.g., \cite{Dupree:1963, Dupree:1964, Tidman:1965, Birmingham:1965, Bekefi_book:1966, Melrose:1972, Ichimaru_book:1973, Akopyan:1975, Akopyan:1976, Sitenko_book:1982, Ginzburg:1990, Tsytovich:1995, Tsytovich_book:1995}). 
This analysis was carried out with the use of various methods, including direct microscopic calculations of the sources of bremsstrahlung radiation \cite{Birmingham:1965} and the method based on the model of pairwise collisions \cite{Bekefi_book:1966}. The fluctuation approach \cite{Tidman:1965, Dupree:1963, Dupree:1964, Ichimaru_book:1973, Sitenko_book:1982} has also been successfully developed based on the interpretation of the bremsstrahlung field as resulting from scattering of fluctuation electromagnetic fields by fluctuations in the charged particle density. In \cite{Tidman:1965, Dupree:1963, Dupree:1964, Ichimaru_book:1973}, such an approach was realized with the use of the equations for the microscopic phase density using perturbation theory methods, while in \cite{Sitenko_book:1982} this approach was realized with the use of the kinetic equations for the waves.\\
\indent Taking into account collective bremsstrahlung effects leads also to the appearance of additional features in the radiation spectra. It is shown that here it takes place important phenomenon such as interference between the radiation corresponding to the ordinary noncoherent Bethe-Heitler radiation and the radiation corresponding to the so-called {\it transition bremsstrahlung} \cite{Ginzburg_book:1990, Platonov:2002} (known also as coherent bremsstrahlung \cite{Platonov:1990} and polarization bremsstrah\-lung \cite{Tsytovich_book:1992}). The latter generated by a relativistic particle in a medium characterized by macroscopic density inhomogeneities, for example, in a plasma with Langmuir waves excited in it, moreover, the spectrum of the Langmuir waves can be arbitrary. It was realized that this effect is fundamental in nature. The transition bremsstrahlung turns out to be significant in electron-ion collisions which produce radiation of the same order as and sometimes also much stronger than ordinary bremsstrahlung. Interference sets in between the Compton and the transition bremsstrahlung and, strictly speaking, the two effects cannot be separated. The interest in bremsstrahlung radiation in an Abelian plasma is mainly due to the fact that it plays a definite role in the generation of radio emission from astrophysical objects and also with the fact that in many cases bremsstrahlung is the main source of fields in thermonuclear plasma and, thus, determines the energy losses of the plasma to radiation. In the papers by Tsytovich {\it et al.} \cite{Akopyan:1975, Akopyan:1976, Ginzburg:1990, Tsytovich:1995, Tsytovich_book:1995} a general technique for calculating collective bremsstrahlung emission and absorption in the QED plasma was suggested.\\
\indent The generalization to a plasma with a non-Abelian type of interaction provides fundamentally new features to bremsstrahlung. As we discussed in detail in \cite{Markov:2024} to the leading order in the strong coupling constant, the basic mechanism of interactions of classical hard color-charged particles and soft gluon excitations is caused not by the spatial oscillations of the charged particle (the normal Thomson scattering), as occurs in an electromagnetic plasma, but it is induced by a precession of the so-called color vector $Q_{\alpha}(t) = (Q^{\hspace{0.03cm}a}_{\alpha}(t))$ of the color-charged particles with the label $\alpha = 1,\,2,\,\ldots$ in a field of soft gluon wave. Particle collisions inside of the QCD medium lead to the rotation of their color charges, as soon as a particle ``hits'' the field of the other particle at the collision point thereby resulting in bremsstrahlung of collective plasma excitations. The QED-like part of bremstrahlung is suppressed and only the dominant non-Abelian contribution survives. In particular, this is manifested in the structure of the corresponding scattering amplitudes and in the form of kinetic equations (see sections \ref{section_7} and \ref{section_8}).\\
\indent The other interaction mechanism is the scattering of a collective wave off the dynamic polarization (transition bremsstrahlung) of the QGP plasma that surrounds the charges in the plasma. The given bremsstrahlung mechanism is possible only in the presence of a medium and cannot exist in vacuum,
i.e. the process of transition bremsstrahlung is a purely collective effect.
The scattering associated with it is sometimes called the {\it nonlinear} one, since the charge of the dynamic polarization of the plasma depends nonlinearly on the gauge fields. In non-Abelian plasma, in contrast to the Abelian plasma \cite{Akopyan:1975, Akopyan:1976, Ginzburg:1990, Tsytovich_book:1995, Tsytovich:1995}, the nonlinear scattering is produced not by the oscillation of the screening polarization cloud (Debye sphere) around the color charge as a result of interaction with the incident scattering wave, but as a consequence of the fact that it is induced by synchronous precession of color vectors of particles forming this cloud in the incident wave field with the frequency $\omega_{\hspace{0.03cm}{\bf k}}$. All color-charged particles within the Debye sphere in the quark-gluon plasma radiate coherently. For calculation of the transition radiation it is necessary to find an effective current produced by the particles. The current depends on both the fast particle field and the background particle spatial distribution, and can be found, for example, from the kinetic equation.\\
\indent The interest in bremsstrahlung in non-abelian plasmas is largely related to  the problem of calculating the radiation energy losses of high-energy color particles passing through a hot QCD medium. As is well known, the energy losses are one of the most important methods for diagnostics of the quark-gluon plasma in ultrarelativistic heavy-ion collisions \cite{Gyulassy:1994, Wang:1995, Baier:1997, Baier:2000, Gyulassy:2001, Zakharov:2001, Zakharov:2007, Qin:2008, Djordjevic:2009}. Here, we are not concerned with this issue in the present study. The radiation energy losses within the framework of the classical Hamilton formalism will be considered in detail in a separate publication.\\
\indent In our previous works \cite{Markov:2020, Markov:2023, Markov:2024} for constructing the kinetic description of collective plasma excitations we have used the approach based on the sequential obtaining of an interconnected chain of kinetic equations which describe the time evolution of hierarchy of irreducible correlation functions (up to the sixth-order correlation functions) of basic dynamical variables: the normal field bosonic $a^{\phantom{\ast}\!\!a}_{\hspace{0.03cm}{\bf k}}$ or fermionic $b^{\phantom{\ast}\!\!i}_{\hspace{0.03cm}{\bf p}}$ variables and the color charge $Q^{\hspace{0.03cm}a}$ of a hard particle. We closed the chain of equations for the correlation functions in the standard way by expressing  sixth-order correlation functions in terms of pair correlation functions. In this paper, we will use a slightly different approach to obtain kinetic equations for the collective excitations of QGP and evolution equations for the (averaged) color charges of two hard particles. This approach known as the {\it multiple-time-scale perturbation expansions} has found wide application in constructing the kinetic equations both for gases and for plasma media \cite{Sandri_1:1963, Sandri_2:1963, Sandri_1965, Frieman:1963, Klimontovich:1965, Benney:1966, Zakharov:1967, Zakharov1968, Davidson:1972, Sitenko:1973, Crawford:1980, Oberman:1983}. It is a modification of the general Bogolubov-Krylov method used in nonlinear mechanics \cite{Krylov:1950}. The multiple time scale approach uses a technique that separates completely the different time components exhibited by the evolution of QGP when an appropriate parameter is small. For the problem being considering it turned out to be more effective. As shown in section \ref{section_7}, it is within the framework of the multiple time scale method that we succeed to correctly reproduce the energy conservation law in the quasiclassical limit in the elementary process of single-plasmon bremsstrahlung in the integrand of kinetic equations for plasmon number densities and in the evolution equations for the averaged color charges.\\
%
%
\indent The paper is organized as follows. In section \ref{section_2}, the general form of the decomposition into plane waves of the gauge field potential is written out. A normal field variable $a^{\phantom{\ast}\!\!a}_{\hspace{0.03cm}{\bf k}}$ included in this decomposition, is presented as a sum of two independent components $a^{\phantom{\ast}\!\!\!(1)\hspace{0.02cm}a}_{\hspace{0.02cm}{\bf k}}$ and $a^{\phantom{\ast}\!\!\!(2)	\hspace{0.02cm}a}_{\hspace{0.02cm}{\bf k}}$.
In the terms of the new variables $a^{\phantom{\ast}\!\!\!(\alpha)\hspace{0.03cm}a}_{\hspace{0.02cm}{\bf k}},\,\alpha = 1, 2$, the free Hamiltonian $H^{(0)}$ of a system of noninteracting plasmons is defined. A generalization of the Lie-Poisson bracket including two Yang-Mills charges $Q^{\hspace{0.03cm}a}_{1}$ and $Q^{\hspace{0.03cm}a}_{2}$, and two components $a^{\phantom{\ast}\!\!\!(\alpha)\hspace{0.03cm}a}_{\hspace{0.02cm}{\bf k}}$ of the normal field variable is given. The corresponding Hamilton equations are defined and the most general structure of the third-order interaction Hamiltonian $H^{(3)}$ with respect to the color charges $Q^{\hspace{0.03cm}a}_{1}$ and $Q^{\hspace{0.03cm}a}_{2}$ of hard particles and the normal variable $a^{\phantom{\ast}\!\!a}_{\hspace{0.03cm}{\bf k}}$ 
of soft bosonic field of a hot quark-gluon plasma is written out. 
In section \ref{section_3}, the canonical transformations including both soft bosonic and color charge degrees of freedom of the quark-gluon plasma are discussed. Two systems of the canonicity conditions for these transformations based on the Lie-Poisson bracket are derived. The most general structure of the canonical transformations in the form of integro-power series in the components $c^{\phantom{\ast}\!\!\!(\alpha)\hspace{0.03cm}a}_{\hspace{0.02cm}{\bf k}},\,\alpha = 1, 2$, of the new normal field variable $c^{\,a}_{\hspace{0.02cm}{\bf k}}$ and the new color charges ${\mathcal Q}^{\hspace{0.03cm}a}_{\alpha}$ up to the terms of sixth order, is presented. The algebraic relations for the second-order coefficient functions of the canonical transformations are provided.
In section~\ref{section_4}, making use of  the aforementioned canonical transformations, the problem of eliminating the third-order interaction Hamiltonian $H^{(3)}$ is considered. The explicit expressions for the coefficient functions linear and quadratic in the variables $c^{\phantom{\ast}\!\!\!(\alpha)\hspace{0.03cm}a}_{\hspace{0.02cm}{\bf k}}$ and ${\mathcal Q}^{\hspace{0.03cm}a}_{\alpha}$ of the canonical transformations are found. An explicit form of the effective amplitude ${T}^{\hspace{0.03cm}(\rho)\hspace{0.03cm}a\,a_{1}\hspace{0.03cm}a_{2}}_{\; {\bf k}}(t)$ describing the plasmon bremsstrahlung occurring upon collision of two hard color-charged particles with one another in the leading tree-level order is given and the corresponding effective fifth-order Hamiltonian ${\mathcal H}^{(5)}$ is written out. A diagrammatic interpretation of the individual terms in the effective amplitude is presented.\\ 
\indent Section \ref{section_5} is concerned with the so-called the multiple-time-scale perturbation expansions approach. According to this approach, the wave field $c^{\hspace{0.02cm}(\alpha)\hspace{0.02cm}a}_{\hspace{0.02cm}{\bf k}}(t)$ and 
color charges $\mathcal{Q}^{\,a}_{\hspace{0.03cm}\alpha}(t)$ of hard particles are taken as a sum of a slowly varying (in time $\tau$) components $C^{\hspace{0.02cm}(\alpha)\hspace{0.02cm}a}_{\hspace{0.02cm}{\bf k}}(\tau)$ and
$\mathcal{Q}^{\,a}_{\hspace{0.03cm}\alpha}(\tau)$ and a small, rapidly varying (in time $t^{\prime}$) components $\widehat{C}^{\hspace{0.02cm}(\alpha)\hspace{0.02cm}a}_{\hspace{0.02cm}{\bf k}}(t')$ and $\widehat{\mathcal{Q}}^{\,d}_{\hspace{0.03cm}\alpha}(t')$. The corresponding decomposition of Hamilton's equations into equations for the fast and slow variables is carried out, and integral nonlinear representations for the fast components as functions of the slow components are explicitly defined.
In section \ref{section_6}, the equation for the slowly varying field components $C^{\hspace{0.02cm}(\alpha)\hspace{0.02cm}a}_{\hspace{0.02cm}{\bf k}}(\tau)$ is written out. The notion of the plasmon number density ${\mathcal N}^{\,(\alpha,\hspace{0.03cm}\alpha^{\prime})\hspace{0.03cm}a\hspace{0.03cm}a^{\prime}_{\phantom{1}}\!}_{\hspace{0.02cm}{\bf k}}\!(\tau)$, which is a nontrivial matrix in effective color and two-particle spaces, is introduced. The higher-order correlation functions of the slowly field components $C^{\hspace{0.02cm}(\alpha)\hspace{0.02cm}a}_{\hspace{0.02cm}{\bf k}}(\tau)$ and the slowly charges $\mathcal{Q}^{\,d}_{\hspace{0.03cm}\alpha}(\tau)$ within the Gaussian approximation for a low nonlinearity level of interacting Bose-excitations are expressed in terms of pair correlation functions. On the basis of this approximation, a matrix kinetic equation for the number density of color plasmons describing bremsstrahlung radiation process of collective gluon excitations from hard test color-charged particles is constructed. The decomposition of the matrix function ${\mathcal N}^{\,(\alpha,\hspace{0.03cm}\alpha^{\prime})\hspace{0.03cm}a\hspace{0.03cm}a^{\prime}_{\phantom{1}}\!}_{\hspace{0.02cm}{\bf k}}\!(\tau)$ into two independent components in the effective two-particle space, i.e. the decomposition with respect to a pair of indices $\alpha$ and $\alpha^{\prime}$ is suggested.\\
\indent In section \ref{section_7}, the color decomposition of two independent matrix functions ${\mathcal N}^{\;a\hspace{0.03cm}a^{\prime}_{\phantom{1}}\!}_{\hspace{0.02cm}
{\bf k}}$ and ${\mathcal W}^{\;a\hspace{0.03cm}a^{\prime}_{\phantom{1}}\!}
_{\hspace{0.02cm}{\bf k}}$ is written out and the first moment about color of two matrix kinetic equations, defining scalar kinetic equations for the colorless parts $N^{\hspace{0.03cm}(1)}_{\bf k}$ and $W^{\hspace{0.03cm}(1)}_{\bf k}$ of this decomposition, is calculated. It is determined a complete structure of the bremsstrahlung plasmon amplitudes ${T}^{\hspace{0.03cm}(\rho)
\hspace{0.03cm}a\,a_{1}\hspace{0.03cm}a_{2}}_{\; {\bf k},\,{\bf q}},\,\rho = 1,2$, each of which is a sum of two terms of a different physical nature. Certain colorless combinations of the mean values of the color charges $\langle{\mathcal Q}^{\hspace{0.03cm}a}_{1}\rangle$ and $\langle{\mathcal Q}^{\hspace{0.03cm}a}_{2}\rangle$. These combinations entering as time-dependent factors into the desired kinetic equations, are introduced. Two special cases $SU(2_{c})$ and $SU(3_{c})$ of the color group are discussed.
Section \ref{section_8} is devoted to determining the second moment about color of the matrix kinetic equations. The moment defines scalar kinetic equations for the second pair of the scalar functions $N^{\hspace{0.03cm}(2)}_{\bf k}$ and $W^{\hspace{0.03cm}(2)}_{\bf k}$ of the decomposition of the matrix plasmon number densities ${\mathcal N}^{\;a\hspace{0.03cm}a^{\prime}_{\phantom{1}}\!}_{\hspace{0.02cm}{\bf k}}$ and 
${\mathcal W}^{\,a\hspace{0.03cm}a^{\prime}_{\phantom{1}}\!}_{\hspace{0.02cm}
{\bf k}}$. The kinetic equations with the relevant conservation law of the elementary bremsstrahlung process are obtained only for the particular value $N_{c} = 3$. The non-closedness of a system of kinetic equations derived in this and in the previous sections is discussed.
In sections \ref{section_9} and \ref{section_10}, the derivation of the equations of motion for the expected value of color charges $Q^{\hspace{0.03cm}a_{1}}_{1}(t)$ and $Q^{\hspace{0.03cm}a_{2}}_{2}(t)$ is considered in details. Based on these equations, the equations determining the time evolution of certain colorless combinations of the second and fourth orders with respect to the averaged color charges $\langle{\mathcal Q}^{\hspace{0.03cm}a_{1}}_{1}\rangle$ and $\langle{\mathcal Q}^{\hspace{0.03cm}a_{2}}_{2}\rangle$ for the specific choice of the color group are defined. It is shown that the resulting system of nonlinear ordinary differential equations, even for the special case of $N_{c}=3$, is not self-contained, since the new higher-order colorless structures inevitably appear. For the colorless structures it is necessary to derive proper equations, etc.\\
\indent Section \ref{section_11} is concerned with the construction in an explicit form of solution of the self-consistent system of kinetic equations describing the evolution of the scalar plasmon number densities $N^{(1)}_{\mathbf k},\, N^{(2)}_{\mathbf k},\,W^{(1)}_{\mathbf k}$ and $W^{(2)}_{\mathbf k}$. In the approximation of the fixed averaged color charges of two interacting particles, these kinetic equations are reduced to a system of four linear ordinary differential equations of first order with constant coefficients. At a certain relation between the averaged color charges, an exact solution of this system is found by standard method of the theory of differential equations.
Finally, in the concluding section \ref{section_12}, three points underlying the construction of the classical Hamiltonian formalism for the one-plasmon bremsstrahlung process in a quark-gluon plasma, are emphasized once again. Also here several interesting issues which are very close to the subject of the present research are briefly discussed, namely, the issues relating to  consideration of higher-order radiation processes: two-plasmon bremsstrahlung and so-called soft ``one-loop''  corrections to tree-level single plasmon bremsstrahlung. In addition, in this section we outline possible ways of a generalization of the radiation processes in question to the fermion sector of hard and soft excitations of the quark-gluon plasma and also possible application of the developed theory to the problem of energy loss.\\
\indent In Appendix \ref{appendix_A}, we present of the basic expression for the effective three-gluon vertex function 
within the framework of the hard thermal loop approximation. 
In Appendix \ref{appendix_B}, the kinetic equation for the second color matrix function ${\mathcal W}_{\hspace{0.02cm}{\bf k}}(\tau) = \bigl({\mathcal W}^{\hspace{0.03cm}a\hspace{0.03cm}a^{\prime}_{\phantom{1}}\!}_{\hspace{0.02cm}{\bf k}}(\tau)\bigr)$ is given. This equation complements a similar kinetic equation for the first color matrix function, i.e. for ${\mathcal N}_{\hspace{0.02cm}{\bf k}}(\tau) = \bigl({\mathcal N}^{\hspace{0.03cm}a\hspace{0.03cm}a^{\prime}_{\phantom{1}}\!}_{\hspace{0.02cm}{\bf k}}(\tau)\bigr)$, introduced in section \ref{section_6}. 
In Appendix \ref{appendix_C}, a general form of evolution equation for the expected value of color charge of the second hard test particle $\langle{\mathcal Q}^{\hspace{0.03cm}a_{2}}_{2}(t)\rangle$ is written out. This equation complements a similar equation for $\langle{\mathcal Q}^{\hspace{0.03cm}a_{1}}_{1}(t)\rangle$, given in the section \ref{section_9}.\\
\indent In Appendix \ref{appendix_D}, the necessary traces for generators in the adjoint representation of the $SU(N_{c})$ color group up to the fifth order and some useful relations between these generators are presented. There are also two additional identities valid only for the special case of $N_{c} = 3$.
In Appendix \ref{appendix_E}, for the special case $N_{c}= 3$, a number of relations and identities are given for certain colorless combinations of two averaged color charges $\langle{\mathcal Q}^{\hspace{0.03cm}a_{1}}_{1}(t)\rangle$ and $\langle{\mathcal Q}^{\hspace{0.03cm}a_{2}}_{2}(t)\rangle$. These relations and identities have been actively used throughout the article.
Appendix \ref{appendix_F} is devoted to a comprehensive analysis of imaginary contributions to an evolution equation for the colorless combination
$\Lambda^{2} = \Lambda^{a}\Lambda^{a}$, where $\Lambda^{a} \equiv f^{\hspace{0.03cm}a\,a_{1}\hspace{0.03cm}a_{2}\hspace{0.03cm}}
\bigl\langle\hspace{0.03cm}\mathcal{Q}^{\,a_{1}}_{\hspace{0.03cm}1}
\hspace{0.03cm}\bigr\rangle\bigl\langle\hspace{0.03cm}
\mathcal{Q}^{\,a_{2}}_{\hspace{0.03cm}2}\hspace{0.03cm}\bigr\rangle$, and to the proof that these contributions vanish at $N_{c} = 3$.

\section{Interaction Hamiltonian of plasmons and two hard particles}
\setcounter{equation}{0}
\label{section_2}

\begin{figure}[t]
\centering
\begin{center}
\begin{tabular*}{0.5\textwidth}{@{}ccc@{}}
\raisebox{-0.26\height}{\resizebox{0.2\textwidth}{!}
{\includegraphics{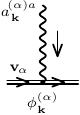}}}
&{ }{ }{ }{ }{ }${\mathbf +}${ }{ }{ }{ }&
\raisebox{-0.26\height}{\resizebox{0.218\textwidth}{!}
{\includegraphics{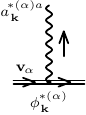}}
}
\end{tabular*}
\end{center}  
\caption{\small Elementary interaction vertices of soft boson excitations with a hard test color-charged particle with a label $\alpha$. The double line denotes a hard particle carrying a color charge $Q^{\hspace{0.03cm}a}_{\alpha}$. The interaction vertices for the incoming and outgoing wave lines (plasmons) are defined by the functions ${\upphi}^{(\alpha)}_{\hspace{0.02cm}{\bf k}}$ and ${\upphi}^{\hspace{0.01cm}\ast\,(\alpha)}_{\hspace{0.02cm}{\bf k}}$ in the Hamiltonian $H^{(3)}$}
\label{fig1-crop}
\end{figure}

The gauge field potentials describing the gluon field in the system are $N_c\times N_c$ matrices in the color space and are defined in terms of $A_{\mu}(x) = A_{\mu}^{a}(x)\, t^{a}$ with $N^{\hspace{0.03cm}2}_c - 1$ Hermitian generators $t^{a}$ of the color $SU(N_c)$ group in the fundamental representation\footnote{\,The color indices $a,\,b,\,c,\,\ldots$ run through values $1,2,\,\ldots\,,N^{\hspace{0.02cm}2}_{c}-1$, while the vector indices $\mu,\,\nu,\,\lambda,\,\ldots$ run through values $0,1,2,3$. Everywhere in this article, we imply summation over repeated indices and use the system of units with $\hbar = c = 1$.}.\\
\indent It is known that there exist two types of the physical soft gluon fields in an equilibrium hot quark-gluon plasma: transverse- and longitudinal-polarized ones \cite{Kalashnikov:1980}. For simplicity, we confine our analysis only to processes involving longitudinally polarized plasma excitations, which are known as {\it plasmons}. These excitations are a purely collective effect of the medium, which has no analogs in the conventional quantum field theory. Let us consider the gauge field potential in the form of the decomposition into plane waves \cite{Blaizot:1994, Hakim:2011}
\begin{equation}
	A^{a}_{\mu}(x) = \int\!d\hspace{0.02cm}{\bf k}\left(\frac{Z_{l}({\bf k})}
	{2\omega^{l}_{{\bf k}}}\right)^{\!\!1/2}\!\!
	\left\{\epsilon^{\ \! l}_{\mu}({\bf k})\, a^{\phantom{\ast}\!\!a}_{{\bf k}}\ \!{\rm e}^{-i\hspace{0.03cm}\omega^{l}_{{\bf k}}\hspace{0.02cm}t\hspace{0.03cm} +\hspace{0.03cm} i\hspace{0.03cm}{\bf k}\hspace{0.02cm}\cdot\hspace{0.02cm} {\bf x}}
	+
	\epsilon^{\ast\, l}_{\mu}({\bf k})\, a^{\ast\ \!\!a}_{{\bf k}}\ \!{\rm e}^{\hspace{0.02cm}i\hspace{0.03cm}\omega^{l}_{{\bf k}}\hspace{0.02cm}t\hspace{0.03cm} -\hspace{0.03cm} i\hspace{0.03cm}{\bf k}\hspace{0.02cm}\cdot\hspace{0.02cm} {\bf x}}
	\right\},
	\label{eq:2q}
\end{equation}
where $\epsilon^{\ \! l}_{\mu} ({\bf k})$ is the polarization vector of a longitudinal mode  (${\bf k}$ is the wave vector). The asterisk $\ast$ denotes the complex conjugation. The factor $Z_{l}({\bf k})$ is the residue of the effective gluon propagator at the longitudinal pole. Finally, $\omega^{\ \! l}_{{\bf k}}$ is the dispersion relation of the longitudinal mode. We consider the amplitude for longitudinal $a^{\phantom{\ast}\!\!a}_{{\bf k}}$ excitations as ordinary (complex) random function. The expectation value of the product of two bosonic amplitudes is
\[
	\bigl\langle\hspace{0.03cm}a^{\ast\hspace{0.03cm}a}_{{\bf k}}\hspace{0.03cm} a^{\phantom{\ast}\!\!b}_{{\bf k}^{\prime}}\bigr\rangle
	=
	\delta^{\hspace{0.03cm}a\hspace{0.02cm}b}\hspace{0.03cm}\delta({\bf k} - {\bf k}^{\prime})\hspace{0.05cm}{\mathcal N}^{\hspace{0.03cm}l}_{\bf k},
\]
where ${\mathcal N}^{\hspace{0.03cm}l}_{\bf k}$ is the number density of the longitudinal plasma waves. Note that the correlation function is written for a hot quark-gluon plasma without external color fields or high-energy color-charged particles penetrating into the plasma from outside. The dispersion relation $\omega^{\ \! l}_{\hspace{0.02cm}{\bf k}}$ for plasmons satisfies the following dispersion equation \cite{Kalashnikov:1980}:
\begin{equation}
{\rm Re}\ \!\varepsilon^{\hspace{0.02cm}l}(\omega,{\bf k})=0\ \!,
\label{eq:2e}
\end{equation}
where
\[
\varepsilon^{\hspace{0.02cm}l}(\omega,{\bf k})=1+\frac{3\hspace{0.02cm}\omega^{\hspace{0.02cm}2}_{pl}}{{\bf k}^{\hspace{0.01cm}2}}
\biggl[1-F\biggl(\frac{\omega}{|{\bf k}|^{2}}\biggr)\biggr],
\quad
F(x) = \frac{x}{2}\left[\hspace{0.03cm}\ln\left|\frac{1+x}{1-x}\right|-i\pi\theta(1-|x|)\right]
\]
is the longitudinal permittivity, $\omega_{\rm pl}^2 = g^2(2N_c+N_f)T^2/18$ is a plasma frequency squared, $T$ is the temperature of the system, $g$ is the strong interaction constant, and $N_{f}$ represents the number of flavors of massless quarks.\\
\indent An important initial step is to represent the original normal variable $a^{\phantom{\ast}\!\!a}_{\hspace{0.02cm}{\bf k}}$ as the following sum of two independent variables:
\begin{equation}
a^{\phantom{\ast}\!\!a}_{\hspace{0.02cm}{\bf k}}
=
\sum_{\alpha = 1,2}a^{\phantom{\ast}\!\!\!(\alpha)
\hspace{0.02cm}a}_{\hspace{0.02cm}{\bf k}}.
\label{eq:2ee}
\end{equation} 
This is the first of our underlying assumptions in constructing the classical theory of plasmon bremsstrahlung within the framework of the Hamiltonian formalism. Next, we want the components $a^{\phantom{\ast}\!\!(\alpha)
\hspace{0.02cm}a}_{\hspace{0.02cm}{\bf k}}$ and $a^{\ast\ \!\!(\alpha)
\hspace{0.02cm}a}_{\hspace{0.02cm}{\bf k}}$ to be satisfied the following Lie-Poisson bracket $({\rm LPB})$ relations
\begin{equation}
\bigl\{a^{(\alpha)
\hspace{0.02cm}a}_{\hspace{0.02cm}{\bf k}},\,a^{(\beta)
\hspace{0.02cm}b}_{\hspace{0.02cm}{\bf k}^{\prime}}\bigr\}_{\rm LPB} = 0,
\quad\!
\bigl\{a^{\ast\ \!\!(\alpha)
\hspace{0.02cm}a}_{\hspace{0.02cm}{\bf k}},\,a^{\ast\ \!\!(\beta)
\hspace{0.02cm}b}_{\hspace{0.02cm}{\bf k}^{\prime}}\bigr\}_{\rm LPB} = 0, 
\quad\!
\bigl\{a^{(\alpha)
\hspace{0.02cm}a}_{\hspace{0.02cm}{\bf k}},\,a^{\ast\ \!\!(\beta)
\hspace{0.02cm}b}_{\hspace{0.02cm}{\bf k}^{\prime}}\bigr\}_{\rm LPB}
=
\delta^{\hspace{0.02cm} \alpha\beta}\hspace{0.02cm}
\delta^{\hspace{0.02cm} ab}\hspace{0.02cm}\delta({\bf k} - {\bf k}^{\prime}).
\label{eq:2r}
\end{equation}
On the other hand, when we consider the color charges $Q^{\hspace{0.03cm}a}_{1}$ and $Q^{\hspace{0.03cm}a}_{2}$ of two hard test particles, the same Lie-Poisson bracket, must have the following form:
\begin{equation}
\hspace{0.04cm}
\bigl\{Q^{\,a}_{\alpha},\hspace{0.03cm}Q^{\,b}_{\beta}\hspace{0.03cm} \bigr\}_{\rm LPB} 
= 
\,\delta_{\hspace{0.02cm} \alpha\beta}\hspace{0.02cm}
f^{\hspace{0.03cm}a\hspace{0.03cm}b\hspace{0.03cm}c}\hspace{0.03cm}Q^{\hspace{0.03cm}c}_{\beta},
\quad (\mbox{no summation over $\beta$\hspace{0.03cm}!}).
\label{eq:2t}
\end{equation}
For the case of continuous media, we take the following expression as the definition of the Lie-Poisson bracket:
\begin{equation}
\bigl\{F,\,G\bigr\}_{\rm LPB} 
=
\sum_{\rho\hspace{0.02cm}=\hspace{0.02cm}1,2}
\int\! d\hspace{0.02cm}{\bf k\hspace{0.01cm}}'\!\hspace{0.02cm}
\left\{\frac{\delta\hspace{0.01cm} F}{\delta\hspace{0.01cm} a^{(\rho)\hspace{0.02cm}c}_{{\bf k}'}}
\hspace{0.03cm}\frac{\delta\hspace{0.01cm}G}{\delta\hspace{0.01cm}a^{\ast\ \!\!(\rho)\hspace{0.02cm}c}_{{\bf k}'}}
\,-\,
\frac{\delta\hspace{0.01cm}F}{\delta\hspace{0.01cm} a^{\ast\ \!\!(\rho)\hspace{0.02cm}c}_{{\bf k}'}}\hspace{0.03cm}
\frac{\delta\hspace{0.01cm}G}{\delta\hspace{0.01cm} a^{(\rho)\hspace{0.02cm}c}_{{\bf k}'}}\right\}
\,+\,
i\hspace{0.01cm}\sum_{\rho\hspace{0.02cm}=\hspace{0.02cm}1,2}
\frac{\partial F}{\,\partial\hspace{0.03cm}Q^{\,a}_{\rho}}\hspace{0.03cm}
\frac{\partial\hspace{0.03cm}G}{\,\partial\hspace{0.03cm} Q^{\hspace{0.03cm}b}_{\rho}}
\,f^{\hspace{0.03cm}a\hspace{0.03cm}b\hspace{0.03cm}c}\hspace{0.03cm}
Q^{\hspace{0.03cm}c}_{\rho}.
\label{eq:2y}
\end{equation}
The first term is the standard canonical bracket. Next, for the sake of simplicity of notation, the abbreviation ${\rm LPB}$ will be omitted, thereby suggesting that by the braces $\{\,,\}$ we always mean the Lie-Poisson bracket (\ref{eq:2y}).\\
\indent Let us write the Hamilton equations for the components  $a^{\phantom{\ast}\!\!\!(\alpha)
\hspace{0.02cm}a}_{\hspace{0.02cm}{\bf k}}$, $a^{\ast\ \!\!\!(\alpha)
\hspace{0.02cm}a}_{\hspace{0.02cm}{\bf k}}$ and color charges $Q^{\,a}_{\alpha}$:  
\begin{align}
&\frac{\partial\hspace{0.02cm}a^{\phantom{\ast}\!\!\!(\alpha)
\hspace{0.02cm}a}_{\hspace{0.02cm}{\bf k}}}{\partial\hspace{0.02cm} t}
=
-\hspace{0.03cm}i\hspace{0.03cm}\bigl\{a^{\phantom{\ast}\!\!\!(\alpha)
\hspace{0.02cm}a}_{\hspace{0.02cm}{\bf k}}, H\bigr\} \equiv  -i\,\frac{\!\!\!\delta H}{\delta\hspace{0.01cm} a^{\!\ast\ \!\!(\alpha)
\hspace{0.02cm}a}_{\hspace{0.02cm}{\bf k}}}\,,
\qquad
\frac{\partial\hspace{0.02cm}a^{\ast\ \!\!(\alpha)
\hspace{0.02cm}a}_{\hspace{0.02cm}{\bf k}}}{\partial\hspace{0.02cm} t}
=
-\hspace{0.03cm}i\hspace{0.03cm}\bigl\{a^{\ast\ \!\!(\alpha)
\hspace{0.02cm}a}_{\hspace{0.02cm}{\bf k}}, H\bigr\} 
\equiv  
i\,\frac{\!\!\delta H}{\delta\hspace{0.01cm} a^{\phantom{\ast}\!\!\!(\alpha)
\hspace{0.02cm}a}_{\hspace{0.02cm}{\bf k}}}\,,
\label{eq:2u}\\[1ex]
&\frac{d\hspace{0.02cm}Q^{\hspace{0.03cm}a}_{\alpha}}{d\hspace{0.02cm} t}
=
-\hspace{0.03cm}i\hspace{0.01cm}
\left\{Q^{\hspace{0.03cm}a}_{\alpha}, H\right\} =  
\frac{\!\partial H}{\partial\hspace{0.03cm} Q^{\hspace{0.03cm}b}_{\alpha}}\,
f^{\hspace{0.03cm}a\hspace{0.03cm}b\hspace{0.03cm}c}
\hspace{0.03cm}Q^{\hspace{0.03cm}c}_{\alpha},
\quad 
Q^{\hspace{0.03cm}a}_{\alpha}|_{t\hspace{0.02cm}=\hspace{0.02cm}t_{0}} = Q^{\hspace{0.03cm}a}_{0\hspace{0.02cm}\alpha},
\label{eq:2i}
\end{align}
where on the right-hand side of the last equation no summation over the ``index'' $\alpha$ is implied. The function $H$ represents a Hamiltonian for the system of plasmons and two hard test particles, which is equal to a sum $H =  H^{(0)} + H_{int}$, where $H^{(0)}$ is the Hamiltonian of noninteracting plasmons and ${H}_{int}$ is the interaction Hamiltonian of plasmons and the hard color-charged particles. We assume that a given (test) particle with the label $\alpha\,(= 1,2)$ moves in the quark-gluon plasma with a constant velocity ${\bf v}_{\alpha}$. We make another of our underlying assumptions, in this case concerning the structure of the free Hamiltonian $H^{(0)}$, namely, we assume it to be equal to 
\begin{equation}
H^{(0)} =  
\sum_{\alpha\hspace{0.02cm}=\hspace{0.02cm}1,2}
\int\!d\hspace{0.02cm}{\bf k}\hspace{0.04cm}(\omega^{\hspace{0.03cm}l}_{\hspace{0.02cm}{\bf k}} - {\mathbf v}_{\alpha\!}\cdot {\mathbf k})\ \!
a^{\ast\ \!\!(\alpha)\hspace{0.02cm}a}_{\hspace{0.02cm}{\bf k}}
\hspace{0.03cm}
a^{\phantom{\ast}\!\!\!(\alpha)	\hspace{0.02cm}a}_{\hspace{0.02cm}{\bf k}}.
\label{eq:2p}
\end{equation}
Owing to the specific character of the dispersion equation for soft (longitudinal) bosonic excitations (\ref{eq:2e}) in a hot quark-gluon plasma, the factor $(\omega^{\hspace{0.03cm}l}_{\hspace{0.02cm}{\bf k}} - {\mathbf v}_{\alpha\!}\cdot {\mathbf k})$ in the integrand (\ref{eq:2p}) will never turn to zero, i.e.,
\begin{equation}
\omega^{\hspace{0.03cm}l}_{\hspace{0.02cm}{\bf k}} - {\mathbf v}_{\alpha\!}\cdot {\mathbf k} \neq 0	
\label{eq:2a}
\end{equation}
for arbitrary values of the wave vector ${\bf k}$. In other words, linear Landau damping (Cherenkov emission) is kinematically forbidden in the hot quark-gluon plasma.\\ 
\indent In the approximation of small amplitudes the interaction Hamiltonian can be presented in the form of a formal integro-power series in the bosonic functions $a^{\phantom{\ast}\!\!a}_{\hspace{0.02cm}{\bf k}}$ and $a^{\ast\ \!\!a}_{\hspace{0.02cm}{\bf k}}$, and in the color charges $Q^{\hspace{0.03cm}a}_{\alpha}$:
\[
H_{int} = H^{(3)} + H^{(4)} + \, \ldots\,,
\]
where the third-order interaction Hamiltonian has the following structure:
\begin{align}
H^{(3)} 
= \sum_{\alpha\hspace{0.02cm}=\hspace{0.02cm}1,2}&\int\!d\hspace{0.02cm}{\bf k}\hspace{0.03cm}
\Bigl[\hspace{0.02cm}{\upphi}^{(\alpha)}_{\hspace{0.03cm} {\bf k}}(t)
\hspace{0.03cm} Q^{\hspace{0.03cm}a}_{\alpha}\,{a}^{\,a}_{\hspace{0.02cm}{\bf k}}\hspace{0.03cm} 
+
{\upphi}^{\hspace{0.01cm}\ast\hspace{0.03cm}(\alpha)}_{\hspace{0.02cm}{\bf k}}(t)\hspace{0.03cm}
Q^{\hspace{0.03cm}a}_{\alpha}\, {a}^{\hspace{0.03cm}\ast\hspace{0.03cm} a}_{\hspace{0.02cm}{\bf k}}\,\hspace{0.03cm}\Bigr]
\label{eq:2s}\\[0.7ex]
+ 
&\int\!d\hspace{0.02cm}{\bf k}\, d\hspace{0.02cm}{\bf k}_{1}\hspace{0.03cm} d\hspace{0.02cm}{\bf k}_{2}\hspace{0.03cm}
\Bigl\{\hspace{0.02cm}{\mathcal V}^{\; a\, a_{1}\hspace{0.03cm} a_{2}}_{{\bf k},\, {\bf k}_{1},\, {\bf k}_{2}}(t)\, 
a^{\ast\hspace{0.03cm} a}_{\hspace{0.02cm}{\bf k}}\,
a^{\,a_{1}}_{\hspace{0.03cm}{\bf k}_{1}}\, a^{\,a_{2}}_{\hspace{0.03cm}{\bf k}_{2}}
\,+\,
{\mathcal V}^{\,*\,a\, a_{1}\hspace{0.03cm} a_{2}}_{\, {\bf k},\, {\bf k}_{1},\, {\bf k}_{2}}(t)\, 
a^{\!\!\phantom{\ast}a}_{\hspace{0.02cm}{\bf k}}\,a^{\ast\,a_{1}}_{\hspace{0.03cm}{\bf k}_{1}}\hspace{0.03cm}a^{\ast\,a_{2}}_{\hspace{0.03cm}{\bf k}_{2}}
\Bigr\}\hspace{0.04cm}
\delta({\bf k} - {\bf k}_{1} - {\bf k}_{2}) 
\notag\\[0.7ex]
+\, \frac{1}{3}&\int\!d\hspace{0.02cm}{\bf k}\, d\hspace{0.02cm}{\bf k}_{1}\hspace{0.03cm} d\hspace{0.02cm}{\bf k}_{2}\hspace{0.03cm}
\Bigl\{\hspace{0.02cm}{\mathcal U}^{\; a\, a_{1}\hspace{0.03cm} a_{2}}_{\,{\bf k},\,{\bf k}_{1},\, {\bf k}_{2}}(t)\, 
a^{a}_{\hspace{0.02cm}{\bf k}}\, a^{a_{1}}_{\hspace{0.03cm}{\bf k}_{1}}\hspace{0.03cm}
a^{a_{2}}_{\hspace{0.03cm}{\bf k}_{2}}
\,+\,
{\mathcal U}^{\,*\,a\, a_{1}\hspace{0.03cm} a_{2}}_{\; {\bf k},\,
{\bf k}_{1},\, {\bf k}_{2}}(t)\, 
a^{\ast\,a}_{\hspace{0.02cm}{\bf k}}\hspace{0.03cm}
a^{\ast\,a_{1}}_{\hspace{0.03cm}{\bf k}_{1}}\hspace{0.03cm}a^{\ast\,a_{2}}_{\hspace{0.03cm}{\bf k}_{2}}
\Bigr\}\hspace{0.04cm}
\delta({\bf k} + {\bf k}_{1} + {\bf k}_{2}). 
\notag
\end{align}
We will not need an explicit form of higher-order interaction Hamiltonians $H^{(4)},\,\ldots$ for deriving the bremsstrahlung radiation amplitude in the leading approximation. We should immediately note that, unlike the free Hamiltonian (\ref{eq:2p}), we used {\it full} amplitudes  $a^{\phantom{\ast}\!\!a}_{\hspace{0.02cm}{\bf k}}$ and $a^{\ast\ \!\!a}_{\hspace{0.02cm}{\bf k}}$ in writing the interaction Hamiltonian (\ref{eq:2s}) rather than their separate components\footnote{It can be postulated that this circumstance will also be true for higher-order interaction Hamiltonians $H^{(n)},\,n\geq 4$.}. Besides, it should be especially explained that in spite of the fact that, for example, the Hamiltonian $H^{(3)}$ contains mixed contributions that are formally quadratic in the variables $a^{\,a}_{\,{\bf k}}$ and $Q^{\,a}_{\alpha}$, and contributions that are cubic in $a^{\,a}_{\,{\bf k}}$, nevertheless, the whole expression (\ref{eq:2s}) is a third-order interaction Hamiltonian. Thus, we assign the degree of nonlinearity {\it two} to the color charges $Q^{\hspace{0.03cm}a}_{\alpha}$, treating them as composite objects. We refer to the two terms in square brackets in the first line in (\ref{eq:2s}) as elementary vertices of the interaction between soft gluon excitations and a hard color-charged particle with a label $\alpha$, as shown in Fig.\,\ref{fig1-crop}. The vertex functions ${\mathcal V}^{\; a\, a_{1}\hspace{0.03cm} a_{2}}_{{\bf k},\, {\bf k}_{1},\, {\bf k}_{2}}(t)$ and ${\mathcal U}^{\; a\, a_{1}\hspace{0.03cm} a_{2}}_{{\bf k},\, {\bf k}_{1},\, {\bf k}_{2}}(t)$ in the remaining contributions in (\ref{eq:2s}) determine the processes of three-plasmon interaction.\\ 
\indent In the vertex functions ${\upphi}^{(\alpha)}_{\hspace{0.03cm} {\bf k}}(t)$, ${\mathcal V}^{\; a\, a_{1}\hspace{0.03cm} a_{2}}_{{\bf k},\, {\bf k}_{1},\, {\bf k}_{2}}(t)$ and ${\mathcal U}^{\; a\, a_{1}\hspace{0.03cm} a_{2}}_{\,{\bf k},\,{\bf k}_{1},\, {\bf k}_{2}}(t)$ we have explicitly introduced time dependence $t$. In further constructions related to the multi-time formalism (see section \ref{section_5}), we will refer this time to the so-called ``fast'' time. For three-plasmon vertices ${\mathcal V}^{\; a\, a_{1}\hspace{0.03cm} a_{2}}_{{\bf k},\, {\bf k}_{1},\, {\bf k}_{2}}(t)$ and ${\mathcal U}^{\; a\, a_{1}\hspace{0.03cm} a_{2}}_{\,{\bf k},\,{\bf k}_{1},\, {\bf k}_{2}}(t)$ the dependence on $t$ is trivially defined in the following form:
\begin{equation}
{\mathcal V}^{\; a\, a_{1}\hspace{0.03cm} a_{2}}_{{\bf k},\, {\bf k}_{1},\, {\bf k}_{2}}(t) 
=
{\mathcal V}^{\; a\, a_{1}\hspace{0.03cm} a_{2}}_{{\bf k},\, {\bf k}_{1},\, {\bf k}_{2}}\,
{\rm e}^{i\hspace{0.03cm}(\omega^{l}_{{\bf k}} - \omega^{l}_{{\bf k}_{1}} - \omega^{l}_{{\bf k}_{2}})\hspace{0.02cm}t},
\qquad
{\mathcal U}^{\; a\, a_{1}\hspace{0.03cm} a_{2}}_{\,{\bf k},\,{\bf k}_{1},\, {\bf k}_{2}}(t)
=\,
{\mathcal U}^{\; a\, a_{1}\hspace{0.03cm} a_{2}}_{\,{\bf k},\,{\bf k}_{1},\, {\bf k}_{2}}\,
{\rm e}^{-\hspace{0.03cm}i\hspace{0.03cm}(\omega^{l}_{{\bf k}} + \omega^{l}_{{\bf k}_{1}} + \omega^{l}_{{\bf k}_{2}})\hspace{0.02cm}t}.
\label{eq:2dd}
\end{equation} 
Dependence of the vertex function ${\upphi}^{(\alpha)}_{\hspace{0.03cm} {\bf k}}(t)$ on the ``fast'' time is defined just below. The vertex functions ${\mathcal V}^{\; a\, a_{1}\hspace{0.03cm} a_{2}}_{{\bf k},\, {\bf k}_{1},\, {\bf k}_{2}}$ and ${\mathcal U}^{\; a\, a_{1}\hspace{0.03cm} a_{2}}_{\hspace{0.03cm}{\bf k},\, {\bf k}_{1},\, {\bf k}_{2}}$ satisfy the ``conditions of natural symmetry'', which specify that the integrals in Eqs.\,(\ref{eq:2s}) are unaffected by relabeling of the dummy color indices and integration variables. These conditions have the following form:
\begin{equation}
{\mathcal V}^{\;a\,a_{1}\hspace{0.03cm}a_{2}}_{{\bf k},\, {\bf k}_{1},\, {\bf k}_{2}} = {\mathcal V}^{\; a\, a_{2}\, a_{1}}_{{\bf k},\, {\bf k}_{2},\, {\bf k}_{1}},
\qquad\;
{\mathcal U}^{\,a\,a_{1}\hspace{0.03cm}a_{2}}_{\hspace{0.03cm}{\bf k},\, {\bf k}_{1},\,{\bf k}_{2}} 
= 
{\mathcal U}^{\,a\,a_{2}\,a_{1}}_{\hspace{0.03cm}{\bf k},\, {\bf k}_{2},\, {\bf k}_{1}}
= 
{\mathcal U}^{\,a_{1}\hspace{0.03cm}a\,a_{2}}_{\hspace{0.03cm}{\bf k}_{1},\, {\bf k},\, {\bf k}_{2}}.
\label{eq:2f}
\end{equation}
The real nature of the Hamiltonian (\ref{eq:2s}) is obvious.\\ 
\indent The vertex functions in the Hamiltonian $H^{(3)}$ are defined by specific properties of the system under study, in our case, by a high-temperature quark-gluon plasma. An explicit form of the three-point amplitudes ${\mathcal V}^{\; a\, a_{1}\hspace{0.03cm} a_{2}}_{{\bf k},\, {\bf k}_{1},\, {\bf k}_{2}}$ and $\,{\mathcal U}^{\; a\, a_{1}\hspace{0.03cm} a_{2}}_{{\bf k},\, {\bf k}_{1},\, {\bf k}_{2}}$ within the hard thermal loop approximation was obtained in \cite{Markov:2020}. They have the following color and momentum structures: 
\begin{equation}
{\mathcal V}^{\ \!a\,a_{1}\hspace{0.03cm}a_{2}}_{{\bf k},\, {\bf k}_{1},\,{\bf k}_{2}}
=
f^{\hspace{0.03cm}a\,a_{1}\hspace{0.03cm}a_{2}\,}\hspace{0.02cm}{\mathcal V}_{\, {\bf k},\, {\bf k}_{1},\, {\bf k}_{2}},
\qquad
{\mathcal U}^{\ \! a\, a_{1}\hspace{0.03cm} a_{2}}_{{\bf k},\, {\bf k}_{1},\, {\bf k}_{2}}
=
f^{\hspace{0.03cm}a\,a_{1}\hspace{0.03cm}a_{2}\,}\hspace{0.02cm}{\mathcal U}_{\, {\bf k},\, {\bf k}_{1},\, {\bf k}_{2}},
\label{eq:2h}
\end{equation}
where
\begin{equation}
{\mathcal V}_{\, {\bf k},\, {\bf k}_{1},\, {\bf k}_{2}} = 
\frac{1}{2^{3/4}}\,g\hspace{0.03cm}
\Biggl(\frac{\epsilon^{\hspace{0.03cm}l}_{\mu}({\bf k})}{\sqrt{2\hspace{0.03cm}\omega^ {l}_{\hspace{0.03cm}
			{\bf k}_{\phantom{1}}}}}\Biggr)\!
\Biggl(\frac{\epsilon^{\hspace{0.03cm}l}_{\mu_{1}}({\bf k}_{1})}{\sqrt{2\hspace{0.03cm}\omega^{\hspace{0.03cm}l}_{\hspace{0.03cm}
			{\bf k}_{1}}}}\Biggr)\!
\Biggl(\frac{\epsilon^{\hspace{0.03cm}l}_{\mu_{2}}({\bf k}_{2})}{\sqrt{2\hspace{0.03cm}\omega^{\hspace{0.03cm}l}_{\hspace{0.03cm}
			{\bf k}_{2}}}}\Biggr)\!
\,^{\ast}\Gamma^{\mu\mu_1\mu_2}(k,- k_{1},- k_{2})\Bigr|_{\rm \,on-shell}
\hspace{0.4cm} 
\label{eq:2j}
\end{equation}
and
\begin{equation}
{\mathcal U}_{\, {\bf k},\, {\bf k}_{1},\, {\bf k}_{2}} =
\frac{1}{2^{3/4}}\,g\hspace{0.03cm}
\Biggl(\frac{ \epsilon^{\hspace{0.03cm}l}_{\mu}({\bf k})}{\sqrt{2\hspace{0.03cm}\omega^{\hspace{0.03cm}l}_{\hspace{0.03cm}
			{\bf k}_{\phantom{1}}}}}\Biggr)\!
\Biggl(\frac{\epsilon^{\hspace{0.03cm}l}_{\mu_{1}}({\bf k}_{1})}{\sqrt{2\hspace{0.03cm}\omega^{\hspace{0.03cm}l}_{\hspace{0.03cm}
			{\bf k}_{1}}}}\Biggr)\!
\Biggl(\frac{\epsilon^{\hspace{0.03cm}l}_{\mu_{2}}({\bf k}_{2})}{\sqrt{2\hspace{0.03cm}\omega^{\hspace{0.03cm}l}_{\hspace{0.03cm}
			{\bf k}_{2}}}}\Biggr)\!
\,^{\ast}\Gamma^{\mu\mu_1\mu_2}(- k,- k_{1},- k_{2})\Bigr|_{\rm \,on-shell}.
\!
\label{eq:2k}
\end{equation}
The explicit form of the effective three-gluon vertex $\,^{\ast}\Gamma^{\mu\mu_1\mu_2}(k, k_{1}, k_{2})$ on the right-hand side of these expressions is defined by formulae (\ref{ap:A1})\,--\,(\ref{ap:A3}) in Appendix \ref{appendix_A}. By virtue of the color decomposition (\ref{eq:2h}) and of the symmetry properties (\ref{eq:2f}) for the effective vertices (\ref{eq:2j}) and (\ref{eq:2k}), the following permutation relations 
\begin{equation}
{\mathcal V}_{\, {\bf k},\, {\bf k}_{1},\, {\bf k}_{2}}
=
-\hspace{0.02cm}{\mathcal V}_{\, {\bf k},\, {\bf k}_{2},\, {\bf k}_{1}},
\qquad
{\mathcal U}_{\, {\bf k},\, {\bf k}_{1},\, {\bf k}_{2}}
=
-\,{\mathcal U}_{\, {\bf k},\, {\bf k}_{2},\, {\bf k}_{1}}
=
-\,{\mathcal U}_{\, {\bf k}_{1},\, {\bf k},\, {\bf k}_{2}}
\label{eq:2l}
\end{equation}
are satisfied.\\
\indent In closing this section we define an explicit form of the vertex functions ${\upphi}^{(\alpha)}_{\,{\bf k}}(t)$, which enters into the third-order interaction Hamiltonian (\ref{eq:2s}). It is simpler to define this function on the basis of the interaction Hamiltonian of two classical color-charged particles with an external gauge field $A^{a}_{\mu}(x)$:
\[
H_{int} = \sum_{\alpha\hspace{0.02cm}=\hspace{0.02cm}1,2} 
\int\!\frac{d\hspace{0.02cm}{\bf x}}{(2\pi)^{3}}\,j^{a\hspace{0.03cm}\mu}_{Q_{\alpha}}(x)A^{a}_{\mu}(x),
\]
where the color currents of the hard particles $j^{a\hspace{0.03cm}\mu}_{Q_{\alpha}}(x)$, considering (\ref{eq:1w}), are defined by the expressions  
\[
j_{Q_{\alpha}}^{\hspace{0.02cm}a\hspace{0.02cm}\mu} = g\hspace{0.03cm}v^{\hspace{0.02cm}\mu}_{\alpha}\hspace{0.02cm}
Q^{\hspace{0.03cm}a}_{\alpha}(t)	\hspace{0.03cm}
{\delta}^{(3)}\bigl({\bf x} - {\bf x}_{0\hspace{0.03cm}\alpha} - {\bf v}_{\alpha}(t - t_{0})\bigr),\quad
v^{\hspace{0.02cm}\mu}_{\alpha} = (1,{\bf v}_{\alpha}).
\]
We substitute these currents in $H_{int}$ and integrate over ${\bf x}$. As the gauge potential $A^{a}_{\mu}(x)$ we use the longitudinal polarized one on the right-hand side of the expansion (\ref{eq:2q}). As a result, we get for the specific choice of $t_{0} = 0$, 
\[
H_{int} = g\hspace{0.03cm}\sum_{\alpha\hspace{0.02cm}=\hspace{0.02cm}1,2}
v^{\mu}_{\alpha}\hspace{0.03cm}Q^{\hspace{0.03cm}a}_{\alpha}(t)\hspace{0.03cm}
A^{a}_{\mu}\bigl(t,\,{\bf x}_{0\hspace{0.03cm}\alpha} + {\bf v}_{\alpha}\hspace{0.03cm}t)\bigr) 
\vspace{-0.2cm}
\]  
\vspace{-0.2cm}
\begin{align}
= \sum_{\alpha\hspace{0.02cm}=\hspace{0.02cm}1,2}
\int\!d\hspace{0.02cm}{\bf k}\,
\biggl[\hspace{0.03cm}&g
\left(\frac{Z_{l}({\bf k})}
{2\hspace{0.02cm}\omega^{l}_{{\bf k}}}\right)^{\!\!1/2}\!\!\!
(v_{\alpha}\cdot\epsilon^{\ \! l}({\bf k}))\, a^{\phantom{\ast}\!\!a}_{{\bf k}}\hspace{0.03cm}Q^{\hspace{0.03cm}a}_{\alpha}\hspace{0.02cm}
\ \!{\rm e}^{-i\hspace{0.03cm}(\omega^{l}_{{\bf k}} - {\bf k}\cdot{\bf v}_{\alpha})\hspace{0.02cm}t}
{\rm e}^{i\hspace{0.03cm}
{\bf k}\hspace{0.02cm}\cdot\hspace{0.02cm} {\bf x}_{0\hspace{0.02cm}\alpha}}
\notag \\[1ex]
+\,&g
\left(\frac{Z_{l}({\bf k})}
{2\hspace{0.02cm}\omega^{l}_{{\bf k}}}\right)^{\!\!1/2}\!\!\!
(v\hspace{0.03cm}\cdot\epsilon^{\ast\,l}({\bf k}))\, 
a^{\ast\,a}_{{\bf k}}\hspace{0.03cm}Q^{\hspace{0.03cm}a}\hspace{0.02cm}
\ \!{\rm e}^{\hspace{0.03cm}i\hspace{0.03cm}(\omega^{l}_{{\bf k}} - {\bf k}
\cdot{\bf v}\hspace{0.03cm})\hspace{0.02cm}t}{\rm e}^{-i\hspace{0.03cm}
{\bf k}\hspace{0.02cm}\cdot\hspace{0.02cm} {\bf x}_{0\hspace{0.02cm}\alpha}}\biggr].
\notag
\end{align}
Comparing this expression with Hamiltonian $H^{(3)}$, Eq.\,(\ref{eq:2s}), we find the following identification for the vertex function ${\upphi}^{(\alpha)}_{\hspace{0.03cm}{\bf k}}(t)$:
\begin{equation}
{\upphi}^{(\alpha)}_{\hspace{0.03cm}{\bf k}}(t)
=
{\upphi}^{(\alpha)}_{\hspace{0.03cm}{\bf k}}
\ \!{\rm e}^{-i\hspace{0.03cm}(\omega^{l}_{{\bf k}} - {\bf k}\cdot{\bf v}_{\alpha})\hspace{0.02cm}t}\hspace{0.03cm}
{\rm e}^{i\hspace{0.03cm}
{\bf k}\hspace{0.02cm}\cdot\hspace{0.02cm} {\bf x}_{0\hspace{0.02cm}\alpha}},
\label{eq:2z}
\end{equation}
where
\[
{\upphi}^{(\alpha)}_{\hspace{0.03cm}{\bf k}}
\equiv
g\left(\frac{Z_{l}({\bf k})}
{2\hspace{0.03cm}\omega^{\hspace{0.03cm}l}_{\hspace{0.03cm}
		{\bf k}}}\right)^{\!\!1/2}\!\!\!
(v_{\alpha}\cdot\epsilon^{\hspace{0.03cm}l}({\bf k})).
\]

%
%

\section{Canonical transformation}
\setcounter{equation}{0}
\label{section_3}

It is intuitively clear that in the case of a nondecay dispersion law, the terms with the vertex functions ${\mathcal V}^{\; a\, a_{1}\hspace{0.03cm} a_{2}}_{{\bf k},\, {\bf k}_{1},\, {\bf k}_{2}}$ and ${\mathcal U}^{\; a\, a_{1}\hspace{0.03cm} a_{2}}_{{\bf k},\, {\bf k}_{1},\, {\bf k}_{2}}$ in the Hamiltonian $H^{(3)}$ (\ref{eq:2s}) describing three-wave processes may turn out to be irrelevant in some respect. Indeed, due to the specific character of the dispersion equation for soft bosonic longitudinal excitation (\ref{eq:2e}) in a hot quark-gluon plasma, the resonance conditions for three-wave processes involving plasmons
\begin{equation}
\left\{
\begin{array}{ll}
	{\bf k} = {\bf k}_{1} + {\bf k}_{2}, \\[1.5ex]
	\omega^{\hspace{0.03cm}l}_{\hspace{0.03cm}{\bf k}} = \omega^{\hspace{0.03cm}l}_{\hspace{0.03cm}{\bf k}_{1}} + \omega^{\hspace{0.03cm}l}_{\hspace{0.03cm}{\bf k}_{2}},
\end{array}
\right.
\quad
\left\{
\begin{array}{ll}
	{\bf k} + {\bf k}_{1} + {\bf k}_{2} = 0, \\[1.5ex]
	\omega^{\hspace{0.03cm}l}_{\hspace{0.03cm}{\bf k}} + \omega^{\hspace{0.03cm}l}_{\hspace{0.03cm}{\bf k}_{1}} + \omega^{\hspace{0.03cm}l}_{\hspace{0.03cm}{\bf k}_{2}} = 0,
\end{array}\
\right.
\label{eq:3qq}
\end{equation}
have no solutions. Furthermore, the terms in $H^{(3)}$ with vertex function ${\upphi}^{(\alpha)}_{\hspace{0.03cm}{\bf k}}$ describe the processes of Cherenkov emission (absorption) by a test particle with the label $\alpha$ moving through the medium under consideration. These processes are kinematically forbidden due to the condition (\ref{eq:2a}). We shall show that in this case it is possible to move to new canonical variables ($c^{(\alpha)\hspace{0.02cm}a}_{\hspace{0.02cm}{\bf k}}$, $c^{\ast\ \!\!(\alpha)
\hspace{0.02cm}a}_{\hspace{0.02cm}{\bf k}}$) and ${\mathcal Q}^{\hspace{0.03cm}a}_{\hspace{0.02cm}\alpha}$ such that in these variables the third-order interaction Hamiltonian $H^{(3)}$ vanishes. In this procedure of excluding the Hamiltonian $H^{(3)}$, a canonical transformation is derived by successively eliminating the terms for which the resonance conditions (\ref{eq:3qq}) and (\ref{eq:2a}) are not satisfied.\\
\indent Let us consider the transformation from the components $a^{\phantom{\ast}\!\!(\alpha)\hspace{0.02cm}a}_{\hspace{0.02cm}{\bf k}}$ of the initial normal boson variable $a^{a}_{\bf k}$ and the classical color charges ${Q}^{\hspace{0.03cm}a}_{\alpha}$ to the new field components $c^{\phantom{\ast}\!\!(\hspace{0.03cm}\alpha)
\hspace{0.02cm}a}_{\hspace{0.02cm}{\bf k}}$ and the new color charges ${\mathcal Q}^{\hspace{0.03cm}a}_{\hspace{0.02cm}\alpha}$:
\begin{align}
&a^{(\alpha)\hspace{0.02cm}a}_{\hspace{0.02cm}{\bf k}} = a^{\phantom{\ast}\!\!(\alpha)\hspace{0.02cm}a}_{\hspace{0.02cm}{\bf k}}\hspace{0.01cm}
(c^{(\hspace{0.03cm}\alpha)
\hspace{0.02cm}a}_{\hspace{0.02cm}{\bf k}}\!, 
c^{\ast\ \!\!(\alpha)
\hspace{0.02cm}a}_{\hspace{0.02cm}{\bf k}}\!, 
{\mathcal Q}^{\hspace{0.03cm}a}_{\hspace{0.02cm}\alpha}),
\label{eq:3q}\\[0.8ex]
&Q^{\,a}_{\hspace{0.01cm}\alpha} = Q^{\,a}_{\hspace{0.01cm}\alpha}\hspace{0.01cm}(\hspace{0.02cm}c^{\phantom{\ast}\!\!(\hspace{0.03cm}\alpha)\hspace{0.02cm}a}_{\hspace{0.02cm}{\bf k}},\, 
c^{\ast\ \!\!(\alpha)\hspace{0.02cm}a}_{\hspace{0.02cm}{\bf k}}\!,\, {\mathcal Q}^{\hspace{0.03cm}a}_{\hspace{0.02cm}\alpha}\hspace{0.02cm}).
\label{eq:3w}
\end{align}
We will demand that the Hamilton equations in terms of the new variables take the form (\ref{eq:2u}) and (\ref{eq:2i}) with the same Hamiltonian $H$. Because we consider those transformations that are not explicitly time  
dependent, the old and the new Hamiltonians are numerically equal, but generally differ in functional form, since they are written in different variables. Straightforward but rather cumbersome calculations result in two systems of integral relations. The first of them has the following form:
\begin{subequations} 
\label{eq:3e}
\begin{align}
\sum_{\rho}
&\int\! d\hspace{0.02cm}{\bf k\hspace{0.01cm}}'\!\hspace{0.01cm}
\left\{\frac{\delta\hspace{0.01cm}  a^{(\alpha)\hspace{0.02cm}a}_{\hspace{0.02cm}{\bf k}}}{\delta\hspace{0.01cm} c^{(\rho)\hspace{0.02cm}c}_{{\bf k}'}}
\hspace{0.03cm}\frac{\delta\hspace{0.01cm}  
a^{\ast\ \!\!(\beta)\hspace{0.02cm}b}_{\hspace{0.02cm}{\bf k}''}}{\delta\hspace{0.01cm}c^{\ast\ \!\!(\rho)\hspace{0.02cm}c}_{{\bf k}'}}
\,-\,
\frac{\delta\hspace{0.01cm}a^{(\alpha)\hspace{0.02cm}a}_{\hspace{0.02cm}{\bf k}}}
{\delta\hspace{0.01cm}c^{\!\ast\ \!\!(\rho)\hspace{0.02cm}c}_{{\bf k}'}}\hspace{0.03cm}
\frac{\delta\hspace{0.01cm}a^{\ast\ \!\!(\beta)\hspace{0.02cm}b}_{\hspace{0.02cm}{\bf k}''}}
{\delta\hspace{0.01cm} c^{(\rho)\hspace{0.02cm}c}_{{\bf k}'}}\right\}
+
i\,\sum_{\rho}
\frac{\partial a^{(\alpha)\hspace{0.02cm}a}_{\hspace{0.02cm}{\bf k}}}{\,\partial\hspace{0.03cm} {\mathcal Q}^{\hspace{0.03cm}c}_{\hspace{0.02cm}\rho}}\hspace{0.03cm}
\frac{\partial\hspace{0.03cm}a^{\ast\ \!\!(\beta)\hspace{0.02cm}b}_{\hspace{0.02cm}{\bf k}''}}{\,\partial\hspace{0.03cm} {\mathcal Q}^{\hspace{0.03cm}c^{\hspace{0.02cm}\prime}}_{\hspace{0.02cm}\rho}}
\,f^{\hspace{0.03cm}c\hspace{0.03cm}c^{\hspace{0.02cm}\prime}\hspace{0.02cm}d}\hspace{0.03cm}{\mathcal Q}^{\hspace{0.03cm}d}_{\hspace{0.02cm}\rho}
\!=
\delta^{\hspace{0.03cm}a\hspace{0.02cm}b}
\delta^{\hspace{0.03cm}\alpha\hspace{0.02cm}\beta}
\delta ({\bf k}-{\bf k}\!\ ''),
\label{eq:3ea}
\\[0.8ex]
\sum_{\rho}
&\int\! d\hspace{0.02cm}{\bf k\hspace{0.01cm}}'\!\hspace{0.01cm}
\left\{\frac{\delta\hspace{0.01cm}  a^{(\alpha)\hspace{0.02cm}a}_{\hspace{0.02cm}{\bf k}}}{\delta\hspace{0.01cm} c^{(\rho)\hspace{0.02cm}c}_{{\bf k}'}}
\hspace{0.03cm}\frac{\delta\hspace{0.01cm}  
	a^{(\beta)\hspace{0.02cm}b}_{\hspace{0.02cm}{\bf k}''}}{\delta\hspace{0.01cm}c^{\ast\ \!\!(\rho)\hspace{0.02cm}c}_{{\bf k}'}}
\,-\,
\frac{\delta\hspace{0.01cm}a^{(\alpha)\hspace{0.02cm}a}_{\hspace{0.02cm}{\bf k}}}
{\delta\hspace{0.01cm}c^{\!\ast\ \!\!(\rho)\hspace{0.02cm}c}_{{\bf k}'}}\hspace{0.03cm}
\frac{\delta\hspace{0.01cm}a^{(\beta)\hspace{0.02cm}b}_{\hspace{0.02cm}{\bf k}''}}
{\delta\hspace{0.01cm} c^{(\rho)\hspace{0.02cm}c}_{{\bf k}'}}\right\}
+
i\,\sum_{\rho}
\frac{\partial a^{(\alpha)\hspace{0.02cm}a}_{\hspace{0.02cm}{\bf k}}}{\,\partial\hspace{0.03cm} {\mathcal Q}^{\hspace{0.03cm}c}_{\hspace{0.02cm}\rho}}\hspace{0.03cm}
\frac{\partial\hspace{0.03cm}a^{(\beta)\hspace{0.02cm}b}_{\hspace{0.02cm}{\bf k}''}}{\,\partial\hspace{0.03cm} {\mathcal Q}^{\hspace{0.03cm}c^{\hspace{0.02cm}\prime}}_{\hspace{0.02cm}\rho}}
\,f^{\hspace{0.03cm}c\hspace{0.03cm}c^{\hspace{0.02cm}\prime}\hspace{0.02cm}d}\hspace{0.03cm}{\mathcal Q}^{\hspace{0.03cm}d}_{\hspace{0.02cm}\rho}
= 0,
\label{eq:3eb}
\\[0.8ex]
\sum_{\rho}
&\int\! d\hspace{0.02cm}{\bf k\hspace{0.01cm}}'\!\hspace{0.01cm}
\left\{\frac{\delta\hspace{0.01cm}  a^{(\alpha)\hspace{0.02cm}a}_{\hspace{0.02cm}{\bf k}}}{\delta\hspace{0.01cm} c^{(\rho)\hspace{0.02cm}c}_{{\bf k}'}}
\hspace{0.03cm}\frac{\delta\hspace{0.01cm}  
	Q^{\,b}_{\beta}}{\delta\hspace{0.01cm}c^{\ast\ \!\!(\rho)\hspace{0.02cm}c}_{{\bf k}'}}
\,-\,
\frac{\delta\hspace{0.01cm}a^{(\alpha)\hspace{0.02cm}a}_{\hspace{0.02cm}{\bf k}}}
{\delta\hspace{0.01cm}c^{\!\ast\ \!\!(\rho)\hspace{0.02cm}c}_{{\bf k}'}}\hspace{0.03cm}
\frac{\delta\hspace{0.01cm}Q^{\,b}_{\beta}}
{\delta\hspace{0.01cm} c^{(\rho)\hspace{0.02cm}c}_{{\bf k}'}}\right\}
+
i\,\sum_{\rho}
\frac{\partial a^{(\alpha)\hspace{0.02cm}a}_{\hspace{0.02cm}{\bf k}}}{\,\partial\hspace{0.03cm} {\mathcal Q}^{\hspace{0.03cm}c}_{\hspace{0.02cm}\rho}}\hspace{0.03cm}
\frac{\partial\hspace{0.03cm}Q^{\,b}_{\beta}}{\,\partial\hspace{0.03cm} {\mathcal Q}^{\hspace{0.03cm}c^{\hspace{0.02cm}\prime}}_{\hspace{0.02cm}\rho}}
\,f^{\hspace{0.03cm}c\hspace{0.03cm}c^{\hspace{0.02cm}\prime}\hspace{0.02cm}d}\hspace{0.03cm}{\mathcal Q}^{\hspace{0.03cm}d}_{\hspace{0.02cm}\rho}
= 0.
\label{eq:3ec}
\end{align}
\end{subequations}
Correspondingly, the second system is
\begin{subequations} 
\label{eq:3r}
\begin{align}
		\sum_{\rho}
		&\int\! d\hspace{0.02cm}{\bf k\hspace{0.01cm}}'\!\hspace{0.01cm}
		\left\{\frac{\delta\hspace{0.01cm}  {Q}^{\phantom{\ast}\!\!a}_{\hspace{0.02cm}\alpha}}
		{\delta\hspace{0.01cm} c^{(\rho)\hspace{0.02cm}c}_{{\bf k}'}}
		\hspace{0.03cm}\frac{\delta\hspace{0.01cm}  
			a^{\ast\ \!\!(\beta)\hspace{0.02cm}b}_{\hspace{0.02cm}{\bf k}''}}{\delta\hspace{0.01cm}c^{\ast\ \!\!(\rho)\hspace{0.02cm}c}_{{\bf k}'}}
		\,-\,
\frac{\delta\hspace{0.01cm}{Q}^{\phantom{\ast}\!\!a}_{\hspace{0.02cm}\alpha}}
		{\delta\hspace{0.01cm}c^{\!\ast\ \!\!(\rho)\hspace{0.02cm}c}_{{\bf k}'}}\hspace{0.03cm}
		\frac{\delta\hspace{0.01cm}a^{\ast\ \!\!(\beta)\hspace{0.02cm}b}_{\hspace{0.02cm}{\bf k}''}}
		{\delta\hspace{0.01cm} c^{(\rho)\hspace{0.02cm}c}_{{\bf k}'}}\right\}
		+
		i\,\sum_{\rho}
		\frac{\partial\hspace{0.03cm} {Q}^{\phantom{\ast}\!\!a}_{\hspace{0.02cm}\alpha}}{\,\partial\hspace{0.03cm} {\mathcal Q}^{\hspace{0.03cm}c}_{\hspace{0.02cm}\rho}}\hspace{0.03cm}
		\frac{\partial\hspace{0.03cm}a^{\ast\ \!\!(\beta)\hspace{0.02cm}b}_{\hspace{0.02cm}{\bf k}''}}{\,\partial\hspace{0.03cm} {\mathcal Q}^{\hspace{0.03cm}c^{\hspace{0.02cm}\prime}}_{\hspace{0.02cm}\rho}}
		\,f^{\hspace{0.03cm}c\hspace{0.03cm}c^{\hspace{0.02cm}\prime}\hspace{0.02cm}d}\hspace{0.03cm}{\mathcal Q}^{\hspace{0.03cm}d}_{\hspace{0.02cm}\rho}
		\!= 0,
		\label{eq:3ra}
		\\[0.8ex]
		\sum_{\rho}
		&\int\! d\hspace{0.02cm}{\bf k\hspace{0.01cm}}'\!\hspace{0.01cm}
		\left\{\frac{\delta\hspace{0.01cm}  {Q}^{\phantom{\ast}\!\!a}_{\hspace{0.02cm}\alpha}}{\delta\hspace{0.01cm} c^{(\rho)\hspace{0.02cm}c}_{{\bf k}'}}
		\hspace{0.03cm}\frac{\delta\hspace{0.01cm}  
			a^{(\beta)\hspace{0.02cm}b}_{\hspace{0.02cm}{\bf k}''}}{\delta\hspace{0.01cm}c^{\ast\ \!\!(\rho)\hspace{0.02cm}c}_{{\bf k}'}}
		\,-\,
\frac{\delta\hspace{0.01cm}{Q}^{\phantom{\ast}\!\!a}_{\hspace{0.02cm}\alpha}}
		{\delta\hspace{0.01cm}c^{\!\ast\ \!\!(\rho)\hspace{0.02cm}c}_{{\bf k}'}}\hspace{0.03cm}
		\frac{\delta\hspace{0.01cm}a^{(\beta)\hspace{0.02cm}b}_{\hspace{0.02cm}{\bf k}''}}
		{\delta\hspace{0.01cm} c^{(\rho)\hspace{0.02cm}c}_{{\bf k}'}}\right\}
		+
		i\,\sum_{\rho}
		\frac{\partial {Q}^{\phantom{\ast}\!\!a}_{\hspace{0.02cm}\alpha}}{\,\partial\hspace{0.03cm} {\mathcal Q}^{\hspace{0.03cm}c}_{\hspace{0.02cm}\rho}}\hspace{0.03cm}
		\frac{\partial\hspace{0.03cm}a^{(\beta)\hspace{0.02cm}b}_{\hspace{0.02cm}{\bf k}''}}{\,\partial\hspace{0.03cm} {\mathcal Q}^{\hspace{0.03cm}c^{\hspace{0.02cm}\prime}}_{\hspace{0.02cm}\rho}}
		\,f^{\hspace{0.03cm}c\hspace{0.03cm}c^{\hspace{0.02cm}\prime}\hspace{0.02cm}d}\hspace{0.03cm}{\mathcal Q}^{\hspace{0.03cm}d}_{\hspace{0.02cm}\rho}
		= 0,
		\label{eq:3rb}
		\\[0.8ex]
		\sum_{\rho}
		&\int\! d\hspace{0.02cm}{\bf k\hspace{0.01cm}}'\!\hspace{0.01cm}
		\left\{\frac{\delta\hspace{0.01cm}  {Q}^{\phantom{\ast}\!\!a}_{\hspace{0.02cm}\alpha}}{\delta\hspace{0.01cm} c^{(\rho)\hspace{0.02cm}c}_{{\bf k}'}}
		\hspace{0.03cm}\frac{\delta\hspace{0.01cm}  
{Q}^{\phantom{\ast}\!\!b}_{\hspace{0.02cm}\beta}}{\delta\hspace{0.01cm}c^{\ast\ \!\!(\rho)\hspace{0.02cm}c}_{{\bf k}'}}
\,-\,
\frac{\delta\hspace{0.01cm}{Q}^{\phantom{\ast}\!\!a}_{\hspace{0.02cm}\alpha}}
		{\delta\hspace{0.01cm}c^{\!\ast\ \!\!(\rho)\hspace{0.02cm}c}_{{\bf k}'}}\hspace{0.03cm}
\frac{\delta\hspace{0.01cm}{Q}^{\phantom{\ast}\!\!b}_{\hspace{0.02cm}\beta}}
		{\delta\hspace{0.01cm} c^{(\rho)\hspace{0.02cm}c}_{{\bf k}'}}\right\}
		+
		i\,\sum_{\rho}
		\frac{\partial\hspace{0.03cm}
{Q}^{\phantom{\ast}\!\!a}_{\hspace{0.02cm}\alpha}}{\,\partial\hspace{0.03cm} {\mathcal Q}^{\hspace{0.03cm}c}_{\hspace{0.02cm}\rho}}\hspace{0.03cm}
\frac{\partial\hspace{0.03cm}{Q}^{\phantom{\ast}\!\!b}_{\hspace{0.02cm}\beta}}{\,\partial\hspace{0.03cm} {\mathcal Q}^{\hspace{0.03cm}c^{\hspace{0.02cm}\prime}}_{\hspace{0.02cm}\rho}}
		\,f^{\hspace{0.03cm}c\hspace{0.03cm}c^{\hspace{0.02cm}\prime}\hspace{0.02cm}d}\hspace{0.03cm}{\mathcal Q}^{\hspace{0.03cm}d}_{\hspace{0.02cm}\rho}
		= i\hspace{0.02cm} 
		f^{\hspace{0.03cm}a\hspace{0.03cm}b\hspace{0.03cm}d}\hspace{0.03cm} {Q}^{\hspace{0.03cm}d}_{\alpha}\,
		\delta_{\alpha\beta}.
\label{eq:3rc}
\end{align}
\end{subequations}
These canonicity conditions can be written in a very compact form if we make use of the definition of the Lie-Poisson bracket (\ref{eq:2y}) and replace the variation variables by the new ones: $a^{(\rho)\hspace{0.02cm}c}_{{\bf k}'}\rightarrow c^{(\rho)\hspace{0.02cm}c}_{{\bf k}'}$ and $Q^{\,c}_{\rho} \rightarrow {\mathcal Q}^{\,c}_{\hspace{0.03cm}\rho}$. In this case the Lie-Poisson bracket for the original variables $a^{(\alpha)\hspace{0.02cm}a}_{\hspace{0.02cm}{\bf k}}$ and $Q^{\,a}_{\alpha}$, Eqs.\,(\ref{eq:2r}) and (\ref{eq:2t}), turns to the canonicity conditions (\ref{eq:3e}) and (\ref{eq:3r}), which impose certain restrictions on the functional dependencies (\ref{eq:3q}) and (\ref{eq:3w}). Let us present the canonical transformations (\ref{eq:3q}) and (\ref{eq:3w}) in the form of integro-power series in the components $c^{(\hspace{0.03cm}\alpha)	\hspace{0.02cm}a}_{\hspace{0.02cm}{\bf k}}$ and the color charges ${\mathcal Q}^{\,a}_{\hspace{0.03cm}\alpha}$. In this case the transformation (\ref{eq:3q}) up to the terms of the third order\footnote{\hspace{0.03cm}Recall again that we consider a degree of nonlinearity of the color charge to be two.} in the new variables $c^{(\hspace{0.03cm}\alpha)	\hspace{0.02cm}a}_{\hspace{0.02cm}{\bf k}}$ and ${\mathcal Q}^{\hspace{0.03cm}a}_{\hspace{0.03cm}\alpha}$ has the following form:
\begin{equation}
a^{(\alpha)\hspace{0.02cm}a}_{\hspace{0.02cm}{\bf k}} = c^{(\alpha)\hspace{0.02cm}a}_{\hspace{0.02cm}{\bf k}}\,
+ 
\sum_{\beta}{F}^{(\hspace{0.02cm}\alpha,\hspace{0.02cm}\beta)}_{\hspace{0.03cm}\bf k}\hspace{0.02cm}
{\mathcal Q}^{\hspace{0.03cm}a}_{\hspace{0.03cm}\beta}
\label{eq:3t}
\end{equation}
\[
+ \sum_{\alpha_{1},\hspace{0.02cm}\alpha_{2}} \int\!d\hspace{0.02cm}{\bf k}_{1}\hspace{0.02cm} d\hspace{0.02cm}{\bf k}_{2} 
\left[\hspace{0.03cm}
V^{\hspace{0.02cm}(1\hspace{0.02cm}|\hspace{0.02cm}\alpha,\alpha_{1},\alpha_{2})\,a\,a_{1}\hspace{0.03cm}a_{2}}_{\ {\bf k},\, {\bf k}_{1},\, 
{\bf k}_{2}}\hspace{0.03cm}
c^{(\alpha_{1})\hspace{0.02cm}a_{1}}_{\hspace{0.02cm}
{\bf k}_{1}}\hspace{0.03cm}
c^{(\alpha_{2})\hspace{0.02cm}a_{2}}_{\hspace{0.02cm}{\bf k}_{2}}
+
V^{\hspace{0.02cm}(2\hspace{0.02cm}|\hspace{0.02cm}\alpha,\alpha_{1},\alpha_{2})\,a\,a_{1}\hspace{0.03cm}a_{2}}_{\ {\bf k},\, {\bf k}_{1},\, {\bf k}_{2}}
\hspace{0.03cm}
c^{\hspace{0.03cm}\ast\,(\alpha_{1}) a_{1}}_{\hspace{0.02cm}{\bf k}_{1}}\hspace{0.03cm}c^{(\alpha_{2})\hspace{0.02cm}a_{2}}_{\hspace{0.02cm}{\bf k}_{2}}
\right.
\]
\[
\hspace{8.8cm}
\left.
+\,
V^{\hspace{0.02cm}(3\hspace{0.02cm}|\hspace{0.02cm}\alpha,\alpha_{1},\alpha_{2})\,a\,a_{1}\hspace{0.03cm}a_{2}}_{\ {\bf k},\, {\bf k}_{1},\, {\bf k}_{2}}\, 
c^{\hspace{0.03cm}\ast\,(\alpha_{1}) a_{1}}_{\hspace{0.02cm}{\bf k}_{1}} c^{\hspace{0.03cm}\ast\,(\alpha_{2}) a_{2}}_{\hspace{0.02cm}{\bf k}_{2}}\hspace{0.03cm}\right] 
\]
\[
\hspace{0.3cm}
+ \sum_{\alpha_{1},\hspace{0.02cm}\alpha_{2}}
\int\!d\hspace{0.02cm}{\bf k}_{1}\! 
\left[\hspace{0.03cm}\widetilde{V}^{\hspace{0.02cm}(1|\hspace{0.02cm}\alpha,\alpha_{1},\alpha_{2})\,a\,a_{1}\hspace{0.03cm}a_{2}}_{\ {\bf k},\, {\bf k}_{1}}\,
c^{\hspace{0.03cm}\ast\,(\alpha_{1})\hspace{0.03cm}a_{1}}_{\hspace{0.02cm}{\bf k}_{1}}\hspace{0.03cm}{\mathcal Q}^{\hspace{0.03cm}a_{2}}_{\hspace{0.03cm}\alpha_{2}}
\,+\,
\widetilde{V}^{\hspace{0.02cm}(2|\hspace{0.02cm}\alpha,\alpha_{1},\alpha_{2})\,a\,a_{1}\hspace{0.03cm}a_{2}}_{\ {\bf k},\, {\bf k}_{1}}\,
c^{\hspace{0.03cm}(\alpha_{1})\hspace{0.03cm}a_{1}}_{\hspace{0.02cm}{\bf k}_{1}}\hspace{0.03cm}{\mathcal Q}^{\hspace{0.03cm}a_{2}}_{\hspace{0.03cm}\alpha_{2}}\hspace{0.03cm}
\right] +\,\ldots\,. 
\]
Similarly, the most common power-series expansion for the transformation (\ref{eq:3w}) up to the terms of the fourth order is
\begin{equation}
Q^{\hspace{0.03cm}a}_{\alpha} = {\mathcal Q}^{\hspace{0.03cm}a}_{\hspace{0.03cm}\alpha}\,
+\! 
\sum_{\alpha_{1},\hspace{0.02cm}\alpha_{2}}
\int\!d\hspace{0.02cm}{\bf k}_{1}
\left[\hspace{0.03cm}M^{\hspace{0.02cm}(\alpha,\alpha_{1},\alpha_{2})\,a\,a_{1}\hspace{0.03cm}a_{2}}_{\; {\bf k}_{1}}\, 
c^{\hspace{0.03cm}(\alpha_{1}) a_{1}}_{\hspace{0.02cm}{\bf k}_{1}}\hspace{0.03cm}{\mathcal Q}^{\hspace{0.03cm}a_{2}}_{\hspace{0.03cm}\alpha_{2}}
\,+\,
M^{\hspace{0.02cm}\ast\hspace{0.02cm}(\alpha,\alpha_{1},\alpha_{2})\,a\,a_{1}\hspace{0.03cm}a_{2}}_{\; {\bf k}_{1}}\, c^{\hspace{0.03cm}\ast\,(\alpha_{1}) a_{1}}_{\hspace{0.02cm}{\bf k}_{1}}\hspace{0.03cm}{\mathcal Q}^{\hspace{0.03cm}a_{2}}_{\hspace{0.03cm}\alpha_{2}}\hspace{0.03cm} 
\right] 
\label{eq:3y}
\end{equation}
\[
+\!\!\!
\sum_{\alpha_{1},\hspace{0.02cm}\alpha_{2},\hspace{0.02cm}\alpha_{3}}\!
\int\!\!d\hspace{0.02cm}{\bf k}_{1}\hspace{0.02cm} d\hspace{0.02cm}{\bf k}_{2}
\Bigl[M^{\hspace{0.03cm}(1\hspace{0.02cm}|\hspace{0.02cm}\alpha,\alpha_{1},\alpha_{2},\alpha_{3})\,a\,a_{1}\hspace{0.03cm}a_{2}\,a_{3}}_{\ {\bf k}_{1},\,{\bf k}_{2}}\hspace{0.02cm}
c^{\hspace{0.03cm}(\alpha_{1}) a_{1}}_{\hspace{0.02cm}{\bf k}_{1}}\hspace{0.01cm} 
c^{\hspace{0.03cm}(\alpha_{2}) a_{2}}_{\hspace{0.02cm}{\bf k}_{2}}\hspace{0.01cm}
{\mathcal Q}^{\hspace{0.03cm}a_{3}}_{\hspace{0.03cm}\alpha_{3}}
+
M^{\hspace{0.03cm}(2\hspace{0.02cm}|\hspace{0.02cm}\alpha,\alpha_{1},\alpha_{2},\alpha_{3})\,a\,a_{1}\hspace{0.03cm}a_{2}\,a_{3}}_{\ {\bf k}_{1},\, {\bf k}_{2}}\hspace{0.01cm}
c^{\hspace{0.03cm}\ast\,(\alpha_{1}) a_{1}}_{\hspace{0.02cm}{\bf k}_{1}}\hspace{0.01cm} 
c^{\hspace{0.03cm}(\alpha_{2}) a_{2}}_{\hspace{0.02cm}{\bf k}_{2}}\hspace{0.01cm}
{\mathcal Q}^{\hspace{0.03cm}a_{3}}_{\hspace{0.03cm}\alpha_{3}}
\]
\[
+\,
M^{\hspace{0.03cm}\ast\,(1\hspace{0.02cm}|\hspace{0.02cm}\alpha,\alpha_{1},\alpha_{2},\alpha_{3})\,a\,a_{1}\hspace{0.03cm}a_{2}\,a_{3}}_{\ {\bf k}_{1},\, {\bf k}_{2}}\hspace{0.01cm} 
c^{\hspace{0.03cm}\ast\,(\alpha_{1}) a_{1}}_{\hspace{0.02cm}{\bf k}_{1}}\hspace{0.01cm} 
c^{\hspace{0.03cm}\ast\,(\alpha_{2}) a_{2}}_{\hspace{0.02cm}{\bf k}_{2}}\hspace{0.01cm}
{\mathcal Q}^{\hspace{0.03cm}a_{3}}_{\hspace{0.03cm}\alpha_{3}}
\hspace{0.03cm}\Bigr]
+\,\ldots\ _{.}
\]
Note that the coefficient functions in (\ref{eq:3t}) and (\ref{eq:3y}) are contracted with the components $c^{(\alpha)\hspace{0.02cm}a}_{\hspace{0.02cm}{\bf k}}$ of the amplitude $c^{\,a}_{\hspace{0.02cm}{\bf k}}$ of the new bosonic field and, thus, can now depend in a rather nontrivial way on the new ``index'' $\alpha$. This fundamentally distinguishes the coefficient functions from the vertex ones entering in the original interaction Hamiltonian $H^{(3)}$, Eq.\,(\ref{eq:2s}), since this Hamiltonian includes only the full amplitude (\ref{eq:2ee}). This nontrivial dependence, as we will see below, is generated solely by the chosen structure of the free Hamiltonian $H^{(0)}$, Eq.\,(\ref{eq:2p}).\\
\indent In the transformation (\ref{eq:3y}) the requirement of a reality of the color charge is taken into account, in particular, it also leads to the condition
\[
M^{\hspace{0.03cm}\ast\hspace{0.03cm}(2\hspace{0.02cm}|\hspace{0.02cm}\alpha,\alpha_{1},\alpha_{2},\alpha_{3})\,a\,a_{1}\hspace{0.03cm}a_{2}\,a_{3}}_{\ {\bf k}_{1},\, {\bf k}_{2}}
=
M^{\hspace{0.03cm}(2\hspace{0.02cm}|\hspace{0.02cm}\alpha,\alpha_{2},\alpha_{1},\alpha_{3})\,a\,a_{2}\hspace{0.03cm}a_{1}\,a_{3}}_{\ {\bf k}_{2},\, {\bf k}_{1}}.
\]
In addition we note that the coefficient functions $V^{\hspace{0.02cm}(1\hspace{0.02cm}|\hspace{0.02cm}\alpha,\alpha_{1},\alpha_{2})\,a\,a_{1}\hspace{0.03cm}a_{2}}_{\ {\bf k},\, {\bf k}_{1},\,{\bf k}_{2}}$, $V^{\hspace{0.02cm}(3\hspace{0.02cm}|\hspace{0.02cm}\alpha,\alpha_{1},\alpha_{2})\,a\,a_{1}\hspace{0.03cm}a_{2}}_{\ {\bf k},\, {\bf k}_{1},\,{\bf k}_{2}}$ in (\ref{eq:3t}) and $M^{\hspace{0.03cm}(1\hspace{0.02cm}|\hspace{0.02cm}\alpha,\alpha_{1},\alpha_{2},\alpha_{3})\,a\,a_{1}\hspace{0.03cm}a_{2}\,a_{3}}_{\ {\bf k}_{1},\,{\bf k}_{2}}$ in (\ref{eq:3y}) must satisfy the natural symmetry conditions:
\begin{align}
&V^{\hspace{0.02cm}(1\hspace{0.02cm}|\hspace{0.02cm}\alpha,\alpha_{1},\alpha_{2})\,a\,a_{1}\hspace{0.03cm}a_{2}}_{\ {\bf k},\, {\bf k}_{1},\,{\bf k}_{2}}
= 
V^{\hspace{0.02cm}(1\hspace{0.02cm}|\hspace{0.02cm}\alpha,\alpha_{2},\alpha_{1})\,a\,a_{2}\hspace{0.03cm}a_{1}}_{\ {\bf k},\, {\bf k}_{2},\,{\bf k}_{1}},
\quad
V^{\hspace{0.02cm}(3\hspace{0.02cm}|\hspace{0.02cm}\alpha,\alpha_{1},\alpha_{2})\,a\,a_{1}\hspace{0.03cm}a_{2}}_{\ {\bf k},\, {\bf k}_{1},\,{\bf k}_{2}}
= 
V^{\hspace{0.02cm}(3\hspace{0.02cm}|\hspace{0.02cm}\alpha,\alpha_{2},\alpha_{1})\,a\,a_{2}\hspace{0.03cm}a_{1}}_{\ {\bf k},\, {\bf k}_{2},\,{\bf k}_{1}},
\notag\\[2.5ex]
&M^{\hspace{0.03cm}(1\hspace{0.02cm}|\hspace{0.02cm}\alpha,\alpha_{1},\alpha_{2},\alpha_{3})\,a\,a_{1}\hspace{0.03cm}a_{2}\,a_{3}}_{\ {\bf k}_{1},\,{\bf k}_{2}}
= 
M^{\hspace{0.03cm}(1\hspace{0.02cm}|\hspace{0.02cm}\alpha,\alpha_{2},\alpha_{1},\alpha_{3})\,a\,a_{2}\hspace{0.03cm}a_{1}\,a_{3}}_{\ {\bf k}_{2},\,{\bf k}_{1}}.
\notag
\end{align}
\indent Furthermore, substituting the expansions (\ref{eq:3t}) and  (\ref{eq:3y}) into a system of the canonicity conditions (\ref{eq:3e}) and (\ref{eq:3r}), we obtain rather nontrivial integral relations connecting various coefficient functions among themselves. Here, we have provided only algebraic relations for the lowest second-order coefficient functions:
\begin{equation}
V^{\hspace{0.02cm}(2\hspace{0.02cm}|\hspace{0.02cm}\alpha,\alpha_{1},\alpha_{2})\,a\,a_{1}\hspace{0.03cm}a_{2}}_{\ {\bf k},\, {\bf k}_{1},\, {\bf k}_{2}} 
= 
-\hspace{0.01cm}2\hspace{0.03cm}V^{\,\ast\hspace{0.03cm}(1\hspace{0.02cm}|\hspace{0.02cm}\alpha_{2},\alpha_{1},\alpha)\, a_{2}\,a_{1}\hspace{0.03cm}a}_{\ {\bf k}_{2},\, {\bf k}_{1},\, {\bf k}},
\quad
V^{\hspace{0.02cm}(3\hspace{0.02cm}|\hspace{0.02cm}\alpha,\alpha_{1},\alpha_{2})\,a\,a_{1}\hspace{0.03cm}a_{2}}_{\ {\bf k},\, {\bf k}_{1},\, {\bf k}_{2}}
= 
V^{\hspace{0.02cm}(3\hspace{0.02cm}|\hspace{0.02cm}\alpha_{1},\alpha,\alpha_{2})\,a_{1}\hspace{0.03cm}a\,a_{2}}_{\ {\bf k}_{1},\, {\bf k},\,{\bf k}_{2}},
\label{eq:3i}
\end{equation}
\begin{equation}
M^{\hspace{0.02cm}(\alpha,\alpha_{1},\alpha_{2})\,a\,a_{1}\hspace{0.03cm}a_{2}}_{\; {\bf k}_{1}}
+ 
i\hspace{0.03cm} f^{\,a\, a_{1}\hspace{0.03cm}a_{2}}\hspace{0.02cm}
\delta^{\hspace{0.03cm}\alpha\hspace{0.03cm}\alpha_{2}} F^{\hspace{0.03cm}\ast\hspace{0.03cm}(\alpha_{1},\hspace{0.03cm}\alpha)}_{\, {\bf k}_{1}} = 0,
\label{eq:3o}
\end{equation}
\begin{align}
&\widetilde{V}^{\hspace{0.02cm}(1|\hspace{0.02cm}\alpha,\alpha_{1},\alpha_{2})\,a\,a_{1}\hspace{0.03cm}a_{2}}_{\ {\bf k},\, {\bf k}_{1}}
-
\widetilde{V}^{\hspace{0.03cm}(1|\hspace{0.02cm}\alpha_{1},\alpha,\alpha_{2})\,a_{1}\hspace{0.03cm}a\,a_{2}}_{\ {\bf k}_{1},\, {\bf k}} 
\,-\,
i\hspace{0.03cm}f^{\,a\, a_{1}\hspace{0.03cm}a_{2}}\hspace{0.02cm}
F^{\hspace{0.02cm}(\hspace{0.02cm}\alpha,\hspace{0.03cm}\alpha_{2})}_{\mathbf k}\hspace{0.02cm}
F^{\hspace{0.02cm}(\hspace{0.02cm}\alpha_{1},\hspace{0.03cm}\alpha_{2})}_{{\mathbf k}_{1}} = 0,\label{eq:3p}\\[1.5ex]
&\widetilde{V}^{\hspace{0.02cm}(2|\hspace{0.02cm}\alpha,\alpha_{1},\alpha_{2})\,a\,a_{1}\hspace{0.03cm}a_{2}}_{\ {\bf k},\, {\bf k}_{1}}
+
\widetilde{V}^{\,\ast\hspace{0.03cm}(2|\hspace{0.02cm}\alpha_{1},\alpha,\alpha_{2})\,a_{1}\hspace{0.03cm}a\,a_{2}}_{\ {\bf k}_{1},\, {\bf k}} 
+
i\hspace{0.02cm}f^{\,a\, a_{1}\hspace{0.03cm}a_{2}}
F^{\hspace{0.02cm}(\alpha,\hspace{0.02cm}\alpha_{2})}_{\mathbf k}
F^{\,\ast\hspace{0.02cm}(\alpha_{1},\hspace{0.02cm}\alpha_{2})}_{{\mathbf k}_{1}}\! = 0.
\label{eq:3a}
\end{align}
We emphasize that there is no summation over indexes $\alpha$ and $\alpha_{2}$ in the relations (\ref{eq:3o})\,--\,(\ref{eq:3a}), and that in (\ref{eq:3o}) we intentionally kept the momentum ${\bf k}_{1}$ with label 1, to show that this momentum is associated with the specific indexes $\alpha_{1}$ and $a_{1}$ in the coefficient function $M^{\hspace{0.02cm}(\alpha,\alpha_{1},\alpha_{2})\,a\,a_{1}\hspace{0.03cm}a_{2}}_{\; {\bf k}_{1}}$.

%
%

\section{Eliminating ``nonessential'' Hamiltonian $H^{(3)}$. Effective fifth-order Hamiltonian}
\setcounter{equation}{0}
\label{section_4}

\begin{figure}[t]
\centering
\begin{center}
\begin{tabular*}{0.8\textwidth}{@{}ccc@{}}
\raisebox{-0.4\height}{\resizebox{0.35\textwidth}{!}
{\includegraphics{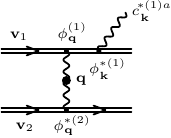}}
}
&$\!\!\!\!\!{\mathbf +}${ }{ }{ }{ }{ }{ }{ }&
\raisebox{-0.55\height}{\resizebox{0.35\textwidth}{!}
{\includegraphics{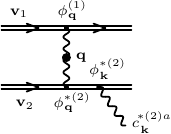}}
}
\end{tabular*}
\end{center}  
\caption{\small Diagrammatic representation of the first and last terms in square brackets in the complex conjugate effective amplitude ${T}^{\,{\rm I}\,(\rho)\hspace{0.03cm}a\,a_{1}\hspace{0.03cm}a_{2}}_{\; {\bf k},\,{\bf q}}$, Eq.\,(\ref{eq:4u}). The blob stands for the HTL-resummation and the double line denotes a hard particle}
\label{fig2-crop}
\end{figure}

\begin{figure}[t]
\centering
\begin{center}
\begin{tabular*}{0.8\textwidth}{@{}ccc@{}}
\raisebox{-0.44\height}{\resizebox{0.33\textwidth}{!}
{\includegraphics{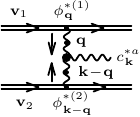}}}
&${ }{ }{ }{ }{ }\hspace{1cm}{\mathbf +}${ }{ }{ }{ }{ }{ }{ }&
\raisebox{-0.44\height}{\resizebox{0.33\textwidth}{!}
{\includegraphics{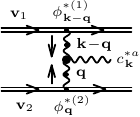}}
}
\\
\raisebox{-0.44\height}{\resizebox{0.33\textwidth}{!}
	{\includegraphics{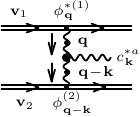}}}
&${ }{ }{ }{ }{ }\hspace{1cm}{\mathbf +}${ }{ }{ }{ }{ }{ }{ }&
\raisebox{-0.44\height}{\resizebox{0.33\textwidth}{!}
	{\includegraphics{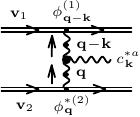}}}
\end{tabular*}
\end{center}  
\caption{\small Diagrammatic representation of contributions related to the three-plasmon vertex function ${\mathcal V}^{\;a\,a_{1}\hspace{0.03cm}a_{2}}_{\;{\bf k},\, {\bf k}_{1},\, {\bf k}_{2}}$ in the complex conjugate expression of the effective amplitude ${T}^{\,{\rm II}\,(\rho)\hspace{0.03cm}a\,a_{1}\hspace{0.03cm}a_{2}}_{\; {\bf k},\,{\bf q}}$, Eq.\,(\ref{eq:4uu}). There are also the contribution connected with another three-plasmon vertex function
${\mathcal U}^{\;a\,a_{1}\hspace{0.03cm}a_{2}}_{\;{\bf k},\, {\bf k}_{1},\, {\bf k}_{2}}$}
\label{fig3-crop}
\end{figure}

The next step in constructing the effective theory is the procedure of eliminating the third-order interaction Hamiltonian $H^{(3)}$, Eq.\,(\ref{eq:2s}), upon switching from the components $a^{(\alpha)\hspace{0.02cm}a}_{\hspace{0.02cm}{\bf k}}$ of the original bosonic function $a^{\hspace{0.02cm}a}_{\hspace{0.02cm}{\bf k}}$ and the color charges $Q^{\hspace{0.03cm}a}_{\alpha}$ to the new components $c^{(\alpha)\hspace{0.02cm}a}_{\hspace{0.02cm}{\bf k}}$ and color charges ${\mathcal Q}^{\hspace{0.03cm}a}_{\hspace{0.03cm}\alpha}$ as a result of the canonical transformations (\ref{eq:3t}) and (\ref{eq:3y}). We have already performed such an elimination in \cite{Markov:2024}, considering the interaction of soft bosonic excitations with one hard test particle. In this paper, a new element is the consideration of the interaction of soft bosonic excitations with two hard test particles. Therefore, here we will focus in more detail on this new aspect.\\
\indent To eliminate the third-order interaction Hamiltonian $H^{(3)}$, we substitute the expansions (\ref{eq:3t}) and (\ref{eq:3y}) into the free-field Hamiltonian $H^{(0)}$ given by expression (\ref{eq:2p}) and keep only the terms that have a quadratic or cubic nonlinearity in the new variables $c^{(\alpha)\hspace{0.02cm}a}_{\hspace{0.02cm}{\bf k}}$ and ${\mathcal Q}^{\hspace{0.03cm}a}_{\hspace{0.03cm}\alpha}$. Then in the third-order Hamiltonian $H^{(3)}$, Eq.\,(\ref{eq:2s}), we perform the simple replacements of variables: $a^{(\alpha)\hspace{0.02cm}a}_{\hspace{0.02cm}{\bf k}}\rightarrow c^{(\alpha)\hspace{0.02cm}a}_{\hspace{0.02cm}{\bf k}}$ and $Q^{\hspace{0.03cm}a}_{\alpha} \rightarrow {\mathcal Q}^{\hspace{0.03cm}a}_{\hspace{0.03cm}\alpha}$. Adding the expression thus obtained to the expression that follows from the free-field Hamiltonian $H^{(0)}$ and collecting similar terms, finally we obtain 
\begin{equation}
H^{(0)} + H^{(3)}
=  
\sum_{\alpha}
\int\!d\hspace{0.02cm}{\bf k}\,
(\omega^{\hspace{0.03cm}l}_{\hspace{0.02cm}{\bf k}} - {\mathbf v}^{\phantom{l}}_{\alpha\!}\cdot {\mathbf k})\ \!
c^{\ast\ \!\!(\alpha)\hspace{0.02cm}a}_{\hspace{0.02cm}{\bf k}}
\hspace{0.03cm}
c^{\phantom{\ast}\!\!\!(\alpha)	\hspace{0.02cm}a}_{\hspace{0.02cm}{\bf k}}
\label{eq:4q}
\end{equation}
\[
+\sum_{\alpha,\beta}
\!\int\!d\hspace{0.02cm}{\bf k}\,
\Bigl\{\left[\hspace{0.03cm}
(\omega^{\hspace{0.03cm}l}_{\hspace{0.02cm}{\bf k}} - {\mathbf v}^{\phantom{l}}_{\alpha\!}\cdot {\mathbf k})\ \!
{F}^{\,\ast\hspace{0.03cm}(\hspace{0.02cm}\alpha,\hspace{0.02cm}\beta)}_{\hspace{0.03cm}\bf k}
+ 
{\upphi}^{(\beta)}_{\hspace{0.03cm}{\bf k}}\hspace{0.03cm}\right]
c^{(\alpha)	\hspace{0.02cm}a}_{\hspace{0.02cm}{\bf k}}\hspace{0.03cm}
{\mathcal Q}^{\hspace{0.03cm}a}_{\hspace{0.02cm}\beta}
\hspace{0.03cm}+\hspace{0.03cm}
\left[\hspace{0.03cm}(\omega^{\hspace{0.03cm}l}_{\hspace{0.02cm}{\bf k}} - {\mathbf v}^{\phantom{l}}_{\alpha\!}\cdot {\mathbf k})\ \!
{F}^{\hspace{0.03cm}(\hspace{0.02cm}\alpha,\hspace{0.02cm}\beta)}_{\hspace{0.03cm}\bf k} 
+ 
{\upphi}^{\ast\hspace{0.03cm}(\beta)}_{\hspace{0.03cm}{\bf k}}
\hspace{0.03cm}\right]
c^{\ast\ \!\!(\alpha)\hspace{0.02cm}a}_{\hspace{0.02cm}{\bf k}}\hspace{0.03cm}
{\mathcal Q}^{\hspace{0.03cm}a}_{\hspace{0.02cm}\beta}\hspace{0.03cm} 
\Bigr\}.
\]
Requiring that the expression in curly brackets on the right-hand side of (\ref{eq:4q}) to be zero, we obtain an explicit form of the coefficient function ${F}^{\hspace{0.03cm}(\hspace{0.02cm}\alpha,\hspace{0.02cm}\beta)}_{\hspace{0.03cm}\bf k}$ in the canonical transformation (\ref{eq:3t}) in terms of the vertex functions ${\upphi}^{(\alpha)}_{\hspace{0.03cm}{\bf k}}$:
\begin{equation}
{F}^{\hspace{0.03cm}(\hspace{0.02cm}\alpha,\hspace{0.02cm}\beta)}_{\hspace{0.03cm}\bf k} 
=
-\hspace{0.03cm}\frac{{\upphi}^{\ast\hspace{0.03cm}(\beta)}_{\hspace{0.03cm}{\bf k}}}
{\omega^{\hspace{0.03cm}l}_{\hspace{0.02cm}{\bf k}} - {\mathbf v}^{\phantom{l}}_{\alpha\!}\cdot {\mathbf k}}\,.
\label{eq:4w}
\end{equation}
The relation (\ref{eq:4w}) has a meaning due to the condition (\ref{eq:2a}). Making use of (\ref{eq:4w}), from (\ref{eq:3o}) we immediately find the explicit form of the coefficient function $M^{\hspace{0.02cm}(\alpha,\alpha_{1},\alpha_{2})\,a\,a_{1}\hspace{0.03cm}a_{2}}_{\; {\bf k}_{1}}$ entering into the canonical transformation of color charges $Q^{\hspace{0.03cm}a}_{\alpha}$, Eq.\,(\ref{eq:3y}):
\begin{equation}
M^{\hspace{0.02cm}(\alpha,\hspace{0.03cm}\alpha_{1},\alpha_{2})\,a\,a_{1}\hspace{0.03cm}a_{2}}_{\; {\bf k}_{1}} 
= 
i\hspace{0.02cm}f^{\,a\, a_{1}\hspace{0.03cm}a_{2}}
\delta^{\hspace{0.03cm}\alpha\hspace{0.03cm}\alpha_{2}}\, \frac{{\upphi}^{\hspace{0.03cm}(\alpha)}_{\hspace{0.03cm}{\bf k}_{1}}}
{\omega^{\hspace{0.03cm}l}_{\hspace{0.02cm}{\bf k}_{1}} - {\mathbf v}^{\phantom{l}}_{\alpha_{1}\!}\cdot {\mathbf k}^{\phantom{l}}_{1}}\,.
\label{eq:4e}
\end{equation}
\indent Furthermore, the requirement to exclude third-order terms in the Hamiltonian $H^{(3)}$, containing the vertex functions ${\mathcal V}^{\,a\,a_{1}\hspace{0.03cm}a_{2}}_{\ {\bf k},\,{\bf k}_{1},\, {\bf k}_{2}}$ and ${\mathcal U}^{\,a\,a_{1}\hspace{0.03cm}a_{2}}_{\ {\bf k},\,{\bf k}_{1},\,{\bf k}_{2}}$ leads to the following expressions for the coefficient functions $V^{\hspace{0.02cm}(1,3\hspace{0.02cm}|\hspace{0.02cm}\alpha,\alpha_{1},\alpha_{2})\,a\,a_{1}\hspace{0.03cm}a_{2}}_{\ {\bf k},\, {\bf k}_{1},\,{\bf k}_{2}}$ in the canonical transformation (\ref{eq:3t}):
\[
\begin{split}
&V^{\hspace{0.02cm}(1\hspace{0.02cm}|\hspace{0.02cm}\alpha,\alpha_{1},\alpha_{2})\,a\,a_{1}\hspace{0.03cm}a_{2}}_{\ {\bf k},\, {\bf k}_{1},\,	{\bf k}_{2}} 
=
-\hspace{0.02cm}\displaystyle\frac{{\mathcal V}^{\,a\,a_{1}\hspace{0.03cm}a_{2}}_{\ {\bf k},\,{\bf k}_{1},\, 
{\bf k}_{2}}}
{\bigl(\omega^{\hspace{0.03cm}l}_{\hspace{0.03cm}{\bf k}} - {\mathbf v}^{\phantom{l}}_{\alpha\!}\cdot{\mathbf k}\bigr) 
- 
\bigl(\omega^{\hspace{0.03cm}l}_{\hspace{0.03cm}{\bf k}_{1}} - {\mathbf v}^{\phantom{l}}_{\alpha_{1}\!}\cdot {\mathbf k}^{\phantom{l}}_{1}\bigr)
- 
\bigl(\omega^{\hspace{0.03cm}l}_{\hspace{0.03cm}{\bf k}_{2}} - {\mathbf v}^{\phantom{l}}_{\alpha_{2}\!}\cdot {\mathbf k}^{\phantom{l}}_{2}\bigr)}\,
\delta({\bf k} - {\bf k}_{1} - {\bf k}_{2}), 
\\[4ex]
&V^{\hspace{0.02cm}(3\hspace{0.02cm}|\hspace{0.02cm}\alpha,\alpha_{1},\alpha_{2})\,a\,a_{1}\hspace{0.03cm}a_{2}}_{\ {\bf k},\, {\bf k}_{1},\,{\bf k}_{2}}
= 
-\hspace{0.02cm}\displaystyle\frac{{\mathcal U}^{\hspace{0.03cm}*\,a\, a_{1}\hspace{0.03cm}a_{2}}_{\ {\bf k},\,{\bf k}_{1},\,{\bf k}_{2}}}
{\bigl(\omega^{\hspace{0.03cm}l}_{\hspace{0.03cm}{\bf k}} - {\mathbf v}^{\phantom{l}}_{\alpha\!}\cdot{\mathbf k}\bigr) 
+ 
\bigl(\omega^{\hspace{0.03cm}l}_{\hspace{0.03cm}{\bf k}_{1}} - {\mathbf v}^{\phantom{l}}_{\alpha_{1}\!}\cdot{\mathbf k}^{\phantom{l}}_{1}\bigr)
+ 
\bigl(\omega^{\hspace{0.03cm}l}_{\hspace{0.03cm}{\bf k}_{2}} - {\mathbf v}^{\phantom{l}}_{\alpha_{2}\!}\cdot {\mathbf k}^{\phantom{l}}_{2}\bigr)}\,
\delta({\bf k} + {\bf k}_{1} + {\bf k}_{2}).
\end{split}
\]
Note that the denominators of these expressions for arbitrary values $\alpha,\,\alpha_{1}$ and $\alpha_{2}$ explicitly contain wave vectors, unlike, for example, similar expressions in \cite{Markov:2020, Markov:2024}. 
The coefficient $V^{\hspace{0.02cm}(2\hspace{0.02cm}|\hspace{0.02cm}\alpha,\alpha_{1},\alpha_{2})\,a\,a_{1}\hspace{0.03cm}a_{2}}_{\ {\bf k},\, {\bf k}_{1},\,{\bf k}_{2}}$ is found from Eq.\,(\ref{eq:3i}).\\ 
\indent Thus, instead of the sum of the initial Hamiltonians $H^{(0)} + H^{(3)}$, Eq.\,(\ref{eq:4q}), we now obtain a new free-field Hamiltonian ${\mathcal H}^{(0)}$ for noninteracting plasmons in terms of the new normal variables $c^{\ast\ \!\!(\alpha)\hspace{0.02cm}a}_{\hspace{0.02cm}{\bf k}}$ and $c^{\phantom{\ast}\!\!\!(\alpha)\hspace{0.02cm}a}_{\hspace{0.02cm}{\bf k}}$:
\begin{equation}
{\mathcal H}^{(0)} =  
\sum_{\alpha}
\int\!d\hspace{0.02cm}{\bf k}\,
(\omega^{\hspace{0.03cm}l}_{\hspace{0.02cm}{\bf k}} - {\mathbf v}^{\phantom{l}}_{\alpha\!}\cdot {\mathbf k})\ \!
c^{\ast\ \!\!(\alpha)\hspace{0.02cm}a}_{\hspace{0.02cm}{\bf k}}
\hspace{0.03cm}
c^{(\alpha)	\hspace{0.02cm}a}_{\hspace{0.02cm}{\bf k}}.
\label{eq:4t}
\end{equation}
Hereinafter,\,the\,Hamiltonians\,in\,the\,new\,variables\,will\,be designated by the calligraphic letter~${\mathcal H}$.\\ 
\indent Furthermore, we can move to the construction of an explicit form of effective {\it fifth-order} Hamilto\-nian ${\mathcal H}^{(5)}$, which describes
plasmon bremsstrahlung radiation produced by the interaction between two hard test color-charged particles\footnote{As part of the formal rule of counting degrees of nonlinearity we have adopted, this Hamiltonian has a higher degree of nonlinearity than all the Hamiltonians we have considered so far \cite{Markov:2020, Markov:2023, Markov:2024} and describes a qualitatively new physical process.}. For this purpose, we need to collect all contributions proportional to the products $c^{\hspace{0.02cm}(\alpha)\hspace{0.02cm}a}_{\hspace{0.02cm}{\bf k}}\hspace{0.02cm}{\mathcal Q}^{\hspace{0.03cm}a_{1}}_{\hspace{0.03cm}1}\hspace{0.02cm}{\mathcal Q}^{\hspace{0.03cm}a_{2}}_{\hspace{0.03cm}2}$ 
and 
$c^{\ast\ \!\!(\alpha)\hspace{0.02cm}a}_{\hspace{0.02cm}{\bf k}}\hspace{0.02cm}{\mathcal Q}^{\hspace{0.03cm}a_{1}}_{\hspace{0.03cm}1}\hspace{0.02cm}{\mathcal Q}^{\hspace{0.03cm}a_{2}}_{\hspace{0.03cm}2}$ 
from the free-field Hamiltonian $H^{(0)}$, Eq.\,(\ref{eq:2p}), and from the interaction Hamiltonian $H^{(3)}$, Eq.\,(\ref{eq:2s}), to be generated by the canonical transformations (\ref{eq:3t}) and (\ref{eq:3y}). Thereby we obtain the effective fifth-order Hamiltonian describing plasmon bremsstrahlung:  
\begin{equation}
{\mathcal H}^{(5)}
=
\sum_{\rho}
\int\!d\hspace{0.02cm}{\bf k}\,
{T}^{\hspace{0.03cm}(\rho)\hspace{0.03cm}a\,a_{1}\hspace{0.03cm}a_{2}}_{\; {\bf k}}(t)\,
c^{\hspace{0.02cm}(\rho)\hspace{0.02cm}a}_{\hspace{0.02cm}{\bf k}}\hspace{0.02cm}{\mathcal Q}^{\hspace{0.03cm}a_{1}}_{\hspace{0.03cm}1}\hspace{0.02cm}{\mathcal Q}^{\hspace{0.03cm}a_{2}}_{\hspace{0.03cm}2}
\,+\,
\sum_{\rho}
\int\!d\hspace{0.02cm}{\bf k}\,
{T}^{\hspace{0.03cm}\ast\hspace{0.03cm}(\rho)\hspace{0.03cm}a\,a_{1}
\hspace{0.03cm}a_{2}}_{\;{\bf k}}(t)\,
c^{\ast\ \!\!(\rho)\hspace{0.02cm}a}_{\hspace{0.02cm}{\bf k}}\hspace{0.02cm}{\mathcal Q}^{\hspace{0.03cm}a_{1}}_{\hspace{0.03cm}1}\hspace{0.02cm}{\mathcal Q}^{\hspace{0.03cm}a_{2}}_{\hspace{0.03cm}2}.
\label{eq:4y}
\end{equation}
Here, the effective amplitude ${T}^{\hspace{0.03cm}(\rho)\hspace{0.03cm}a\,a_{1}\hspace{0.03cm}a_{2}}_{\; {\bf k}}(t)$ has the following structure:
\begin{equation}
{T}^{\hspace{0.03cm}(\rho)\hspace{0.03cm}a\,a_{1}\hspace{0.03cm}a_{2}}_{\; {\bf k}}(t)
=
\int\!d\hspace{0.02cm}{\bf q}\,
{T}^{\hspace{0.03cm}(\rho)\hspace{0.03cm}a\,a_{1}\hspace{0.03cm}a_{2}}_{\; {\bf k},\,{\bf q}}(t),
\label{eq:4_1u}
\end{equation}
where, in turn, we represent the ``density'' of the amplitude ${T}^{\hspace{0.03cm}(\rho)\hspace{0.03cm}a\,a_{1}\hspace{0.03cm}a_{2}}_{\; {\bf k},\,{\bf q}}$ as the sum of two parts that differ in physical content
\begin{equation}
{T}^{\hspace{0.03cm}(\rho)\hspace{0.03cm}a\,a_{1}\hspace{0.03cm}a_{2}}_{\; {\bf k},\,{\bf q}}(t)
=
{T}^{\,{\rm I}\,
(\rho)\hspace{0.03cm}a\,a_{1}\hspace{0.03cm}a_{2}}_{\; {\bf k},\,{\bf q}}(t)
+
{T}^{\,{\rm II}\,
(\rho)\hspace{0.03cm}a\,a_{1}\hspace{0.03cm}a_{2}}_{\; {\bf k},\,{\bf q}}(t).
\label{eq:4_1uu}
\end{equation}
The first part, taking into account (\ref{eq:4w}) and (\ref{eq:4e}), has the following structure:
\begin{equation}
{T}^{\,{\rm I}\,(\rho)\hspace{0.03cm}a\,a_{1}\hspace{0.03cm}a_{2}}_{\; {\bf k},\,{\bf q}}(t)
=
-\hspace{0.03cm}i\hspace{0.02cm}
f^{\,a\hspace{0.03cm}a_{1}\hspace{0.03cm}a_{2}}\hspace{0.03cm}
\frac{1}{\omega^{\hspace{0.02cm} l}_{\hspace{0.03cm}{\bf k}} - {\bf v}^{\phantom{l}}_{\rho}\cdot {\bf k}}\,
\biggl(\frac{1}{\omega^{\hspace{0.02cm}l}_{\hspace{0.03cm}{\bf q}} - {\bf v}^{\phantom{l}}_{1}\cdot {\bf q}}
\,+\,
\frac{1}{\omega^{\hspace{0.02cm} l}_{\hspace{0.03cm}{\bf q}} - {\bf v}^{\phantom{l}}_{2}\cdot {\bf q}}
\biggr)\times
\label{eq:4u}
\vspace{-0.3cm}
\end{equation}
\begin{align}
\Bigl[\hspace{0.03cm}&{\upphi}^{\hspace{0.02cm}\ast\hspace{0.03cm}(1)}_{\,{\bf q}}(t)
\hspace{0.03cm}
{\upphi}^{\hspace{0.03cm}(2)}_{\,{\bf q}}(t)
\hspace{0.03cm}
{\upphi}^{\hspace{0.03cm}(1)}_{\,{\bf k}}(t)
\,+\,
{\upphi}^{(1)}_{\,{\bf q}}(t)
\hspace{0.03cm}
{\upphi}^{\hspace{0.02cm}\ast\hspace{0.03cm}(2)}_{\,{\bf q}}(t)
\hspace{0.03cm}
{\upphi}^{\hspace{0.03cm}(1)}_{\,{\bf k}}(t)
\,-\,\notag\\[1ex]
&{\upphi}^{\hspace{0.02cm}\ast\hspace{0.03cm}(2)}_{\,{\bf q}}(t)
\hspace{0.03cm}
{\upphi}^{\hspace{0.03cm}(1)}_{\,{\bf q}}(t)
\hspace{0.03cm}
{\upphi}^{\hspace{0.03cm}(2)}_{\,{\bf k}}(t)
\,-\,
{\upphi}^{\hspace{0.03cm}(2)}_{\,{\bf q}}(t)
\hspace{0.03cm}
{\upphi}^{\hspace{0.02cm}\ast\hspace{0.03cm}(1)}_{\,{\bf q}}(t)
\hspace{0.03cm}
{\upphi}^{\hspace{0.03cm}(2)}_{\,{\bf k}}(t)
\Bigr],
\notag
\end{align}
while the second part can be represented as 
\begin{align}
&\hspace{2.5cm}{T}^{\,{\rm II}\,(\rho)\hspace{0.03cm}a\,a_{1}\hspace{0.03cm}a_{2}}_{\; {\bf k},\,{\bf q}}(t)
=
{\mathcal V}^{\,\ast\,a\,a_{1}\hspace{0.03cm}a_{2}}_{\ {\bf k},\,{\bf q},\, 
{\bf k} - {\bf q}}(t)\hspace{0.03cm}
\biggl(\frac{1}{\omega^{\hspace{0.02cm}l}_{\hspace{0.03cm}{\bf q}} - {\bf v}^{\phantom{l}}_{1}\cdot {\bf q}}
\,+\,
\frac{1}{\omega^{\hspace{0.02cm} l}_{\hspace{0.03cm}{\bf q}} - {\bf v}^{\phantom{l}}_{2}\cdot {\bf q}}
\biggr)\hspace{0.03cm}\times
\label{eq:4uu}\\[1ex]
&\biggl(\frac{1}{\omega^{\hspace{0.02cm}l}_{\hspace{0.03cm}{\bf k} - {\bf q}}\! - {\bf v}^{\phantom{l}}_{1}\cdot ({\bf k} - {\bf q})}
\,+\,
\frac{1}{\omega^{\hspace{0.02cm} l}_{\hspace{0.03cm}{\bf k} - {\bf q}}\! - {\bf v}^{\phantom{l}}_{2}\cdot ({\bf k} - {\bf q})}
\biggr)
\Bigl[\hspace{0.03cm}{\upphi}^{\hspace{0.03cm}(1)}_{\,{\bf q}}(t)
\hspace{0.03cm}
{\upphi}^{\hspace{0.03cm}(2)}_{\,{\bf k} - {\bf q}}(t)
-
{\upphi}^{\hspace{0.03cm}(2)}_{\,{\bf q}}(t)
\hspace{0.03cm}
{\upphi}^{\hspace{0.03cm}(1)}_{\,{\bf k} - {\bf q}}(t)
\Bigr]\,+
\notag\\[1ex]
&\hspace{5.3cm}2\hspace{0.03cm}{\mathcal V}^{\;a_{1}\hspace{0.03cm}a_{2}\,a}_{{\bf q},\,{\bf q} - {\bf k},\,{\bf k}}(t)\hspace{0.03cm}
\biggl(\frac{1}{\omega^{\hspace{0.02cm}l}_{\hspace{0.03cm}{\bf q}} - {\bf v}^{\phantom{l}}_{1}\cdot {\bf q}}
\,+\,
\frac{1}{\omega^{\hspace{0.02cm} l}_{\hspace{0.03cm}{\bf q}} - {\bf v}^{\phantom{l}}_{2}\cdot {\bf q}}
\biggr)\hspace{0.03cm}\times
\notag\\[1ex]
&\biggl(\frac{1}{\omega^{\hspace{0.02cm}l}_{\hspace{0.03cm}{\bf q} - {\bf k}}\! - {\bf v}^{\phantom{l}}_{1}\cdot ({\bf q} - {\bf k})}
\,+\,
\frac{1}{\omega^{\hspace{0.02cm} l}_{\hspace{0.03cm}{\bf q} - {\bf k}}\! - {\bf v}^{\phantom{l}}_{2}\cdot ({\bf q} - {\bf k})}
\biggr)
\Bigl[\hspace{0.03cm}{\upphi}^{\hspace{0.03cm}(1)}_{\,{\bf q}}(t)
\hspace{0.03cm}
{\upphi}^{\ast\,(2)}_{\,{\bf q} - {\bf k}}(t)
-
{\upphi}^{\hspace{0.03cm}(2)}_{\,{\bf q}}(t)
\hspace{0.03cm}
{\upphi}^{\ast\,(1)}_{\,{\bf q} - {\bf k}}(t)
\Bigr]\,+
\notag\\[1ex]
&\hspace{5.6cm}{\mathcal U}^{\;a\,a_{1}\hspace{0.03cm}a_{2}}_{\ {\bf k},\,{\bf q},\, 
- {\bf k} - {\bf q}}(t)\hspace{0.03cm}
\biggl(\frac{1}{\omega^{\hspace{0.02cm}l}_{\hspace{0.03cm}{\bf q}} - {\bf v}^{\phantom{l}}_{1}\cdot {\bf q}}
\,+\,
\frac{1}{\omega^{\hspace{0.02cm} l}_{\hspace{0.03cm}{\bf q}} - {\bf v}^{\phantom{l}}_{2}\cdot {\bf q}}
\biggr)\hspace{0.03cm}\times
\notag\\[1ex]
&\biggl(\frac{1}{\omega^{\hspace{0.02cm}l}_{\hspace{0.03cm}- {\bf k} - {\bf q}}\! + {\bf v}^{\phantom{l}}_{1}\cdot ({\bf k} + {\bf q})}
\,+\,
\frac{1}{\omega^{\hspace{0.02cm} l}_{\hspace{0.03cm}- {\bf k} - {\bf q}}\! 
+ {\bf v}^{\phantom{l}}_{2}\cdot ({\bf k} + {\bf q})}
\biggr)
\Bigl[\hspace{0.03cm}{\upphi}^{\hspace{0.03cm}\ast\,(1)}_{\,{\bf q}}(t)
\hspace{0.03cm}
{\upphi}^{\hspace{0.03cm}\ast\,(2)}_{\,- {\bf k} - {\bf q}}(t)
-
{\upphi}^{\hspace{0.03cm}\ast\,(2)}_{\,{\bf q}}(t)
\hspace{0.03cm}
{\upphi}^{\hspace{0.03cm}\ast\,(1)}_{\,- {\bf k} - {\bf q}}(t)
\Bigr].
\notag
\end{align}
\indent We see that only the first part of the total amplitude ${T}^{\,(\rho)\hspace{0.03cm}a\,a_{1}\hspace{0.03cm}a_{2}}_{\; {\bf k},\,{\bf q}}$, namely ${T}^{\,{\rm I}\,(\rho)\hspace{0.03cm}a\,a_{1}\hspace{0.03cm}a_{2}}_{\; {\bf k},\,{\bf q}}$, actually depends on the ``index'' $\rho$. The effective amplitude ${T}^{\,{\rm I}\,(\rho)\hspace{0.03cm}a\,a_{1}\hspace{0.03cm}a_{2}}_{\; {\bf k},\,{\bf q}}$ describes the Compton-like plasmon bremsstrahlung. Figure\,\ref{fig2-crop} gives the diagrammatic interpretation of different terms in (\ref{eq:4u}) (more exactly, for the complex conjugate expression ${T}^{\hspace{0.03cm}\ast\hspace{0.03cm}{\rm I}\,(\rho)\hspace{0.03cm}a\,a_{1}\hspace{0.03cm}a_{2}}_{\; {\bf k},\,{\bf q}}$\hspace{0.03cm}).
The plasmon bremsstrahlung here comes from either hard particle 1 or 2. This is reflected in the multiplier with the index $\rho = 1,2$ in (\ref{eq:4u}).
The outgoing wave lines in fig.\,\ref{fig2-crop} correspond to the components $c^{\hspace{0.02cm}(\rho)\hspace{0.02cm}a}_{\hspace{0.02cm}{\bf k}}$ of the total normal variable $c^{\,a}_{\hspace{0.03cm}{\bf k}}$, the horizontal double line between two interaction vertices corresponds to the ``propagator'' of the hard particles
\begin{equation}
\frac{1}{\omega^{\hspace{0.02cm} l}_{\hspace{0.03cm}{\bf k}}\, -\, {\bf v}^{\phantom{l}}_{\rho}\!\cdot {\bf k}}
\label{eq:4aa}
\end{equation}
and the interaction vertices, defining plasmon bremsstrahlung, correspond to the vertex functions ${\upphi}^{\hspace{0.02cm}\ast\hspace{0.03cm}(1)}_{\,{\bf k}}(t)$ or ${\upphi}^{\hspace{0.02cm}\ast\hspace{0.03cm}(2)}_{\,{\bf k}}(t)$. The remaining vertex functions of the form ${\upphi}^{\hspace{0.03cm}(1)}_{\,{\bf q}}(t)$ and ${\upphi}^{\hspace{0.03cm}(2)}_{\,{\bf q}}(t)$ are related to the exchange process of an intermediate ``virtual'' oscillation to which the sum in parentheses in (\ref{eq:4u}) corresponds.\\
\indent The effective amplitude ${T}^{\,{\rm II}\,(\rho)\hspace{0.03cm}a\,a_{1}\hspace{0.03cm}a_{2}}_{\; {\bf k},\,{\bf q}}$, Eq.\,(\ref{eq:4uu}), determines the processes, shown in fig.\,\ref{fig3-crop}. It is these processes that represent the transition bremsstrahlung, which was discussed in the Introduction.
The graphs in fig.\,\ref{fig3-crop} are connected with the interaction of hard particles with plasmons through the three-plasmon vertex functions ${\mathcal V}^{\;a\,a_{1}\hspace{0.03cm}a_{2}}_{\;{\bf k},\, {\bf k}_{1},\, {\bf k}_{2}}$, ${\mathcal V}^{\,\ast\,a\,a_{1}\hspace{0.03cm}a_{2}}_{\;{\bf k},\, {\bf k}_{1},\, {\bf k}_{2}}$ and  ${\mathcal U}^{\;a\,a_{1}\hspace{0.03cm}a_{2}}_{\;{\bf k},\, {\bf k}_{1},\, {\bf k}_{2}}$ with the intermediate ``virtual'' oscillations to which the factors  
in parentheses in (\ref{eq:4uu}) correspond. We have depicted the contributions of the various terms in (\ref{eq:4uu}), with different directions of momenta in the internal lines. Unlike the previous amplitude ${T}^{\,{\rm I}\,(\rho)\hspace{0.03cm}a\,a_{1}\hspace{0.03cm}a_{2}}_{\; {\bf k},\,{\bf q}}$ here the outgoing wave lines correspond to the full normal variable $c^{\!\!\!\!\!\!\phantom{(1)}a}_{\hspace{0.03cm}{\bf k}} = c^{\,(1)\,a}_{\hspace{0.03cm}{\bf k}} +
c^{\,(2)\,a}_{\hspace{0.03cm}{\bf k}}$. It is clear why the second effective amplitude ${T}^{\,{\rm II}\,(\rho)\hspace{0.03cm}a\,a_{1}
\hspace{0.03cm}a_{2}}_{\;{\bf k},\,{\bf q}}$ does not depend on the $\rho$ index: particles 1 and 2 enter symmetrically into this radiation process. Finally, the interaction vertices ${\upphi}^{\hspace{0.03cm}(1)}_{\,{\bf q}}(t)$, ${\upphi}^{\hspace{0.03cm}(1)}_{\,{\bf k} - {\bf q}}(t)$ etc. correspond to interaction of ``virtual'' plasmons with hard particles.

%
%

\section{\bf Multiple-time-scale perturbation expansion}
\label{section_5}
\setcounter{equation}{0}

Now we turn to the construction of a kinetic equation describing the process of plasmon bremsstrahlung produced by the color charge rotation of hard particles passing through the hot QCD matter. As the interaction Hamiltonian we consider the effective Hamiltonian ${\mathcal H}^{(5)}$, Eq.\,(\ref{eq:4y}). The equations of motion for the components of the bosonic normal variables $c^{\phantom{\hspace{0.03cm}\ast} \!\!a}_{\hspace{0.02cm}{\bf k}}$ and $c^{\hspace{0.03cm}\ast\,a}_{\hspace{0.02cm}{\bf k}}$ and the color charges $\mathcal{Q}^{\,a}_{\hspace{0.03cm}\alpha}$ are defined by the corresponding Hamilton equations. For the soft Bose-excitations we find
\begin{align}
&\frac{\partial \hspace{0.02cm}c^{\hspace{0.02cm}(\alpha)\hspace{0.02cm}a}_{\hspace{0.02cm}{\bf k}}}{\partial\hspace{0.03cm} t}
=
-\hspace{0.03cm}i\hspace{0.05cm}\Bigl\{c^{\hspace{0.02cm}(\alpha)\hspace{0.02cm}a}_{\hspace{0.02cm}{\bf k}}\hspace{0.03cm},\hspace{0.03cm}
{\mathcal H}^{(5)}\Bigr\}
=
-\hspace{0.03cm}i\,
{T}^{\hspace{0.03cm}\ast\hspace{0.03cm}(\alpha)\hspace{0.03cm}a\,a_{1}
\hspace{0.03cm}a_{2}}_{\;{\bf k}}(t)\,
{\mathcal Q}^{\hspace{0.03cm}a_{1}}_{\hspace{0.03cm}1}
\hspace{0.02cm}
{\mathcal Q}^{\hspace{0.03cm}a_{2}}_{\hspace{0.03cm}2},
\label{eq:5q}\\[1ex]	
&\frac{\partial \hspace{0.02cm}c^{\ast\ \!\!(\alpha)\hspace{0.02cm}a}_{\hspace{0.02cm}{\bf k}}}{\partial\hspace{0.03cm} t}
=
-\hspace{0.03cm}i\hspace{0.05cm}\Bigl\{c^{\ast\ \!\!(\alpha)\hspace{0.02cm}a}_{\hspace{0.02cm}{\bf k}}\hspace{0.03cm},\hspace{0.03cm}{\mathcal H}^{(5)}\Bigr\}
=
i\,{T}^{\hspace{0.03cm}(\alpha)\hspace{0.03cm}a\,a_{1}\hspace{0.03cm}a_{2}}_{\; {\bf k}}(t)\,
\hspace{0.02cm}{\mathcal Q}^{\hspace{0.03cm}a_{1}}_{\hspace{0.03cm}1}
\hspace{0.02cm}
{\mathcal Q}^{\hspace{0.03cm}a_{2}}_{\hspace{0.03cm}2}, 
\label{eq:5w}	
\end{align}
and, respectively, for the classical color charges we get  
\[
\frac{d \hspace{0.01cm}\mathcal{Q}^{\,d}_{\hspace{0.03cm}\alpha}}{d\hspace{0.03cm}t}
=
-\hspace{0.03cm}i\hspace{0.05cm}
\Bigl\{\mathcal{Q}^{\,d}_{\hspace{0.03cm}\alpha}\hspace{0.03cm},\hspace{0.03cm} 
{\mathcal H}^{(5)}\Bigr\}
=
\frac{\partial\hspace{0.03cm} 
{\mathcal H}^{(5)}}{\,\partial\hspace{0.02cm} {\mathcal Q}^{\hspace{0.03cm}d^{\hspace{0.02cm}\prime}}_{\hspace{0.03cm}\alpha}}
\,f^{\hspace{0.03cm}d\hspace{0.03cm}d^{\hspace{0.02cm}\prime}
\hspace{0.01cm}e}\hspace{0.03cm} {\mathcal Q}^{\hspace{0.03cm}e}_{\hspace{0.03cm}\alpha},
\]
or
\begin{align}
&\frac{d \hspace{0.01cm}\mathcal{Q}^{\,a_{1}}_{\hspace{0.03cm}1}}{d\hspace{0.03cm} t}
=
f^{\hspace{0.03cm}a_{1}\hspace{0.03cm}c^{\hspace{0.02cm}\prime}\hspace{0.01cm}e_{1}}
\,\sum_{\rho}\,\biggl[\,
\int\!d\hspace{0.02cm}{\bf k}\,
{T}^{\hspace{0.03cm}(\rho)\hspace{0.03cm}a\,c^{\hspace{0.02cm}\prime}
\hspace{0.03cm}a_{2}}_{\; {\bf k}}(t)\,
c^{\hspace{0.02cm}(\rho)\hspace{0.02cm}a}_{\hspace{0.02cm}{\bf k}}\hspace{0.02cm}{\mathcal Q}^{\hspace{0.03cm}e_{1}}_{\hspace{0.03cm}1}\hspace{0.02cm}{\mathcal Q}^{\hspace{0.03cm}a_{2}}_{\hspace{0.03cm}2}
\,+\,
\int\!d\hspace{0.02cm}{\bf k}\,
{T}^{\hspace{0.03cm}\ast\hspace{0.03cm}(\rho)\hspace{0.03cm}a\,
c^{\hspace{0.02cm}\prime}\hspace{0.03cm}a_{2}}_{\; {\bf k}}(t)\,
c^{\ast\ \!\!(\rho)\hspace{0.02cm}a}_{\hspace{0.02cm}{\bf k}}\hspace{0.02cm}{\mathcal Q}^{\hspace{0.03cm}e_{1}}_{\hspace{0.03cm}1}\hspace{0.02cm}{\mathcal Q}^{\hspace{0.03cm}a_{2}}_{\hspace{0.03cm}2}
\biggr],
\label{eq:5e}\\[1.5ex]
&\frac{d \hspace{0.01cm}\mathcal{Q}^{\,a_{2}}_{\hspace{0.03cm}2}}{d\hspace{0.03cm}t}
=
f^{\hspace{0.03cm}a_{2}\hspace{0.03cm}c^{\hspace{0.02cm}\prime}\hspace{0.01cm}e_{2}}
\,\sum_{\rho}\,\biggl[\,
\int\!d\hspace{0.02cm}{\bf k}\,
{T}^{\hspace{0.03cm}(\rho)\hspace{0.03cm}a\hspace{0.03cm}a_{1}\,c^{\hspace{0.02cm}\prime}}_{\;{\bf k}}(t)\,
c^{\hspace{0.02cm}(\rho)\hspace{0.02cm}a}_{\hspace{0.02cm}{\bf k}}\hspace{0.02cm}{\mathcal Q}^{\hspace{0.03cm}a_{1}}_{\hspace{0.03cm}1}\hspace{0.02cm}{\mathcal Q}^{\hspace{0.03cm}e_{2}}_{\hspace{0.03cm}2}
\,+\,
\int\!d\hspace{0.02cm}{\bf k}\,
{T}^{\hspace{0.03cm}\ast\hspace{0.03cm}(\rho)\hspace{0.03cm}a\hspace{0.03cm}a_{1}\,c^{\hspace{0.02cm}\prime}}_{\;{\bf k}}(t)\,
c^{\ast\ \!\!(\rho)\hspace{0.02cm}a}_{\hspace{0.02cm}{\bf k}}\hspace{0.02cm}{\mathcal Q}^{\hspace{0.03cm}a_{1}}_{\hspace{0.03cm}1}\hspace{0.02cm}{\mathcal Q}^{\hspace{0.03cm}e_{2}}_{\hspace{0.03cm}2}
\biggr].
\label{eq:5ee}
\end{align}
In the equations (\ref{eq:5q}) and (\ref{eq:5w}) for the bosonic normal variables $c^{\phantom{\hspace{0.03cm}\ast} \!\!a}_{\hspace{0.02cm}{\bf k}}$ and $c^{\hspace{0.03cm}\ast\,a}_{\hspace{0.02cm}{\bf k}}$ instead of the total Hamiltonian ${\mathcal H}^{(0)\!} + {\mathcal H}^{(5)}$ we have used only the interaction Hamiltonian ${\mathcal H}^{(5)}$. The contribution of the new free Hamiltonian ${\mathcal H}^{(0)}$, Eq.\,(\ref{eq:4t}), to an evolution of soft bosonic excitations was actually taken into account in the definition of the vertex functions ${\mathcal V}^{\; a\, a_{1}\hspace{0.03cm} a_{2}}_{{\bf k},\, {\bf k}_{1},\, {\bf k}_{2}}(t)$, ${\mathcal U}^{\; a\, a_{1}\hspace{0.03cm} a_{2}}_{\,{\bf k},\,{\bf k}_{1},\, {\bf k}_{2}}(t)$ and  ${\upphi}^{(\alpha)}_{\hspace{0.03cm} {\bf k}}(t)$, Eqs.\,(\ref{eq:2dd}) and (\ref{eq:2z}), respectively. Therefore, accounting ${\mathcal H}^{(0)}$ on the right-hand sides of (\ref{eq:5q}) and (\ref{eq:5w}), as it is not difficult to see, result in a kind of ``double-counting''.\\
\indent To construct the required kinetic equation it is convenient to use the multiple-time-scale perturbation expansions \cite{Frieman:1963, Zakharov:1967,  Zakharov1968, Davidson:1972, Crawford:1980}, which was already mentioned in the Introduction. According to this method  the wave field $c^{\hspace{0.02cm}(\alpha)\hspace{0.02cm}a}_{\hspace{0.02cm}{\bf k}}(t)$ can be divided into a slowly varying (in time) components $C^{\hspace{0.02cm}(\alpha)\hspace{0.02cm}a}_{\hspace{0.02cm}{\bf k}}(\tau)$ with a characteristic variation time much larger than $1/\omega_{\rm pl}$ and a small, rapidly varying components $\widehat{C}^{\hspace{0.02cm}(\alpha)\hspace{0.02cm}a}_{\hspace{0.02cm}{\bf k}}(t')$ (the value of $\widehat{C}^{\hspace{0.02cm}(\alpha)\hspace{0.02cm}a}_{\hspace{0.02cm}{\bf k}}(t')$ changes appreciably within a time on the order of $1/\omega_{\rm pl}$). Most of the energy in the wave field is contained in the slowly component $C^{\hspace{0.02cm}(\alpha)\hspace{0.02cm}a}_{\hspace{0.02cm}{\bf k}}(\tau)$. Let us assume that a similar separation into slow and fast components is valid for color charges $\mathcal{Q}^{\,a}_{\hspace{0.03cm}\alpha}(t)$ of hard particles. These assumptions permits us to write 
\begin{subequations} 
\label{eq:5r}
\begin{align}
&c^{\hspace{0.02cm}(\alpha)\hspace{0.02cm}a}_{\hspace{0.02cm}{\bf k}}(t)
=
\varepsilon^{\alpha_{1}}\hspace{0.02cm}
C^{\hspace{0.02cm}(\alpha)\hspace{0.02cm}a}_{\hspace{0.02cm}{\bf k}}(\tau)
\,+\,
\varepsilon^{\alpha_{2}}\hspace{0.03cm}
\widehat{C}^{\hspace{0.02cm}(\alpha)\hspace{0.02cm}a}_{\hspace{0.02cm}{\bf k}}(t'),
\label{eq:5ra}\\[1.5ex]
&\mathcal{Q}^{\,a}_{\hspace{0.03cm}\alpha}(t)
=
\varepsilon^{\beta_{1}}\hspace{0.02cm}\mathcal{Q}^{\,a}_{\hspace{0.03cm}\alpha}
(\tau)
\,+\,
\varepsilon^{\beta_{2}}\hspace{0.02cm}\widehat{\mathcal{Q}}^{\,a}_{\hspace{0.03cm}\alpha}(t'). 
\label{eq:5rb}
\end{align}
\end{subequations}
Here, $\varepsilon$ is a dimensionless small parameter describing the magnitude of the nonlinearity. The fast time scale is $t'$, and the slow time scale is $\tau$. We consider them to be independent variables. The total time derivative is to be expanded according to
\begin{equation}
\partial/\partial\hspace{0.03cm}t 
=
\varepsilon^{\gamma_{1}}\hspace{0.02cm}(\partial/\partial\hspace{0.03cm}t') 
\,+\, 
\varepsilon^{\gamma_{2}}\hspace{0.02cm}(\partial/\partial\hspace{0.03cm}\tau).
\label{eq:5t}
\end{equation}
Unknown exponents $\alpha_{1,2},\,\beta_{1,2}$ and $\gamma_{1,2}$ on the right-hand sides (\ref{eq:5r}) and (\ref{eq:5t}) represent non-negative integers\footnote{\hspace{0.03cm}The condition that the exponents must be integers is most likely unnecessary.}, moreover, the natural requirement here is the fulfillment of inequalities
\[
\alpha_{1}<\alpha_{2}, \quad \beta_{1}<\beta_{2}, \quad \gamma_{1}<\gamma_{2}. 
\]	
We will determine these exponents while finding a self-consistent solution to the equations for fast and slow components of the original system (\ref{eq:5q})\,--\,(\ref{eq:5ee}).\\
\indent Substituting expansions (\ref{eq:5r}) and (\ref{eq:5t}) into (\ref{eq:5q}) and (\ref{eq:5e}) we decompose the Hamilton equations into the ones for the fast and slow variables as follows:    
\begin{align}
&\varepsilon^{\alpha_{2} + \gamma_{1}}\,
\frac{\partial \hspace{0.04cm}\hat{C}^{\hspace{0.02cm}(\alpha)\hspace{0.02cm}a}_{\hspace{0.02cm}{\bf k}}(t')}{\partial\hspace{0.03cm}t'}
\hspace{0.03cm}=\hspace{0.03cm}
\varepsilon^{2\beta_{1}}\hspace{0.03cm}(-\hspace{0.03cm}i)\,
{T}^{\hspace{0.03cm}\ast\hspace{0.03cm}(\alpha)\hspace{0.03cm}a\,a_{1}
\hspace{0.03cm}a_{2}}_{\;{\bf k}}(t')\,
{\mathcal Q}^{\hspace{0.03cm}a_{1}}_{\hspace{0.03cm}1}(\tau)
\hspace{0.02cm}
{\mathcal Q}^{\hspace{0.03cm}a_{2}}_{\hspace{0.03cm}2}(\tau),
\label{eq:5y}\\[1ex]
&\varepsilon^{\alpha_{1} + \gamma_{2}}\,
\frac{\partial \hspace{0.04cm}C^{\hspace{0.02cm}(\alpha)\hspace{0.02cm}a}_{\hspace{0.02cm}{\bf k}}(\tau)}{\partial\hspace{0.03cm}\tau}
\,=\,
\varepsilon^{\beta_{1} + \beta_{2}}\hspace{0.03cm}
(-\hspace{0.03cm}i)\,
{T}^{\hspace{0.03cm}\ast\hspace{0.03cm}(\alpha)\hspace{0.03cm}a\,a_{1}
\hspace{0.03cm}a_{2}}_{\;{\bf k}}(t')\,
\Bigl[{\mathcal Q}^{\hspace{0.03cm}a_{1}}_{\hspace{0.03cm}1}(\tau)
\hspace{0.02cm}
\widehat{\mathcal{Q}}^{\hspace{0.03cm}a_{2}}_{\hspace{0.03cm}2}(t')
+
\widehat{\mathcal{Q}}^{\hspace{0.03cm}a_{1}}_{\hspace{0.03cm}1 }(t')
\hspace{0.02cm}
{\mathcal Q}^{\hspace{0.03cm}a_{2}}_{\hspace{0.03cm}2}(\tau)
\Bigr],
\label{eq:5u}\\[2ex]
&\varepsilon^{\beta_{2} + \gamma_{1}}\,
\frac{d \hspace{0.01cm}\widehat{\mathcal{Q}}^{\,a_{1}}_{\hspace{0.03cm}1}(t')}{d\hspace{0.03cm} t'}
=
\varepsilon^{\alpha_{1} + 2\beta_{1}}\hspace{0.03cm}
f^{\hspace{0.03cm}a_{1}\hspace{0.03cm}c^{\hspace{0.02cm}\prime}\hspace{0.01cm}e_{1}}
\,\sum_{\rho}\,\biggl[\,
\int\!d\hspace{0.02cm}{\bf k}\,
{T}^{\hspace{0.03cm}(\rho)\hspace{0.03cm}a\,c^{\hspace{0.02cm}\prime}
\hspace{0.03cm}a_{2}}_{\; {\bf k}}(t')\,
C^{\hspace{0.02cm}(\rho)\hspace{0.02cm}a}_{\hspace{0.02cm}{\bf k}}(\tau)
\hspace{0.02cm}
{\mathcal Q}^{\hspace{0.03cm}e_{1}}_{\hspace{0.03cm}1}(\tau)
\hspace{0.02cm}
{\mathcal Q}^{\hspace{0.03cm}a_{2}}_{\hspace{0.03cm}2}(\tau)
\label{eq:5i}\\[1ex]
&\hspace{8.5cm}+\!
\int\!d\hspace{0.02cm}{\bf k}\,
{T}^{\hspace{0.03cm}\ast\hspace{0.03cm}(\rho)\hspace{0.03cm}a\,
c^{\hspace{0.02cm}\prime}\hspace{0.03cm}a_{2}}_{\; {\bf k}}(t')\,
C^{\,\ast\hspace{0.03cm}(\rho)\hspace{0.02cm}a}_{\hspace{0.02cm}{\bf k}}(\tau)
\hspace{0.02cm}
{\mathcal Q}^{\hspace{0.03cm}e_{1}}_{\hspace{0.03cm}1}(\tau)
\hspace{0.02cm}
{\mathcal Q}^{\hspace{0.03cm}a_{2}}_{\hspace{0.03cm}2}(\tau)
\biggr],
\notag\\[1ex]
&
\varepsilon^{\beta_{1} + \gamma_{2}}\,
\frac{d \hspace{0.01cm}\mathcal{Q}^{\,a_{1}}_{\hspace{0.03cm}1}(\tau)}{d\hspace{0.03cm} \tau}
\,=
\varepsilon^{\alpha_{2} + 2\beta_{1}}\hspace{0.03cm}
f^{\hspace{0.03cm}a_{1}\hspace{0.03cm}c^{\hspace{0.02cm}\prime}\hspace{0.01cm}e_{1}}
\,\sum_{\rho}\,\biggl[\,
\int\!d\hspace{0.02cm}{\bf k}\,
{T}^{\hspace{0.03cm}(\rho)\hspace{0.03cm}a\,c^{\hspace{0.02cm}\prime}
	\hspace{0.03cm}a_{2}}_{\; {\bf k}}(t')\,
\widehat{C}^{\hspace{0.02cm}(\rho)\hspace{0.02cm}a}_{\hspace{0.02cm}{\bf k}}(t')
\hspace{0.02cm}
{\mathcal Q}^{\hspace{0.03cm}e_{1}}_{\hspace{0.03cm}1}(\tau)
\hspace{0.02cm}
{\mathcal Q}^{\hspace{0.03cm}a_{2}}_{\hspace{0.03cm}2}(\tau)
\label{eq:5o}\\[1ex]
&\hspace{8.5cm}
+\!
\int\!d\hspace{0.02cm}{\bf k}\,
{T}^{\hspace{0.03cm}\ast\hspace{0.03cm}(\rho)\hspace{0.03cm}a\,
	c^{\hspace{0.02cm}\prime}\hspace{0.03cm}a_{2}}_{\; {\bf k}}(t')\,
\widehat{C}^{\,\ast\hspace{0.02cm}(\rho)\hspace{0.02cm}a}_{\hspace{0.02cm}{\bf k}}(t')
\hspace{0.02cm}
{\mathcal Q}^{\hspace{0.03cm}e_{1}}_{\hspace{0.03cm}1}(\tau)
\hspace{0.02cm}
{\mathcal Q}^{\hspace{0.03cm}a_{2}}_{\hspace{0.03cm}2}(\tau)
\biggr]
\notag\\
&+\,
\varepsilon^{\alpha_{1} + \beta_{1} + \beta_{2}}\hspace{0.03cm}
f^{\hspace{0.03cm}a_{1}\hspace{0.03cm}c^{\hspace{0.02cm}\prime}\hspace{0.01cm}e_{1}}
\,\sum_{\rho}\,\biggl[\,
\int\!d\hspace{0.02cm}{\bf k}\,
{T}^{\hspace{0.03cm}(\rho)\hspace{0.03cm}a\,c^{\hspace{0.02cm}\prime}
	\hspace{0.03cm}a_{2}}_{\; {\bf k}}(t')\,
C^{\hspace{0.02cm}(\rho)\hspace{0.02cm}a}_{\hspace{0.02cm}{\bf k}}(\tau)
\hspace{0.02cm}
\Bigl(
{\mathcal Q}^{\hspace{0.03cm}a_{2}}_{\hspace{0.03cm}2}(\tau)
\hspace{0.02cm}
\widehat{\mathcal{Q}}^{\hspace{0.03cm}e_{1}}_{\hspace{0.03cm}1}(t')
+
\widehat{\mathcal{Q}}^{\hspace{0.03cm}a_{2}}_{\hspace{0.03cm}2}(t')
\hspace{0.02cm}
{\mathcal Q}^{\hspace{0.03cm}e_{1}}_{\hspace{0.03cm}1}(\tau)
\Bigr)
\notag\\[1ex]
&\hspace{5cm}+\!
\int\!d\hspace{0.02cm}{\bf k}\,
{T}^{\hspace{0.03cm}\ast\hspace{0.03cm}(\rho)\hspace{0.03cm}a\,
	c^{\hspace{0.02cm}\prime}\hspace{0.03cm}a_{2}}_{\; {\bf k}}(t')\,
C^{\,\ast\hspace{0.03cm}(\rho)\hspace{0.02cm}a}_{\hspace{0.02cm}{\bf k}}(\tau)
\hspace{0.02cm}
\Bigl(
{\mathcal Q}^{\hspace{0.03cm}a_{2}}_{\hspace{0.03cm}2}(\tau)
\hspace{0.02cm}
\widehat{\mathcal{Q}}^{\hspace{0.03cm}e_{1}}_{\hspace{0.03cm}1}(t')
+
\widehat{\mathcal{Q}}^{\hspace{0.03cm}a_{2}}_{\hspace{0.03cm}2}(t')
\hspace{0.02cm}
{\mathcal Q}^{\hspace{0.03cm}e_{1}}_{\hspace{0.03cm}1}(\tau)
\Bigr)\biggr].
\notag
\end{align}
Let us make a few comments. The effective amplitude ${T}^{\hspace{0.03cm}(\rho)\hspace{0.03cm}a\,a_{1}\hspace{0.03cm}a_{2}}_{\; {\bf k}}$, that is included in the initial dynamic equations 
(\ref{eq:5q})\,--\,(\ref{eq:5ee}), is itself a function of time. We attributed this dependence to the ``fast'' time $t'$. The most problematic in our decomposition is the  first term on the right-hand side Eq.\,(\ref{eq:5q}) (and, accordingly, in the conjugate equation (\ref{eq:5w})). The term of such a kind did not occur in the earlier works on the application of the multiple-time-scale formalism to the construction of kinetic equations. We have placed this contribution only to the equation for the rapidly varying component $\widehat{C}^{\hspace{0.02cm}(\alpha)
\hspace{0.02cm}a}_{\hspace{0.02cm}{\bf k}}(t')$. As will be shown later, this leads to correct final expressions.\\
\indent The requirement of equality of the left- and right-hand sides of the system (\ref{eq:5y})\,--\,(\ref{eq:5o}) with respect to the small parameter $\varepsilon$ leads to the following system of algebraic equations for the exponents: 
\[
	\begin{array}{lll}
		(1) \quad\; &\alpha_{2} + \gamma_{1} = 2\hspace{0.02cm}\beta_{1}, \qquad\; &\beta_{2} + \gamma_{1} = \alpha_{1} + 2\hspace{0.02cm}\beta_{1}, \\[2ex]
		(2) \quad\; &\alpha_{1} + \gamma_{2} = \beta_{1} + \beta_{2}, \qquad\; &\beta_{1} + \gamma_{2} = \alpha_{2} + 2\beta_{1},\\[2ex]
		(3) \quad\; &\beta_{1} + \gamma_{2} = \alpha_{1} + \beta_{1} + \beta_{2}. \qquad\; &
	\end{array}
\]
We easily express the solution of the equations in lines (1) and (2) in terms of the exponents $\alpha_{1}$ and $\alpha_{2}$:
\begin{equation}
\beta_{1} = 2\hspace{0.02cm}\alpha_{1},
\quad
\beta_{2} = \alpha_{1} + \alpha_{2},
\quad
\gamma_{1} = 4\hspace{0.02cm}\alpha_{1} - \alpha_{2},
\quad
\gamma_{2} = 2\hspace{0.02cm}\alpha_{1} + \alpha_{2}.
\label{eq:5oo}
\end{equation}
The equation in line (3), which is related to the last contribution to (\ref{eq:5o}), turns into an identity for these solutions. 
If one fix values for the exponents $\alpha_{1}$ and $\gamma_{1}$, putting, for example,
\[
\alpha_{1} = 1,\quad \gamma_{1} = 0,
\]
to be as close as possible to the early work \cite{Crawford:1980}, we further 
get  
\[
\alpha_{2} = 4
\quad
\beta_{1} = 2,
\quad
\beta_{2} = 5,
\quad
\gamma_{2} = 6.
\] 
Hence, the representations (\ref{eq:5r}) and (\ref{eq:5t}) take the final form  	
\begin{align}
&c^{\hspace{0.02cm}(\alpha)\hspace{0.02cm}a}_{\hspace{0.02cm}{\bf k}}(t)
=
\varepsilon\,
C^{\hspace{0.02cm}(\alpha)\hspace{0.02cm}a}_{\hspace{0.02cm}{\bf k}}(\tau)
\,+\,
\varepsilon^{4}\hspace{0.03cm}
\widehat{C}^{\hspace{0.02cm}(\alpha)\hspace{0.02cm}a}_{\hspace{0.02cm}{\bf k}}(t'),
\notag\\[1.5ex]
&\mathcal{Q}^{\,d}_{\hspace{0.03cm}\alpha}(t)
=
\varepsilon^{2}\hspace{0.02cm}\mathcal{Q}^{\,d}_{\hspace{0.03cm}\alpha}(\tau)
\,+\,
\varepsilon^{5}\hspace{0.02cm}\widehat{\mathcal{Q}}^{\,d}_{\hspace{0.03cm}\alpha}(t'), 
\notag\\[1.5ex]
&\partial/\partial\hspace{0.03cm}t 
=
(\partial/\partial\hspace{0.03cm}t') 
\,+\, 
\varepsilon^{6}\hspace{0.02cm}(\partial/\partial\hspace{0.03cm}\tau).
\notag
\end{align}
\indent In equations (\ref{eq:5y})\,--\,(\ref{eq:5o}) now we can set $\varepsilon = 1$. Since the terms involving $C^{\hspace{0.02cm}(\rho)\hspace{0.02cm}a}_{\hspace{0.02cm}{\bf k}}(\tau)$
and ${\mathcal Q}^{\hspace{0.03cm}d}_{\hspace{0.03cm}\alpha}(\tau)$ on the right-hand side of equation (\ref{eq:5i}) do not depend on $t'$, we can integrate with respect to $t'$ to give, to leading order,
\begin{align}
	\widehat{\mathcal{Q}}^{\,a_{1}}_{\hspace{0.03cm}1}(t')
	=
	f^{\hspace{0.03cm}a_{1}\hspace{0.03cm}c^{\hspace{0.02cm}\prime}\hspace{0.01cm}e_{1}}
	\,\sum_{\rho}\,\biggl[\,
	&\int\!d\hspace{0.02cm}{\bf k}\,
	\biggl(\int{T}^{\hspace{0.03cm}(\rho)\hspace{0.03cm}a\,c^{\hspace{0.02cm}\prime}\hspace{0.03cm}a^{\prime}_{2}}_{\; {\bf k}}(t')\hspace{0.03cm}dt'\biggr)\,
	C^{\hspace{0.02cm}(\rho)\hspace{0.02cm}a}_{\hspace{0.02cm}{\bf k}}(\tau)
	\hspace{0.02cm}
	{\mathcal Q}^{\hspace{0.03cm}e_{1}}_{\hspace{0.03cm}1}(\tau)
	\hspace{0.02cm}
	{\mathcal Q}^{\hspace{0.03cm}a^{\prime}_{2}}_{\hspace{0.03cm}2}(\tau)
	\label{eq:5a}\\[1ex]
	+\!
	&\int\!d\hspace{0.02cm}{\bf k}\,
	\biggl(\int{T}^{\hspace{0.03cm}\ast\hspace{0.03cm}(\rho)\hspace{0.03cm}a\,
		c^{\hspace{0.02cm}\prime}\hspace{0.03cm}a^{\prime}_{2}}_{\; {\bf k}}(t')\hspace{0.03cm}dt'\biggr)\,
	C^{\,\ast\hspace{0.03cm}(\rho)\hspace{0.02cm}a}_{\hspace{0.02cm}{\bf k}}(\tau)
	\hspace{0.02cm}
	{\mathcal Q}^{\hspace{0.03cm}e_{1}}_{\hspace{0.03cm}1}(\tau)
	\hspace{0.02cm}
	{\mathcal Q}^{\hspace{0.03cm}a^{\prime}_{2}}_{\hspace{0.03cm}2}(\tau)
	\biggr]
	\notag
\end{align}	
and similarly	
\begin{align}
	\widehat{\mathcal{Q}}^{\,a_{2}}_{\hspace{0.03cm}2}(t')
	=
	f^{\hspace{0.03cm}a_{2}\hspace{0.03cm}c^{\hspace{0.02cm}\prime}\hspace{0.01cm}e_{2}}
	\,\sum_{\rho}\,\biggl[\,
	&\int\!d\hspace{0.02cm}{\bf k}\,
	\biggl(\int{T}^{\hspace{0.03cm}(\rho)\hspace{0.03cm}a\hspace{0.03cm}a^{\prime}_{1}\hspace{0.03cm}c^{\hspace{0.02cm}\prime}}_{\; {\bf k}}(t')\hspace{0.03cm}dt'\biggr)\,
	C^{\hspace{0.02cm}(\rho)\hspace{0.02cm}a}_{\hspace{0.02cm}{\bf k}}(\tau)
	\hspace{0.02cm}
	{\mathcal Q}^{\hspace{0.03cm}e_{2}}_{\hspace{0.03cm}2}(\tau)
	\hspace{0.02cm}
	{\mathcal Q}^{\hspace{0.03cm}a^{\prime}_{1}}_{\hspace{0.03cm}1}(\tau)
	\label{eq:5s}\\[1ex]
	+\!
	&\int\!d\hspace{0.02cm}{\bf k}\,
	\biggl(\int{T}^{\hspace{0.03cm}\ast\hspace{0.03cm}(\rho)\hspace{0.03cm}a\hspace{0.03cm}a^{\prime}_{1}\hspace{0.03cm}c^{\hspace{0.02cm}\prime}}_{\; {\bf k}}(t')\hspace{0.03cm}dt'\biggr)\,
	C^{\,\ast\hspace{0.03cm}(\rho)\hspace{0.02cm}a}_{\hspace{0.02cm}{\bf k}}(\tau)
	\hspace{0.02cm}
	{\mathcal Q}^{\hspace{0.03cm}e_{2}}_{\hspace{0.03cm}2}(\tau)
	\hspace{0.02cm}
	{\mathcal Q}^{\hspace{0.03cm}a^{\prime}_{1}}_{\hspace{0.03cm}1}(\tau)
	\biggr].
	\notag
\end{align}	
The integral over $t'$ is understood as indefinite. We assume the integration constant to be zero without loss of generality. Further, the integration of the equation Eq.\,(\ref{eq:5y}) with respect to $t'$ leads us to the following representation for the rapidly varying components $\widehat{C}^{\hspace{0.02cm}(\alpha)\hspace{0.02cm}a}_{\hspace{0.02cm}{\bf k}}(t')$:
\begin{equation}
\widehat{C}^{\hspace{0.02cm}(\alpha)\hspace{0.02cm}a}_{\hspace{0.02cm}{\bf k}}(t')
=
-\hspace{0.03cm}i\hspace{0.04cm}
\biggl(\int
{T}^{\hspace{0.03cm}\ast\hspace{0.03cm}(\alpha)\hspace{0.03cm}a\,
a_{1}\hspace{0.03cm}a_{2}}_{\; {\bf k}}(t')\hspace{0.03cm}dt'\biggr)
{\mathcal Q}^{\hspace{0.03cm}a_{1}}_{\hspace{0.03cm}1}(\tau)
\hspace{0.02cm}
{\mathcal Q}^{\hspace{0.03cm}a_{2}}_{\hspace{0.03cm}2}(\tau). 
\label{eq:5d} 	
\end{equation}

%
%

\section{\bf Kinetic equation for soft gluon excitations}
\label{section_6}
\setcounter{equation}{0}

Substituting the solutions obtained (\ref{eq:5a}) and (\ref{eq:5s}) in Eq.\,(\ref{eq:5u}), we derive the following equation for the slowly varying field component $C^{\hspace{0.02cm}(\alpha)\hspace{0.02cm}a}_{\hspace{0.02cm}{\bf k}}(\tau)$:
\begin{equation}
\frac{\partial \hspace{0.04cm}C^{\hspace{0.02cm}(\alpha)\hspace{0.02cm}a}_{\hspace{0.02cm}{\bf k}}(\tau)}{\partial\hspace{0.03cm}\tau}
\,=\,
(-\hspace{0.03cm}i)\,
{T}^{\hspace{0.03cm}\ast\hspace{0.03cm}(\alpha)\hspace{0.03cm}a\,a_{1}
	\hspace{0.03cm}a_{2}}_{\;{\bf k}}(t')\,\times
\vspace{-0.5cm}
\label{eq:6q}
\end{equation}
\begin{align}
\biggl\{ {\mathcal Q}^{\hspace{0.03cm}a_{1}}_{\hspace{0.03cm}1}(\tau)
\hspace{0.02cm}	
f^{\hspace{0.03cm}a_{2}\hspace{0.03cm}c^{\hspace{0.02cm}\prime}\hspace{0.01cm}e_{2}}
\,\sum_{\rho}\,\biggl[\,
&\int\!d\hspace{0.02cm}{\bf k}^{\prime}\,
\biggl(\int{T}^{\hspace{0.03cm}(\rho)\hspace{0.03cm}a^{\prime}\hspace{0.03cm}a^{\prime}_{1}\hspace{0.03cm}c^{\hspace{0.02cm}\prime}}_{\; {\bf k}^{\prime}}(t')\hspace{0.03cm}dt'\biggr)\,
C^{\hspace{0.02cm}(\rho)\hspace{0.02cm}a^{\prime}}_{\hspace{0.02cm}{\bf k}^{\prime}}(\tau)
\hspace{0.02cm}
{\mathcal Q}^{\hspace{0.03cm}e_{2}}_{\hspace{0.03cm}2}(\tau)
\hspace{0.02cm}
{\mathcal Q}^{\hspace{0.03cm}a^{\prime}_{1}}_{\hspace{0.03cm}1}(\tau)
\notag\\[1ex]
+\!
&\int\!d\hspace{0.02cm}{\bf k}^{\prime}\,
\biggl(\int{T}^{\hspace{0.03cm}\ast\hspace{0.03cm}(\rho)\hspace{0.03cm}a^{\prime}\hspace{0.03cm}a^{\prime}_{1}\hspace{0.03cm}c^{\hspace{0.02cm}\prime}}_{\; {\bf k}^{\prime}}(t')\hspace{0.03cm}dt'\biggr)\,
C^{\,\ast\hspace{0.03cm}(\rho)\hspace{0.02cm}a^{\prime}}_{\hspace{0.02cm}{\bf k}^{\prime}}(\tau)
\hspace{0.02cm}
{\mathcal Q}^{\hspace{0.03cm}e_{2}}_{\hspace{0.03cm}2}(\tau)
\hspace{0.02cm}
{\mathcal Q}^{\hspace{0.03cm}a^{\prime}_{1}}_{\hspace{0.03cm}1}(\tau)
\biggr]
\notag\\[1ex]
+\,
f^{\hspace{0.03cm}a_{1}\hspace{0.03cm}c^{\hspace{0.02cm}\prime}\hspace{0.01cm}e_{1}}
\,\sum_{\rho}\,\biggl[\,
&\int\!d\hspace{0.02cm}{\bf k}^{\prime}\,
\biggl(\int{T}^{\hspace{0.03cm}(\rho)\hspace{0.03cm}a^{\prime}\,c^{\hspace{0.02cm}\prime}\hspace{0.03cm}a^{\prime}_{2}}_{\; {\bf k}^{\prime}}(t')\hspace{0.03cm}dt'\biggr)\,
C^{\hspace{0.02cm}(\rho)\hspace{0.02cm}a^{\prime}}_{\hspace{0.02cm}{\bf k}^{\prime}}(\tau)
\hspace{0.02cm}
{\mathcal Q}^{\hspace{0.03cm}e_{1}}_{\hspace{0.03cm}1}(\tau)
\hspace{0.02cm}
{\mathcal Q}^{\hspace{0.03cm}a^{\prime}_{2}}_{\hspace{0.03cm}2}(\tau)
\notag\\[1ex]
+\!
&\int\!d\hspace{0.02cm}{\bf k}^{\prime}\,
\biggl(\int{T}^{\hspace{0.03cm}\ast\hspace{0.03cm}(\rho)\hspace{0.03cm}a^{\prime}
\,
c^{\hspace{0.02cm}\prime}\hspace{0.03cm}a^{\prime}_{2}}_{\; {\bf k}^{\prime}}(t')\hspace{0.03cm}dt'\biggr)\,
C^{\,\ast\hspace{0.03cm}(\rho)\hspace{0.02cm}a^{\prime}}_{\hspace{0.02cm}{\bf k}^{\prime}}(\tau)
\hspace{0.02cm}
{\mathcal Q}^{\hspace{0.03cm}e_{1}}_{\hspace{0.03cm}1}(\tau)
\hspace{0.02cm}
{\mathcal Q}^{\hspace{0.03cm}a^{\prime}_{2}}_{\hspace{0.03cm}2}(\tau)
\biggr]{\mathcal Q}^{\hspace{0.03cm}a_{2}}_{\hspace{0.03cm}2}(\tau)\biggr\}.
\notag
\end{align}	
Similar equation  is valid for the slowly varying conjugate component $C^{\,\ast\hspace{0.02cm}(\alpha)\hspace{0.02cm}a}_{\hspace{0.02cm}{\bf k}}(\tau)$ also. If the ensemble of interacting Bose-excitations at a low nonlinearity level has random phases, then it can be statistically described by introducing the bosonic one-time, two-component correlation function of the following form:
\begin{equation}
\bigl\langle\hspace{0.01cm}C^{\,\ast\hspace{0.02cm}(\alpha)\hspace{0.02cm}a}_{\hspace{0.02cm}{\bf k}}(\tau)
\hspace{0.03cm}
C^{\hspace{0.02cm}(\alpha^{\prime})\hspace{0.02cm}a^{\prime}}_{\hspace{0.02cm}{\bf k}'}(\tau)\bigr\rangle
=
{\mathcal N}^{\,(\alpha,\hspace{0.03cm}\alpha^{\prime})\hspace{0.03cm}a\hspace{0.03cm}a^{\prime}_{\phantom{1}}\!}_{\hspace{0.02cm}{\bf k}}\!(\tau)
\hspace{0.03cm}
\delta({\bf k} - {\bf k}^{\hspace{0.02cm}\prime}),
\label{eq:6w}
\end{equation}
where $\langle\cdot\rangle$ denotes ensemble averaging and the matrix function ${\mathcal N}^{\,(\alpha,\alpha^{\prime})\hspace{0.03cm}a\hspace{0.03cm}a^{\prime}_{\phantom{1}}\!}_{\hspace{0.02cm}{\bf k}}\!(\tau)$ we interpret as the plasmon number density.\\
\indent Let us define a kinetic equation for the  plasmon number density  employing the equation (\ref{eq:6q}) and the definition (\ref{eq:6w}).
The time evolution of the ${\mathcal N}^{\,(\alpha,\hspace{0.03cm}\alpha^{\prime})\hspace{0.03cm}a\hspace{0.03cm}a^{\prime}_{\phantom{1}}\!}_{\hspace{0.02cm}{\bf k}}\!(\tau)$ is obtained by multiplying Eq.\,(\ref{eq:6q}) for $C^{\hspace{0.02cm}(\alpha^{\prime})\hspace{0.02cm}a^{\prime}}_{\hspace{0.02cm}{\bf k}}(\tau)$ by $C^{\,\ast\hspace{0.02cm}(\alpha)\hspace{0.02cm}a}_{\hspace{0.02cm}{\bf k}}(\tau)$, ensemble averaging and adding the Hermitian transpose.  The right-hand side of the resulting expression will contain eighth-order correlation functions of the following form:
\begin{equation}
\begin{split}
&\bigl\langle\hspace{0.01cm}C^{\,\ast\hspace{0.02cm}(\alpha)\hspace{0.02cm}a}_{\hspace{0.02cm}{\bf k}}(\tau)
\hspace{0.03cm}
C^{\hspace{0.02cm}(\alpha^{\prime})\hspace{0.02cm}a^{\prime}}_{\hspace{0.02cm}{\bf k}'}(\tau)
\hspace{0.03cm}
\mathcal{Q}^{\hspace{0.03cm}a^{\prime}_{2}}_{2}(\tau)
\hspace{0.03cm}
\mathcal{Q}^{\hspace{0.03cm}e_{1}}_{1}(\tau)
\hspace{0.03cm}
\mathcal{Q}^{\hspace{0.03cm}a_{2}}_{2}(\tau)\bigr\rangle,\\[1.5ex]
&\bigl\langle\hspace{0.01cm}C^{\,\ast\hspace{0.02cm}(\alpha)\hspace{0.02cm}a}_{\hspace{0.02cm}{\bf k}}(\tau)
\hspace{0.03cm}
C^{\hspace{0.02cm}(\alpha^{\prime})\hspace{0.02cm}a^{\prime}}_{\hspace{0.02cm}{\bf k}'}(\tau)
\hspace{0.03cm}
\mathcal{Q}^{\hspace{0.03cm}a^{\prime}_{1}}_{1}(\tau)
\hspace{0.03cm}
\mathcal{Q}^{\hspace{0.03cm}e_{2}}_{2}(\tau)
\hspace{0.03cm}
\mathcal{Q}^{\hspace{0.03cm}a_{1}}_{1}(\tau)\bigr\rangle
\end{split}
\label{eq:6ww}
\end{equation}
etc. Further, by differentiating the eighth-order correlation functions with respect to $\tau$ with allowance made for (\ref{eq:6q}), (\ref{eq:5o}), (\ref{eq:5a})\,--\,(\ref{eq:5d}) we derive the equation the right-hand side of which will contain the even higher order correlation functions in the variables $C^{\,\ast\hspace{0.02cm}(\alpha)\hspace{0.02cm}a}_{\hspace{0.02cm}{\bf k}}(\tau),\,C^{\hspace{0.02cm}(\alpha^{\prime})\hspace{0.02cm}a^{\prime}}_{\hspace{0.02cm}{\bf k}'}(\tau)$ and $\mathcal{Q}^{\,d}_{\hspace{0.03cm}\alpha}$.
However, we simplify the task as much as possible. We close the chain of equations at the first stage by expressing the eighth-order correlation functions (\ref{eq:6ww}) in terms of the pair correlation functions for the slowly varying components $C^{\,\ast\hspace{0.02cm}(\alpha)\hspace{0.02cm}a}_{\hspace{0.02cm}{\bf k}}(\tau)$ and  $C^{\hspace{0.02cm}(\alpha^{\prime})\hspace{0.02cm}a^{\prime}}_{\hspace{0.02cm}{\bf k}'}(\tau)$, and for the mean value of the slowly varying color charges $\mathcal{Q}^{\,d}_{\hspace{0.03cm}\alpha}(\tau)$, i.e. we put
\begin{align}
&\bigl\langle\hspace{0.01cm}C^{\,\ast\hspace{0.02cm}(\alpha)\hspace{0.02cm}a}_{\hspace{0.02cm}{\bf k}}(\tau)
\hspace{0.03cm}
C^{\hspace{0.02cm}(\alpha^{\prime})\hspace{0.02cm}a^{\prime}}_{\hspace{0.02cm}{\bf k}'}(\tau)
\hspace{0.03cm}
\mathcal{Q}^{\hspace{0.03cm}a^{\prime}_{2}}_{2}(\tau)
\hspace{0.03cm}
\mathcal{Q}^{\hspace{0.03cm}e_{1}}_{1}(\tau)
\hspace{0.03cm}
\mathcal{Q}^{\hspace{0.03cm}a_{2}}_{2}(\tau)\bigr\rangle
\simeq\notag\\[1ex]
&\bigl\langle\hspace{0.01cm}C^{\,\ast\hspace{0.02cm}(\alpha)\hspace{0.02cm}a}_{\hspace{0.02cm}{\bf k}}(\tau)
\hspace{0.03cm}
C^{\hspace{0.02cm}(\alpha^{\prime})\hspace{0.02cm}a^{\prime}}_{\hspace{0.02cm}{\bf k}'}(\tau)\bigr\rangle
\hspace{0.03cm}
\bigl\langle\hspace{0.03cm}\mathcal{Q}^{\hspace{0.03cm}a^{\prime}_{2}}_{2}(\tau)
\hspace{0.03cm}\mathcal{Q}^{\hspace{0.03cm}e_{1}}_{1}(\tau)
\hspace{0.03cm}\mathcal{Q}^{\hspace{0.03cm}a_{2}}_{2}(\tau)
\hspace{0.03cm}\bigr\rangle
\simeq
\label{eq:6www}\\[1ex]
&\delta({\bf k} - {\bf k}^{\hspace{0.02cm}\prime})
\hspace{0.03cm}
{\mathcal N}^{\,(\alpha,\hspace{0.03cm}\alpha^{\prime})\hspace{0.03cm}a\hspace{0.03cm}a^{\prime}_{\phantom{1}}\!}_{\hspace{0.02cm}{\bf k}}\!(\tau)
\hspace{0.03cm}
\bigl\langle\hspace{0.03cm}\mathcal{Q}^{\hspace{0.03cm}a^{\prime}_{2}}_{2}(\tau)
\hspace{0.03cm}\bigr\rangle
\hspace{0.03cm}\bigl\langle\hspace{0.03cm}\mathcal{Q}^{\hspace{0.03cm}e_{1}}_{1}(\tau)
\hspace{0.03cm}\bigr\rangle
\hspace{0.01cm}\bigl\langle\hspace{0.03cm}\mathcal{Q}^{\hspace{0.03cm}a_{2}}_{2}(\tau)
\hspace{0.03cm}\bigr\rangle
\notag
\end{align}
and so on.\\
\indent The kinetic equation can be presented in a more visual form if we set for the effective amplitude ${T}^{\hspace{0.03cm}(\alpha)\hspace{0.03cm}a\,a_{1}\hspace{0.03cm}a_{2}}_{\; {\bf k}}(t')$ the following color and momentum decomposition
\begin{equation}
{T}^{\hspace{0.03cm}(\alpha)\hspace{0.03cm}a\,a_{1}\hspace{0.03cm}a_{2}}_{\; {\bf k}}(t') 
=
f^{\hspace{0.03cm}a\,a_{1}\hspace{0.03cm}a_{2}\hspace{0.03cm}} {T}^{\hspace{0.03cm}(\alpha)}_{\; {\bf k}}(t').
\label{eq:6wwww}
\end{equation}
After simple algebraic transformations we get the desired matrix-valued kinetic equation
\begin{equation}
\delta({\bf k} - {\bf k}\!\ ')\,
\frac{\partial\hspace{0.02cm}{\mathcal N}^{\,(\alpha,\hspace{0.03cm}\alpha^{\prime})\hspace{0.03cm}a\hspace{0.03cm}a^{\prime}_{\phantom{1}}\!}_{\hspace{0.02cm}{\bf k}}\!(\tau)}{\partial\hspace{0.03cm}\tau}
=
\label{eq:6e}
\end{equation}
\begin{align}
-\,&{T}^{\hspace{0.03cm}(\alpha)}_{\; {\bf k}}(t')
\,\sum_{\rho}\,
\biggl(\int{T}^{\,\ast\hspace{0.03cm}(\rho)}_{\; {\bf k}'}(t')\hspace{0.03cm}dt'\biggr)
\bigl(\hspace{0.03cm}T^{\,a_{2}}\hspace{0.03cm}T^{\,e_{1}}\hspace{0.03cm}T^{\,a^{\prime}_{2}}{\mathcal N}^{\,(\rho,\hspace{0.03cm}\alpha^{\prime})}_{\hspace{0.02cm}{\bf k}'}\hspace{0.01cm}
\bigr)^{\hspace{0.01cm}a\hspace{0.03cm}a^{\prime}_{\phantom{1}}\!}
\hspace{0.01cm}
\bigl\langle\hspace{0.03cm}\mathcal{Q}^{\hspace{0.03cm}a^{\prime}_{2}}_{2}(\tau)
\hspace{0.03cm}\bigr\rangle
\hspace{0.03cm}\bigl\langle\hspace{0.03cm}\mathcal{Q}^{\hspace{0.03cm}e_{1}}_{1}(\tau)
\hspace{0.03cm}\bigr\rangle
\hspace{0.01cm}\bigl\langle\hspace{0.03cm}\mathcal{Q}^{\hspace{0.03cm}a_{2}}_{2}(\tau)
\hspace{0.03cm}\bigr\rangle
\notag\\[1ex]
-\,&{T}^{\hspace{0.03cm}(\alpha)}_{\; {\bf k}}(t')
\,\sum_{\rho}\,
\biggl(\int{T}^{\,\ast\hspace{0.03cm}(\rho)}_{\; {\bf k}'}(t')\hspace{0.03cm}dt'\biggr)
\bigl(\hspace{0.03cm}T^{\,a_{1}}\hspace{0.03cm}T^{\,e_{2}}\hspace{0.03cm}T^{\,a^{\prime}_{1}}{\mathcal N}^{\,(\rho,\hspace{0.03cm}\alpha^{\prime})}_{\hspace{0.02cm}{\bf k}'}\hspace{0.01cm}
\bigr)^{\hspace{0.01cm}a\hspace{0.03cm}a^{\prime}_{\phantom{1}}\!}
\hspace{0.01cm}
\bigl\langle\hspace{0.03cm}\mathcal{Q}^{\hspace{0.03cm}a^{\prime}_{1}}_{1}(\tau)
\hspace{0.03cm}\bigr\rangle
\hspace{0.03cm}\bigl\langle\hspace{0.03cm}\mathcal{Q}^{\hspace{0.03cm}e_{2}}_{2}(\tau)
\hspace{0.03cm}\bigr\rangle
\hspace{0.01cm}\bigl\langle\hspace{0.03cm}\mathcal{Q}^{\hspace{0.03cm}a_{1}}_{1}(\tau)
\hspace{0.03cm}\bigr\rangle
\notag\\[1ex]
-\,&{T}^{\,\ast\hspace{0.03cm}(\alpha^{\prime})}_{\; {\bf k}'}(t')
\,\sum_{\rho}\,
\biggl(\int{T}^{\hspace{0.03cm}(\rho)}_{\; {\bf k}}(t')\hspace{0.03cm}dt'\biggr)
\bigl(\hspace{0.03cm}{\mathcal N}^{\,(\alpha,\hspace{0.03cm}\rho)}_{\hspace{0.02cm}{\bf k}}
\hspace{0.03cm}
T^{\,a^{\prime}_{2}}\hspace{0.03cm}T^{\,e_{1}}\hspace{0.03cm}
T^{\,a_{2}}\hspace{0.01cm}
\bigr)^{\hspace{0.01cm}a\hspace{0.03cm}a^{\prime}_{\phantom{1}}\!}
\hspace{0.01cm}
\bigl\langle\hspace{0.03cm}\mathcal{Q}^{\hspace{0.03cm}a^{\prime}_{2}}_{2}(\tau)
\hspace{0.03cm}\bigr\rangle
\hspace{0.03cm}\bigl\langle\hspace{0.03cm}\mathcal{Q}^{\hspace{0.03cm}e_{1}}_{1}(\tau)
\hspace{0.03cm}\bigr\rangle
\hspace{0.01cm}\bigl\langle\hspace{0.03cm}\mathcal{Q}^{\hspace{0.03cm}a_{2}}_{2}(\tau)
\hspace{0.03cm}\bigr\rangle
\notag\\[1ex]
-\,&{T}^{\,\ast\hspace{0.03cm}(\alpha^{\prime})}_{\; {\bf k}'}(t')
\,\sum_{\rho}\,
\biggl(\int{T}^{\hspace{0.03cm}(\rho)}_{\; {\bf k}}(t')\hspace{0.03cm}dt'\biggr)
\bigl(\hspace{0.03cm}{\mathcal N}^{\,(\alpha,\hspace{0.03cm}\rho)}_{\hspace{0.02cm}{\bf k}}
\hspace{0.03cm}
T^{\,a^{\prime}_{1}}\hspace{0.03cm}T^{\,e_{2}}\hspace{0.03cm}
T^{\,a_{1}}\hspace{0.01cm}
\bigr)^{\hspace{0.01cm}a\hspace{0.03cm}a^{\prime}_{\phantom{1}}\!}
\hspace{0.01cm}
\bigl\langle\hspace{0.03cm}\mathcal{Q}^{\hspace{0.03cm}a^{\prime}_{1}}_{1}(\tau)
\hspace{0.03cm}\bigr\rangle
\hspace{0.03cm}\bigl\langle\hspace{0.03cm}\mathcal{Q}^{\hspace{0.03cm}e_{2}}_{2}(\tau)
\hspace{0.03cm}\bigr\rangle
\hspace{0.01cm}\bigl\langle\hspace{0.03cm}\mathcal{Q}^{\hspace{0.03cm}a_{1}}_{1}(\tau)
\hspace{0.03cm}\bigr\rangle,
\notag
\end{align}
where $(\hspace{0.01cm}T^{\,a})^{\hspace{0.01cm}b\hspace{0.03cm}c} \equiv -i\hspace{0.03cm}f^{\hspace{0.03cm}a\hspace{0.02cm}b\hspace{0.03cm}c}$ are generators in the adjoint representation and we have introduced matrix notation with respect to the color indices for the plasmon number density ${\mathcal N}^{\,(\alpha,\hspace{0.03cm}\alpha^{\prime})}_{\hspace{0.02cm}{\bf k}} = \bigl({\mathcal N}^{\,(\alpha,\hspace{0.03cm}\alpha^{\prime})\hspace{0.03cm}a\hspace{0.03cm}a^{\prime}_{\phantom{1}}\!}_{\hspace{0.02cm}{\bf k}}\bigr)$. It is recalled that the plasmon number density ${\mathcal N}^{\,(\alpha,\hspace{0.03cm}\alpha^{\prime})\hspace{0.03cm}a\hspace{0.03cm}a^{\prime}_{\phantom{1}}\!}_{\hspace{0.02cm}{\bf k}}\!(\tau)$ is a matrix function of both the ``indices'' $\alpha$ and $\alpha^{\prime}$, labeling hard particles under consideration, and the color indices $a$ and $a^{\prime}$. Let us consider first the following decomposition of the plasmon number density with respect to $\alpha$ and $\alpha^{\prime}$:
\begin{equation}
{\mathcal N}^{\,(\alpha,\hspace{0.03cm}\alpha^{\prime})\hspace{0.03cm}a\hspace{0.03cm}a^{\prime}_{\phantom{1}}\!}_{\hspace{0.02cm}{\bf k}} 
= 
\frac{1}{2}\,\delta^{\,\alpha\hspace{0.02cm}\alpha^{\prime}}\!\hspace{0.01cm} 
{\mathcal N}^{\,a\hspace{0.02cm}a^{\prime}}_{\bf k} 
+\, 
\frac{1}{2}\,\frac{(\Delta {\mathbf v}\cdot
\boldsymbol{\sigma})^{\alpha\hspace{0.03cm}\alpha^{\prime}}}{|\Delta {\mathbf v}|}\,{\mathcal W}^{\,a\hspace{0.02cm}a^{\prime}}_{\bf k},
\label{eq:6r}
\end{equation}
where $\Delta{\mathbf v} = {\mathbf v}_{1} - {\mathbf v}_{2}$; $\boldsymbol{\sigma} = (\sigma_{1},\sigma_{2},\sigma_{3})$ are the Pauli matrices. However, even such a representation in the case of an arbitrary direction of the velocity difference $\Delta{\mathbf v}$ of two colliding hard particles leads to highly complicated expressions. Therefore, in order to simplify task as much as possible, we consider that $\Delta{\mathbf v}$ is directed along the axis $OZ$ and then instead of (\ref{eq:6r}) we have
\begin{equation}
{\mathcal N}^{\,(\alpha,\hspace{0.03cm}\alpha^{\prime})\hspace{0.03cm}a\hspace{0.03cm}a^{\prime}_{\phantom{1}}\!}_{\hspace{0.02cm}{\bf k}} 
= 
\frac{1}{2}\,\delta^{\,\alpha\hspace{0.02cm}\alpha^{\prime}}\!\hspace{0.01cm} 
{\mathcal N}^{\,a\hspace{0.02cm}a^{\prime}}_{\bf k} 
+\, 
\frac{1}{2}\,({\sigma}_{3})^{\alpha\hspace{0.03cm}\alpha^{\prime}}\hspace{0.03cm} 
{\mathcal W}^{\,a\hspace{0.02cm}a^{\prime}}_{\bf k}.
\label{eq:6t}
\end{equation}
\indent The first step is to define an equation for the ``trivial'' part of the plasmon number density, i.e. for ${\mathcal N}^{\,a\hspace{0.02cm}a^{\prime}}_{\bf k}(\tau)$ in (\ref{eq:6e}). For this purpose, we take the trace of the left- and right-hand sides of equation (\ref{eq:6e}) with respect to the indices $\alpha$ and $\alpha^{\prime}$, i.e. we put $\alpha = \alpha^{\prime}$ and perform the summation over $\alpha$. Setting ${\bf k} = {\bf k}^{\hspace{0.02cm}\prime}$, we arrive at the following kinetic equation for the plasmon number density ${\mathcal N}^{\,a\hspace{0.02cm}a^{\prime}}_{\bf k}(\tau)$, instead of (\ref{eq:6e}):
\begin{align}
	\frac{\partial\hspace{0.02cm}{\mathcal N}^{\,a\hspace{0.03cm}a^{\prime}_{\phantom{1}}\!}_{\hspace{0.02cm}{\bf k}}\!(\tau)}{\partial\hspace{0.03cm}\tau}
	=
	-\,\frac{1}{2}\,\sum_{\rho}\,&{T}^{\hspace{0.03cm}(\rho)}_{\; {\bf k}}(t')
	\,\biggl(\int{T}^{\,\ast\hspace{0.03cm}(\rho)}_{\; {\bf k}}(t')\hspace{0.03cm}dt'\biggr)
	\Bigl\{
	\bigl(\hspace{0.03cm}T^{\,a_{2}}\hspace{0.03cm}T^{\,e_{1}}\hspace{0.03cm}T^{\,a^{\prime}_{2}}{\mathcal N}_{\hspace{0.02cm}{\bf k}}\hspace{0.01cm}
	\bigr)^{\hspace{0.01cm}a\hspace{0.03cm}a^{\prime}_{\phantom{1}}\!}\!
	\hspace{0.01cm}\bigl\langle\hspace{0.03cm}\mathcal{Q}^{\hspace{0.03cm}a^{\prime}_{2}}_{2}
	\hspace{0.03cm}\bigr\rangle
	\hspace{0.03cm}\bigl\langle\hspace{0.03cm}\mathcal{Q}^{\hspace{0.03cm}e_{1}}_{1}
	\hspace{0.03cm}\bigr\rangle
	\hspace{0.01cm}\bigl\langle\hspace{0.03cm}\mathcal{Q}^{\hspace{0.03cm}a_{2}}_{2}
	\hspace{0.03cm}\bigr\rangle\,+
	\notag\\[1ex]
&\hspace{4cm}
\bigl(\hspace{0.03cm}T^{\,a_{1}}\hspace{0.03cm}T^{\,e_{2}}\hspace{0.03cm}T^{\,a^{\prime}_{1}}{\mathcal N}_{\hspace{0.02cm}{\bf k}}\hspace{0.01cm}
\bigr)^{\hspace{0.01cm}a\hspace{0.03cm}a^{\prime}_{\phantom{1}}\!}\!
\hspace{0.01cm}\bigl\langle\hspace{0.03cm}\mathcal{Q}^{\hspace{0.03cm}a^{\prime}_{1}}_{1}
\hspace{0.03cm}\bigr\rangle
\hspace{0.03cm}\bigl\langle\hspace{0.03cm}\mathcal{Q}^{\hspace{0.03cm}e_{2}}_{2}
\hspace{0.03cm}\bigr\rangle
\hspace{0.01cm}\bigl\langle\hspace{0.03cm}\mathcal{Q}^{\hspace{0.03cm}a_{1}}_{1}
\hspace{0.03cm}\bigr\rangle	
\Bigr\}\,-
\notag\\[1ex]
\frac{1}{2}\,\sum_{\rho}\,&{T}^{\,\ast\hspace{0.03cm}({\rho})}_{\; {\bf k}}(t')
\,
\biggl(\int{T}^{\hspace{0.03cm}(\rho)}_{\; {\bf k}}(t')\hspace{0.03cm}dt'\biggr)
\Bigl\{
\bigl(\hspace{0.03cm}{\mathcal N}_{\hspace{0.02cm}{\bf k}}
\hspace{0.03cm}
T^{\,a^{\prime}_{2}}\hspace{0.03cm}T^{\,e_{1}}\hspace{0.03cm}
T^{\,a_{2}}\hspace{0.01cm}
\bigr)^{\hspace{0.01cm}a\hspace{0.03cm}a^{\prime}_{\phantom{1}}\!}
\hspace{0.01cm}
\bigl\langle\hspace{0.03cm}\mathcal{Q}^{\hspace{0.03cm}a^{\prime}_{2}}_{2}
\hspace{0.03cm}\bigr\rangle
\hspace{0.03cm}\bigl\langle\hspace{0.03cm}\mathcal{Q}^{\hspace{0.03cm}e_{1}}_{1}
\hspace{0.03cm}\bigr\rangle
\hspace{0.01cm}\bigl\langle\hspace{0.03cm}\mathcal{Q}^{\hspace{0.03cm}a_{2}}_{2}
\hspace{0.03cm}\bigr\rangle\,+
\notag\\[1ex]
&\hspace{4cm}
\bigl(\hspace{0.03cm}{\mathcal N}_{\hspace{0.02cm}{\bf k}}
\hspace{0.03cm}
T^{\,a^{\prime}_{1}}\hspace{0.03cm}T^{\,e_{2}}\hspace{0.03cm}
T^{\,a_{1}}\hspace{0.01cm}
\bigr)^{\hspace{0.01cm}a\hspace{0.03cm}a^{\prime}_{\phantom{1}}\!}
\hspace{0.01cm}
\bigl\langle\hspace{0.03cm}\mathcal{Q}^{\hspace{0.03cm}a^{\prime}_{1}}_{1}
\hspace{0.03cm}\bigr\rangle
\hspace{0.03cm}\bigl\langle\hspace{0.03cm}\mathcal{Q}^{\hspace{0.03cm}e_{2}}_{2}
\hspace{0.03cm}\bigr\rangle
\hspace{0.01cm}\bigl\langle\hspace{0.03cm}\mathcal{Q}^{\hspace{0.03cm}a_{1}}_{1}
\hspace{0.03cm}\bigr\rangle
\Bigr\}\,-
\label{eq:6y}
\end{align}
\begin{align}
&\frac{1}{2}\,\sum_{\rho}\hspace{0.02cm}(-1)^{\rho + 1}\,
{T}^{\hspace{0.03cm}(\rho)}_{\; {\bf k}}(t')
\,\biggl(\int{T}^{\,\ast\hspace{0.03cm}(\rho)}_{\; {\bf k}}(t')\hspace{0.03cm}dt'\biggr)
\Bigl\{
\bigl(\hspace{0.03cm}T^{\,a_{2}}\hspace{0.03cm}T^{\,e_{1}}\hspace{0.03cm}T^{\,a^{\prime}_{2}}{\mathcal W}_{\hspace{0.02cm}{\bf k}}\hspace{0.01cm}
\bigr)^{\hspace{0.01cm}a\hspace{0.03cm}a^{\prime}_{\phantom{1}}\!}\!
\hspace{0.01cm}\bigl\langle\hspace{0.03cm}\mathcal{Q}^{\hspace{0.03cm}a^{\prime}_{2}}_{2}
\hspace{0.03cm}\bigr\rangle
\hspace{0.03cm}\bigl\langle\hspace{0.03cm}\mathcal{Q}^{\hspace{0.03cm}e_{1}}_{1}
\hspace{0.03cm}\bigr\rangle
\hspace{0.01cm}\bigl\langle\hspace{0.03cm}\mathcal{Q}^{\hspace{0.03cm}a_{2}}_{2}
\hspace{0.03cm}\bigr\rangle\,
+
\notag\\[1ex]
&{\hspace{8.5cm}}\bigl(\hspace{0.03cm}T^{\,a_{1}}\hspace{0.03cm}T^{\,e_{2}}\hspace{0.03cm}T^{\,a^{\prime}_{1}}{\mathcal W}_{\hspace{0.02cm}{\bf k}}\hspace{0.01cm}
\bigr)^{\hspace{0.01cm}a\hspace{0.03cm}a^{\prime}_{\phantom{1}}\!}\!
\hspace{0.01cm}\bigl\langle\hspace{0.03cm}\mathcal{Q}^{\hspace{0.03cm}a^{\prime}_{1}}_{1}
\hspace{0.03cm}\bigr\rangle
\hspace{0.03cm}\bigl\langle\hspace{0.03cm}\mathcal{Q}^{\hspace{0.03cm}e_{2}}_{2}
\hspace{0.03cm}\bigr\rangle
\hspace{0.01cm}\bigl\langle\hspace{0.03cm}\mathcal{Q}^{\hspace{0.03cm}a_{1}}_{1}
\hspace{0.03cm}\bigr\rangle	
\Bigr\}\,-
\notag\\[1ex]
&\frac{1}{2}\,\sum_{\rho}\hspace{0.02cm}(-1)^{\rho + 1}\,
{T}^{\,\ast\hspace{0.03cm}({\rho})}_{\; {\bf k}}(t')
\,
\biggl(\int{T}^{\hspace{0.03cm}(\rho)}_{\; {\bf k}}(t')\hspace{0.03cm}dt'\biggr)
\Bigl\{
\bigl(\hspace{0.03cm}{\mathcal W}_{\hspace{0.02cm}{\bf k}}
\hspace{0.03cm}
T^{\,a^{\prime}_{2}}\hspace{0.03cm}T^{\,e_{1}}\hspace{0.03cm}
T^{\,a_{2}}\hspace{0.01cm}
\bigr)^{\hspace{0.01cm}a\hspace{0.03cm}a^{\prime}_{\phantom{1}}\!}
\hspace{0.01cm}
\bigl\langle\hspace{0.03cm}\mathcal{Q}^{\hspace{0.03cm}a^{\prime}_{2}}_{2}
\hspace{0.03cm}\bigr\rangle
\hspace{0.03cm}\bigl\langle\hspace{0.03cm}\mathcal{Q}^{\hspace{0.03cm}e_{1}}_{1}
\hspace{0.03cm}\bigr\rangle
\hspace{0.01cm}\bigl\langle\hspace{0.03cm}\mathcal{Q}^{\hspace{0.03cm}a_{2}}_{2}
\hspace{0.03cm}\bigr\rangle\,
+
\notag\\[1ex]
&{\hspace{8.5cm}}\bigl(\hspace{0.03cm}{\mathcal W}_{\hspace{0.02cm}{\bf k}}
\hspace{0.03cm}
T^{\,a^{\prime}_{1}}\hspace{0.03cm}T^{\,e_{2}}\hspace{0.03cm}
T^{\,a_{1}}\hspace{0.01cm}
\bigr)^{\hspace{0.01cm}a\hspace{0.03cm}a^{\prime}_{\phantom{1}}\!}
\hspace{0.01cm}
\bigl\langle\hspace{0.03cm}\mathcal{Q}^{\hspace{0.03cm}a^{\prime}_{1}}_{1}
\hspace{0.03cm}\bigr\rangle
\hspace{0.03cm}\bigl\langle\hspace{0.03cm}\mathcal{Q}^{\hspace{0.03cm}e_{2}}_{2}
\hspace{0.03cm}\bigr\rangle
\hspace{0.01cm}\bigl\langle\hspace{0.03cm}\mathcal{Q}^{\hspace{0.03cm}a_{1}}_{1}
\hspace{0.03cm}\bigr\rangle
\Bigr\}.
\notag
\end{align}
Here, on the right-hand side we have introduced matrix notations  ${\mathcal N}^{\phantom{a_{1}}\!\!\! }_{\hspace{0.02cm}{\bf k}} = \bigl({\mathcal N}^{\;a\hspace{0.03cm}a_{1}}_{{\bf k}^{\phantom{\prime}}}\bigr)$ and
${\mathcal W}^{\phantom{a_{1}}\!\!\! }_{\hspace{0.02cm}{\bf k}} = \bigl({\mathcal W}^{\;a\hspace{0.03cm}a_{1}}_{{\bf k}^{\phantom{\prime}}}\bigr)$ 
and we suppressed the slow time $\tau$-dependence of these functions, as well as the average values of the slowly varying components of the color charges $\bigl\langle\hspace{0.03cm}\mathcal{Q}^{\hspace{0.03cm}d}_{\hspace{0.03cm}\alpha}(\tau)\hspace{0.03cm}\bigr\rangle,\,\alpha = 1,2$.\\
\indent Next, to obtain the kinetic equation for the second part of the plasmon number density, i.e. for ${\mathcal W}^{\,a\hspace{0.02cm}a^{\prime}}_{\bf k}(\tau)$ we need now to contract the left- and right-hand sides of the equation (\ref{eq:6e}) with the Pauli matrix $({\sigma}_{3})^{\alpha^{\prime}\hspace{0.01cm}\alpha}$ taking into account that ${\rm tr}(\sigma_{3})^{2} = 2$. An explicit form of the required equation for the color matrix function ${\mathcal W}^{\,a\hspace{0.02cm}a^{\prime}}_{\bf k}(\tau)$ is given in Appendix \ref{appendix_B}, Eq.\,(\ref{ap:B1}).

%
%

\section{First moment about color of the kinetic equations (\ref{eq:6y}) and (\ref{ap:B1})}
\label{section_7}
\setcounter{equation}{0}

Let us consider the following color decomposition of the matrix functions ${\mathcal N}^{\;a\hspace{0.03cm}a^{\prime}_{\phantom{1}}\!}_{\hspace{0.02cm}
{\bf k}}$ and ${\mathcal W}^{\;a\hspace{0.03cm}a^{\prime}_{\phantom{1}}\!}_{\hspace{0.02cm}
{\bf k}}$:
\begin{equation}
\begin{split}
&{\mathcal N}^{\;a\hspace{0.03cm}a^{\prime}_{\phantom{1}}\!}_{\hspace{0.02cm}{\bf k}} 
= 
\delta^{\,a\hspace{0.02cm}a^{\prime}}\hspace{0.01cm} 
N^{\hspace{0.03cm}(1)}_{\bf k} 
\hspace{0.03cm}+\, 
\bigl(T^{\,c_{1}}\bigr)^{a\hspace{0.02cm}a^{\prime}}\!
\bigl\langle\hspace{0.03cm}\mathcal{Q}^{\hspace{0.03cm}c_{1}}_{1}
\hspace{0.03cm}\bigr\rangle
\hspace{0.03cm} N^{\hspace{0.03cm}(2)}_{\bf k},\\[1.5ex]
&{\mathcal W}^{\;a\hspace{0.03cm}a^{\prime}_{\phantom{1}}\!}_{\hspace{0.02cm}{\bf k}} 
= 
\delta^{\,a\hspace{0.02cm}a^{\prime}}\hspace{0.01cm} 
W^{\hspace{0.03cm}(1)}_{\bf k} 
\!+\, 
\bigl(T^{\,c_{2}}\bigr)^{a\hspace{0.02cm}a^{\prime}}\!
\bigl\langle\hspace{0.03cm}\mathcal{Q}^{\hspace{0.03cm}c_{2}}_{2}
\hspace{0.03cm}\bigr\rangle
\,W^{\hspace{0.03cm}(2)}_{\bf k},
\end{split}
\label{eq:7q}
\end{equation}
where $N^{\hspace{0.03cm}(i)}_{\bf k}$ and $W^{\hspace{0.03cm}(i)}_{\bf k},\,i = 1,2$, are ordinary numeric functions of the variables $\tau$ and ${\bf k}$. In this section we define equations for the colorless parts of the plasmon number density, i.e. for $N^{\hspace{0.03cm}(1)}_{\bf k}$ and $W^{\hspace{0.03cm}(1)}_{\bf k}$. For concreteness, we first define the equation for the function $N^{\hspace{0.03cm}(1)}_{\bf k}$. For this purpose, we take the trace of the left- and right-hand sides of equation (\ref{eq:6y}) with respect to the color indices, i.e. we set $a = a^{\prime}$ and perform the summation over $a$. Using the color expansion (\ref{eq:7q}) and the obvious formula 
\[
{\rm tr}\,{\mathcal N}_{\bf k} = (N^{\hspace{0.02cm}2}_{c} - 1)N^{\hspace{0.03cm}(1)}_{\bf k}, 
\]
we easily find from (\ref{eq:6y})
\begin{align}
d_{A}\hspace{0.04cm}\frac{\partial\hspace{0.01cm} N^{\hspace{0.03cm}(1)}_{\bf k}}{\!\!\partial\hspace{0.03cm}\tau}
=
-\,\frac{1}{2}\,\sum_{\rho}\,&{T}^{\hspace{0.03cm}(\rho)}_{\; {\bf k}}(t)
\,\biggl(\int{T}^{\,\ast\hspace{0.03cm}(\rho)}_{\; {\bf k}}(t)\hspace{0.03cm}dt\biggr)
\Bigl\{{\rm tr}\hspace{0.03cm}
\bigl(\hspace{0.03cm}T^{\,a_{2}}\hspace{0.03cm}T^{\,e_{1}}\hspace{0.03cm}T^{\,a^{\prime}_{2}}{\mathcal N}_{\hspace{0.02cm}{\bf k}}\hspace{0.01cm}
\bigr)
\hspace{0.01cm}\bigl\langle\hspace{0.03cm}\mathcal{Q}^{\hspace{0.03cm}a^{\prime}_{2}}_{2}
\hspace{0.03cm}\bigr\rangle
\hspace{0.03cm}\bigl\langle\hspace{0.03cm}\mathcal{Q}^{\hspace{0.03cm}e_{1}}_{1}
\hspace{0.03cm}\bigr\rangle
\hspace{0.01cm}\bigl\langle\hspace{0.03cm}\mathcal{Q}^{\hspace{0.03cm}a_{2}}_{2}
\hspace{0.03cm}\bigr\rangle\,+
\notag\\[1ex]
	&\hspace{4.5cm}{\rm tr}\hspace{0.03cm}
	\bigl(\hspace{0.03cm}T^{\,a_{1}}\hspace{0.03cm}T^{\,e_{2}}\hspace{0.03cm}T^{\,a^{\prime}_{1}}{\mathcal N}_{\hspace{0.02cm}{\bf k}}\hspace{0.01cm}
	\bigr)
	\hspace{0.01cm}\bigl\langle\hspace{0.03cm}\mathcal{Q}^{\hspace{0.03cm}a^{\prime}_{1}}_{1}
	\hspace{0.03cm}\bigr\rangle
	\hspace{0.03cm}\bigl\langle\hspace{0.03cm}\mathcal{Q}^{\hspace{0.03cm}e_{2}}_{2}
	\hspace{0.03cm}\bigr\rangle
	\hspace{0.01cm}\bigl\langle\hspace{0.03cm}\mathcal{Q}^{\hspace{0.03cm}a_{1}}_{1}
	\hspace{0.03cm}\bigr\rangle	
	\Bigr\}\,-
	\notag\\[1ex]
	\frac{1}{2}\,\sum_{\rho}\,&{T}^{\,\ast\hspace{0.03cm}({\rho})}_{\; {\bf k}}(t)
	\,
	\biggl(\int{T}^{\hspace{0.03cm}(\rho)}_{\; {\bf k}}(t)\hspace{0.03cm}dt\biggr)
	\Bigl\{{\rm tr}\hspace{0.03cm}
	\bigl(\hspace{0.03cm}{\mathcal N}_{\hspace{0.02cm}{\bf k}}
	\hspace{0.03cm}
	T^{\,a^{\prime}_{2}}\hspace{0.03cm}T^{\,e_{1}}\hspace{0.03cm}
	T^{\,a_{2}}\hspace{0.01cm}
	\bigr)
	\hspace{0.01cm}
	\bigl\langle\hspace{0.03cm}\mathcal{Q}^{\hspace{0.03cm}a^{\prime}_{2}}_{2}
	\hspace{0.03cm}\bigr\rangle
	\hspace{0.03cm}\bigl\langle\hspace{0.03cm}\mathcal{Q}^{\hspace{0.03cm}e_{1}}_{1}
	\hspace{0.03cm}\bigr\rangle
	\hspace{0.01cm}\bigl\langle\hspace{0.03cm}\mathcal{Q}^{\hspace{0.03cm}a_{2}}_{2}
	\hspace{0.03cm}\bigr\rangle\,+
	\notag\\[1ex]
	&\hspace{4.5cm}{\rm tr}\hspace{0.03cm}
	\bigl(\hspace{0.03cm}{\mathcal N}_{\hspace{0.02cm}{\bf k}}
	\hspace{0.03cm}
	T^{\,a^{\prime}_{1}}\hspace{0.03cm}T^{\,e_{2}}\hspace{0.03cm}
	T^{\,a_{1}}\hspace{0.01cm}
	\bigr)
	\hspace{0.01cm}
	\bigl\langle\hspace{0.03cm}\mathcal{Q}^{\hspace{0.03cm}a^{\prime}_{1}}_{1}
	\hspace{0.03cm}\bigr\rangle
	\hspace{0.03cm}\bigl\langle\hspace{0.03cm}\mathcal{Q}^{\hspace{0.03cm}e_{2}}_{2}
	\hspace{0.03cm}\bigr\rangle
	\hspace{0.01cm}\bigl\langle\hspace{0.03cm}\mathcal{Q}^{\hspace{0.03cm}a_{1}}_{1}
	\hspace{0.03cm}\bigr\rangle
	\Bigr\}\,-
	\label{eq:7e}
\end{align}
\begin{align}
	&\frac{1}{2}\,\sum_{\rho}\hspace{0.02cm}(-1)^{\rho + 1}\,{T}^{\hspace{0.03cm}(\rho)}_{\; {\bf k}}(t)
	\,\biggl(\int{T}^{\,\ast\hspace{0.03cm}(\rho)}_{\; {\bf k}}(t)\hspace{0.03cm}dt\biggr)
	\Bigl\{{\rm tr}\hspace{0.03cm}
	\bigl(\hspace{0.03cm}T^{\,a_{2}}\hspace{0.03cm}T^{\,e_{1}}\hspace{0.03cm}T^{\,a^{\prime}_{2}}{\mathcal W}_{\hspace{0.02cm}{\bf k}}\hspace{0.01cm}
	\bigr)
	\hspace{0.01cm}\bigl\langle\hspace{0.03cm}\mathcal{Q}^{\hspace{0.03cm}a^{\prime}_{2}}_{2}
	\hspace{0.03cm}\bigr\rangle
	\hspace{0.03cm}\bigl\langle\hspace{0.03cm}\mathcal{Q}^{\hspace{0.03cm}e_{1}}_{1}
	\hspace{0.03cm}\bigr\rangle
	\hspace{0.01cm}\bigl\langle\hspace{0.03cm}\mathcal{Q}^{\hspace{0.03cm}a_{2}}_{2}
	\hspace{0.03cm}\bigr\rangle\,
	+
	\notag\\[1ex]
	&{\hspace{7.9cm}}
	{\rm tr}\hspace{0.03cm}\bigl(\hspace{0.03cm}T^{\,a_{1}}\hspace{0.03cm}T^{\,e_{2}}\hspace{0.03cm}T^{\,a^{\prime}_{1}}{\mathcal W}_{\hspace{0.02cm}{\bf k}}\hspace{0.01cm}
	\bigr)
	\hspace{0.01cm}\bigl\langle\hspace{0.03cm}\mathcal{Q}^{\hspace{0.03cm}a^{\prime}_{1}}_{1}
	\hspace{0.03cm}\bigr\rangle
	\hspace{0.03cm}\bigl\langle\hspace{0.03cm}\mathcal{Q}^{\hspace{0.03cm}e_{2}}_{2}
	\hspace{0.03cm}\bigr\rangle
	\hspace{0.01cm}\bigl\langle\hspace{0.03cm}\mathcal{Q}^{\hspace{0.03cm}a_{1}}_{1}
	\hspace{0.03cm}\bigr\rangle	
	\Bigr\}\,-
	\notag\\[1ex]
	&\frac{1}{2}\,\sum_{\rho}\hspace{0.02cm}(-1)^{\rho + 1}\,{T}^{\,\ast\hspace{0.03cm}({\rho})}_{\; {\bf k}}(t)
	\,
	\biggl(\int{T}^{\hspace{0.03cm}(\rho)}_{\; {\bf k}}(t)\hspace{0.03cm}dt\biggr)
	\Bigl\{{\rm tr}\hspace{0.03cm}
	\bigl(\hspace{0.03cm}{\mathcal W}_{\hspace{0.02cm}{\bf k}}
	\hspace{0.03cm}
	T^{\,a^{\prime}_{2}}\hspace{0.03cm}T^{\,e_{1}}\hspace{0.03cm}
	T^{\,a_{2}}\hspace{0.01cm}
	\bigr)
	\hspace{0.01cm}
	\bigl\langle\hspace{0.03cm}\mathcal{Q}^{\hspace{0.03cm}a^{\prime}_{2}}_{2}
	\hspace{0.03cm}\bigr\rangle
	\hspace{0.03cm}\bigl\langle\hspace{0.03cm}\mathcal{Q}^{\hspace{0.03cm}e_{1}}_{1}
	\hspace{0.03cm}\bigr\rangle
	\hspace{0.01cm}\bigl\langle\hspace{0.03cm}\mathcal{Q}^{\hspace{0.03cm}a_{2}}_{2}
	\hspace{0.03cm}\bigr\rangle\,
	+
	\notag\\[1ex]
    &{\hspace{7.9cm}}
	{\rm tr}\hspace{0.03cm}\bigl(\hspace{0.03cm}{\mathcal W}_{\hspace{0.02cm}{\bf k}}
	\hspace{0.03cm}
	T^{\,a^{\prime}_{1}}\hspace{0.03cm}T^{\,e_{2}}\hspace{0.03cm}
	T^{\,a_{1}}\hspace{0.01cm}
	\bigr)
	\hspace{0.01cm}
	\bigl\langle\hspace{0.03cm}\mathcal{Q}^{\hspace{0.03cm}a^{\prime}_{1}}_{1}
	\hspace{0.03cm}\bigr\rangle
	\hspace{0.03cm}\bigl\langle\hspace{0.03cm}\mathcal{Q}^{\hspace{0.03cm}e_{2}}_{2}
	\hspace{0.03cm}\bigr\rangle
	\hspace{0.01cm}\bigl\langle\hspace{0.03cm}\mathcal{Q}^{\hspace{0.03cm}a_{1}}_{1}
	\hspace{0.03cm}\bigr\rangle
	\Bigr\}.
	\notag
\end{align}
Hereinafter, for simplicity, we omit the prime at the fast time $t'$. At the beginning we analyze the coefficient functions related to the product of effective amplitudes ${T}^{\hspace{0.03cm}(\rho)}_{\; {\bf k}}(t)$ and ${T}^{\,\ast\hspace{0.03cm}(\rho)}_{\; {\bf k}}(t)$. Here we are interested in the dependence on the fast time $t$.\\
\indent The first step is to write out again the representation for ${T}^{\hspace{0.03cm}(\rho)}_{\; {\bf k}}(t)$ in terms of the integral 
over the momentum transfer ${\mathbf q}$, Eq.\,(\ref{eq:4_1u}),
\begin{equation}
{T}^{\hspace{0.03cm}(\rho)\hspace{0.03cm}a\,a_{1}\hspace{0.03cm}a_{2}}_{\; {\bf k}}(t)
=
\int\!d\hspace{0.02cm}{\bf q}\,
{T}^{\hspace{0.03cm}(\rho)\hspace{0.03cm}a\,a_{1}\hspace{0.03cm}a_{2}}_{\; {\bf k},\,{\bf q}}(t).
\label{eq:7r}
\end{equation}
Due to the structure of the effective amplitude ${T}^{\hspace{0.03cm}(\rho)\hspace{0.03cm}a\,a_{1}\hspace{0.03cm}a_{2}}_{\; {\bf k},\,{\bf q}}(t)$, as it was defined by expressions (\ref{eq:4_1uu})\,--\,(\ref{eq:4uu}), dependence on the fast time is included in this amplitude through the vertex functions. Let us consider in the first part of the effective amplitude ${T}^{\,{\rm I}\,(\rho)\hspace{0.03cm}a\,a_{1}\hspace{0.03cm}a_{2}}_{\; {\bf k},\,{\bf q}}(t)$ the products of three vertex functions ${\upphi}^{\hspace{0.03cm}(1,2)}(t)$ in the square brackets. Using the representation (\ref{eq:2z}) these products can be written in the following form:
\begin{subequations} 
\label{eq:7t}
\begin{align}
&{\upphi}^{\hspace{0.02cm}\ast\hspace{0.03cm}(1)}_{\,{\bf q}}(t)
\hspace{0.03cm}
{\upphi}^{\hspace{0.03cm}(2)}_{\,{\bf q}}(t)
\hspace{0.03cm}
{\upphi}^{\hspace{0.03cm}(1)}_{\,{\bf k}}(t)
=
{\upphi}^{\hspace{0.02cm}\ast\hspace{0.03cm}(1)}_{\,{\bf q}}
\hspace{0.03cm}
{\upphi}^{\hspace{0.03cm}(2)}_{\,{\bf q}}
\hspace{0.03cm}
{\upphi}^{\hspace{0.03cm}(1)}_{\,{\bf k}}
\,
{\rm e}^{-i\hspace{0.03cm}(\omega^{l}_{{\bf k}} - {\bf k}\cdot{\bf v}_{1}
+ ({\mathbf v}_{1} - {\mathbf v}_{2})\cdot{\mathbf q})\hspace{0.02cm}t}
\hspace{0.04cm}
{\rm e}^{-i\hspace{0.03cm}{\bf q}\hspace{0.02cm}\cdot\hspace{0.02cm}({\mathbf x}_{0\hspace{0.02cm}1} - {\mathbf x}_{0\hspace{0.02cm}2})
+	
i\hspace{0.03cm}{\bf k}\hspace{0.02cm}\cdot\hspace{0.02cm} 
{\bf x}_{0\hspace{0.02cm}1}},
\label{eq:7ta}\\[1.5ex]
&{\upphi}^{(1)}_{\,{\bf q}}(t)
\hspace{0.03cm}
{\upphi}^{\hspace{0.02cm}\ast\hspace{0.03cm}(2)}_{\,{\bf q}}(t)
\hspace{0.03cm}
{\upphi}^{\hspace{0.03cm}(1)}_{\,{\bf k}}(t)
=
{\upphi}^{(1)}_{\,{\bf q}}
\hspace{0.03cm}
{\upphi}^{\hspace{0.02cm}\ast\hspace{0.03cm}(2)}_{\,{\bf q}}
\hspace{0.03cm}
{\upphi}^{\hspace{0.03cm}(1)}_{\,{\bf k}}
\,
{\rm e}^{-i\hspace{0.03cm}(\omega^{l}_{{\bf k}} - {\bf k}\cdot{\bf v}_{1}
- ({\mathbf v}_{1} - {\mathbf v}_{2})\cdot{\mathbf q})\hspace{0.02cm}t}
\hspace{0.04cm}
{\rm e}^{i\hspace{0.03cm}{\bf q}\hspace{0.02cm}\cdot\hspace{0.02cm}({\mathbf x}_{0\hspace{0.02cm}1} - {\mathbf x}_{0\hspace{0.02cm}2})
+	
i\hspace{0.03cm}{\bf k}\hspace{0.02cm}\cdot\hspace{0.02cm} 
{\bf x}_{0\hspace{0.02cm}1}},
\label{eq:7tb}\\[1.5ex]
&{\upphi}^{\hspace{0.02cm}\ast\hspace{0.03cm}(2)}_{\,{\bf q}}(t)
\hspace{0.03cm}
{\upphi}^{\hspace{0.03cm}(1)}_{\,{\bf q}}(t)
\hspace{0.03cm}
{\upphi}^{\hspace{0.03cm}(2)}_{\,{\bf k}}(t)
=
{\upphi}^{\hspace{0.02cm}\ast\hspace{0.03cm}(2)}_{\,{\bf q}}
\hspace{0.03cm}
{\upphi}^{\hspace{0.03cm}(1)}_{\,{\bf q}}
\hspace{0.03cm}
{\upphi}^{\hspace{0.03cm}(2)}_{\,{\bf k}}
\,
{\rm e}^{-i\hspace{0.03cm}(\omega^{l}_{{\bf k}} - {\bf k}\cdot{\bf v}_{2}
- ({\mathbf v}_{1} - {\mathbf v}_{2})\cdot{\mathbf q})\hspace{0.02cm}t}
\hspace{0.04cm}
{\rm e}^{i\hspace{0.03cm}{\bf q}\hspace{0.02cm}\cdot\hspace{0.02cm}({\mathbf x}_{0\hspace{0.02cm}1} - {\mathbf x}_{0\hspace{0.02cm}2})
+	
i\hspace{0.03cm}{\bf k}\hspace{0.02cm}\cdot\hspace{0.02cm} 
{\bf x}_{0\hspace{0.02cm}2}},
\label{eq:7tc}\\[1.5ex]
&{\upphi}^{\hspace{0.03cm}(2)}_{\,{\bf q}}(t)
\hspace{0.03cm}
{\upphi}^{\hspace{0.02cm}\ast\hspace{0.03cm}(1)}_{\,{\bf q}}(t)
\hspace{0.03cm}
{\upphi}^{\hspace{0.03cm}(2)}_{\,{\bf k}}(t)
=
{\upphi}^{\hspace{0.03cm}(2)}_{\,{\bf q}}
\hspace{0.03cm}
{\upphi}^{\hspace{0.02cm}\ast\hspace{0.03cm}(1)}_{\,{\bf q}}
\hspace{0.03cm}
{\upphi}^{\hspace{0.03cm}(2)}_{\,{\bf k}}
\,
{\rm e}^{-i\hspace{0.03cm}(\omega^{l}_{{\bf k}} - {\bf k}\cdot{\bf v}_{2}
+ ({\mathbf v}_{1} - {\mathbf v}_{2})\cdot{\mathbf q})\hspace{0.02cm}t}
\hspace{0.04cm}
{\rm e}^{-i\hspace{0.03cm}{\bf q}\hspace{0.02cm}\cdot\hspace{0.02cm}({\mathbf x}_{0\hspace{0.02cm}1} - {\mathbf x}_{0\hspace{0.02cm}2})
+	
i\hspace{0.03cm}{\bf k}\hspace{0.02cm}\cdot\hspace{0.02cm} 
{\bf x}_{0\hspace{0.02cm}2}}.
\label{eq:7td}
\end{align}
\end{subequations}
From the expressions obtained, we see that they have different exponential multipliers. By performing a change of the integration variable ${\bf q}$, they can be brought to the same form. This replacement depends on the value of the ``index'' $\rho = 1,2$. For the amplitude ${T}^{\,{\rm I}\,(1)\hspace{0.03cm}a\,a_{1}\hspace{0.03cm}a_{2}}_{\; {\bf k},\,{\bf q}}(t)$, as a primary expression we take (\ref{eq:7ta}), and in all terms in (\ref{eq:4u}) related to the products (\ref{eq:7tb})\,--\,(\ref{eq:7td}), we perform the following substitutions: 
\begin{align}
&{\mathbf q} \rightarrow -	{\mathbf q}\quad \mbox{for (\ref{eq:7tb})},\notag\\[1ex]
&{\mathbf q} \rightarrow {\mathbf k} -	{\mathbf q}\quad \mbox{for (\ref{eq:7tc})},\label{eq:7y}\\[1ex]
&{\mathbf q} \rightarrow -({\mathbf k} -	{\mathbf q})\quad \mbox{for (\ref{eq:7td})}.\notag
\end{align} 
On the other hand, for the amplitude ${T}^{\,{\rm I}\,(2)\hspace{0.03cm}a\,a_{2}\hspace{0.03cm}a_{2}}_{\; {\bf k},\,{\bf q}}(t)$, as a primary expression we will use (\ref{eq:7tc}), and in the remaining terms in (\ref{eq:4u}) related to the products (\ref{eq:7ta}), (\ref{eq:7tb}) and (\ref{eq:7td}), we perform the substitutions similar to the previous ones:
\begin{align}
	&{\mathbf q} \rightarrow {\mathbf k} -	{\mathbf q}\quad \mbox{for (\ref{eq:7ta})},\notag\\[1ex]
	&{\mathbf q} \rightarrow -({\mathbf k} -	{\mathbf q})\quad \mbox{for (\ref{eq:7tb})},\label{eq:7u}\\[1ex]
	&{\mathbf q} \rightarrow -	{\mathbf q}\quad \mbox{for (\ref{eq:7td})}.\notag
\end{align}
\indent We apply completely similar reasoning for the second part of the effective amplitude: ${T}^{\,{\rm II}\,(\rho)\hspace{0.03cm}a\,a_{1}\hspace{0.03cm}
a_{2}}_{\; {\bf k},\,{\bf q}}(t)$, Eq.\,(\ref{eq:4uu}). Using the representations for the vertex functions (\ref{eq:2dd}) and (\ref{eq:2z}), we obtain the following expressions for the products in (\ref{eq:4uu}):  
\begin{align}
&{\mathcal V}^{\,\ast\,a\,a_{1}\hspace{0.03cm}a_{2}}_{\ {\bf k},\,{\bf q},\, 
{\bf k} - {\bf q}}(t)
\hspace{0.03cm}
{\upphi}^{\,(1)}_{\,{\bf q}}(t)
\hspace{0.03cm}
{\upphi}^{\,(2)}_{\,{\bf k} - {\bf q}}(t)
=
{\mathcal V}^{\,\ast\,a\,a_{1}\hspace{0.03cm}a_{2}}_{\ {\bf k},\,{\bf q},\, 
{\bf k} - {\bf q}}
\hspace{0.03cm}
{\upphi}^{\,(1)}_{\,{\bf q}}
\hspace{0.03cm}
{\upphi}^{\,(2)}_{\,{\bf k} - {\bf q}}
\,
{\rm e}^{-i\hspace{0.03cm}(\omega^{l}_{{\bf k}} - {\bf k}\cdot{\bf v}_{2}
- ({\mathbf v}_{1} - {\mathbf v}_{2})\cdot{\mathbf q})\hspace{0.02cm}t}
\hspace{0.04cm}
{\rm e}^{i\hspace{0.03cm}{\bf q}\hspace{0.02cm}\cdot\hspace{0.02cm}({\mathbf x}_{0\hspace{0.02cm}1} - {\mathbf x}_{0\hspace{0.02cm}2})
+	
i\hspace{0.03cm}{\bf k}\hspace{0.02cm}\cdot\hspace{0.02cm} 
{\bf x}_{0\hspace{0.02cm}2}},
\notag\\[1.5ex]
&{\mathcal V}^{\,\ast\,a\,a_{1}\hspace{0.03cm}a_{2}}_{\ {\bf k},\,{\bf q},\, 
	{\bf k} - {\bf q}}(t)
\hspace{0.03cm}
{\upphi}^{\,(2)}_{\,{\bf q}}(t)
\hspace{0.03cm}
{\upphi}^{\,(1)}_{\,{\bf k} - {\bf q}}(t)
=
{\mathcal V}^{\,\ast\,a\,a_{1}\hspace{0.03cm}a_{2}}_{\ {\bf k},\,{\bf q},\, 
{\bf k} - {\bf q}}
\hspace{0.03cm}
{\upphi}^{\,(2)}_{\,{\bf q}}
\hspace{0.03cm}
{\upphi}^{\,(1)}_{\,{\bf k} - {\bf q}}
\,
{\rm e}^{-i\hspace{0.03cm}(\omega^{l}_{{\bf k}} - {\bf k}\cdot{\bf v}_{1}
+ ({\mathbf v}_{1} - {\mathbf v}_{2})\cdot{\mathbf q})\hspace{0.02cm}t}
\hspace{0.04cm}
{\rm e}^{-i\hspace{0.03cm}{\bf q}\hspace{0.02cm}\cdot\hspace{0.02cm}({\mathbf x}_{0\hspace{0.02cm}1} - {\mathbf x}_{0\hspace{0.02cm}2})
+	
i\hspace{0.03cm}{\bf k}\hspace{0.02cm}\cdot\hspace{0.02cm} 
{\bf x}_{0\hspace{0.02cm}1}},
\notag\\[1.5ex]
&\hspace{9cm}\ldots\notag\\[1ex]
&{\mathcal U}^{\;a\,a_{1}\hspace{0.03cm}a_{2}}_{\hspace{0.03cm}{\bf k},\,{\bf q},\, 
- {\bf k} - {\bf q}}(t)
\hspace{0.03cm}
{\upphi}^{\hspace{0.03cm}\ast\,(1)}_{\,{\bf q}}(t)
\hspace{0.03cm}
{\upphi}^{\hspace{0.03cm}\ast\,(2)}_{- {\bf k} - {\bf q}}(t)
=
{\mathcal U}^{\;a\,a_{1}\hspace{0.03cm}a_{2}}_{\hspace{0.03cm}{\bf k},\,{\bf q},\, 
- {\bf k} - {\bf q}}
\hspace{0.03cm}
{\upphi}^{\hspace{0.03cm}\ast\,(1)}_{\,{\bf q}}
\hspace{0.03cm}
{\upphi}^{\hspace{0.03cm}\ast\,(2)}_{- {\bf k} - {\bf q}}
\,
{\rm e}^{-i\hspace{0.03cm}(\omega^{l}_{{\bf k}} - {\bf k}\cdot{\bf v}_{2}
+ ({\mathbf v}_{1} - {\mathbf v}_{2})\cdot{\mathbf q})\hspace{0.02cm}t}
\hspace{0.04cm}
{\rm e}^{-i\hspace{0.03cm}{\bf q}\hspace{0.02cm}\cdot\hspace{0.02cm}({\mathbf x}_{0\hspace{0.02cm}1} - {\mathbf x}_{0\hspace{0.02cm}2})
+	
i\hspace{0.03cm}{\bf k}\hspace{0.02cm}\cdot\hspace{0.02cm} 
{\bf x}_{0\hspace{0.02cm}2}},
\notag\\[1.5ex]
&{\mathcal U}^{\;a\,a_{1}\hspace{0.03cm}a_{2}}_{\hspace{0.03cm}{\bf k},\,{\bf q},\, 
	- {\bf k} - {\bf q}}(t)
\hspace{0.03cm}
{\upphi}^{\hspace{0.03cm}\ast\,(2)}_{\,{\bf q}}(t)
\hspace{0.03cm}
{\upphi}^{\hspace{0.03cm}\ast\,(1)}_{- {\bf k} - {\bf q}}(t)
=
{\mathcal U}^{\;a\,a_{1}\hspace{0.03cm}a_{2}}_{\hspace{0.03cm}{\bf k},\,{\bf q},\, 
	- {\bf k} - {\bf q}}
\hspace{0.03cm}
{\upphi}^{\hspace{0.03cm}\ast\,(2)}_{\,{\bf q}}
\hspace{0.03cm}
{\upphi}^{\hspace{0.03cm}\ast\,(1)}_{- {\bf k} - {\bf q}}
\,
{\rm e}^{-i\hspace{0.03cm}(\omega^{l}_{{\bf k}} - {\bf k}\cdot{\bf v}_{1}
- ({\mathbf v}_{1} - {\mathbf v}_{2})\cdot{\mathbf q})\hspace{0.02cm}t}
\hspace{0.04cm}
{\rm e}^{i\hspace{0.03cm}{\bf q}\hspace{0.02cm}\cdot\hspace{0.02cm}({\mathbf x}_{0\hspace{0.02cm}1} - {\mathbf x}_{0\hspace{0.02cm}2})
+	
i\hspace{0.03cm}{\bf k}\hspace{0.02cm}\cdot\hspace{0.02cm} 
{\bf x}_{0\hspace{0.02cm}1}}.
\notag
\end{align}
The exponential factors here are exactly the same as in (\ref{eq:7t}). Using the same reasoning as above, these factors can be reduced to the same form by the replacements of the integration variable ${\bf q}$, Eqs.\,(\ref{eq:7y}) or (\ref{eq:7u}), depending on the value $\rho$.\\
\indent Taking into account all of the above, we can now explicitly single out the dependence on the fast time $t$ (as well as on the initial coordinates of the position of particles 1 and 2) from the amplitude ${T}^{\hspace{0.03cm}(\rho)\hspace{0.03cm}a\,a_{1}\hspace{0.03cm}a_{2}}_{\; {\bf k},\,{\bf q}}(t)$ in the following form: 
\begin{equation}
{T}^{\hspace{0.03cm}(\rho)\hspace{0.03cm}a\,a_{1}\hspace{0.03cm}a_{2}}_{\; {\bf k},\,{\bf q}}(t)
\,=\,
{\rm e}^{-i\hspace{0.03cm}(\omega^{l}_{{\bf k}} - {\bf k}\cdot{\bf v}_{\rho}
\,-\, (-1)^{\rho}\, ({\mathbf v}_{1} - {\mathbf v}_{2})\cdot{\mathbf q})\hspace{0.02cm}t}
\hspace{0.04cm}
{\rm e}^{(-1)^{\rho}\,i\hspace{0.03cm}{\bf q}\hspace{0.02cm}\cdot\hspace{0.02cm}({\mathbf x}_{0\hspace{0.02cm}1} - {\mathbf x}_{0\hspace{0.02cm}2})
\,+\,	
i\hspace{0.03cm}{\bf k}\hspace{0.02cm}\cdot\hspace{0.02cm} 
{\bf x}_{0\hspace{0.02cm}\rho}}\,
{T}^{\hspace{0.03cm}(\rho)\hspace{0.03cm}a\,a_{1}\hspace{0.03cm}a_{2}}_{\; {\bf k},\,{\bf q}},
\label{eq:7i}
\end{equation}
where now the amplitude ${T}^{\hspace{0.03cm}(\rho)\hspace{0.03cm}a\,a_{1}\hspace{0.03cm}a_{2}}_{\; {\bf k},\,{\bf q}}$ on the right side is independent of $t$ and have the following structure, instead of (\ref{eq:4_1uu})\,--\,(\ref{eq:4uu}),
\[
	{T}^{\hspace{0.03cm}(\rho)\hspace{0.03cm}a\,a_{1}\hspace{0.03cm}a_{2}}_{\; {\bf k},\,{\bf q}}
	=
	{T}^{\,{\rm I}\,
	(\rho)\hspace{0.03cm}a\,a_{1}\hspace{0.03cm}a_{2}}_{\; {\bf k},\,{\bf q}}
	+
	{T}^{\,{\rm II}\,
	(\rho)\hspace{0.03cm}a\,a_{1}\hspace{0.03cm}a_{2}}_{\; {\bf k},\,{\bf q}},
\quad
\rho = 1,2.
\]
Here, for the first part ${T}^{\,{\rm I}\,(\rho)\hspace{0.03cm}a\,a_{1}
\hspace{0.03cm}a_{2}}_{\; {\bf k},\,{\bf q}},\,\rho = 1,2,$ we have the final expressions:
\begin{align}
	&{T}^{\,{\rm I}\,
	(1)\hspace{0.03cm}a\,a_{1}\hspace{0.03cm}a_{2}}_{\; {\bf k},\,{\bf q}}
	=
	-\hspace{0.03cm}i\hspace{0.02cm}
	f^{\,a\hspace{0.03cm}a_{1}\hspace{0.03cm}a_{2}}\hspace{0.03cm}
	\frac{1}{\omega^{\hspace{0.02cm} l}_{\hspace{0.03cm}{\bf k}} - {\bf v}^{\phantom{l}}_{1}\cdot {\bf k}}\,
	\times\notag\\[1ex]
	&\biggl\{
	\biggl(\frac{1}{\omega^{\hspace{0.02cm}l}_{\hspace{0.03cm}{\bf q}} - {\bf v}^{\phantom{l}}_{1}\cdot {\bf q}}
	\,+\,
	\frac{1}{\omega^{\hspace{0.02cm} l}_{\hspace{0.03cm}{\bf q}} - {\bf v}^{\phantom{l}}_{2}\cdot {\bf q}}
	\biggr)
	\hspace{0.03cm}{\upphi}^{\ast\hspace{0.03cm}(1)}_{\,{\bf q}}
	\hspace{0.03cm}
	{\upphi}^{(2)}_{\,{\bf q}}
	\hspace{0.03cm}
	{\upphi}^{(1)}_{\,{\bf k}}
	\,+
	\notag\\[1ex]
	&\biggl(\frac{1}{\omega^{\hspace{0.02cm}l}_{-\hspace{0.03cm}{\bf q}} + {\bf v}^{\phantom{l}}_{1}\cdot {\bf q}}
	\,+\,
	\frac{1}{\omega^{\hspace{0.02cm} l}_{-\hspace{0.03cm}{\bf q}} + {\bf v}^{\phantom{l}}_{2}\cdot {\bf q}}
	\biggr)
	\hspace{0.03cm}{\upphi}^{(1)}_{\,-{\bf q}}
	\hspace{0.03cm}
	{\upphi}^{\hspace{0.02cm}\ast\hspace{0.03cm}(2)}_{\,-{\bf q}}
	\hspace{0.03cm}
	{\upphi}^{(1)}_{\,{\bf k}}
	\,-\,
	\notag\\[1ex]
	&\biggl(\frac{1}{\omega^{\hspace{0.02cm}l}_{\hspace{0.03cm}{\bf k} - {\bf q}} - {\bf v}^{\phantom{l}}_{1}\cdot({\bf k} - {\bf q})}
	\,+\,
	\frac{1}{\omega^{\hspace{0.02cm} l}_{\hspace{0.03cm}{\bf k} - {\bf q}} - {\bf v}^{\phantom{l}}_{2}\cdot({\bf k} - {\bf q})}
	\biggr)
	{\upphi}^{\hspace{0.02cm}\ast\hspace{0.03cm}(2)}_{\,{\bf k} - {\bf q}}
	\hspace{0.03cm}
	{\upphi}^{(1)}_{\,{\bf k} - {\bf q}}
	\hspace{0.03cm}
	{\upphi}^{(2)}_{\,{\bf k}}
	\,-
	\notag\\[1ex]
	&\biggl(\frac{1}{\omega^{\hspace{0.02cm}l}_{-\hspace{0.03cm}({\bf k} - {\bf q})} + {\bf v}^{\phantom{l}}_{1}\cdot({\bf k} - {\bf q})}
	\,+\,
	\frac{1}{\omega^{\hspace{0.02cm} l}_{-\hspace{0.03cm}({\bf k} - {\bf q})} 
		+ {\bf v}^{\phantom{l}}_{2}\cdot({\bf k} - {\bf q})}
	\biggr)
	{\upphi}^{(2)}_{\,-({\bf k} - {\bf q})}
	\hspace{0.03cm}
	{\upphi}^{\hspace{0.02cm}\ast\hspace{0.03cm}(1)}_{\,-({\bf k} - {\bf q})}
	\hspace{0.03cm}
	{\upphi}^{(2)}_{\,{\bf k}}
	\biggr\}
	\notag
\end{align}
and
\begin{align}
	&{T}^{\,{\rm I}\,
	(2)\hspace{0.03cm}a\,a_{1}\hspace{0.03cm}a_{2}}_{\; {\bf k},\,{\bf q}}
	=
	i\hspace{0.02cm}
	f^{\,a\hspace{0.03cm}a_{1}\hspace{0.03cm}a_{2}}\hspace{0.03cm}
	\frac{1}{\omega^{\hspace{0.02cm} l}_{\hspace{0.03cm}{\bf k}} - {\bf v}^{\phantom{l}}_{2}\cdot {\bf k}}\,
	\times\notag\\[1ex]
	&\biggl\{
	\biggl(\frac{1}{\omega^{\hspace{0.02cm}l}_{\hspace{0.03cm}{\bf q}} - {\bf v}^{\phantom{l}}_{1}\cdot {\bf q}}
	\,+\,
	\frac{1}{\omega^{\hspace{0.02cm} l}_{\hspace{0.03cm}{\bf q}} - {\bf v}^{\phantom{l}}_{2}\cdot {\bf q}}
	\biggr)
	\hspace{0.03cm}{\upphi}^{\ast\hspace{0.03cm}(2)}_{\,{\bf q}}
	\hspace{0.03cm}
	{\upphi}^{(1)}_{\,{\bf q}}
	\hspace{0.03cm}
	{\upphi}^{(2)}_{\,{\bf k}}
	\,+
	\notag\\[1ex]
	&\biggl(\frac{1}{\omega^{\hspace{0.02cm}l}_{-\hspace{0.03cm}{\bf q}} + {\bf v}^{\phantom{l}}_{1}\cdot {\bf q}}
	\,+\,
	\frac{1}{\omega^{\hspace{0.02cm} l}_{-\hspace{0.03cm}{\bf q}} + {\bf v}^{\phantom{l}}_{2}\cdot {\bf q}}
	\biggr)
	\hspace{0.03cm}{\upphi}^{(2)}_{\,-{\bf q}}
	\hspace{0.03cm}
	{\upphi}^{\hspace{0.02cm}\ast\hspace{0.03cm}(1)}_{\,-{\bf q}}
	\hspace{0.03cm}
	{\upphi}^{(2)}_{\,{\bf k}}
	\,-
	\notag\\[1ex]
	&\biggl(\frac{1}{\omega^{\hspace{0.02cm}l}_{\hspace{0.03cm}{\bf k} - {\bf q}} - {\bf v}^{\phantom{l}}_{1}\cdot({\bf k} - {\bf q})}
	\,+\,
	\frac{1}{\omega^{\hspace{0.02cm} l}_{\hspace{0.03cm}{\bf k} - {\bf q}} - {\bf v}^{\phantom{l}}_{2}\cdot({\bf k} - {\bf q})}
	\biggr)
	{\upphi}^{\hspace{0.02cm}\ast\hspace{0.03cm}(1)}_{\,{\bf k} - {\bf q}}
	\hspace{0.03cm}
	{\upphi}^{(2)}_{\,{\bf k} - {\bf q}}
	\hspace{0.03cm}
	{\upphi}^{(1)}_{\,{\bf k}}
	\,-
	\notag\\[1ex]
	&\biggl(\frac{1}{\omega^{\hspace{0.02cm}l}_{-\hspace{0.03cm}({\bf k} - {\bf q})} + {\bf v}^{\phantom{l}}_{1}\cdot({\bf k} - {\bf q})}
	\,+\,
	\frac{1}{\omega^{\hspace{0.02cm} l}_{-\hspace{0.03cm}({\bf k} - {\bf q})} 
		+ {\bf v}^{\phantom{l}}_{2}\cdot({\bf k} - {\bf q})}
	\biggr)
	{\upphi}^{(1)}_{\,-({\bf k} - {\bf q})}
	\hspace{0.03cm}
	{\upphi}^{\hspace{0.02cm}\ast\hspace{0.03cm}(2)}_{\,-({\bf k} - {\bf q})}
	\hspace{0.03cm}
	{\upphi}^{(1)}_{\,{\bf k}}
	\biggr\}.
	\notag
\end{align}
The second part ${T}^{\,{\rm II}\,(\rho)\hspace{0.03cm}a\,a_{1}
\hspace{0.03cm}a_{2}}_{\; {\bf k},\,{\bf q}}$, connected with the effective three-plasmon vertices, can be presented in a similar form:
\begin{align}
T^{\,{\rm II}\,(1)\hspace{0.03cm}a\,a_{1}\hspace{0.03cm}a_{2}}_{\;{\bf k},\,{\bf q}}
= -\hspace{-6cm}&\hspace{6cm}2\hspace{0.03cm}
{\mathcal V}^{\,\ast\,a\,a_{1}\hspace{0.03cm}a_{2}}_{\ {\bf k},\,{\bf q},\, 
{\bf k} - {\bf q}}
\hspace{0.03cm}
{\upphi}^{\,(2)}_{\,{\bf q}}
\hspace{0.03cm}
{\upphi}^{\,(1)}_{\,{\bf k} - {\bf q}}
\,\times
\notag\\[1.5ex]
&\biggl(\frac{1}{\omega^{\hspace{0.02cm}l}_{\hspace{0.03cm}{\bf q}} - {\bf v}^{\phantom{l}}_{1}\cdot {\bf q}}
\,+\,
\frac{1}{\omega^{\hspace{0.02cm}l}_{\hspace{0.03cm}{\bf q}} - {\bf v}^{\phantom{l}}_{2}\cdot {\bf q}}
\biggr)
\biggl(\frac{1}{\omega^{\hspace{0.02cm}l}_{\hspace{0.03cm}{\bf k} - {\bf q}}\! - {\bf v}^{\phantom{l}}_{1}\cdot ({\bf k} - {\bf q})}
\,+\,
\frac{1}{\omega^{\hspace{0.02cm}l}_{\hspace{0.03cm}{\bf k} - {\bf q}}\! - {\bf v}^{\phantom{l}}_{2}\cdot ({\bf k} - {\bf q})}
\biggr)\,-
\notag\\[1.5ex]
&\hspace{6cm}2\hspace{0.03cm}{\mathcal V}^{\;a_{1}\hspace{0.03cm}a_{2}\,a}_{\ {\bf q},\,
{\bf q} - {\bf k},\,{\bf k}} 
\hspace{0.03cm}
{\upphi}^{\,(2)}_{\,{\bf q}}
\hspace{0.03cm}
{\upphi}^{\hspace{0.03cm}\ast\,(1)}_{\,{\bf q} - {\bf k}}
\,\times
\notag\\[1.5ex]
&\biggl(\frac{1}{\omega^{\hspace{0.02cm}l}_{\hspace{0.03cm}{\bf q}} - {\bf v}^{\phantom{l}}_{1}\cdot {\bf q}}
\,+\,
\frac{1}{\omega^{\hspace{0.02cm} l}_{\hspace{0.03cm}{\bf q}} - {\bf v}^{\phantom{l}}_{2}\cdot {\bf q}}
\biggr)
\biggl(\frac{1}{\omega^{\hspace{0.02cm}l}_{\hspace{0.03cm}{\bf q} - {\bf k}}\! - {\bf v}^{\phantom{l}}_{1}\cdot ({\bf q} - {\bf k})}
\,+\,
\frac{1}{\omega^{\hspace{0.02cm} l}_{\hspace{0.03cm}{\bf q} - {\bf k}}\! 
- {\bf v}^{\phantom{l}}_{2}\cdot ({\bf q} - {\bf k})}
\biggr)\,+
\notag\\[1.5ex]
&\hspace{6cm}2\hspace{0.03cm}{\mathcal V}^{\;a_{1}\,a_{2}\,a}_{\ {\bf k} - {\bf q},\,
- {\bf q},\,{\bf k}} 
\hspace{0.03cm}
{\upphi}^{\,(1)}_{\,{\bf k} - {\bf q}}
\hspace{0.03cm}
{\upphi}^{\hspace{0.03cm}\ast\,(2)}_{\,- {\bf q}}
\,\times
\notag\\[1.5ex]
&\biggl(\frac{1}{\omega^{\hspace{0.02cm}l}_{-\hspace{0.03cm}{\bf q}} + 
{\bf v}^{\phantom{l}}_{1}\cdot {\bf q}}
\,+\,
\frac{1}{\omega^{\hspace{0.02cm} l}_{-\hspace{0.03cm}{\bf q}} + 
{\bf v}^{\phantom{l}}_{2}\cdot {\bf q}}
\biggr)
\biggl(\frac{1}{\omega^{\hspace{0.02cm}l}_{\hspace{0.03cm}{\bf k} - {\bf q}}\! - {\bf v}^{\phantom{l}}_{1}\cdot ({\bf k} - {\bf q})}
\,+\,
\frac{1}{\omega^{\hspace{0.02cm} l}_{\hspace{0.03cm}{\bf k} - {\bf q}}\! 
	- {\bf v}^{\phantom{l}}_{2}\cdot ({\bf k} - {\bf q})}
\biggr)\,+
\notag\\[1.5ex]
&\hspace{6cm}2\,{\mathcal U}^{\;a\,a_{1}\hspace{0.03cm}a_{2}}_{\ {\bf k},\,{\bf q} - {\bf k},\,-{\bf q}} 
\hspace{0.03cm}
{\upphi}^{\hspace{0.03cm}\ast\,(1)}_{\,{\bf q} - {\bf k}}
\hspace{0.03cm}
{\upphi}^{\hspace{0.03cm}\ast\,(2)}_{\,-{\bf q}}
\,\times
\notag\\[1.5ex]
&\biggl(\frac{1}{\omega^{\hspace{0.02cm}l}_{-\hspace{0.03cm}{\bf q}} 
+ {\bf v}^{\phantom{l}}_{1}\cdot {\bf q}}
\,+\,
\frac{1}{\omega^{\hspace{0.02cm} l}_{-\hspace{0.03cm}{\bf q}} 
+ {\bf v}^{\phantom{l}}_{2}\cdot {\bf q}}
\biggr)
\biggl(\frac{1}{\omega^{\hspace{0.02cm}l}_{\hspace{0.03cm}{\bf q} - {\bf k}}\! - {\bf v}^{\phantom{l}}_{1}\cdot ({\bf q} - {\bf k})}
\,+\,
\frac{1}{\omega^{\hspace{0.02cm} l}_{\hspace{0.03cm}{\bf q} - {\bf k}}\! - 
{\bf v}^{\phantom{l}}_{2}\cdot ({\bf q} - {\bf k})}
\biggr)
\notag
\end{align}
and
\begin{align}
	T^{\,{\rm II}\,(2)\hspace{0.03cm}a\,a_{1}\hspace{0.03cm}a_{2}}_{\;{\bf k},\,{\bf q}}
	=\; \hspace{-6cm}&\hspace{6cm}2\hspace{0.03cm}
	{\mathcal V}^{\,\ast\,a\,a_{1}\hspace{0.03cm}a_{2}}_{\ {\bf k},\,{\bf q},\, 
	{\bf k} - {\bf q}}
	\hspace{0.03cm}
	{\upphi}^{\,(1)}_{\,{\bf q}}
	\hspace{0.03cm}
	{\upphi}^{\,(2)}_{\,{\bf k} - {\bf q}}
	\,\times
	\notag\\[1.5ex]
	&\biggl(\frac{1}{\omega^{\hspace{0.02cm}l}_{\hspace{0.03cm}{\bf q}} - {\bf v}^{\phantom{l}}_{1}\cdot {\bf q}}
	\,+\,
	\frac{1}{\omega^{\hspace{0.02cm}l}_{\hspace{0.03cm}{\bf q}} - {\bf v}^{\phantom{l}}_{2}\cdot {\bf q}}
	\biggr)
	\biggl(\frac{1}{\omega^{\hspace{0.02cm}l}_{\hspace{0.03cm}{\bf k} - {\bf q}}\! - {\bf v}^{\phantom{l}}_{1}\cdot ({\bf k} - {\bf q})}
	\,+\,
	\frac{1}{\omega^{\hspace{0.02cm}l}_{\hspace{0.03cm}{\bf k} - {\bf q}}\! - {\bf v}^{\phantom{l}}_{2}\cdot ({\bf k} - {\bf q})}
	\biggr)\,+
	\notag\\[1.5ex]
	&\hspace{6cm}2\hspace{0.03cm}{\mathcal V}^{\;a_{1}\hspace{0.03cm}a_{2}\,a}_{\ {\bf q},\,
		{\bf q} - {\bf k},\,{\bf k}} 
	\hspace{0.03cm}
	{\upphi}^{\,(1)}_{\,{\bf q}}
	\hspace{0.03cm}
	{\upphi}^{\hspace{0.03cm}\ast\,(2)}_{\,{\bf q} - {\bf k}}
	\,\times
	\notag\\[1.5ex]
	&\biggl(\frac{1}{\omega^{\hspace{0.02cm}l}_{\hspace{0.03cm}{\bf q}} - {\bf v}^{\phantom{l}}_{1}\cdot {\bf q}}
	\,+\,
	\frac{1}{\omega^{\hspace{0.02cm} l}_{\hspace{0.03cm}{\bf q}} - {\bf v}^{\phantom{l}}_{2}\cdot {\bf q}}
	\biggr)
	\biggl(\frac{1}{\omega^{\hspace{0.02cm}l}_{\hspace{0.03cm}{\bf q} - {\bf k}}\! - {\bf v}^{\phantom{l}}_{1}\cdot ({\bf q} - {\bf k})}
	\,+\,
	\frac{1}{\omega^{\hspace{0.02cm} l}_{\hspace{0.03cm}{\bf q} - {\bf k}}\! 
		- {\bf v}^{\phantom{l}}_{2}\cdot ({\bf q} - {\bf k})}
	\biggr)\,-
	\notag\\[1.5ex]
	&\hspace{6cm}2\hspace{0.03cm}{\mathcal V}^{\;a_{1}\,a_{2}\,a}_{\ {\bf k} - {\bf q},\,
		- {\bf q},\,{\bf k}} 
	\hspace{0.03cm}
	{\upphi}^{\,(2)}_{\,{\bf k} - {\bf q}}
	\hspace{0.03cm}
	{\upphi}^{\hspace{0.03cm}\ast\,(1)}_{\,- {\bf q}}
	\,\times
	\notag\\[1.5ex]
	&\biggl(\frac{1}{\omega^{\hspace{0.02cm}l}_{-\hspace{0.03cm}{\bf q}} + 
		{\bf v}^{\phantom{l}}_{1}\cdot {\bf q}}
	\,+\,
	\frac{1}{\omega^{\hspace{0.02cm} l}_{-\hspace{0.03cm}{\bf q}} + 
		{\bf v}^{\phantom{l}}_{2}\cdot {\bf q}}
	\biggr)
	\biggl(\frac{1}{\omega^{\hspace{0.02cm}l}_{\hspace{0.03cm}{\bf k} - {\bf q}}\! - {\bf v}^{\phantom{l}}_{1}\cdot ({\bf k} - {\bf q})}
	\,+\,
	\frac{1}{\omega^{\hspace{0.02cm} l}_{\hspace{0.03cm}{\bf k} - {\bf q}}\! 
		- {\bf v}^{\phantom{l}}_{2}\cdot ({\bf k} - {\bf q})}
	\biggr)\,+
	\notag\\[1.5ex]
	&\hspace{6cm}2\,{\mathcal U}^{\;a\,a_{1}\hspace{0.03cm}a_{2}}_{\ {\bf k},\,-{\bf q},\,{\bf q} - {\bf k}} 
	\hspace{0.03cm}
	{\upphi}^{\hspace{0.03cm}\ast\,(1)}_{\,-{\bf q}}
	\hspace{0.03cm}
	{\upphi}^{\hspace{0.03cm}\ast\,(2)}_{\,{\bf q} - {\bf k}}
	\,\times
	\notag\\[1.5ex]
	&\biggl(\frac{1}{\omega^{\hspace{0.02cm}l}_{-\hspace{0.03cm}{\bf q}} 
		+ {\bf v}^{\phantom{l}}_{1}\cdot {\bf q}}
	\,+\,
	\frac{1}{\omega^{\hspace{0.02cm} l}_{-\hspace{0.03cm}{\bf q}} 
		+ {\bf v}^{\phantom{l}}_{2}\cdot {\bf q}}
	\biggr)
	\biggl(\frac{1}{\omega^{\hspace{0.02cm}l}_{\hspace{0.03cm}{\bf q} - {\bf k}}\! - {\bf v}^{\phantom{l}}_{1}\cdot ({\bf q} - {\bf k})}
	\,+\,
	\frac{1}{\omega^{\hspace{0.02cm} l}_{\hspace{0.03cm}{\bf q} - {\bf k}}\! - {\bf v}^{\phantom{l}}_{2}\cdot ({\bf q} - {\bf k})}
	\biggr).
	\notag
\end{align}
Here we used the antisymmetry property of vertex functions when rearranging arguments, Eq.\,(\ref{eq:2l}). By employing the color and momentum factorization for the vertex functions, Eq.\,(\ref{eq:2h}), we can write in fully analogy with (\ref{eq:6wwww})
\[
{T}^{\hspace{0.03cm}(\rho)\hspace{0.03cm}a\,a_{1}\hspace{0.03cm}a_{2}}_{\; {\bf k},\,{\bf q}}
=
f^{\hspace{0.03cm}a\,a_{1}\hspace{0.03cm}a_{2}\hspace{0.03cm}} {T}^{\hspace{0.03cm}(\rho)}_{\; {\bf k},\,{\bf q}}.
\]
\indent Let us now return to our original kinetic equation (\ref{eq:7e}). As was mentioned above the fast time integrals on the right-hand side are considered as the indefinite ones. Taking into account the representations (\ref{eq:7r}) and (\ref{eq:7i}), we find that 
\begin{equation}
{T}^{\hspace{0.03cm}(\rho)}_{\; {\bf k}}(t)
\,\biggl(\int{T}^{\,\ast\hspace{0.03cm}(\rho)}_{\; {\bf k}}(t)\hspace{0.03cm}dt\biggr)
=
\label{eq:7a}
\end{equation}
\[
-\!\int\!d\hspace{0.02cm}{\bf q}\,d\hspace{0.02cm}{\bf q}'\;
{T}^{\hspace{0.03cm}(\rho)}_{\; {\bf k},\,{\bf q}}\,
{T}^{\,\ast\hspace{0.03cm}(\rho)}_{\; {\bf k},\,{\bf q}'}\,
{\rm e}^{i\hspace{0.03cm}\,(-1)^{\rho}\,\Delta{\mathbf v}\cdot({\mathbf q} - {\mathbf q}')\hspace{0.02cm}t}
\hspace{0.04cm}
{\rm e}^{i\,(-1)^{\rho}\,({\mathbf x}_{0\hspace{0.02cm}1} - {\mathbf x}_{0\hspace{0.02cm}2})\cdot({\bf q} - {\mathbf q}')}\,
\frac{i}{\omega^{l}_{{\bf k}} - {\bf k}\cdot{\bf v}_{\rho}
-(-1)^{\rho}\, \Delta{\mathbf v}\cdot{\mathbf q}' + i0}.
\]  
Here, there is no summation over $\rho$ is implied and for brevity we introduced the notation
\[
\Delta{\mathbf v} \equiv {\mathbf v}_{1} - {\mathbf v}_{2}.
\] 
For convenience, we also provide the complex conjugate expression for (\ref{eq:7a}), in which, however, we exchange the integration variables ${\mathbf q}\rightleftharpoons {\mathbf q}'$:
\begin{equation}
{T}^{\,\ast\hspace{0.03cm}({\rho})}_{\; {\bf k}}(t)
\,
\biggl(\int{T}^{\hspace{0.03cm}(\rho)}_{\; {\bf k}}(t)\hspace{0.03cm}dt\biggr)
=
\label{eq:7s}
\end{equation}
\[
\int\!d\hspace{0.02cm}{\bf q}\,d\hspace{0.02cm}{\bf q}'\;
{T}^{\hspace{0.03cm}(\rho)}_{\; {\bf k},\,{\bf q}}\,
{T}^{\,\ast\hspace{0.03cm}(\rho)}_{\; {\bf k},\,{\bf q}'}\,
{\rm e}^{i\hspace{0.03cm}\,(-1)^{\rho}\,\Delta{\mathbf v}\cdot({\mathbf q} - {\mathbf q}')\hspace{0.02cm}t}
\hspace{0.04cm}
{\rm e}^{i\,(-1)^{\rho}\,({\mathbf x}_{0\hspace{0.02cm}1} - {\mathbf x}_{0\hspace{0.02cm}2})\cdot({\bf q} - {\mathbf q}')}\,
\frac{i}{\omega^{l}_{{\bf k}} - {\bf k}\cdot{\bf v}_{\rho}
-(-1)^{\rho}\, \Delta{\mathbf v}\cdot{\mathbf q} - i0}.
\]  
With this replacement, the integrand remains unchanged up to the sign (cp. with (\ref{eq:7a})), except for the expression in the denominator. For concreteness, let us consider the first two terms on the right-hand side of (\ref{eq:7e}). Now we consider the expressions for color traces. We assume that the color matrix function ${\mathcal N}_{\bf k}$ is Hermitian, i.e.
\[
{\mathcal N}^{\,\dagger}_{\bf k} = {\mathcal N}_{\bf k}.
\]
By virtue of the decomposition (\ref{eq:7q}), this means that the scalar functions $N^{\hspace{0.03cm}(1)}_{\bf k}$ and $N^{\hspace{0.03cm}(2)}_{\bf k}$ are real. In this case, for the traces in the second term on the right-hand side of (\ref{eq:7e}), due to the Hermitian nature of the adjoint representation matrices $T^{\,a}$, the following relations hold
\begin{equation}
{\rm tr}\hspace{0.03cm}
\bigl(\hspace{0.03cm}{\mathcal N}_{\hspace{0.02cm}{\bf k}}
\hspace{0.03cm}
T^{\,a^{\prime}_{2}}\hspace{0.03cm}T^{\,e_{1}}\hspace{0.03cm}
T^{\,a_{2}}\hspace{0.01cm}\bigr)^{\ast}
=
{\rm tr}\hspace{0.03cm}
\bigl(\hspace{0.03cm}T^{\,a_{2}}\hspace{0.03cm}T^{\,e_{1}}\hspace{0.03cm}T^{\,a^{\prime}_{2}}{\mathcal N}_{\hspace{0.02cm}{\bf k}}\hspace{0.01cm}
\bigr),
\;
{\rm tr}\hspace{0.03cm}
\bigl(\hspace{0.03cm}{\mathcal N}_{\hspace{0.02cm}{\bf k}}
\hspace{0.03cm}
T^{\,a^{\prime}_{1}}\hspace{0.03cm}T^{\,e_{2}}\hspace{0.03cm}
T^{\,a_{1}}\hspace{0.01cm}\bigr)^{\ast}
=
{\rm tr}\hspace{0.03cm}
\bigl(\hspace{0.03cm}T^{\,a_{1}}\hspace{0.03cm}T^{\,e_{2}}\hspace{0.03cm}T^{\,a^{\prime}_{1}}{\mathcal N}_{\hspace{0.02cm}{\bf k}}\hspace{0.01cm}
\bigr).
\label{eq:7d}
\end{equation}
In general, when using the color decomposition (\ref{eq:7q}), the traces on the left- and right-hand sides here are complex, i.e. they do not satisfy the relations of the form
\begin{equation}
{\rm tr}\hspace{0.03cm}
\bigl(\hspace{0.03cm}T^{\,a_{2}}\hspace{0.03cm}T^{\,e_{1}}\hspace{0.03cm}T^{\,a^{\prime}_{2}}{\mathcal N}_{\hspace{0.02cm}{\bf k}}\hspace{0.01cm}
\bigr)^{\ast}
=
{\rm tr}\hspace{0.03cm}
\bigl(\hspace{0.03cm}T^{\,a_{2}}\hspace{0.03cm}T^{\,e_{1}}\hspace{0.03cm}T^{\,a^{\prime}_{2}}{\mathcal N}_{\hspace{0.02cm}{\bf k}}\hspace{0.01cm}
\bigr),
\;	
{\rm tr}\hspace{0.03cm}
\bigl(\hspace{0.03cm}{\mathcal N}_{\hspace{0.02cm}{\bf k}}
\hspace{0.03cm}
T^{\,a^{\prime}_{2}}\hspace{0.03cm}T^{\,e_{1}}\hspace{0.03cm}
T^{\,a_{2}}\hspace{0.01cm}\bigr)^{\ast}
=
{\rm tr}\hspace{0.03cm}
\bigl(\hspace{0.03cm}{\mathcal N}_{\hspace{0.02cm}{\bf k}}
\hspace{0.03cm}
T^{\,a^{\prime}_{2}}\hspace{0.03cm}T^{\,e_{1}}\hspace{0.03cm}
T^{\,a_{2}}\hspace{0.01cm}\bigr)	
\label{eq:7dd}
\end{equation}	
etc. However, as we will see below, when contracting these traces with the averaged color charges, as is the case in the kinetic equation (\ref{eq:7e}), we get real-valued expressions. This leads to the fact that, by virtue of the properties (\ref{eq:7d}) the following equalities are valid:
\[
{\rm tr}\hspace{0.03cm}
\bigl(\hspace{0.03cm}{\mathcal N}_{\hspace{0.02cm}{\bf k}}
\hspace{0.03cm}
T^{\,a^{\prime}_{2}}\hspace{0.03cm}T^{\,e_{1}}\hspace{0.03cm}
T^{\,a_{2}}\hspace{0.01cm}
\bigr)
\hspace{0.01cm}
\bigl\langle\hspace{0.03cm}\mathcal{Q}^{\hspace{0.03cm}a^{\prime}_{2}}_{2}
\hspace{0.03cm}\bigr\rangle
\hspace{0.03cm}\bigl\langle\hspace{0.03cm}\mathcal{Q}^{\hspace{0.03cm}e_{1}}_{1}
\hspace{0.03cm}\bigr\rangle
\hspace{0.01cm}\bigl\langle\hspace{0.03cm}\mathcal{Q}^{\hspace{0.03cm}a_{2}}_{2}
\hspace{0.03cm}\bigr\rangle
=
{\rm tr}\hspace{0.03cm}
\bigl(\hspace{0.03cm}T^{\,a_{2}}\hspace{0.03cm}T^{\,e_{1}}\hspace{0.03cm}T^{\,a^{\prime}_{2}}{\mathcal N}_{\hspace{0.02cm}{\bf k}}\hspace{0.01cm}
\bigr)
\hspace{0.01cm}\bigl\langle\hspace{0.03cm}\mathcal{Q}^{\hspace{0.03cm}a^{\prime}_{2}}_{2}
\hspace{0.03cm}\bigr\rangle
\hspace{0.03cm}\bigl\langle\hspace{0.03cm}\mathcal{Q}^{\hspace{0.03cm}e_{1}}_{1}
\hspace{0.03cm}\bigr\rangle
\hspace{0.01cm}\bigl\langle\hspace{0.03cm}\mathcal{Q}^{\hspace{0.03cm}a_{2}}_{2}
\hspace{0.03cm}\bigr\rangle,
\]
\[
{\rm tr}\hspace{0.03cm}
\bigl(\hspace{0.03cm}{\mathcal N}_{\hspace{0.02cm}{\bf k}}
\hspace{0.03cm}
T^{\,a^{\prime}_{1}}\hspace{0.03cm}T^{\,e_{2}}\hspace{0.03cm}
T^{\,a_{1}}\hspace{0.01cm}
\bigr)
\hspace{0.01cm}
\bigl\langle\hspace{0.03cm}\mathcal{Q}^{\hspace{0.03cm}a^{\prime}_{1}}_{1}
\hspace{0.03cm}\bigr\rangle
\hspace{0.03cm}\bigl\langle\hspace{0.03cm}\mathcal{Q}^{\hspace{0.03cm}e_{2}}_{2}
\hspace{0.03cm}\bigr\rangle
\hspace{0.01cm}\bigl\langle\hspace{0.03cm}\mathcal{Q}^{\hspace{0.03cm}a_{1}}_{1}
\hspace{0.03cm}\bigr\rangle
=
{\rm tr}\hspace{0.03cm}
\bigl(\hspace{0.03cm}T^{\,a_{1}}\hspace{0.03cm}T^{\,e_{2}}\hspace{0.03cm}T^{\,a^{\prime}_{1}}{\mathcal N}_{\hspace{0.02cm}{\bf k}}\hspace{0.01cm}
\bigr)
\hspace{0.01cm}\bigl\langle\hspace{0.03cm}\mathcal{Q}^{\hspace{0.03cm}a^{\prime}_{1}}_{1}
\hspace{0.03cm}\bigr\rangle
\hspace{0.03cm}\bigl\langle\hspace{0.03cm}\mathcal{Q}^{\hspace{0.03cm}e_{2}}_{2}
\hspace{0.03cm}\bigr\rangle
\hspace{0.01cm}\bigl\langle\hspace{0.03cm}\mathcal{Q}^{\hspace{0.03cm}a_{1}}_{1}
\hspace{0.03cm}\bigr\rangle.	
\]
Similar relations hold for the expressions in (\ref{eq:7e}) with another color matrix function ${\mathcal W}_{\hspace{0.02cm}{\bf k}}$. This in turn with accounting made for representations (\ref{eq:7a}), (\ref{eq:7s}) and the relation (\ref{eq:7d}) enables us to bring the first two terms on the right-hand side of (\ref{eq:7e}) to the following form:
\begin{equation}
\sum_{\rho}\,\biggl[{T}^{\hspace{0.03cm}(\rho)}_{\; {\bf k}}(t)
\,\biggl(\int{T}^{\,\ast\hspace{0.03cm}(\rho)}_{\; {\bf k}}(t)\hspace{0.03cm}dt\biggr)
+
{T}^{\,\ast\hspace{0.03cm}({\rho})}_{\; {\bf k}}(t)
\,
\biggl(\int{T}^{\hspace{0.03cm}(\rho)}_{\; {\bf k}}(t)\hspace{0.03cm}dt\biggr)
\biggr]\hspace{0.03cm}\times
\label{eq:7f}
\end{equation}
\[
\Bigl[
{\rm tr}\hspace{0.03cm}
\bigl(\hspace{0.03cm}T^{\,a_{2}}\hspace{0.03cm}T^{\,e_{1}}\hspace{0.03cm}T^{\,a^{\prime}_{2}}{\mathcal N}_{\hspace{0.02cm}{\bf k}}\hspace{0.01cm}
\bigr)
\hspace{0.01cm}\bigl\langle\hspace{0.03cm}\mathcal{Q}^{\hspace{0.03cm}a^{\prime}_{2}}_{2}
\hspace{0.03cm}\bigr\rangle
\hspace{0.03cm}\bigl\langle\hspace{0.03cm}\mathcal{Q}^{\hspace{0.03cm}e_{1}}_{1}
\hspace{0.03cm}\bigr\rangle
\hspace{0.01cm}\bigl\langle\hspace{0.03cm}\mathcal{Q}^{\hspace{0.03cm}a_{2}}_{2}
\hspace{0.03cm}\bigr\rangle
+
{\rm tr}\hspace{0.03cm}
\bigl(\hspace{0.03cm}T^{\,a_{1}}\hspace{0.03cm}T^{\,e_{2}}\hspace{0.03cm}T^{\,a^{\prime}_{1}}{\mathcal N}_{\hspace{0.02cm}{\bf k}}\hspace{0.01cm}
\bigr)
\hspace{0.01cm}\bigl\langle\hspace{0.03cm}\mathcal{Q}^{\hspace{0.03cm}a^{\prime}_{1}}_{1}
\hspace{0.03cm}\bigr\rangle
\hspace{0.03cm}\bigl\langle\hspace{0.03cm}\mathcal{Q}^{\hspace{0.03cm}e_{2}}_{2}
\hspace{0.03cm}\bigr\rangle
\hspace{0.01cm}\bigl\langle\hspace{0.03cm}\mathcal{Q}^{\hspace{0.03cm}a_{1}}_{1}
\hspace{0.03cm}\bigr\rangle	
\Bigr]
=
\]
\[
-\!\int\!d\hspace{0.02cm}{\bf q}\,d\hspace{0.02cm}{\bf q}'\;
{T}^{\hspace{0.03cm}(\rho)}_{\; {\bf k},\,{\bf q}}\,
{T}^{\,\ast\hspace{0.03cm}(\rho)}_{\; {\bf k},\,{\bf q}'}\,
{\rm e}^{i\hspace{0.03cm}\,(-1)^{\rho}\,\Delta{\mathbf v}\cdot({\mathbf q} - {\mathbf q}')\hspace{0.02cm}t}
\hspace{0.04cm}
{\rm e}^{i\,(-1)^{\rho}\,({\mathbf x}_{0\hspace{0.02cm}1} - {\mathbf x}_{0\hspace{0.02cm}2})\cdot({\bf q} - {\mathbf q}')}\,\times
\]
\[
\biggl(
\frac{i}{\omega^{l}_{{\bf k}} - {\bf k}\cdot{\bf v}_{\rho}
-(-1)^{\rho}\, \Delta{\mathbf v}\cdot{\mathbf q}' + i0}
\,-\,
\frac{i}{\omega^{l}_{{\bf k}} - {\bf k}\cdot{\bf v}_{\rho}
-(-1)^{\rho}\, \Delta{\mathbf v}\cdot{\mathbf q} - i0}
\biggr)\times
\]
\[
\Bigl[
{\rm tr}\hspace{0.03cm}
\bigl(\hspace{0.03cm}T^{\,a_{2}}\hspace{0.03cm}T^{\,e_{1}}\hspace{0.03cm}T^{\,a^{\prime}_{2}}{\mathcal N}_{\hspace{0.02cm}{\bf k}}\hspace{0.01cm}
\bigr)
\hspace{0.01cm}\bigl\langle\hspace{0.03cm}\mathcal{Q}^{\hspace{0.03cm}a^{\prime}_{2}}_{2}
\hspace{0.03cm}\bigr\rangle
\hspace{0.03cm}\bigl\langle\hspace{0.03cm}\mathcal{Q}^{\hspace{0.03cm}e_{1}}_{1}
\hspace{0.03cm}\bigr\rangle
\hspace{0.01cm}\bigl\langle\hspace{0.03cm}\mathcal{Q}^{\hspace{0.03cm}a_{2}}_{2}
\hspace{0.03cm}\bigr\rangle
+
{\rm tr}\hspace{0.03cm}
\bigl(\hspace{0.03cm}T^{\,a_{1}}\hspace{0.03cm}T^{\,e_{2}}\hspace{0.03cm}T^{\,a^{\prime}_{1}}{\mathcal N}_{\hspace{0.02cm}{\bf k}}\hspace{0.01cm}
\bigr)
\hspace{0.01cm}\bigl\langle\hspace{0.03cm}\mathcal{Q}^{\hspace{0.03cm}a^{\prime}_{1}}_{1}
\hspace{0.03cm}\bigr\rangle
\hspace{0.03cm}\bigl\langle\hspace{0.03cm}\mathcal{Q}^{\hspace{0.03cm}e_{2}}_{2}
\hspace{0.03cm}\bigr\rangle
\hspace{0.01cm}\bigl\langle\hspace{0.03cm}\mathcal{Q}^{\hspace{0.03cm}a_{1}}_{1}
\hspace{0.03cm}\bigr\rangle	
\Bigr].
\]
A completely similar expression is true for terms in (\ref{eq:7e}) with the matrix function ${\mathcal W}_{\hspace{0.02cm}{\bf k}}$.\\
\indent For the difference of two terms in parentheses on the right-hand side of (\ref{eq:7f}), we cannot directly use Sokhotsky's formula
\begin{equation}
\frac{i}{\omega^{l}_{{\bf k}} - {\bf k}\cdot{\bf v}_{\rho}
	-(-1)^{\rho}\, \Delta{\mathbf v}\cdot{\mathbf q} + i0}
\,-\,
\frac{i}{\omega^{l}_{{\bf k}} - {\bf k}\cdot{\bf v}_{\rho}
	-(-1)^{\rho}\, \Delta{\mathbf v}\cdot{\mathbf q} - i0}
=
\label{eq:7g}
\end{equation}
\[
2\hspace{0.02cm}\pi\hspace{0.02cm}\delta(\omega^{l}_{{\bf k}} - {\bf k}\cdot{\bf v}_{\rho}
-(-1)^{\rho}\, \Delta{\mathbf v}\cdot{\mathbf q}),
\]
since the first term there depends on the variable ${\mathbf q}'$, while the second one depends on ${\mathbf q}$. However, the relation (\ref{eq:7g}) is necessary to obtain the desired kinetic equation. The bremsstrahlung probability must contain a $\delta$-function which describes the conservation of energy in an elementary bremsstrahlung event in the quasiclassical  limit\footnote{\hspace{0.03cm}The conservation law
\begin{equation}
\omega^{l}_{{\bf k}} - {\bf k}\cdot{\bf v}_{\rho}
-(-1)^{\rho}\, \Delta{\mathbf v}\cdot{\mathbf q} = 0
\label{eq:7h} 
\end{equation}
can be derived from the following simple reasoning. Indeed, let the particles 1 and 2 have had momenta ${\mathbf p}_{1}$ and ${\mathbf p}_{2}$ before the collision. As a result of the collision, the momentum ${\mathbf q}$ was transferred from particle 2 to particle 1, and in addition, 
a wave with energy $\omega^{l}_{\mathbf k}$ and momentum ${\mathbf k}$ was emitted (this wave corresponds to the bremsstrahlung quantum). 	
The final momenta after the collision of the particles are ${\mathbf p}^{\prime}_{1}$ and ${\mathbf p}^{\prime}_{2}$:
\[
{\mathbf p}^{\prime}_{1} = {\mathbf p}_{1} + {\mathbf q} - {\mathbf k},
\quad
{\mathbf p}^{\prime}_{2} = {\mathbf p}_{2} - {\mathbf q}.
\] 
The conservation energy law has the form
\[
\varepsilon({\mathbf p}_{1}) + \varepsilon({\mathbf p}_{2})
=
\varepsilon({\mathbf p}^{\prime}_{1}) + \varepsilon({\mathbf p}^{\prime}_{2})
+
\omega^{l}_{\mathbf k},
\] 
where $\varepsilon({\mathbf p})$ is energy of the incoming or outgoing hard particles. From here in the limit $|{\mathbf p}_{\rho}|\gg |{\mathbf k}|,\,|{\mathbf q}|,\,\rho = 1,2$, we result in (\ref{eq:7h}) for the value $\rho = 1$. If the radiation of the bremsstrahlung quantum was associated with particle 2, then we would arrive at (\ref{eq:7h}) for the value $\rho = 2$. 
}.\\
\indent To obtain relation (\ref{eq:7g}) we use the product of two exponential factors in the integrand on the right-hand side of (\ref{eq:7f}). The plasmon number density ${\mathcal N}_{\hspace{0.02cm}{\bf k}}$ and the averaged color charges generally change slowly compared with the exponential oscillatory factors. We should take the average over these oscillatory factors.
The first step is to integrate over $t$ in the first exponential factor that gives
\begin{equation}
\int\limits^{+\infty}_{-\infty}\!d\hspace{0.02cm}t\,
{\rm e}^{i\hspace{0.03cm}\,(-1)^{\rho}\,\Delta{\mathbf v}\cdot({\mathbf q} - {\mathbf q}')\hspace{0.02cm}t}
=
2\hspace{0.02cm}\pi\hspace{0.02cm}\delta(\Delta{\mathbf v}\cdot({\mathbf q} - {\mathbf q}'))
=
2\hspace{0.02cm}\pi\,\frac{1}{|\Delta{\mathbf v}|}\,
\delta({q}_{\hspace{0.02cm}\parallel} - {q}^{\prime}_{\hspace{0.02cm}\parallel}),
\label{eq:7j} 
\end{equation}
where ${q}_{\hspace{0.02cm}\parallel}$ (and ${q}^{\prime}_{\hspace{0.02cm}\parallel}$) is the longitudinal component of the momentum transfer.\\
Further in the second exponential factor without loss of generality one can set ${\mathbf x}_{0\hspace{0.02cm}1}$ = 0, and choose the vector ${\mathbf x}_{0\hspace{0.02cm}2} = ({\mathbf b},z_{0\hspace{0.02cm}2})$, where two-dimensional vector ${\mathbf b}$ is orthogonal to the relative velocity $\Delta{\mathbf v} = {\mathbf v}_1 - {\mathbf v}_{2}$. Besides in the subsequent discussion the longitudinal component $z_{0\hspace{0.02cm}2}$ also does not play any role and thus it can be set equal to zero. As a result, here we have	 
\begin{equation}
{\rm e}^{\hspace{0.02cm}i\,(-1)^{\rho}\,({\mathbf x}_{0\hspace{0.02cm}1} - {\mathbf x}_{0\hspace{0.02cm}2})\cdot({\bf q} - {\mathbf q}')}
=
{\rm e}^{\hspace{0.02cm}i\,(-1)^{\rho}\,{\mathbf b}\cdot({\bf q} - {\mathbf q}')_{\perp}}.
\label{eq:7jj} 
\end{equation}
The vector ${\mathbf b}$ plays the role of an impact parameter. Thus the scalar functions $N^{\hspace{0.03cm}(i)}_{\bf k}$ and $W^{\hspace{0.03cm}(i)}_{\bf k},\,i = 1,2$, in the decompositions (\ref{eq:7q}) do not depend only on the slow time $\tau$ and the wave vector ${\mathbf k}$, but also on the two-dimensional vector ${\mathbf b}$. Instead of these functions, we introduce their ``averaged'' versions over the vector ${\mathbf b}$, while keeping the same notations
\begin{equation}
N^{\hspace{0.03cm}(i)}_{\bf k}(\tau) 
\equiv 
\int\!d\hspace{0.03cm}{\mathbf b}\,
N^{\hspace{0.03cm}(i)}_{\bf k}(\tau; {\mathbf b}),
\qquad
W^{\hspace{0.03cm}(i)}_{\bf k}(\tau) 
\equiv 
\int\!d\hspace{0.03cm}{\mathbf b}\,
W^{\hspace{0.03cm}(i)}_{\bf k}(\tau; {\mathbf b}).
\label{eq:7k} 
\end{equation}
We perform integration over the impact parameter ${\mathbf b}$, taking into account the expression (\ref{eq:7jj}) and the definition (\ref{eq:7k}) and using the approximations 
\begin{equation}
\begin{split}
&\int\!d\hspace{0.03cm}{\mathbf b}\,
{\rm e}^{\hspace{0.02cm}i\,(-1)^{\rho}\,{\mathbf b}\cdot({\bf q} - {\mathbf q}')_{\perp}}
\hspace{0.03cm}
{\mathcal N}_{{\bf k}}(\tau;{\mathbf b})
\simeq
(2\hspace{0.02cm}\pi)^{2}\,
\delta^{(2)}(({\mathbf q} - {\mathbf q}^{\prime})_{\hspace{0.02cm}\perp})\,
{\mathcal N}_{{\bf k}}(\tau),\\[1ex]
&\int\!d\hspace{0.03cm}{\mathbf b}\,
{\rm e}^{\hspace{0.02cm}i\,(-1)^{\rho}\,{\mathbf b}\cdot({\bf q} - {\mathbf q}')_{\perp}}
\hspace{0.03cm}
{\mathcal W}_{{\bf k}}(\tau;{\mathbf b})
\simeq
(2\hspace{0.02cm}\pi)^{2}\,
\delta^{(2)}(({\mathbf q} - {\mathbf q}^{\prime})_{\hspace{0.02cm}\perp})\,
{\mathcal W}_{{\bf k}}(\tau).
\end{split}
\label{eq:7l} 
\end{equation}
Eqs.\,(\ref{eq:7j}) and (\ref{eq:7l}) enables us to perform complete integration with respect to ${\bf q}'$ in (\ref{eq:7f}). In the end, we obtain instead of (\ref{eq:7e}), 
\begin{align}
d_{A}\hspace{0.04cm}\frac{\partial\hspace{0.01cm}N^{\hspace{0.03cm}(1)}_{\bf k}}{\!\!\partial\hspace{0.03cm}\tau}
=
&\frac{(2\hspace{0.02cm}\pi)^{3}}{2\hspace{0.03cm}|\Delta{\mathbf v}|}\,\sum_{\rho}\,
\int\!d\hspace{0.02cm}{\bf q}\;
\bigl|\hspace{0.03cm}{T}^{\hspace{0.03cm}(\rho)}_{\; {\bf k},\,{\bf q}}\bigr|^{\hspace{0.02cm}2}\,
2\hspace{0.02cm}\pi\hspace{0.02cm}
\delta(\omega^{l}_{{\bf k}} - {\bf k}\cdot{\bf v}_{\rho}
-(-1)^{\rho}\, \Delta{\mathbf v}\cdot{\mathbf q}) 
\,\times
\vspace{-0.5cm}
\label{eq:7z}\\[1ex] 
&{\hspace{-1.8cm}}\Bigl[\hspace{0.03cm}
{\rm tr}\hspace{0.03cm}
\bigl(\hspace{0.03cm}T^{\,a_{2}}\hspace{0.03cm}T^{\,e_{1}}\hspace{0.03cm}T^{\,a^{\prime}_{2}}{\mathcal N}_{\hspace{0.02cm}{\bf k}}\hspace{0.01cm}
\bigr)
\hspace{0.01cm}\bigl\langle\hspace{0.03cm}\mathcal{Q}^{\hspace{0.03cm}a^{\prime}_{2}}_{2}
\hspace{0.03cm}\bigr\rangle
\hspace{0.03cm}\bigl\langle\hspace{0.03cm}\mathcal{Q}^{\hspace{0.03cm}e_{1}}_{1}
\hspace{0.03cm}\bigr\rangle
\hspace{0.01cm}\bigl\langle\hspace{0.03cm}\mathcal{Q}^{\hspace{0.03cm}a_{2}}_{2}
\hspace{0.03cm}\bigr\rangle
\,+\,
{\rm tr}\hspace{0.03cm}
\bigl(\hspace{0.03cm}T^{\,a_{1}}\hspace{0.03cm}T^{\,e_{2}}\hspace{0.03cm}T^{\,a^{\prime}_{1}}{\mathcal N}_{\hspace{0.02cm}{\bf k}}\hspace{0.01cm}
\bigr)
\hspace{0.01cm}\bigl\langle\hspace{0.03cm}\mathcal{Q}^{\hspace{0.03cm}a^{\prime}_{1}}_{1}
\hspace{0.03cm}\bigr\rangle
\hspace{0.03cm}\bigl\langle\hspace{0.03cm}\mathcal{Q}^{\hspace{0.03cm}e_{2}}_{2}
\hspace{0.03cm}\bigr\rangle
\hspace{0.01cm}\bigl\langle\hspace{0.03cm}\mathcal{Q}^{\hspace{0.03cm}a_{1}}_{1}
\hspace{0.03cm}\bigr\rangle	
\Bigr]\,+
\notag\\[1.5ex]
&\frac{(2\hspace{0.02cm}\pi)^{3}}{2\hspace{0.03cm}|\Delta{\mathbf v}|}\,\sum_{\rho}\hspace{0.02cm}(-1)^{\rho + 1}\!
\int\!d\hspace{0.02cm}{\bf q}\;
\bigl|\hspace{0.03cm}{T}^{\hspace{0.03cm}(\rho)}_{\; {\bf k},\,{\bf q}}\bigr|^{\hspace{0.02cm}2}\,
2\hspace{0.02cm}\pi\hspace{0.02cm}
\delta(\omega^{l}_{{\bf k}} - {\bf k}\cdot{\bf v}_{\rho}
-(-1)^{\rho}\, \Delta{\mathbf v}\cdot{\mathbf q}) 
\,\times
\notag\\[1.5ex]
&{\hspace{-1.75cm}}
\Bigl[\hspace{0.03cm}
{\rm tr}\hspace{0.03cm}
\bigl(\hspace{0.03cm}T^{\,a_{2}}\hspace{0.03cm}T^{\,e_{1}}\hspace{0.03cm}T^{\,a^{\prime}_{2}}\hspace{0.03cm}{\mathcal W}_{\hspace{0.02cm}{\bf k}}\hspace{0.01cm}
\bigr)
\hspace{0.01cm}\bigl\langle\hspace{0.03cm}\mathcal{Q}^{\hspace{0.03cm}a^{\prime}_{2}}_{2}
\hspace{0.03cm}\bigr\rangle
\hspace{0.03cm}\bigl\langle\hspace{0.03cm}\mathcal{Q}^{\hspace{0.03cm}e_{1}}_{1}
\hspace{0.03cm}\bigr\rangle
\hspace{0.01cm}\bigl\langle\hspace{0.03cm}\mathcal{Q}^{\hspace{0.03cm}a_{2}}_{2}
\hspace{0.03cm}\bigr\rangle
\,+\,
{\rm tr}\hspace{0.03cm}
\bigl(\hspace{0.03cm}T^{\,a_{1}}\hspace{0.03cm}T^{\,e_{2}}\hspace{0.03cm}T^{\,a^{\prime}_{1}}\hspace{0.03cm}{\mathcal W}_{\hspace{0.02cm}{\bf k}}\hspace{0.01cm}
\bigr)
\hspace{0.01cm}\bigl\langle\hspace{0.03cm}\mathcal{Q}^{\hspace{0.03cm}a^{\prime}_{1}}_{1}
\hspace{0.03cm}\bigr\rangle
\hspace{0.03cm}\bigl\langle\hspace{0.03cm}\mathcal{Q}^{\hspace{0.03cm}e_{2}}_{2}
\hspace{0.03cm}\bigr\rangle
\hspace{0.01cm}\bigl\langle\hspace{0.03cm}\mathcal{Q}^{\hspace{0.03cm}a_{1}}_{1}
\hspace{0.03cm}\bigr\rangle	
\Bigr].
\notag
\end{align}
\indent All that is left to do, is to explicitly calculate the color traces. Using the color decomposition (\ref{eq:7q}) and the formulae for the traces of the product of three and four color matrices (generators) in the adjoint representation in Appendix \ref{appendix_D}, Eqs.\,(\ref{ap:D3}) and (\ref{ap:D4}), we easily find
\begin{equation}
{\rm tr}\hspace{0.03cm}
\bigl(\hspace{0.03cm}T^{\,a_{2}}\hspace{0.03cm}T^{\,e_{1}}\hspace{0.03cm}T^{\,a^{\prime}_{2}}{\mathcal N}_{\hspace{0.02cm}{\bf k}}\hspace{0.01cm}\bigr)
=
\label{eq:7_1z} 
\end{equation}
\[
N^{\hspace{0.03cm}(1)}_{\bf k}\hspace{0.03cm}
{\rm tr}\hspace{0.03cm}
\bigl(\hspace{0.03cm}T^{\,a_{2}}\hspace{0.03cm}T^{\,e_{1}}\hspace{0.03cm}T^{\,a^{\prime}_{2}}\hspace{0.01cm}\bigr)
+
\bigl\langle\hspace{0.03cm}\mathcal{Q}^{\hspace{0.03cm}c_{1}}_{1}
\hspace{0.03cm}\bigr\rangle	N^{\hspace{0.03cm}(2)}_{\bf k}\hspace{0.03cm}
{\rm tr}\hspace{0.03cm}
\bigl(\hspace{0.03cm}T^{\,a_{2}}\hspace{0.03cm}T^{\,e_{1}}\hspace{0.03cm}T^{\,a^{\prime}_{2}}\hspace{0.03cm}T^{\,c_{1}}\hspace{0.01cm}\bigr)
=
\frac{i}{2}\,N^{\hspace{0.03cm}(1)}_{\bf k}\hspace{0.03cm}N_{c}\hspace{0.03cm}
f^{\hspace{0.03cm}a_{2}\,e_{1}\hspace{0.03cm}a^{\prime}_{2}}
\,+
\]
\[
\bigl\langle\hspace{0.03cm}\mathcal{Q}^{\hspace{0.03cm}c_{1}}_{1}
\hspace{0.03cm}\bigr\rangle	N^{\hspace{0.03cm}(2)}_{\bf k}\hspace{0.03cm}
\Bigl\{\delta^{\hspace{0.02cm}a_{2}\hspace{0.02cm}c_{1}}\hspace{0.03cm}\delta^{\hspace{0.02cm}e_{1}\hspace{0.03cm}a^{\prime}_{2}}
+
\frac{1}{2}\,\bigl(\hspace{0.02cm}
\delta^{\hspace{0.02cm}a_{2}\hspace{0.02cm}e_{1}}\hspace{0.03cm}\delta^{\hspace{0.02cm}a^{\prime}_{2}\hspace{0.03cm}c_{1}}
+
\delta^{\hspace{0.02cm}a_{2}\hspace{0.02cm}a^{\prime}_{2}}\hspace{0.03cm}
\delta^{\hspace{0.02cm}e_{1}\hspace{0.03cm}c_{1}}\hspace{0.02cm}\bigr)
+
\frac{1}{4}\,N_{c}\hspace{0.03cm}\bigl(\hspace{0.02cm}
f^{\hspace{0.03cm}a_{2}\hspace{0.03cm}c_{1}\hspace{0.03cm}e}
\hspace{0.01cm}f^{\hspace{0.03cm}e_{1}\hspace{0.03cm}a^{\prime}_{2}\hspace{0.03cm}e}
\!+
d^{\hspace{0.04cm}a_{2}\hspace{0.03cm}c_{1}\hspace{0.03cm}e}
\hspace{0.02cm}d^{\hspace{0.04cm}e_{1}\hspace{0.03cm}a^{\prime}_{2}\hspace{0.03cm}e}
\hspace{0.02cm}\bigr)\Bigr\},
\]
where $d^{\hspace{0.04cm}a\hspace{0.03cm}b\hspace{0.03cm}c}$ are the totally symmetric structure constants in the $\mathfrak{su}(N_{c})$ Lie algebra. As mentioned above, this trace is a complex quantity, i.e. it does not satisfy the first relation in (\ref{eq:7dd}). However, this trace enters into the kinetic equation (\ref{eq:7z}) in the form of contraction with the color charges
\[
{\rm tr}\hspace{0.03cm}
\bigl(\hspace{0.03cm}T^{\,a_{2}}\hspace{0.03cm}T^{\,e_{1}}\hspace{0.03cm}T^{\,a^{\prime}_{2}}{\mathcal N}_{\hspace{0.02cm}{\bf k}}\hspace{0.01cm}
\bigr)
\hspace{0.01cm}\bigl\langle\hspace{0.03cm}\mathcal{Q}^{\hspace{0.03cm}a^{\prime}_{2}}_{2}
\hspace{0.03cm}\bigr\rangle
\hspace{0.03cm}\bigl\langle\hspace{0.03cm}\mathcal{Q}^{\hspace{0.03cm}e_{1}}_{1}
\hspace{0.03cm}\bigr\rangle
\hspace{0.01cm}\bigl\langle\hspace{0.03cm}\mathcal{Q}^{\hspace{0.03cm}a_{2}}_{2}
\hspace{0.03cm}\bigr\rangle,
\]
and therefore, considering the antisymmetry of the structure constants $f^{\hspace{0.03cm}a_{2}\,e_{1}\hspace{0.03cm}a^{\prime}_{2}}$, we see that the imaginary part of (\ref{eq:7_1z}) vanishes. For the real part, we get  
\begin{equation}
{\rm tr}\hspace{0.03cm}
\bigl(\hspace{0.03cm}T^{\,a_{2}}\hspace{0.03cm}T^{\,e_{1}}\hspace{0.03cm}T^{\,a^{\prime}_{2}}{\mathcal N}_{\hspace{0.02cm}{\bf k}}\hspace{0.01cm}
\bigr)
\hspace{0.01cm}\bigl\langle\hspace{0.03cm}\mathcal{Q}^{\hspace{0.03cm}a^{\prime}_{2}}_{2}
\hspace{0.03cm}\bigr\rangle
\hspace{0.03cm}\bigl\langle\hspace{0.03cm}\mathcal{Q}^{\hspace{0.03cm}e_{1}}_{1}
\hspace{0.03cm}\bigr\rangle
\hspace{0.01cm}\bigl\langle\hspace{0.03cm}\mathcal{Q}^{\hspace{0.03cm}a_{2}}_{2}
\hspace{0.03cm}\bigr\rangle
=
\biggl\{\frac{3}{2}\,{\mathfrak q}^{\hspace{0.03cm}2}_{12} 
\,+\, 
\frac{1}{2}\,{\mathfrak q}_{1}\hspace{0.03cm}{\mathfrak q}_{2} 
\,+\, 
\frac{1}{4}\,N_{c}\hspace{0.03cm}\bigl(\hspace{0.02cm} - 
\Lambda^{2} + \Omega^{\hspace{0.03cm}2}_{12}\bigl)\biggr\}
N^{\hspace{0.03cm}(2)}_{\bf k}.
\label{eq:7zz} 
\end{equation}
Here, we set $\Lambda^{2} \equiv \Lambda^{e}\Lambda^{e},\, \Omega^{\hspace{0.03cm}2}_{12}
\equiv \Omega^{\hspace{0.03cm}e}_{12}\,\Omega^{\hspace{0.03cm}e}_{12}$.
Similarly, for the second trace in (\ref{eq:7z}) we obtain
\begin{equation}
{\rm tr}\hspace{0.03cm}
\bigl(\hspace{0.03cm}T^{\,a_{1}}\hspace{0.03cm}T^{\,e_{2}}\hspace{0.03cm}T^{\,a^{\prime}_{1}}{\mathcal N}_{\hspace{0.02cm}{\bf k}}\hspace{0.01cm}
\bigr)
\hspace{0.01cm}\bigl\langle\hspace{0.03cm}\mathcal{Q}^{\hspace{0.03cm}a^{\prime}_{1}}_{1}
\hspace{0.03cm}\bigr\rangle
\hspace{0.03cm}\bigl\langle\hspace{0.03cm}\mathcal{Q}^{\hspace{0.03cm}e_{2}}_{2}
\hspace{0.03cm}\bigr\rangle
\hspace{0.01cm}\bigl\langle\hspace{0.03cm}\mathcal{Q}^{\hspace{0.03cm}a_{1}}_{1}
\hspace{0.03cm}\bigr\rangle	
=
\biggl\{2\hspace{0.03cm}{\mathfrak q}_{1}\hspace{0.03cm}{\mathfrak q}_{12} + \frac{1}{4}\,N_{c}\,\Omega^{\hspace{0.03cm}e}_{11}
\hspace{0.03cm}\Omega^{\hspace{0.03cm}e}_{12}\bigl)\biggr\}
N^{\hspace{0.03cm}(2)}_{\bf k}.
\hspace{2.2cm}
\label{eq:7zzz} 
\end{equation}
On the right-hand sides of Eqs.\,(\ref{eq:7zz}) and (\ref{eq:7zzz}) we have introduced, by the definition, the following functions of the slow time $\tau$:
\begin{equation}
{\mathfrak q}_{1} \equiv 
\hspace{0.03cm}\bigl\langle\hspace{0.03cm}\mathcal{Q}^{\hspace{0.03cm}e}_{1}
\hspace{0.03cm}\bigr\rangle
\bigl\langle\hspace{0.03cm}\mathcal{Q}^{\hspace{0.03cm}e}_{1}
\hspace{0.03cm}\bigr\rangle,
\quad
{\mathfrak q}_{2} \equiv 
\hspace{0.03cm}\bigl\langle\hspace{0.03cm}\mathcal{Q}^{\hspace{0.03cm}e}_{2}
\hspace{0.03cm}\bigr\rangle
\bigl\langle\hspace{0.03cm}\mathcal{Q}^{\hspace{0.03cm}e}_{2}
\hspace{0.03cm}\bigr\rangle,
\quad
{\mathfrak q}_{12} \equiv 
\hspace{0.03cm}\bigl\langle\hspace{0.03cm}\mathcal{Q}^{\hspace{0.03cm}e}_{1}
\hspace{0.03cm}\bigr\rangle
\bigl\langle\hspace{0.03cm}\mathcal{Q}^{\hspace{0.03cm}e}_{2}
\hspace{0.03cm}\bigr\rangle,
\vspace{-0.3cm}
\label{eq:7w}
\end{equation}
\begin{align}
&\Lambda^{c} =  \Lambda^{c}(\tau)
\equiv 
f^{\hspace{0.03cm}c\,b_{1}\hspace{0.03cm}b_{2}\hspace{0.03cm}}
\bigl\langle\hspace{0.03cm}\mathcal{Q}^{\,b_{1}}_{\hspace{0.03cm}1}
(\tau)
\hspace{0.03cm}\bigr\rangle
\bigl\langle\hspace{0.03cm}\mathcal{Q}^{\,b_{2}}_{\hspace{0.03cm}2}
(\tau)
\hspace{0.03cm}\bigr\rangle,
\notag\\[1ex]
&\Omega^{\hspace{0.03cm}c}_{11} = \Omega^{\hspace{0.03cm}c}_{11}(\tau)
\equiv 
d^{\,c\,b_{1}\hspace{0.03cm}b^{\prime}_{1}\hspace{0.03cm}}
\bigl\langle\hspace{0.03cm}\mathcal{Q}^{\,b_{1}}_{\hspace{0.03cm}1}
(\tau)
\hspace{0.03cm}\bigr\rangle
\bigl\langle\hspace{0.03cm}\mathcal{Q}^{\,b^{\prime}_{1}}_{\hspace{0.03cm}1}
(\tau)
\hspace{0.03cm}\bigr\rangle,
\notag\\[1ex]
&\Omega^{\hspace{0.03cm}c}_{22} = \Omega^{\hspace{0.03cm}c}_{22}(\tau)
\equiv 
d^{\,c\,b_{2}\hspace{0.03cm}b^{\prime}_{2}\hspace{0.03cm}}
\bigl\langle\hspace{0.03cm}\mathcal{Q}^{\,b_{2}}_{\hspace{0.03cm}2}
(\tau)
\hspace{0.03cm}\bigr\rangle
\bigl\langle\hspace{0.03cm}\mathcal{Q}^{\,b^{\prime}_{2}}_{\hspace{0.03cm}2}
(\tau)
\hspace{0.03cm}\bigr\rangle,
\notag\\[1ex]
&\Omega^{\hspace{0.03cm}c}_{12} = \Omega^{\hspace{0.03cm}c}_{12}(\tau)
\equiv 
d^{\,c\,b_{1}\hspace{0.03cm}b^{\prime}_{2}\hspace{0.03cm}}
\bigl\langle\hspace{0.03cm}\mathcal{Q}^{\,b_{1}}_{\hspace{0.03cm}1}
(\tau)
\hspace{0.03cm}\bigr\rangle
\bigl\langle\hspace{0.03cm}\mathcal{Q}^{\,b^{\prime}_{2}}_{\hspace{0.03cm}2}
(\tau)
\hspace{0.03cm}\bigr\rangle.
\notag
\end{align}
We see that in the case of an arbitrary value of $N_{c}$, the number of color structures associated with the color charges is too large. To minimize their number, we will consider specific values for $N_{c}$. For the ``trivial'' case $\mathfrak{su}(2_{\hspace{0.02cm}c})$, when $d^{\,a\,b\hspace{0.03cm}c}\! \equiv 0$ the last three color structure in (\ref{eq:7w}) vanish. However, as is shown in Appendix \ref{appendix_E}, it is possible to completely get rid of these color structures (or rather, from their colorless combinations such as $\Omega^{\hspace{0.03cm}2}_{12}
= \Omega^{\hspace{0.03cm}e}_{12}\hspace{0.03cm}\Omega^{\hspace{0.03cm}e}_{12}$ and $\Omega^{\hspace{0.03cm}e}_{11}\hspace{0.03cm}\Omega^{\hspace{0.03cm}e}_{12}$)  for another, more physically important case of the $\mathfrak{su}(3_{c})$ Lie algebra. For $N_{c} = 3$, by virtue of the relations  (\ref{ap:E3}) and (\ref{ap:E5}), the following two equalities are true
\[
\Omega^{\hspace{0.03cm}2}_{12}
=
\frac{1}{3}\,
\bigl(
{\mathfrak q}_{1}\hspace{0.02cm}{\mathfrak q}_{2} - \Lambda^{2}
\hspace{0.03cm}\bigr)
\quad\mbox{and}\quad
\Omega^{\hspace{0.03cm}e}_{11}\hspace{0.03cm}\Omega^{\hspace{0.03cm}e}_{12}	
=
\frac{1}{3}\,{\mathfrak q}_{1}\hspace{0.03cm}{\mathfrak q}_{12}.
\]
With these equalities at hand we have, instead of the expressions (\ref{eq:7zz}) and (\ref{eq:7zzz}), 
\begin{equation}
\begin{split}
&{\rm tr}\hspace{0.03cm}
\bigl(\hspace{0.03cm}T^{\,a_{2}}\hspace{0.03cm}T^{\,e_{1}}\hspace{0.03cm}T^{\,a^{\prime}_{2}}{\mathcal N}_{\hspace{0.02cm}{\bf k}}\hspace{0.01cm}
\bigr)
\hspace{0.01cm}\bigl\langle\hspace{0.03cm}\mathcal{Q}^{\hspace{0.03cm}a^{\prime}_{2}}_{2}
\hspace{0.03cm}\bigr\rangle
\hspace{0.03cm}\bigl\langle\hspace{0.03cm}\mathcal{Q}^{\hspace{0.03cm}e_{1}}_{1}
\hspace{0.03cm}\bigr\rangle
\hspace{0.01cm}\bigl\langle\hspace{0.03cm}\mathcal{Q}^{\hspace{0.03cm}a_{2}}_{2}
\hspace{0.03cm}\bigr\rangle
=
\biggl\{\frac{3}{2}\,{\mathfrak q}^{\hspace{0.03cm}2}_{12} \,+\, \frac{3}{4}\,{\mathfrak q}_{1}\hspace{0.03cm}{\mathfrak q}_{2} 
\,-\,\Lambda^{2}\biggr\}
N^{\hspace{0.03cm}(2)}_{\bf k},\\[1ex]
%
&{\rm tr}\hspace{0.03cm}
\bigl(\hspace{0.03cm}T^{\,a_{1}}\hspace{0.03cm}T^{\,e_{2}}\hspace{0.03cm}T^{\,a^{\prime}_{1}}{\mathcal N}_{\hspace{0.02cm}{\bf k}}\hspace{0.01cm}
\bigr)
\hspace{0.01cm}\bigl\langle\hspace{0.03cm}\mathcal{Q}^{\hspace{0.03cm}a^{\prime}_{1}}_{1}
\hspace{0.03cm}\bigr\rangle
\hspace{0.03cm}\bigl\langle\hspace{0.03cm}\mathcal{Q}^{\hspace{0.03cm}e_{2}}_{2}
\hspace{0.03cm}\bigr\rangle
\hspace{0.01cm}\bigl\langle\hspace{0.03cm}\mathcal{Q}^{\hspace{0.03cm}a_{1}}_{1}
\hspace{0.03cm}\bigr\rangle	
=
\frac{9}{4}\,{\mathfrak q}_{1}\hspace{0.03cm}{\mathfrak q}_{12}
N^{\hspace{0.03cm}(2)}_{\bf k}.
\end{split}
\label{eq:7bb}
\end{equation}
The relative simplicity of these latter expressions stems from our choice of the color decomposition (\ref{eq:7q}). With any other, a more sophisticated choice of the representations (\ref{eq:7q}), expressions for the color traces become noticeably more complicated.\\ 
\indent For the traces in (\ref{eq:7z}) with the matrix function ${\mathcal W}_{\hspace{0.02cm}{\bf k}}$ the equalities (\ref{eq:7bb}) take the form
\[
	\begin{split}
		&{\rm tr}\hspace{0.03cm}
		\bigl(\hspace{0.03cm}T^{\,a_{2}}\hspace{0.03cm}T^{\,e_{1}}\hspace{0.03cm}T^{\,a^{\prime}_{2}}{\mathcal W}_{\hspace{0.02cm}{\bf k}}\hspace{0.01cm}
		\bigr)
		\hspace{0.01cm}\bigl\langle\hspace{0.03cm}\mathcal{Q}^{\hspace{0.03cm}a^{\prime}_{2}}_{2}
		\hspace{0.03cm}\bigr\rangle
		\hspace{0.03cm}\bigl\langle\hspace{0.03cm}\mathcal{Q}^{\hspace{0.03cm}e_{1}}_{1}
		\hspace{0.03cm}\bigr\rangle
		\hspace{0.01cm}\bigl\langle\hspace{0.03cm}\mathcal{Q}^{\hspace{0.03cm}a_{2}}_{2}
		\hspace{0.03cm}\bigr\rangle
		=
		\frac{9}{4}\,{\mathfrak q}_{2}\hspace{0.03cm}{\mathfrak q}_{12}
		W^{\hspace{0.03cm}(2)}_{\bf k},\\[1ex]
		%
		&{\rm tr}\hspace{0.03cm}
		\bigl(\hspace{0.03cm}T^{\,a_{1}}\hspace{0.03cm}T^{\,e_{2}}\hspace{0.03cm}T^{\,a^{\prime}_{1}}{\mathcal N}_{\hspace{0.02cm}{\bf k}}\hspace{0.01cm}
		\bigr)
		\hspace{0.01cm}\bigl\langle\hspace{0.03cm}\mathcal{Q}^{\hspace{0.03cm}a^{\prime}_{1}}_{1}
		\hspace{0.03cm}\bigr\rangle
		\hspace{0.03cm}\bigl\langle\hspace{0.03cm}\mathcal{Q}^{\hspace{0.03cm}e_{2}}_{2}
		\hspace{0.03cm}\bigr\rangle
		\hspace{0.01cm}\bigl\langle\hspace{0.03cm}\mathcal{Q}^{\hspace{0.03cm}a_{1}}_{1}
		\hspace{0.03cm}\bigr\rangle	
		=
		\biggl\{\frac{3}{2}\,{\mathfrak q}^{\hspace{0.03cm}2}_{12} \,+\, \frac{3}{4}\,{\mathfrak q}_{1}\hspace{0.03cm}{\mathfrak q}_{2} 
		\,-\,\Lambda^{2}\biggr\}
		W^{\hspace{0.03cm}(2)}_{\bf k}.
	\end{split}
\]
Let us now write the final kinetic equation for the function $N^{\hspace{0.03cm}(1)}_{\bf k}$, instead of (\ref{eq:7z}),
\begin{align}
d_{A}\hspace{0.04cm}\frac{\partial\hspace{0.01cm}N^{\hspace{0.03cm}(1)}_{\bf k}}{\!\!\partial\hspace{0.03cm}\tau}
=\;
&\frac{(2\hspace{0.02cm}\pi)^{3}}{2\hspace{0.03cm}|\Delta{\mathbf v}|}\,\sum_{\rho}\,
\int\!d\hspace{0.02cm}{\bf q}\;
\bigl|\hspace{0.03cm}{T}^{\hspace{0.03cm}(\rho)}_{\; {\bf k},\,{\bf q}}\bigr|^{\hspace{0.02cm}2}\,
2\hspace{0.02cm}\pi\hspace{0.02cm}\delta(\omega^{l}_{{\bf k}} - {\bf k}\cdot{\bf v}_{\rho}
-(-1)^{\rho}\, \Delta{\mathbf v}\cdot{\mathbf q}) 
\,\times
\vspace{-0.5cm}
\label{eq:7x}\\[1ex] 
&\biggl\{\frac{3}{2}\,{\mathfrak q}^{\hspace{0.03cm}2}_{12} \,+\, \frac{3}{4}\,{\mathfrak q}_{1}\hspace{0.03cm}{\mathfrak q}_{2}
+ \frac{9}{4}\,{\mathfrak q}_{1}\hspace{0.03cm}{\mathfrak q}_{12} 
\,-\,\Lambda^{2}\biggr\}\hspace{0.02cm}
N^{\hspace{0.03cm}(2)}_{\bf k}\hspace{0.03cm} \,+
\notag\\[1.5ex]
&\frac{(2\hspace{0.02cm}\pi)^{3}}{2\hspace{0.03cm}|\Delta{\mathbf v}|}\,\sum_{\rho}\hspace{0.03cm}(-1)^{\rho + 1}\!\!
\int\!d\hspace{0.02cm}{\bf q}\;
\bigl|\hspace{0.03cm}{T}^{\hspace{0.03cm}(\rho)}_{\; {\bf k},\,{\bf q}}\bigr|^{\hspace{0.02cm}2}\,
2\hspace{0.02cm}\pi\hspace{0.02cm}
\delta(\omega^{l}_{{\bf k}} - {\bf k}\cdot{\bf v}_{\rho}
-(-1)^{\rho}\, \Delta{\mathbf v}\cdot{\mathbf q}) 
\,\times
\notag\\[1.5ex]
&\biggl\{\frac{3}{2}\,{\mathfrak q}^{\hspace{0.03cm}2}_{12} \,+\, \frac{3}{4}\,{\mathfrak q}_{1}\hspace{0.03cm}{\mathfrak q}_{2}
+ \frac{9}{4}\,{\mathfrak q}_{2 }\hspace{0.03cm}{\mathfrak q}_{12} 
\,-\,\Lambda^{2}\biggr\}\hspace{0.02cm}
W^{\hspace{0.03cm}(2)}_{\bf k}.
\notag
\end{align}
Reasoning fully analogous with those used above, leads us to the kinetic equation for the second ``colorless'' part $W^{\hspace{0.03cm}(1)}_{\bf k}$ in the representation (\ref{eq:7q}):
\begin{align}
	d_{A}\hspace{0.04cm}\frac{\partial\hspace{0.01cm}W^{\hspace{0.03cm}(1)}_{\bf k}}{\!\!\partial\hspace{0.03cm}\tau}
	=\,
	&\frac{(2\hspace{0.02cm}\pi)^{3}}{2\hspace{0.03cm}|\Delta{\mathbf v}|}\,\sum_{\rho}\,
	\int\!d\hspace{0.02cm}{\bf q}\;
	\bigl|\hspace{0.03cm}{T}^{\hspace{0.03cm}(\rho)}_{\; {\bf k},\,{\bf q}}\bigr|^{\hspace{0.02cm}2}\,
	2\hspace{0.02cm}\pi\hspace{0.02cm}\delta(\omega^{l}_{{\bf k}} - {\bf k}\cdot{\bf v}_{\rho}
	-(-1)^{\rho}\, \Delta{\mathbf v}\cdot{\mathbf q}) 
	\,\times
	\vspace{-0.5cm}
\label{eq:7c}\\[1ex]
&\biggl\{\frac{3}{2}\,{\mathfrak q}^{\hspace{0.03cm}2}_{12} \,+\, \frac{3}{4}\,{\mathfrak q}_{1}\hspace{0.03cm}{\mathfrak q}_{2}
+ \frac{9}{4}\,{\mathfrak q}_{2 }\hspace{0.03cm}{\mathfrak q}_{12} 
\,-\,\Lambda^{2}\biggr\}\hspace{0.02cm}
W^{\hspace{0.03cm}(2)}_{\bf k}\hspace{0.03cm} \,+
\notag\\[1.5ex]
&\frac{(2\hspace{0.02cm}\pi)^{3}}{2\hspace{0.03cm}|\Delta{\mathbf v}|}\,\sum_{\rho}\hspace{0.03cm}(-1)^{\rho + 1}\!\!
\int\!d\hspace{0.02cm}{\bf q}\;
\bigl|\hspace{0.03cm}{T}^{\hspace{0.03cm}(\rho)}_{\; {\bf k},\,{\bf q}}\bigr|^{\hspace{0.02cm}2}\,
2\hspace{0.02cm}\pi\hspace{0.02cm}
\delta(\omega^{l}_{{\bf k}} - {\bf k}\cdot{\bf v}_{\rho}
-(-1)^{\rho}\, \Delta{\mathbf v}\cdot{\mathbf q}) 
\,\times
\notag\\[1.5ex]
&\biggl\{\frac{3}{2}\,{\mathfrak q}^{\hspace{0.03cm}2}_{12} \,+\, \frac{3}{4}\,{\mathfrak q}_{1}\hspace{0.03cm}{\mathfrak q}_{2}
+ \frac{9}{4}\,{\mathfrak q}_{1}\hspace{0.03cm}{\mathfrak q}_{12} 
\,-\,\Lambda^{2}\biggr\}\hspace{0.02cm}
N^{\hspace{0.03cm}(2)}_{\bf k}.
\notag
\end{align}
It follows from the equation for the color matrix function ${\mathcal W}^{\,a\hspace{0.02cm}a^{\prime}}_{\bf k}(\tau)$, Eq.\,(\ref{ap:B1}).\\
\indent We can propose a slightly different representation for the plasmon density matrix as an alternative to existing representations (\ref{eq:6t}) and (\ref{eq:7q}), without using the matrix $\sigma_{3}$, namely
\[
{\mathcal N}_{\hspace{0.02cm}{\bf k}}(\tau) 
= 
N^{(0)}_{\bf k}(\tau)\hspace{0.03cm}I^{(1)}\otimes I^{(2)}
+
N^{(1)}_{\bf k}(\tau)\hspace{0.03cm}\bigl(T^{(1)\hspace{0.03cm}a_{1}}\otimes I^{(2)}\hspace{0.03cm}\bigr)\bigl\langle\hspace{0.03cm}
\mathcal{Q}^{\,a_{1}}_{\hspace{0.03cm}1}(\tau)
\hspace{0.03cm}\bigr\rangle
+
N^{(2)}_{\bf k}(\tau)\hspace{0.03cm}\bigl(\hspace{0.03cm}I^{(1)}\otimes T^{(2)\hspace{0.03cm}a_{2}}\bigr)
\bigl\langle\hspace{0.03cm}\mathcal{Q}^{\,a_{2}}_{\hspace{0.03cm}2}
(\tau)\hspace{0.03cm}\bigr\rangle\,
+
\]
\begin{equation}
	N^{(12)}_{\bf k}(\tau)\hspace{0.03cm}\bigl(T^{(1)\hspace{0.03cm}a_{1}}\otimes T^{(2)\hspace{0.03cm}a_{2}}\bigr)
	f^{\hspace{0.03cm}a_{1}\hspace{0.03cm}a_{2}\hspace{0.03cm}c}
	\Lambda^{c}(\tau),
	\label{eq:7ww}
\end{equation}
where $I^{(1,\hspace{0.03cm}2)}$ are the identity matrices of dimension $(N^{2}_{c} - 1)\times (N^{2}_{c} - 1)$ and $\otimes$ is the tensor product symbol. We associate own color space with each of two color particles, as indicated by the labels 1 or 2 in parentheses. On the left-hand side of (\ref{eq:7ww}) the matrix function ${\mathcal N}_{\hspace{0.02cm}{\bf k}}$ has the four-indices color structure ${\mathcal N}_{\hspace{0.02cm}{\bf k}} = (\hspace{0.03cm}{\mathcal N}^{\,a_{1}a_{2},\,a^{\prime}_{1}a^{\prime}_{2}}_{\hspace{0.02cm}{\bf k}}\hspace{0.03cm})$. It can be related in a certain way to the plasmon density matrix ${\mathcal N}^{\,(\alpha,\hspace{0.03cm}\alpha^{\prime})\hspace{0.03cm}a
\hspace{0.03cm}a^{\prime}_{\phantom{1}}\!}_{\hspace{0.02cm}{\bf k}}$, introduced by us because of the definition of the correlation function (\ref{eq:6w}) and, thereby, connect the four scalar functions included in the color decomposition (\ref{eq:7q}) with the four functions from the representation (\ref{eq:7ww}).

%
%

\section{Second moment about color of the kinetic equation (\ref{eq:6y})}
\label{section_8}
\setcounter{equation}{0}

Let us return to the matrix kinetic equation (\ref{eq:6y}). Now we contract the left- and right-hand sides of this equation with the color matrix $(\hspace{0.02cm}T^{\,s_{1}})^{\hspace{0.02cm}a^{\prime}\hspace{0.01cm}a}$.
Using the color decomposition (\ref{eq:7q}) and the formula 
\[
{\rm tr}\hspace{0.03cm}\bigl(T^{\,s_{1}} {\mathcal N}_{{\bf k}} \bigr) =
N_{c}\,\bigl\langle\hspace{0.03cm}
\mathcal{Q}^{\,s_{1}}_{\hspace{0.03cm}1}
\hspace{0.03cm}\bigr\rangle N^{\hspace{0.03cm}(2)}_{\bf k}, 
\]
as a result, from (\ref{eq:6y}) we get 
\begin{equation}
N_{c}\hspace{0.04cm}\frac{\partial\hspace{0.01cm}\bigl(\bigl\langle\hspace{0.03cm}
\mathcal{Q}^{\,s_{1}}_{\hspace{0.03cm}1}
\hspace{0.03cm}\bigr\rangle N^{\hspace{0.03cm}(2)}_{\bf k}\bigr)}{\!\!\partial\hspace{0.03cm}\tau}
=
\label{eq:8q}
\end{equation}
\begin{align}
-\,&\frac{1}{2}\,\sum_{\rho}\,{T}^{\hspace{0.03cm}(\rho)}_{\; {\bf k}}(t)
\,\biggl(\int{T}^{\,\ast\hspace{0.03cm}(\rho)}_{\; {\bf k}}(t)\hspace{0.03cm}dt\biggr)
\Bigl\{{\rm tr}\hspace{0.03cm}
\bigl(\hspace{0.03cm}T^{\,s_{1}}\hspace{0.03cm}T^{\,a_{2}}\hspace{0.03cm}T^{\,e_{1}}\hspace{0.03cm}T^{\,a^{\prime}_{2}}{\mathcal N}_{\hspace{0.02cm}{\bf k}}\hspace{0.01cm}
\bigr)
\hspace{0.01cm}\bigl\langle\hspace{0.03cm}\mathcal{Q}^{\hspace{0.03cm}a^{\prime}_{2}}_{2}
\hspace{0.03cm}\bigr\rangle
\hspace{0.03cm}\bigl\langle\hspace{0.03cm}\mathcal{Q}^{\hspace{0.03cm}e_{1}}_{1}
\hspace{0.03cm}\bigr\rangle
\hspace{0.01cm}\bigl\langle\hspace{0.03cm}\mathcal{Q}^{\hspace{0.03cm}a_{2}}_{2}
\hspace{0.03cm}\bigr\rangle\,+
\notag\\[1ex]
&\hspace{5.7cm}{\rm tr}\hspace{0.03cm}
\bigl(\hspace{0.03cm}T^{\,s_{1}}\hspace{0.03cm}T^{\,a_{1}}\hspace{0.03cm}T^{\,e_{2}}\hspace{0.03cm}T^{\,a^{\prime}_{1}}{\mathcal N}_{\hspace{0.02cm}{\bf k}}\hspace{0.01cm}
\bigr)
\hspace{0.01cm}\bigl\langle\hspace{0.03cm}\mathcal{Q}^{\hspace{0.03cm}a^{\prime}_{1}}_{1}
\hspace{0.03cm}\bigr\rangle
\hspace{0.03cm}\bigl\langle\hspace{0.03cm}\mathcal{Q}^{\hspace{0.03cm}e_{2}}_{2}
\hspace{0.03cm}\bigr\rangle
\hspace{0.01cm}\bigl\langle\hspace{0.03cm}\mathcal{Q}^{\hspace{0.03cm}a_{1}}_{1}
\hspace{0.03cm}\bigr\rangle	
\Bigr\}\,-
\notag\\[1ex]
&\frac{1}{2}\,\sum_{\rho}\,{T}^{\,\ast\hspace{0.03cm}({\rho})}_{\; {\bf k}}(t)
\,
\biggl(\int{T}^{\hspace{0.03cm}(\rho)}_{\; {\bf k}}(t)\hspace{0.03cm}dt\biggr)
\Bigl\{{\rm tr}\hspace{0.03cm}
\bigl(\hspace{0.03cm}{\mathcal N}_{\hspace{0.02cm}{\bf k}}
\hspace{0.03cm}
T^{\,a^{\prime}_{2}}\hspace{0.03cm}T^{\,e_{1}}\hspace{0.03cm}
T^{\,a_{2}}\hspace{0.03cm}T^{\,s_{1}}\hspace{0.01cm}
\bigr)
\hspace{0.01cm}
\bigl\langle\hspace{0.03cm}\mathcal{Q}^{\hspace{0.03cm}a^{\prime}_{2}}_{2}
\hspace{0.03cm}\bigr\rangle
\hspace{0.03cm}\bigl\langle\hspace{0.03cm}\mathcal{Q}^{\hspace{0.03cm}e_{1}}_{1}
\hspace{0.03cm}\bigr\rangle
\hspace{0.01cm}\bigl\langle\hspace{0.03cm}\mathcal{Q}^{\hspace{0.03cm}a_{2}}_{2}
\hspace{0.03cm}\bigr\rangle\,+
\notag\\[1ex]
&\hspace{5.7cm}{\rm tr}\hspace{0.03cm}
\bigl(\hspace{0.03cm}{\mathcal N}_{\hspace{0.02cm}{\bf k}}
\hspace{0.03cm}
T^{\,a^{\prime}_{1}}\hspace{0.03cm}T^{\,e_{2}}\hspace{0.03cm}
T^{\,a_{1}}\hspace{0.03cm}T^{\,s_{1}}\hspace{0.01cm}
\bigr)
\hspace{0.01cm}
\bigl\langle\hspace{0.03cm}\mathcal{Q}^{\hspace{0.03cm}a^{\prime}_{1}}_{1}
\hspace{0.03cm}\bigr\rangle
\hspace{0.03cm}\bigl\langle\hspace{0.03cm}\mathcal{Q}^{\hspace{0.03cm}e_{2}}_{2}
\hspace{0.03cm}\bigr\rangle
\hspace{0.01cm}\bigl\langle\hspace{0.03cm}\mathcal{Q}^{\hspace{0.03cm}a_{1}}_{1}
\hspace{0.03cm}\bigr\rangle
\Bigr\}\,-
\notag\\[2ex]
&\frac{1}{2}\,\sum_{\rho}\hspace{0.03cm}(-1)^{\rho + 1}\,{T}^{\hspace{0.03cm}(\rho)}_{\; {\bf k}}(t)
\,\biggl(\int{T}^{\,\ast\hspace{0.03cm}(\rho)}_{\; {\bf k}}(t)\hspace{0.03cm}dt\biggr)
\Bigl\{{\rm tr}\hspace{0.03cm}
\bigl(\hspace{0.03cm}T^{\,s_{1}}\hspace{0.03cm}T^{\,a_{2}}\hspace{0.03cm}T^{\,e_{1}}\hspace{0.03cm}T^{\,a^{\prime}_{2}}{\mathcal W}_{\hspace{0.02cm}{\bf k}}\hspace{0.01cm}
\bigr)
\hspace{0.01cm}\bigl\langle\hspace{0.03cm}\mathcal{Q}^{\hspace{0.03cm}a^{\prime}_{2}}_{2}
\hspace{0.03cm}\bigr\rangle
\hspace{0.03cm}\bigl\langle\hspace{0.03cm}\mathcal{Q}^{\hspace{0.03cm}e_{1}}_{1}
\hspace{0.03cm}\bigr\rangle
\hspace{0.01cm}\bigl\langle\hspace{0.03cm}\mathcal{Q}^{\hspace{0.03cm}a_{2}}_{2}
\hspace{0.03cm}\bigr\rangle
\,+\,
\notag\\[1ex]
&\hspace{5.7cm}{\rm tr}\hspace{0.03cm}
\bigl(\hspace{0.03cm}T^{\,s_{1}}\hspace{0.03cm}T^{\,a_{1}}\hspace{0.03cm}
T^{\,e_{2}}\hspace{0.03cm}T^{\,a^{\prime}_{1}}{\mathcal W}_{\hspace{0.02cm}
{\bf k}}\hspace{0.01cm}\bigr)
\hspace{0.01cm}\bigl\langle\hspace{0.03cm}\mathcal{Q}^{\hspace{0.03cm}a^{\prime}_{1}}_{1}
\hspace{0.03cm}\bigr\rangle
\hspace{0.03cm}\bigl\langle\hspace{0.03cm}\mathcal{Q}^{\hspace{0.03cm}e_{2}}_{2}
\hspace{0.03cm}\bigr\rangle
\hspace{0.01cm}\bigl\langle\hspace{0.03cm}\mathcal{Q}^{\hspace{0.03cm}a_{1}}_{1}
\hspace{0.03cm}\bigr\rangle	
\Bigr\}\,-
\notag\\[1ex]
&\frac{1}{2}\,\sum_{\rho}\hspace{0.03cm}(-1)^{\rho + 1}\,{T}^{\,\ast\hspace{0.03cm}({\rho})}_{\; {\bf k}}(t)
\,
\biggl(\int{T}^{\hspace{0.03cm}(\rho)}_{\; {\bf k}}(t)\hspace{0.03cm}dt\biggr)
\Bigl\{{\rm tr}\hspace{0.03cm}
\bigl(\hspace{0.03cm}{\mathcal W}_{\hspace{0.02cm}{\bf k}}
\hspace{0.03cm}
T^{\,a^{\prime}_{2}}\hspace{0.03cm}T^{\,e_{1}}\hspace{0.03cm}
T^{\,a_{2}}\hspace{0.03cm}T^{\,s_{1}}\hspace{0.01cm}\bigr)
\hspace{0.01cm}
\bigl\langle\hspace{0.03cm}\mathcal{Q}^{\hspace{0.03cm}a^{\prime}_{2}}_{2}
\hspace{0.03cm}\bigr\rangle
\hspace{0.03cm}\bigl\langle\hspace{0.03cm}\mathcal{Q}^{\hspace{0.03cm}e_{1}}_{1}
\hspace{0.03cm}\bigr\rangle
\hspace{0.01cm}\bigl\langle\hspace{0.03cm}\mathcal{Q}^{\hspace{0.03cm}a_{2}}_{2}
\hspace{0.03cm}\bigr\rangle
\,+\,
\notag\\[1ex]
&\hspace{5.7cm}{\rm tr}\hspace{0.03cm}\bigl(\hspace{0.03cm}{\mathcal W}_{\hspace{0.02cm}{\bf k}}
\hspace{0.03cm}
T^{\,a^{\prime}_{1}}\hspace{0.03cm}T^{\,e_{2}}\hspace{0.03cm}
T^{\,a_{1}}\hspace{0.03cm}T^{\,s_{1}}\hspace{0.01cm}\bigr)
\hspace{0.01cm}
\bigl\langle\hspace{0.03cm}\mathcal{Q}^{\hspace{0.03cm}a^{\prime}_{1}}_{1}
\hspace{0.03cm}\bigr\rangle
\hspace{0.03cm}\bigl\langle\hspace{0.03cm}\mathcal{Q}^{\hspace{0.03cm}e_{2}}_{2}
\hspace{0.03cm}\bigr\rangle
\hspace{0.01cm}\bigl\langle\hspace{0.03cm}\mathcal{Q}^{\hspace{0.03cm}a_{1}}_{1}
\hspace{0.03cm}\bigr\rangle
\Bigr\}.
\notag
\end{align}
The completely analogous reasoning that brought us from equation (\ref{eq:7e}) to equation (\ref{eq:7z}) is also true here. Therefore, omitting the intermediate considerations, we immediately give the necessary equation, which follows from (\ref{eq:8q})
\begin{equation}
N_{c}\hspace{0.04cm}\frac{\partial\hspace{0.01cm}\bigl(\bigl\langle\hspace{0.03cm}
\mathcal{Q}^{\,s_{1}}_{\hspace{0.03cm}1}\hspace{0.03cm}\bigr\rangle 
N^{\hspace{0.03cm}(2)}_{\bf k}\bigr)}{\!\!\partial\hspace{0.03cm}\tau}
=
\frac{(2\hspace{0.02cm}\pi)^{3}}{2\hspace{0.03cm}|\Delta{\mathbf v}|}\,\sum_{\rho}\,
\int\!d\hspace{0.02cm}{\bf q}\;
\bigl|\hspace{0.03cm}{T}^{\hspace{0.03cm}(\rho)}_{\; {\bf k},\,{\bf q}}\bigr|^{\hspace{0.02cm}2}\,
2\hspace{0.02cm}\pi\hspace{0.02cm}\delta(\omega^{l}_{{\bf k}} - {\bf k}\cdot{\bf v}_{\rho}-(-1)^{\rho}\,\Delta{\mathbf v}\cdot{\mathbf q}) 
\,\times
\vspace{-0.5cm}
\label{eq:8w} 
\end{equation}
\begin{align}
	&\Bigl[\hspace{0.03cm}
	{\rm tr}\hspace{0.03cm}
	\bigl(\hspace{0.03cm}T^{\,s_{1}}\hspace{0.03cm}T^{\,a_{2}}\hspace{0.03cm}T^{\,e_{1}}\hspace{0.03cm}T^{\,a^{\prime}_{2}}{\mathcal N}_{\hspace{0.02cm}{\bf k}}\hspace{0.01cm}
	\bigr)
	\hspace{0.01cm}\bigl\langle\hspace{0.03cm}\mathcal{Q}^{\hspace{0.03cm}a^{\prime}_{2}}_{2}
	\hspace{0.03cm}\bigr\rangle
	\hspace{0.03cm}\bigl\langle\hspace{0.03cm}\mathcal{Q}^{\hspace{0.03cm}e_{1}}_{1}
	\hspace{0.03cm}\bigr\rangle
	\hspace{0.01cm}\bigl\langle\hspace{0.03cm}\mathcal{Q}^{\hspace{0.03cm}a_{2}}_{2}
	\hspace{0.03cm}\bigr\rangle
	\,+\,
	{\rm tr}\hspace{0.03cm}
	\bigl(\hspace{0.03cm}T^{\,s_{1}}\hspace{0.03cm}T^{\,a_{1}}\hspace{0.03cm}T^{\,e_{2}}\hspace{0.03cm}T^{\,a^{\prime}_{1}}{\mathcal N}_{\hspace{0.02cm}{\bf k}}\hspace{0.01cm}
	\bigr)
	\hspace{0.01cm}\bigl\langle\hspace{0.03cm}\mathcal{Q}^{\hspace{0.03cm}a^{\prime}_{1}}_{1}
	\hspace{0.03cm}\bigr\rangle
	\hspace{0.03cm}\bigl\langle\hspace{0.03cm}\mathcal{Q}^{\hspace{0.03cm}e_{2}}_{2}
	\hspace{0.03cm}\bigr\rangle
	\hspace{0.01cm}\bigl\langle\hspace{0.03cm}\mathcal{Q}^{\hspace{0.03cm}a_{1}}_{1}
	\hspace{0.03cm}\bigr\rangle	
	\Bigr]\,+
	\notag\\[1.5ex]
	&{\hspace{4.7cm}}\frac{(2\hspace{0.02cm}\pi)^{3}}{2\hspace{0.03cm}|\Delta{\mathbf v}|}\,\sum_{\rho}\hspace{0.03cm}(-1)^{\rho + 1}\!\!
	\int\!d\hspace{0.02cm}{\bf q}\;
	\bigl|\hspace{0.03cm}{T}^{\hspace{0.03cm}(\rho)}_{\; {\bf k},\,{\bf q}}\bigr|^{\hspace{0.02cm}2}\,
	2\hspace{0.02cm}\pi\hspace{0.02cm}\delta(\omega^{l}_{{\bf k}} - {\bf k}\cdot{\bf v}_{\rho}-(-1)^{\rho}\,\Delta{\mathbf v}\cdot{\mathbf q}) 
	\,\times
	\notag\\[1.5ex]
	&\Bigl[\hspace{0.03cm}
	{\rm tr}\hspace{0.03cm}
	\bigl(\hspace{0.03cm}T^{\,s_{1}}\hspace{0.03cm}T^{\,a_{2}}\hspace{0.03cm}T^{\,e_{1}}\hspace{0.03cm}T^{\,a^{\prime}_{2}}\hspace{0.03cm}{\mathcal W}_{\hspace{0.02cm}{\bf k}}\hspace{0.01cm}
	\bigr)
	\hspace{0.01cm}\bigl\langle\hspace{0.03cm}\mathcal{Q}^{\hspace{0.03cm}a^{\prime}_{2}}_{2}
	\hspace{0.03cm}\bigr\rangle
	\hspace{0.03cm}\bigl\langle\hspace{0.03cm}\mathcal{Q}^{\hspace{0.03cm}e_{1}}_{1}
	\hspace{0.03cm}\bigr\rangle
	\hspace{0.01cm}\bigl\langle\hspace{0.03cm}\mathcal{Q}^{\hspace{0.03cm}a_{2}}_{2}
	\hspace{0.03cm}\bigr\rangle
	\,+\,
	{\rm tr}\hspace{0.03cm}
	\bigl(\hspace{0.03cm}T^{\,s_{1}}\hspace{0.03cm}T^{\,a_{1}}\hspace{0.03cm}T^{\,e_{2}}\hspace{0.03cm}T^{\,a^{\prime}_{1}}\hspace{0.03cm}{\mathcal W}_{\hspace{0.02cm}{\bf k}}\hspace{0.01cm}
	\bigr)
	\hspace{0.01cm}\bigl\langle\hspace{0.03cm}\mathcal{Q}^{\hspace{0.03cm}a^{\prime}_{1}}_{1}
	\hspace{0.03cm}\bigr\rangle
	\hspace{0.03cm}\bigl\langle\hspace{0.03cm}\mathcal{Q}^{\hspace{0.03cm}e_{2}}_{2}
	\hspace{0.03cm}\bigr\rangle
	\hspace{0.01cm}\bigl\langle\hspace{0.03cm}\mathcal{Q}^{\hspace{0.03cm}a_{1}}_{1}
	\hspace{0.03cm}\bigr\rangle	
	\Bigr].
	\notag
\end{align}
\indent Just like in case of deriving kinetic equations (\ref{eq:7x}) and (\ref{eq:7c}), from (\ref{eq:8w}) we can obtain the desired kinetic equation for the second scalar function $N^{\hspace{0.03cm}(2)}_{\bf k}$ with the correct conservation law only if the contractions of the type 
\begin{equation}
\begin{split}
&{\rm tr}\hspace{0.03cm}
\bigl(\hspace{0.03cm}T^{\,s_{1}}\hspace{0.03cm}T^{\,a_{2}}\hspace{0.03cm}T^{\,e_{1}}\hspace{0.03cm}T^{\,a^{\prime}_{2}}{\mathcal N}_{\hspace{0.02cm}{\bf k}}\hspace{0.01cm}
\bigr)
\hspace{0.01cm}\bigl\langle
\hspace{0.03cm}\mathcal{Q}^{\hspace{0.03cm}a^{\prime}_{2}}_{2}
\hspace{0.03cm}\bigr\rangle
\hspace{0.03cm}\bigl\langle\hspace{0.03cm}\mathcal{Q}^{\hspace{0.03cm}e_{1}}_{1}
\hspace{0.03cm}\bigr\rangle
\hspace{0.01cm}\bigl\langle\hspace{0.03cm}\mathcal{Q}^{\hspace{0.03cm}a_{2}}_{2}
\hspace{0.03cm}\bigr\rangle,\\[1ex]
&{\rm tr}\hspace{0.03cm}
\bigl(\hspace{0.03cm}T^{\,s_{1}}\hspace{0.03cm}T^{\,a_{1}}\hspace{0.03cm}T^{\,e_{2}}\hspace{0.03cm}T^{\,a^{\prime}_{1}}{\mathcal N}_{\hspace{0.02cm}{\bf k}}\hspace{0.01cm}
\bigr)
\hspace{0.01cm}\bigl\langle\hspace{0.03cm}\mathcal{Q}^{\hspace{0.03cm}a^{\prime}_{1}}_{1}
\hspace{0.03cm}\bigr\rangle
\hspace{0.03cm}\bigl\langle\hspace{0.03cm}\mathcal{Q}^{\hspace{0.03cm}e_{2}}_{2}
\hspace{0.03cm}\bigr\rangle
\hspace{0.01cm}\bigl\langle\hspace{0.03cm}\mathcal{Q}^{\hspace{0.03cm}a_{1}}_{1}
\hspace{0.03cm}\bigr\rangle	
\end{split}
\label{eq:8e}
\end{equation}	
and so on, are real functions. Let us consider the color traces for the first contraction in (\ref{eq:8e}). By taking into account the representation (\ref{eq:7q}), we have
\[
{\rm tr}\hspace{0.03cm}
\bigl(\hspace{0.03cm}T^{\,s_{1}}\hspace{0.03cm}T^{\,a_{2}}\hspace{0.03cm}
T^{\,e_{1}}\hspace{0.03cm}T^{\,a^{\prime}_{2}}{\mathcal N}_{\hspace{0.02cm}
{\bf k}}\hspace{0.01cm}\bigr)
=
N^{\hspace{0.03cm}(1)}_{\bf k}\hspace{0.03cm}
{\rm tr}\hspace{0.03cm}
\bigl(\hspace{0.03cm}T^{\,s_{1}}\hspace{0.03cm}T^{\,a_{2}}\hspace{0.03cm}T^{\,e_{1}}\hspace{0.03cm}T^{\,a^{\prime}_{2}}\hspace{0.01cm}\bigr)
+
\bigl\langle\hspace{0.03cm}
\mathcal{Q}^{\,c_{1}}_{\hspace{0.03cm}1}\hspace{0.03cm}\bigr\rangle N^{\hspace{0.03cm}(2)}_{\bf k}\hspace{0.03cm}
{\rm tr}\hspace{0.03cm}
\bigl(\hspace{0.03cm}T^{\,s_{1}}\hspace{0.03cm}T^{\,a_{2}}\hspace{0.03cm}T^{\,e_{1}}\hspace{0.03cm}T^{\,a^{\prime}_{2}}\hspace{0.03cm}T^{\,c_{1}}\hspace{0.01cm}\bigr).
\]
Here, the fourth-order trace of the adjoint representation matrices $T^{a}$ is real, and the fifth-order trace is purely imaginary. We will show that the contraction of the last trace with the averaged color charges, as occurs in (\ref{eq:8e}) does indeed vanish. The easiest way to show this is to use the relation (\ref{ap:D10}) for the fifth-order trace. After relabel of the dummy summation indices, where it is needed, this gives us
\[
{\rm tr}\hspace{0.03cm}
\bigl(\hspace{0.03cm}T^{\,c_{1}}\hspace{0.03cm}T^{\,s_{1}}\hspace{0.03cm}T^{\,a_{2}}\hspace{0.03cm}T^{\,e_{1}}\hspace{0.03cm}T^{\,a^{\prime}_{2}}\hspace{0.01cm}\bigr)
\hspace{0.01cm}
\bigl\langle\hspace{0.03cm}
\mathcal{Q}^{\,c_{1}}_{\hspace{0.03cm}1}\hspace{0.03cm}\bigr\rangle
\hspace{0.03cm}\bigl\langle
\hspace{0.03cm}\mathcal{Q}^{\hspace{0.03cm}a^{\prime}_{2}}_{2}
\hspace{0.03cm}\bigr\rangle
\hspace{0.03cm}\bigl\langle\hspace{0.03cm}\mathcal{Q}^{\hspace{0.03cm}e_{1}}_{1}
\hspace{0.03cm}\bigr\rangle
\hspace{0.01cm}\bigl\langle\hspace{0.03cm}\mathcal{Q}^{\hspace{0.03cm}a_{2}}_{2}
\hspace{0.03cm}\bigr\rangle
=
\]
\[
-\,
{\rm tr}\hspace{0.03cm}
\bigl(\hspace{0.03cm}T^{\,c_{1}}\hspace{0.03cm}T^{\,s_{1}}\hspace{0.03cm}
T^{\,a_{2}}\hspace{0.03cm}T^{\,e_{1}}\hspace{0.03cm}T^{\,a^{\prime}_{2}}
\hspace{0.01cm}\bigr)
\hspace{0.01cm}
\bigl\langle\hspace{0.03cm}
\mathcal{Q}^{\,c_{1}}_{\hspace{0.03cm}1}\hspace{0.03cm}\bigr\rangle
\hspace{0.03cm}\bigl\langle
\hspace{0.03cm}\mathcal{Q}^{\hspace{0.03cm}a^{\prime}_{2}}_{2}
\hspace{0.03cm}\bigr\rangle
\hspace{0.03cm}\bigl\langle\hspace{0.03cm}\mathcal{Q}^{\hspace{0.03cm}e_{1}}_{1}
\hspace{0.03cm}\bigr\rangle
\hspace{0.01cm}\bigl\langle\hspace{0.03cm}\mathcal{Q}^{\hspace{0.03cm}a_{2}}_{2}
\hspace{0.03cm}\bigr\rangle
\,-\,
\]
\[
i\hspace{0.03cm}f^{\hspace{0.03cm}s_{1}\,c_{1}\hspace{0.03cm}b}
\bigl\langle\hspace{0.03cm}
\mathcal{Q}^{\,c_{1}}_{\hspace{0.03cm}1}\hspace{0.03cm}\bigr\rangle\hspace{0.03cm}
{\rm tr}\hspace{0.03cm}
\bigl(\hspace{0.03cm}T^{\,b}\hspace{0.03cm}T^{\,a_{2}}\hspace{0.03cm}T^{\,e_{1}}\hspace{0.03cm}T^{\,a^{\prime}_{2}}\hspace{0.01cm}\bigr)
\hspace{0.01cm}\bigl\langle
\hspace{0.03cm}\mathcal{Q}^{\hspace{0.03cm}a^{\prime}_{2}}_{2}
\hspace{0.03cm}\bigr\rangle
\hspace{0.03cm}\bigl\langle\hspace{0.03cm}\mathcal{Q}^{\hspace{0.03cm}e_{1}}_{1}
\hspace{0.03cm}\bigr\rangle
\hspace{0.01cm}\bigl\langle\hspace{0.03cm}\mathcal{Q}^{\hspace{0.03cm}a_{2}}_{2}
\hspace{0.03cm}\bigr\rangle
\]
or
\begin{equation}
{\rm tr}\hspace{0.03cm}
\bigl(\hspace{0.03cm}T^{\,c_{1}}\hspace{0.03cm}T^{\,s_{1}}\hspace{0.03cm}
T^{\,a_{2}}\hspace{0.03cm}T^{\,e_{1}}\hspace{0.03cm}T^{\,a^{\prime}_{2}}
\hspace{0.01cm}\bigr)
\bigl\langle\hspace{0.03cm}
\mathcal{Q}^{\,c_{1}}_{\hspace{0.03cm}1}\hspace{0.03cm}\bigr\rangle
\hspace{0.01cm}\bigl\langle
\hspace{0.03cm}\mathcal{Q}^{\hspace{0.03cm}a^{\prime}_{2}}_{2}
\hspace{0.03cm}\bigr\rangle
\hspace{0.03cm}\bigl\langle\hspace{0.03cm}\mathcal{Q}^{\hspace{0.03cm}e_{1}}_{1}
\hspace{0.03cm}\bigr\rangle
\hspace{0.01cm}\bigl\langle\hspace{0.03cm}\mathcal{Q}^{\hspace{0.03cm}a_{2}}_{2}
\hspace{0.03cm}\bigr\rangle
=
\label{eq:8r}
\end{equation}
\[
-\hspace{0.03cm}\frac{i}{2}\,
f^{\hspace{0.03cm}s_{1}\,c_{1}\hspace{0.03cm}b}
\bigl\langle\hspace{0.03cm}
\mathcal{Q}^{\,c_{1}}_{\hspace{0.03cm}1}\hspace{0.03cm}\bigr\rangle\hspace{0.03cm}
{\rm tr}\hspace{0.03cm}
\bigl(\hspace{0.03cm}T^{\,b}\hspace{0.03cm}T^{\,a_{2}}\hspace{0.03cm}T^{\,e_{1}}\hspace{0.03cm}T^{\,a^{\prime}_{2}}\hspace{0.01cm}\bigr)
\hspace{0.01cm}\bigl\langle
\hspace{0.03cm}\mathcal{Q}^{\hspace{0.03cm}a^{\prime}_{2}}_{2}
\hspace{0.03cm}\bigr\rangle
\hspace{0.03cm}\bigl\langle\hspace{0.03cm}\mathcal{Q}^{\hspace{0.03cm}e_{1}}_{1}
\hspace{0.03cm}\bigr\rangle
\hspace{0.01cm}\bigl\langle\hspace{0.03cm}\mathcal{Q}^{\hspace{0.03cm}a_{2}}_{2}
\hspace{0.03cm}\bigr\rangle.
\]
From the last equality, we see that unlike (\ref{eq:7_1z}), (\ref{eq:7zz}) here the fifth-order trace of the $T^{\,a}$ matrices doesn't automatically disappear when contracting with the averaged color charges, since
\[
{\rm tr}\hspace{0.03cm}
\bigl(\hspace{0.03cm}T^{\,b}\hspace{0.03cm}T^{\,a_{2}}\hspace{0.03cm}T^{\,e_{1}}\hspace{0.03cm}T^{\,a^{\prime}_{2}}\hspace{0.01cm}\bigr)
=
{\rm tr}\hspace{0.03cm}
\bigl(\hspace{0.03cm}T^{\,b}\hspace{0.03cm}T^{\,a^{\prime}_{2}}\hspace{0.03cm}T^{\,e_{1}}\hspace{0.03cm}T^{\,a_{2}}\hspace{0.01cm}\bigr)
\]
by virtue of the property (\ref{ap:D8}). It is evident that the expression (\ref{eq:8r}) can be turned to zero if one contracts it with the $\bigl\langle\hspace{0.03cm}
\mathcal{Q}^{\,s_{1}}_{\hspace{0.03cm}1}\hspace{0.03cm}\bigr\rangle$, i.e.,
\begin{equation}
\bigl\langle\hspace{0.03cm}
\mathcal{Q}^{\,s_{1}}_{\hspace{0.03cm}1}\hspace{0.03cm}\bigr\rangle\hspace{0.04cm}
{\rm tr}\hspace{0.03cm}
\bigl(\hspace{0.03cm}T^{\,c_{1}}\hspace{0.03cm}T^{\,s_{1}}\hspace{0.03cm}
T^{\,a_{2}}\hspace{0.03cm}T^{\,e_{1}}\hspace{0.03cm}T^{\,a^{\prime}_{2}}
\hspace{0.01cm}\bigr)
\bigl\langle\hspace{0.03cm}
\mathcal{Q}^{\,c_{1}}_{\hspace{0.03cm}1}\hspace{0.03cm}\bigr\rangle
\hspace{0.01cm}\bigl\langle
\hspace{0.03cm}\mathcal{Q}^{\hspace{0.03cm}a^{\prime}_{2}}_{2}
\hspace{0.03cm}\bigr\rangle
\hspace{0.03cm}\bigl\langle\hspace{0.03cm}\mathcal{Q}^{\hspace{0.03cm}e_{1}}_{1}
\hspace{0.03cm}\bigr\rangle
\hspace{0.01cm}\bigl\langle\hspace{0.03cm}\mathcal{Q}^{\hspace{0.03cm}a_{2}}_{2}
\hspace{0.03cm}\bigr\rangle
= 0.
\label{eq:8t} 
\end{equation}
By virtue of the above, equation (\ref{eq:8w}) will hold true only when contracting it with the averaged color charge $\bigl\langle\hspace{0.03cm}
\mathcal{Q}^{\,s_{1}}_{\hspace{0.03cm}1}\hspace{0.03cm}\bigr\rangle$. Therefore, instead of (\ref{eq:8w}) we should write
\begin{equation}
N_{c}\hspace{0.04cm}\bigl\langle\hspace{0.03cm}
\mathcal{Q}^{\,s_{1}}_{\hspace{0.03cm}1}\hspace{0.03cm}\bigr\rangle\,\frac{\partial\hspace{0.01cm}\bigl(\bigl\langle\hspace{0.03cm}
\mathcal{Q}^{\,s_{1}}_{\hspace{0.03cm}1}\hspace{0.03cm}\bigr\rangle N^{\hspace{0.03cm}(2)}_{\bf k}\bigr)}{\!\!\partial\hspace{0.03cm}\tau}
=
\vspace{-0.3cm}
\label{eq:8y} 
\end{equation}
\begin{align}
&\frac{(2\hspace{0.02cm}\pi)^{3}}{2\hspace{0.03cm}|\Delta{\mathbf v}|}\,\sum_{\rho}\,
\int\!d\hspace{0.02cm}{\bf q}\;
\bigl|\hspace{0.03cm}{T}^{\hspace{0.03cm}(\rho)}_{\; {\bf k},\,{\bf q}}\bigr|^{\hspace{0.02cm}2}\,
2\hspace{0.02cm}\pi\hspace{0.02cm}\delta(\omega^{l}_{{\bf k}} - {\bf k}\cdot{\bf v}_{\rho}-(-1)^{\rho}\,\Delta{\mathbf v}\cdot{\mathbf q})
\,\times\notag\\
\Bigl[\hspace{0.03cm}
&{\rm tr}\hspace{0.03cm}
\bigl(\hspace{0.03cm}T^{\,s_{1}}\hspace{0.03cm}T^{\,a_{2}}\hspace{0.03cm}T^{\,e_{1}}\hspace{0.03cm}T^{\,a^{\prime}_{2}}{\mathcal N}_{\hspace{0.02cm}{\bf k}}\hspace{0.01cm}
\bigr)\bigl\langle\hspace{0.03cm}
\mathcal{Q}^{\,s_{1}}_{\hspace{0.03cm}1}\hspace{0.03cm}\bigr\rangle
\hspace{0.01cm}\bigl\langle
\hspace{0.03cm}\mathcal{Q}^{\hspace{0.03cm}a^{\prime}_{2}}_{2}
\hspace{0.03cm}\bigr\rangle
\hspace{0.03cm}\bigl\langle\hspace{0.03cm}\mathcal{Q}^{\hspace{0.03cm}e_{1}}_{1}
\hspace{0.03cm}\bigr\rangle
\hspace{0.01cm}\bigl\langle\hspace{0.03cm}\mathcal{Q}^{\hspace{0.03cm}a_{2}}_{2}
\hspace{0.03cm}\bigr\rangle
\,+\notag\\[1ex]
&{\rm tr}\hspace{0.03cm}
\bigl(\hspace{0.03cm}T^{\,s_{1}}\hspace{0.03cm}T^{\,a_{1}}\hspace{0.03cm}T^{\,e_{2}}\hspace{0.03cm}T^{\,a^{\prime}_{1}}{\mathcal N}_{\hspace{0.02cm}{\bf k}}\hspace{0.01cm}
\bigr)\bigl\langle\hspace{0.03cm}
\mathcal{Q}^{\,s_{1}}_{\hspace{0.03cm}1}\hspace{0.03cm}\bigr\rangle
\hspace{0.01cm}\bigl\langle
\hspace{0.03cm}\mathcal{Q}^{\hspace{0.03cm}a^{\prime}_{1}}_{1}
\hspace{0.03cm}\bigr\rangle
\hspace{0.03cm}\bigl\langle\hspace{0.03cm}\mathcal{Q}^{\hspace{0.03cm}e_{2}}_{2}
\hspace{0.03cm}\bigr\rangle
\hspace{0.01cm}\bigl\langle\hspace{0.03cm}\mathcal{Q}^{\hspace{0.03cm}a_{1}}_{1}
\hspace{0.03cm}\bigr\rangle	
\Bigr]\,+
\notag\\[1.5ex]
	&\frac{(2\hspace{0.02cm}\pi)^{3}}{2\hspace{0.03cm}|\Delta{\mathbf v}|}\,\sum_{\rho}\hspace{0.03cm}(-1)^{\rho + 1}\!\!
	\int\!d\hspace{0.02cm}{\bf q}\;
	\bigl|\hspace{0.03cm}{T}^{\hspace{0.03cm}(\rho)}_{\; {\bf k},\,{\bf q}}\bigr|^{\hspace{0.02cm}2}\,
	2\hspace{0.02cm}\pi\hspace{0.02cm}\delta(\omega^{l}_{{\bf k}} - {\bf k}\cdot{\bf v}_{\rho}-(-1)^{\rho}\,\Delta{\mathbf v}\cdot{\mathbf q})
	\,\times
	\notag\\[1.5ex]
	\Bigl[\hspace{0.03cm}&{\rm tr}\hspace{0.03cm}
	\bigl(\hspace{0.03cm}T^{\,s_{1}}\hspace{0.03cm}T^{\,a_{2}}\hspace{0.03cm}T^{\,e_{1}}\hspace{0.03cm}T^{\,a^{\prime}_{2}}\hspace{0.03cm}{\mathcal W}_{\hspace{0.02cm}{\bf k}}\hspace{0.01cm}
	\bigr)\bigl\langle\hspace{0.03cm}
	\mathcal{Q}^{\,s_{1}}_{\hspace{0.03cm}1}\hspace{0.03cm}\bigr\rangle
	\hspace{0.01cm}\bigl\langle
	\hspace{0.03cm}\mathcal{Q}^{\hspace{0.03cm}a^{\prime}_{2}}_{2}
	\hspace{0.03cm}\bigr\rangle
	\hspace{0.03cm}\bigl\langle\hspace{0.03cm}\mathcal{Q}^{\hspace{0.03cm}e_{1}}_{1}
	\hspace{0.03cm}\bigr\rangle
	\hspace{0.01cm}\bigl\langle\hspace{0.03cm}\mathcal{Q}^{\hspace{0.03cm}a_{2}}_{2}
	\hspace{0.03cm}\bigr\rangle
	\,+\notag\\[1ex]
	&{\rm tr}\hspace{0.03cm}
	\bigl(\hspace{0.03cm}T^{\,s_{1}}\hspace{0.03cm}T^{\,a_{1}}\hspace{0.03cm}T^{\,e_{2}}\hspace{0.03cm}T^{\,a^{\prime}_{1}}\hspace{0.03cm}{\mathcal W}_{\hspace{0.02cm}{\bf k}}\hspace{0.01cm}
	\bigr)\bigl\langle\hspace{0.03cm}
	\mathcal{Q}^{\,s_{1}}_{\hspace{0.03cm}1}\hspace{0.03cm}\bigr\rangle
	\hspace{0.01cm}\bigl\langle
	\hspace{0.03cm}\mathcal{Q}^{\hspace{0.03cm}a^{\prime}_{1}}_{1}
	\hspace{0.03cm}\bigr\rangle
	\hspace{0.03cm}\bigl\langle\hspace{0.03cm}\mathcal{Q}^{\hspace{0.03cm}e_{2}}_{2}
	\hspace{0.03cm}\bigr\rangle
	\hspace{0.01cm}\bigl\langle\hspace{0.03cm}\mathcal{Q}^{\hspace{0.03cm}a_{1}}_{1}
	\hspace{0.03cm}\bigr\rangle	
	\Bigr].
	\notag
\end{align}
Taking into account the color decomposition (\ref{eq:7q}), the equality (\ref{eq:8t}) and the formula for the trace of fourth order (\ref{ap:D4}) it is easy to find that for the sum of the traces in the first square bracket on the right-hand side of the kinetic equation (\ref{eq:8y}), the following representation holds
\begin{align}
	&{\rm tr}\hspace{0.03cm}
	\bigl(\hspace{0.03cm}T^{\,s_{1}}\hspace{0.03cm}T^{\,a_{2}}\hspace{0.03cm}T^{\,e_{1}}\hspace{0.03cm}T^{\,a^{\prime}_{2}}{\mathcal N}_{\hspace{0.02cm}{\bf k}}\hspace{0.01cm}
	\bigr)\bigl\langle\hspace{0.03cm}
	\mathcal{Q}^{\,s_{1}}_{\hspace{0.03cm}1}\hspace{0.03cm}\bigr\rangle
	\hspace{0.01cm}\bigl\langle
	\hspace{0.03cm}\mathcal{Q}^{\hspace{0.03cm}a^{\prime}_{2}}_{2}
	\hspace{0.03cm}\bigr\rangle
	\hspace{0.03cm}\bigl\langle\hspace{0.03cm}\mathcal{Q}^{\hspace{0.03cm}e_{1}}_{1}
	\hspace{0.03cm}\bigr\rangle
	\hspace{0.01cm}\bigl\langle\hspace{0.03cm}\mathcal{Q}^{\hspace{0.03cm}a_{2}}_{2}
	\hspace{0.03cm}\bigr\rangle
	\,+\notag\\[1ex]
	&{\rm tr}\hspace{0.03cm}
	\bigl(\hspace{0.03cm}T^{\,s_{1}}\hspace{0.03cm}T^{\,a_{1}}\hspace{0.03cm}T^{\,e_{2}}\hspace{0.03cm}T^{\,a^{\prime}_{1}}{\mathcal N}_{\hspace{0.02cm}{\bf k}}\hspace{0.01cm}
	\bigr)\bigl\langle\hspace{0.03cm}
	\mathcal{Q}^{\,s_{1}}_{\hspace{0.03cm}1}\hspace{0.03cm}\bigr\rangle
	\hspace{0.01cm}\bigl\langle
	\hspace{0.03cm}\mathcal{Q}^{\hspace{0.03cm}a^{\prime}_{1}}_{1}
	\hspace{0.03cm}\bigr\rangle
	\hspace{0.03cm}\bigl\langle\hspace{0.03cm}\mathcal{Q}^{\hspace{0.03cm}e_{2}}_{2}
	\hspace{0.03cm}\bigr\rangle
	\hspace{0.01cm}\bigl\langle\hspace{0.03cm}\mathcal{Q}^{\hspace{0.03cm}a_{1}}_{1}
	\hspace{0.03cm}\bigr\rangle	
	=
	\notag\\[1.5ex]
	&\biggl\{\frac{3}{2}\,{\mathfrak q}^{\hspace{0.03cm}2}_{12} \,+\, \frac{3}{4}\,{\mathfrak q}_{1}\hspace{0.03cm}{\mathfrak q}_{2}
	+ \frac{9}{4}\,{\mathfrak q}_{1}\hspace{0.03cm}{\mathfrak q}_{12} 
	\,-\,\Lambda^{2}\biggr\}\hspace{0.02cm}
	N^{\hspace{0.03cm}(1)}_{\bf k}.
	\notag
\end{align}
We get a completely similar expression for the sum of the traces in the last two lines in (\ref{eq:8y}) with the replacement:
$N^{\hspace{0.03cm}(1)}_{\bf k}\rightarrow W^{\hspace{0.03cm}(1)}_{\bf k}$.
Thus, we can write out in addition to (\ref{eq:7x}) and (\ref{eq:7c}) the third kinetic equation for the scalar function $N^{\hspace{0.03cm}(2)}_{\bf k}$, which we present as follows:
\begin{equation}
	N_{c}\hspace{0.04cm}{\mathfrak q}_{1}\,\frac{\partial\hspace{0.01cm}
	N^{\hspace{0.03cm}(2)}_{\bf k}}{\!\!\partial\hspace{0.03cm}\tau}
	\,+\,
	\frac{1}{2}\,N_{c}\,N^{\hspace{0.03cm}(2)}_{\bf k}
	\frac{\partial\hspace{0.03cm}{\mathfrak q}_{1}}{\!\!\partial\hspace{0.03cm}\tau}
	=
\hspace{1cm}
\label{eq:8u} 
\end{equation}
\vspace{-0.7cm}
\begin{align}	
	&\frac{(2\hspace{0.02cm}\pi)^{3}}{2\hspace{0.03cm}|\Delta{\mathbf v}|}\,\sum_{\rho}\,
	\int\!d\hspace{0.02cm}{\bf q}\;
	\bigl|\hspace{0.03cm}{T}^{\hspace{0.03cm}(\rho)}_{\; {\bf k},\,{\bf q}}\bigr|^{\hspace{0.02cm}2}\,
	2\hspace{0.02cm}\pi\hspace{0.02cm}\delta(\omega^{l}_{{\bf k}} - {\bf k}\cdot{\bf v}_{\rho}
	-(-1)^{\rho}\, \Delta{\mathbf v}\cdot{\mathbf q}) 
	\,\times
	\notag\\[1ex]
	&\biggl\{\frac{3}{2}\,{\mathfrak q}^{\hspace{0.03cm}2}_{12} \,+\, \frac{3}{4}\,{\mathfrak q}_{1}\hspace{0.03cm}{\mathfrak q}_{2}
	+ \frac{9}{4}\,{\mathfrak q}_{1}\hspace{0.03cm}{\mathfrak q}_{12} 
	\,-\,\Lambda^{2}\biggr\}\hspace{0.02cm}
	N^{\hspace{0.03cm}(1)}_{\bf k}\hspace{0.03cm} \,+
	\notag\\[1.5ex]
	&\frac{(2\hspace{0.02cm}\pi)^{3}}{2\hspace{0.03cm}|\Delta{\mathbf v}|}
	\,\sum_{\rho}\hspace{0.02cm}(-1)^{\rho + 1}\!\!
	\int\!d\hspace{0.02cm}{\bf q}\;
	\bigl|\hspace{0.03cm}{T}^{\hspace{0.03cm}(\rho)}_{\; {\bf k},\,{\bf q}}\bigr|^{\hspace{0.02cm}2}\,
	2\hspace{0.02cm}\pi\hspace{0.02cm}\delta(\omega^{l}_{{\bf k}} - {\bf k}\cdot{\bf v}_{\rho}
	-(-1)^{\rho}\, \Delta{\mathbf v}\cdot{\mathbf q}) 
	\,\times
	\notag\\[1.5ex]
	&\biggl\{\frac{3}{2}\,{\mathfrak q}^{\hspace{0.03cm}2}_{12} \,+\, \frac{3}{4}\,{\mathfrak q}_{1}\hspace{0.03cm}{\mathfrak q}_{2}
	+ \frac{9}{4}\,{\mathfrak q}_{1}\hspace{0.03cm}{\mathfrak q}_{12} 
	\,-\,\Lambda^{2}\biggr\}\hspace{0.02cm}
	W^{\hspace{0.03cm}(1)}_{\bf k}.
	\notag
\end{align}
We just need to write out the equation for the four and final scalar function $W^{\hspace{0.03cm}(2)}_{\bf k}$. Using the same reasoning as above, here we get from Eq.\,(\ref{ap:B1})
\begin{equation}
N_{c}\hspace{0.04cm}{\mathfrak q}_{2}\,\frac{\partial\hspace{0.01cm}
W^{\hspace{0.03cm}(2)}_{\bf k}}{\!\!\partial\hspace{0.03cm}\tau}
\,+\,
\frac{1}{2}\,N_{c}\,W^{\hspace{0.03cm}(2)}_{\bf k}
\frac{\partial\hspace{0.03cm}{\mathfrak q}_{2}}{\!\!\partial\hspace{0.03cm}\tau}
=
\hspace{1cm}
\label{eq:8i} 
\end{equation}
\vspace{-0.7cm}
\begin{align}	
	&\frac{(2\hspace{0.02cm}\pi)^{3}}{2\hspace{0.03cm}|\Delta{\mathbf v}|}\,\sum_{\rho}\,
	\int\!d\hspace{0.02cm}{\bf q}\;
	\bigl|\hspace{0.03cm}{T}^{\hspace{0.03cm}(\rho)}_{\; {\bf k},\,{\bf q}}\bigr|^{\hspace{0.02cm}2}\,
	2\hspace{0.02cm}\pi\hspace{0.03cm}\delta(\omega^{l}_{{\bf k}} - {\bf k}\cdot{\bf v}_{\rho}
	-(-1)^{\rho}\, \Delta{\mathbf v}\cdot{\mathbf q})
	\,\times 
\notag\\[1ex]	
	&\biggl\{\frac{3}{2}\,{\mathfrak q}^{\hspace{0.03cm}2}_{12} \,+\, \frac{3}{4}\,{\mathfrak q}_{1}\hspace{0.03cm}{\mathfrak q}_{2}
	+ \frac{9}{4}\,{\mathfrak q}_{2}\hspace{0.03cm}{\mathfrak q}_{12} 
	\,-\,\Lambda^{2}\biggr\}\hspace{0.02cm}
	W^{\hspace{0.03cm}(1)}_{\bf k}\hspace{0.03cm} \,+
	\notag\\[1.5ex]
	&\frac{(2\hspace{0.02cm}\pi)^{3}}{2\hspace{0.03cm}|\Delta{\mathbf v}|}\,\sum_{\rho}\hspace{0.02cm}(-1)^{\rho + 1}\!\!
	\int\!d\hspace{0.02cm}{\bf q}\;
	\bigl|\hspace{0.03cm}{T}^{\hspace{0.03cm}(\rho)}_{\; {\bf k},\,{\bf q}}\bigr|^{\hspace{0.02cm}2}\,
	2\hspace{0.02cm}\pi\hspace{0.03cm}\delta(\omega^{l}_{{\bf k}} - {\bf k}\cdot{\bf v}_{\rho}
	-(-1)^{\rho}\, \Delta{\mathbf v}\cdot{\mathbf q}) 
	\,\times
	\notag\\[1.5ex]
	&\biggl\{\frac{3}{2}\,{\mathfrak q}^{\hspace{0.03cm}2}_{12} \,+\, \frac{3}{4}\,{\mathfrak q}_{1}\hspace{0.03cm}{\mathfrak q}_{2}
	+ \frac{9}{4}\,{\mathfrak q}_{2}\hspace{0.03cm}{\mathfrak q}_{12} 
	\,-\,\Lambda^{2}\biggr\}\hspace{0.02cm}
	N^{\hspace{0.03cm}(1)}_{\bf k}.
	\notag
\end{align}
\indent However, the system of kinetic equations obtained in this and the preceding sections is still unclosed. It contains four unknown functions: ${\mathfrak q}_{1}(\tau),\,{\mathfrak q}_{2}(\tau),\,{\mathfrak q}_{12}(\tau)$ and $\Lambda^{2}(\tau)$, as they were defined by the relations (\ref{eq:7w}). In addition, another circumstance should be mentioned. The right-hand side of the equations (\ref{eq:7x}), (\ref{eq:7c}), (\ref{eq:8u}) and (\ref{eq:8i}) according to the construction within the framework of the multi-time formalism,  contains terms linear in the scalar plasmon number densities $N^{\hspace{0.03cm}(1,2)}_{\bf k}$ and $W^{\hspace{0.03cm}(1,2)}_{\bf k}$ and it does not contain any terms that are free from them, i.e. proportional only to the products of the averaged color charges. It is not entirely clear why such contributions do not occur here. Indeed, let us consider the correlation function of the original normal variables $c^{\hspace{0.02cm}(\alpha)\hspace{0.02cm}a}_{\hspace{0.02cm}{\bf k}}(t)$ and $c^{\ast\ \!\!(\alpha)\hspace{0.02cm}a}_{\hspace{0.02cm}{\bf k}}(t)$ before their decomposition into slow and fast variables. Taking into account the representation (\ref{eq:5ra}), we have
\[
\bigl\langle\hspace{0.01cm}c^{\,\ast\hspace{0.02cm}(\alpha)\hspace{0.02cm}a}_{\hspace{0.02cm}{\bf k}}(t)
\hspace{0.03cm}
c^{\hspace{0.02cm}(\alpha^{\prime})\hspace{0.02cm}a^{\prime}}_{\hspace{0.02cm}{\bf k}'}(t)\bigr\rangle
=
\varepsilon^{2\hspace{0.02cm}\alpha_{1}}
\bigl\langle\hspace{0.01cm}C^{\,\ast\hspace{0.02cm}(\alpha)\hspace{0.02cm}a}_{\hspace{0.02cm}{\bf k}}(\tau)
\hspace{0.03cm}
C^{\hspace{0.02cm}(\alpha^{\prime})\hspace{0.02cm}a^{\prime}}_{\hspace{0.02cm}{\bf k}'}(\tau)\bigr\rangle
\,+\,
\varepsilon^{2\hspace{0.02cm}\alpha_{2}}
\bigl\langle\hspace{0.01cm}\widehat{C}^{\,\ast\hspace{0.02cm}(\alpha)\hspace{0.02cm}a}_{\hspace{0.02cm}{\bf k}}(t')
\hspace{0.03cm}
\widehat{C}^{\hspace{0.02cm}(\alpha^{\prime})\hspace{0.02cm}a^{\prime}}_{\hspace{0.02cm}{\bf k}'}(t')\bigr\rangle.
\]
Here, on the right-hand side we have assumed that all the correlation functions mixed the slow and fast components, are equal to zero. Define the derivative of
of this equality with respect to the original time $t$. Using the representation (\ref{eq:5t}) for the total time variation $\partial/ \partial\hspace{0.02cm}t$, we derive
\begin{equation}
\frac{\partial}{\partial\hspace{0.02cm}t}\,
\bigl\langle\hspace{0.01cm}c^{\,\ast\hspace{0.02cm}(\alpha)\hspace{0.02cm}a}_{\hspace{0.02cm}{\bf k}}(t)
\hspace{0.03cm}
c^{\hspace{0.02cm}(\alpha^{\prime})\hspace{0.02cm}a^{\prime}}_{\hspace{0.02cm}{\bf k}'}(t)\bigr\rangle
=
\label{eq:8o} 
\end{equation}
\[
\varepsilon^{2\hspace{0.02cm}\alpha_{1} + \gamma_{2}}\,
\frac{\partial}{\partial\hspace{0.02cm}\tau}\,
\bigl\langle\hspace{0.01cm}C^{\,\ast\hspace{0.02cm}(\alpha)\hspace{0.02cm}a}_{\hspace{0.02cm}{\bf k}}(\tau)
\hspace{0.03cm}
C^{\hspace{0.02cm}(\alpha^{\prime})\hspace{0.02cm}a^{\prime}}_{\hspace{0.02cm}{\bf k}'}(\tau)\bigr\rangle
\,+\,
\varepsilon^{2\hspace{0.02cm}\alpha_{2} + \gamma_{1}}\,
\frac{\partial}{\partial\hspace{0.03cm}t'}\,
\bigl\langle\hspace{0.01cm}\widehat{C}^{\,\ast\hspace{0.02cm}(\alpha)\hspace{0.02cm}a}_{\hspace{0.02cm}{\bf k}}(t')
\hspace{0.03cm}
\widehat{C}^{\hspace{0.02cm}(\alpha^{\prime})\hspace{0.02cm}a^{\prime}}_{\hspace{0.02cm}{\bf k}'}(t')\bigr\rangle.
\]
When choosing solutions (\ref{eq:5oo}) for the exponents of the two contributions, on the right-hand side of the previous expression we have
\[
2\hspace{0.02cm}\alpha_{1} + \gamma_{2} = 
4\hspace{0.02cm}\alpha_{1} + \alpha_{2},
\qquad
2\hspace{0.02cm}\alpha_{2} + \gamma_{1}
= 4\hspace{0.02cm}\alpha_{1} + \alpha_{2},
\]
that is, these contributions have the same power of the small parameter $\varepsilon$. The first term on the right-hand side of (\ref{eq:8o}) with the slow time derivative gives us a system of kinetic equations (\ref{eq:7x}), (\ref{eq:7c}), (\ref{eq:8u}) and (\ref{eq:8i}). The second term with the fast time derivative is easiest to determine using first the initial equation (\ref{eq:5y}), which gives us
\begin{align}
&\frac{\partial}{\partial\hspace{0.03cm}t'}\,
\bigl\langle\hspace{0.01cm}\widehat{C}^{\,\ast\hspace{0.02cm}(\alpha)\hspace{0.02cm}a}_{\hspace{0.02cm}{\bf k}}(t')
\hspace{0.03cm}
\widehat{C}^{\hspace{0.02cm}(\alpha^{\prime})\hspace{0.02cm}a^{\prime}}_{\hspace{0.02cm}{\bf k}'}(t')\bigr\rangle
\,=\,
i\hspace{0.03cm}\bigl[\hspace{0.03cm}
\omega^{\hspace{0.03cm}l}_{\hspace{0.02cm}{\bf k}} - {\mathbf v}^{\phantom{l}}_{\alpha}\!\cdot {\mathbf k}
-
(\omega^{\hspace{0.03cm}l}_{\hspace{0.02cm}{\bf k}'} - {\mathbf v}^{\phantom{l}}_{\alpha'}\!\cdot {\mathbf k}')
\hspace{0.03cm}\bigr]
\bigl\langle\hspace{0.01cm}\widehat{C}^{\,\ast\hspace{0.02cm}(\alpha)\hspace{0.02cm}a}_{\hspace{0.02cm}{\bf k}}(t')
\hspace{0.03cm}
\widehat{C}^{\hspace{0.02cm}(\alpha^{\prime})\hspace{0.02cm}a^{\prime}}_{\hspace{0.02cm}{\bf k}'}(t')\bigr\rangle
\,+
\label{eq:8p}\\[1.5ex] 
&i\hspace{0.03cm}
{T}^{\hspace{0.03cm}(\alpha)\hspace{0.03cm}a\,a_{1}
\hspace{0.03cm}a_{2}}_{\;{\mathbf k}}(t')\,
\bigl\langle\hspace{0.01cm}
{\mathcal Q}^{\hspace{0.03cm}a_{1}}_{\hspace{0.03cm}1}(\tau)
\hspace{0.02cm}
{\mathcal Q}^{\hspace{0.03cm}a_{2}}_{\hspace{0.03cm}2}(\tau)
\hspace{0.03cm}
\widehat{C}^{\hspace{0.02cm}(\alpha^{\prime})\hspace{0.02cm}a^{\prime}}_{\hspace{0.02cm}{\bf k}'}(t')\bigr\rangle
-
i\hspace{0.03cm}{T}^{\hspace{0.03cm}\ast\hspace{0.03cm}(\alpha')\hspace{0.03cm}a'\,a^{\prime}_{1}\hspace{0.03cm}a^{\prime}_{2}}_{\;{\mathbf k}'}(t')\,
\bigl\langle\hspace{0.01cm}\widehat{C}^{\,\ast\hspace{0.02cm}(\alpha)
\hspace{0.02cm}a}_{\hspace{0.02cm}{\bf k}}(t')
\hspace{0.03cm}
{\mathcal Q}^{\hspace{0.03cm}a^{\prime}_{1}}_{\hspace{0.03cm}1}(\tau)
\hspace{0.02cm}
{\mathcal Q}^{\hspace{0.03cm}a^{\prime}_{2}}_{\hspace{0.03cm}2}(\tau)
\bigr\rangle
\notag
\end{align}
and then substituting the explicit form of the rapidly varying component $\widehat{C}^{\hspace{0.02cm}(\alpha)\hspace{0.02cm}a}_{\hspace{0.02cm}{\bf k}}(t')$, as defined by Eq.\,(\ref{eq:5d}). Formally, the fast time derivative (\ref{eq:8p}) will be proportional to the product of the four averaged slowly varying color charges, just how we wanted it. However, the physical meaning of the additional contribution (\ref{eq:8p}) remains very vague for us and we will not take it into account in the kinetic equations obtained above.

%
%

\section{\bf Equations for the expected value of color charges ${\mathcal Q}^{\hspace{0.03cm}a_{1}}_{\hspace{0.03cm}1}$ and ${\mathcal Q}^{\hspace{0.03cm}a_{2}}_{\hspace{0.03cm}2}$}
\label{section_9}
\setcounter{equation}{0}

In this section we consider the derivation of the equations of motion for the expected value of the slowly varying color charges ${\mathcal Q}^{\hspace{0.03cm}a_{1}}_{\hspace{0.03cm}1}$ and ${\mathcal Q}^{\hspace{0.03cm}a_{2}}_{\hspace{0.03cm}2}$. To be more specific, consider the detailed derivation of the equation for the averaged color charge $\langle\hspace{0.01cm}\mathcal{Q}^{\,a_{1}}_{\hspace{0.03cm}1}(\tau)\hspace{0.03cm}\rangle$.\\
\indent Our first step is to average equation (\ref{eq:5o}). Thus, we obtain
\begin{align}
\frac{d \hspace{0.01cm}\bigl\langle\hspace{0.01cm}
\mathcal{Q}^{\,a_{1}}_{\hspace{0.03cm}1}(\tau)\hspace{0.02cm}\bigr\rangle}{d\hspace{0.03cm}\tau}
\,=
\hspace{0.03cm}
f^{\hspace{0.03cm}a_{1}\hspace{0.03cm}c^{\hspace{0.02cm}\prime}\hspace{0.01cm}e_{1}}
\,\sum_{\rho}\,\biggl[\,
&\int\!d\hspace{0.02cm}{\bf k}\,
{T}^{\hspace{0.03cm}(\rho)\hspace{0.03cm}a\,c^{\hspace{0.02cm}\prime}
	\hspace{0.03cm}a_{2}}_{\; {\bf k}}(t')\,
\bigl\langle\hspace{0.01cm}\widehat{C}^{\hspace{0.02cm}(\rho)\hspace{0.02cm}a}_{\hspace{0.02cm}{\bf k}}(t')
\hspace{0.02cm}
{\mathcal Q}^{\hspace{0.03cm}e_{1}}_{\hspace{0.03cm}1}(\tau)
\hspace{0.02cm}
{\mathcal Q}^{\hspace{0.03cm}a_{2}}_{\hspace{0.03cm}2}(\tau)
\hspace{0.02cm}\bigr\rangle
\label{eq:9q}\\[1ex]
%
+\!
&\int\!d\hspace{0.02cm}{\bf k}\,
{T}^{\hspace{0.03cm}\ast\hspace{0.03cm}(\rho)\hspace{0.03cm}a\,
	c^{\hspace{0.02cm}\prime}\hspace{0.03cm}a_{2}}_{\; {\bf k}}(t')\,
\bigl\langle\hspace{0.01cm}\widehat{C}^{\,\ast\hspace{0.02cm}(\rho)\hspace{0.02cm}a}_{\hspace{0.02cm}{\bf k}}(t')
\hspace{0.02cm}
{\mathcal Q}^{\hspace{0.03cm}e_{1}}_{\hspace{0.03cm}1}(\tau)
\hspace{0.02cm}
{\mathcal Q}^{\hspace{0.03cm}a_{2}}_{\hspace{0.03cm}2}(\tau)
\hspace{0.02cm}\bigr\rangle
\biggr]
\notag\\
+\,
f^{\hspace{0.03cm}a_{1}\hspace{0.03cm}c^{\hspace{0.02cm}\prime}\hspace{0.01cm}e_{1}}
\,\sum_{\rho}\,\biggl[\,
&\int\!d\hspace{0.02cm}{\bf k}\,
{T}^{\hspace{0.03cm}(\rho)\hspace{0.03cm}a\,c^{\hspace{0.02cm}\prime}
	\hspace{0.03cm}a_{2}}_{\; {\bf k}}(t')\,
\bigl\langle\hspace{0.01cm}C^{\hspace{0.02cm}(\rho)\hspace{0.02cm}a}_{\hspace{0.02cm}{\bf k}}(\tau)
\hspace{0.02cm}
\bigl(
{\mathcal Q}^{\hspace{0.03cm}a_{2}}_{\hspace{0.03cm}2}(\tau)
\hspace{0.02cm}
\widehat{\mathcal{Q}}^{\hspace{0.03cm}e_{1}}_{\hspace{0.03cm}1}(t')
+
\widehat{\mathcal{Q}}^{\hspace{0.03cm}a_{2}}_{\hspace{0.03cm}2}(t')
\hspace{0.02cm}
{\mathcal Q}^{\hspace{0.03cm}e_{1}}_{\hspace{0.03cm}1}(\tau)
\bigr)\hspace{0.02cm}\bigr\rangle
\notag\\[1ex]
+\!
&\int\!d\hspace{0.02cm}{\bf k}\,
{T}^{\hspace{0.03cm}\ast\hspace{0.03cm}(\rho)\hspace{0.03cm}a\,
	c^{\hspace{0.02cm}\prime}\hspace{0.03cm}a_{2}}_{\; {\bf k}}(t')\,
\bigl\langle\hspace{0.01cm}C^{\,\ast\hspace{0.03cm}(\rho)\hspace{0.02cm}a}_{\hspace{0.02cm}{\bf k}}(\tau)
\hspace{0.02cm}
\bigl(
{\mathcal Q}^{\hspace{0.03cm}a_{2}}_{\hspace{0.03cm}2}(\tau)
\hspace{0.02cm}
\widehat{\mathcal{Q}}^{\hspace{0.03cm}e_{1}}_{\hspace{0.03cm}1}(t')
+
\widehat{\mathcal{Q}}^{\hspace{0.03cm}a_{2}}_{\hspace{0.03cm}2}(t')
\hspace{0.02cm}
{\mathcal Q}^{\hspace{0.03cm}e_{1}}_{\hspace{0.03cm}1}(\tau)
\bigr)\hspace{0.02cm}\bigr\rangle\biggr].
\notag
\end{align}
The next step is to put the representations for the rapidly varying field component $\widehat{C}^{\hspace{0.02cm}(\alpha)\hspace{0.02cm}a}_{\hspace{0.02cm}{\bf k}}(t')$, Eq.\,(\ref{eq:5d}), and for the rapidly varying color charges $\widehat{\mathcal{Q}}^{\hspace{0.03cm}d}_{\hspace{0.03cm}\alpha}(t')$, Eqs.\,(\ref{eq:5a}) and (\ref{eq:5s}), into the right-hand side of equation (\ref{eq:9q}). Finally, the third and last step is to use approximations of the form (\ref{eq:6www}) for the correlation functions of the slowly varying variables. Taking into account the color and momentum decomposition of the effective amplitude ${T}^{\hspace{0.03cm}(\alpha)\hspace{0.03cm}a\,a_{1}\hspace{0.03cm}a_{2}}_{\; {\bf k}}(t')$, Eq.\,(\ref{eq:6wwww}), as a result we have, instead of (\ref{eq:9q}),
\begin{equation}
\frac{d\hspace{0.03cm}\bigl\langle \mathcal{Q}^{\,a_{1}}_{\hspace{0.03cm}1}(\tau)\hspace{0.02cm}\bigr\rangle}{d\hspace{0.03cm}\tau}
=
\label{eq:9w}
\end{equation}
\begin{align}
-\,\sum_{\rho,\,\rho^{\prime}}\,
\int\!d\hspace{0.02cm}{\bf k}\,
{T}^{\hspace{0.03cm}(\rho)}_{\; {\bf k}}(t')
\biggl(\int{T}^{\,\ast\hspace{0.03cm}(\rho^{\prime})}_{\; {\bf k}}(t') \hspace{0.03cm}dt'\biggr)
\Bigl[&{\rm tr}\hspace{0.03cm}
\bigl(\hspace{0.03cm}T^{\,a^{\prime}_{2}}\hspace{0.03cm}T^{\,a_{1}}
\hspace{0.03cm}T^{\,e^{\prime}_{1}}\hspace{0.03cm}
T^{\,a^{\prime\prime}_{2}}{\mathcal N}^{\,(\rho^{\prime},\hspace{0.03cm}\rho)}_{\hspace{0.02cm}
{\bf k}}\hspace{0.01cm}
\bigr)
\hspace{0.01cm}
\bigl\langle\hspace{0.03cm}\mathcal{Q}^{\hspace{0.03cm}a^{\prime}_{2}}_{2}(\tau)\hspace{0.03cm}\bigr\rangle\,-
\notag\\
i\hspace{0.03cm}f^{\hspace{0.03cm}a_{1}\hspace{0.02cm}c^{\prime}\hspace{0.03cm}e_{1}}\,
&{\rm tr}\hspace{0.03cm}
\bigl(\hspace{0.03cm}T^{\,c^{\prime}}\hspace{0.03cm}T^{\,a^{\prime\prime}_{2}}
\hspace{0.03cm}T^{\,e^{\prime}_{1}}
{\mathcal N}^{\,(\rho^{\prime},\hspace{0.03cm}\rho)}_{\hspace{0.02cm}{\bf k}}\hspace{0.01cm}
\bigr)
\hspace{0.03cm}\bigl\langle\hspace{0.03cm}\mathcal{Q}^{\hspace{0.03cm}e_{1}}_{1}(\tau)
\hspace{0.03cm}\bigr\rangle
\Bigr]
\hspace{0.01cm}
\bigl\langle\hspace{0.03cm}\mathcal{Q}^{\hspace{0.03cm}a^{\prime\prime}_{2}}_{2}(\tau)\hspace{0.03cm}\bigr\rangle
\hspace{0.03cm}\bigl\langle\hspace{0.03cm}\mathcal{Q}^{\hspace{0.03cm}e^{\prime}_{1}}_{1}(\tau)
\hspace{0.03cm}\bigr\rangle\,-
\notag\\[1ex]
-\,\sum_{\rho,\,\rho^{\prime}}\,
\int\!d\hspace{0.02cm}{\bf k}\,
{T}^{\,\ast\hspace{0.03cm}(\rho)}_{\; {\bf k}}(t')
\biggl(\int{T}^{\hspace{0.03cm}(\rho^{\prime})}_{\;{\bf k}}(t')\hspace{0.03cm}dt'\biggr)
\Bigl[&{\rm tr}\hspace{0.03cm}\bigl({\mathcal N}^{\,(\rho,\hspace{0.03cm}\rho^{\prime})}_{\hspace{0.02cm}{\bf k}}
\hspace{0.03cm}T^{\,a^{\prime\prime}_{2}}\hspace{0.03cm}T^{\,e^{\prime}_{1}}\hspace{0.03cm}T^{\,a_{1}}\hspace{0.03cm}T^{\,a^{\prime}_{2}}\hspace{0.01cm}
\bigr)
\hspace{0.01cm}
\bigl\langle\hspace{0.03cm}\mathcal{Q}^{\hspace{0.03cm}a^{\prime}_{2}}_{2}(\tau)\hspace{0.03cm}\bigr\rangle\,+
\notag\\
i\hspace{0.03cm}f^{\hspace{0.03cm}a_{1}\hspace{0.02cm}c^{\prime}\hspace{0.03cm}e_{1}}\,
&{\rm tr}\hspace{0.03cm}\bigl({\mathcal N}^{\,(\rho,\hspace{0.03cm}\rho^{\prime})}_{\hspace{0.02cm}{\bf k}}
\hspace{0.03cm}T^{\,e^{\prime}_{1}}\hspace{0.03cm}T^{\,a^{\prime\prime}_{2}}\hspace{0.03cm}T^{\,c^{\prime}}\hspace{0.01cm}
\bigr)
\hspace{0.03cm}\bigl\langle\hspace{0.03cm}\mathcal{Q}^{\hspace{0.03cm}e_{1}}_{1}(\tau)
\hspace{0.03cm}\bigr\rangle
\Bigr]
\hspace{0.01cm}
\bigl\langle\hspace{0.03cm}\mathcal{Q}^{\hspace{0.03cm}a^{\prime\prime}_{2}}_{2}(\tau)\hspace{0.03cm}\bigr\rangle
\hspace{0.03cm}\bigl\langle\hspace{0.03cm}\mathcal{Q}^{\hspace{0.03cm}e^{\prime}_{1}}_{1}(\tau)
\hspace{0.03cm}\bigr\rangle\,-
\notag
\end{align}
\begin{align}
\sum_{\rho}\,\biggl[\,
&\int\!d\hspace{0.02cm}{\bf k}\,
{T}^{\hspace{0.03cm}(\rho)}_{\; {\bf k}}(t')
\biggl(\int {T}^{\hspace{0.03cm}\ast\hspace{0.03cm}(\rho)}_{\; {\bf k}}(t')\hspace{0.03cm}dt'\biggr)
-
\int\!d\hspace{0.02cm}{\bf k}\,
\hspace{0.03cm}
{T}^{\,\ast\hspace{0.03cm}(\rho)}_{\; {\bf k}}(t')
\biggl(\int
%
{T}^{\hspace{0.03cm}(\rho)}_{\; {\bf k}}(t')\hspace{0.03cm}dt'\biggr)\biggr]
\hspace{0.03cm}\times
\notag\\[1ex]	
&\bigl(\hspace{0.03cm}T^{\,e_{1}}\hspace{0.03cm}T^{\,a^{\prime}_{2}}\hspace{0.03cm}T^{\,a^{\prime}_{1}}\hspace{0.01cm}
\bigr)^{\hspace{0.01cm}a^{\phantom{\prime}}_{1}\hspace{0.03cm}a^{\prime\prime}_{2}}\hspace{0.03cm}
\bigl\langle\hspace{0.03cm}\mathcal{Q}^{\hspace{0.03cm}a^{\prime}_{1}}_{1}(\tau)
\hspace{0.03cm}\bigr\rangle
\hspace{0.03cm}\bigl\langle\hspace{0.03cm}\mathcal{Q}^{\hspace{0.03cm}a^{\prime\prime}_{2}}_{2}(\tau)
\hspace{0.03cm}\bigr\rangle
\hspace{0.03cm}\bigl\langle\hspace{0.03cm}\mathcal{Q}^{\hspace{0.03cm}a^{\prime}_{2}}_{2}(\tau)
\hspace{0.03cm}\bigr\rangle
\hspace{0.01cm}\bigl\langle\hspace{0.03cm}\mathcal{Q}^{\hspace{0.03cm}e_{1}}_{1}(\tau)
\hspace{0.03cm}\bigr\rangle.
\notag
\end{align}
Let us make a special note of the fact that the last contribution on the right-hand side of (\ref{eq:9w}) does not depend at all on the plasmon number density ${\mathcal N}^{\,(\rho^{\prime},\hspace{0.03cm}\rho)}_{\hspace{0.02cm}{\bf k}}(\tau)$.
The physical meaning of this contribution is not entirely clear to us. However, as will be shown in the following discussion, the terms of such a kind vanish in all the dynamic equations relevant to us for the colorless combinations (\ref{eq:7w}). The reason for this lies in the fact that the color factors of these contributions, which is not related to the dynamics of the system, become zero. \\
\indent Now we use for the plasmon number density ${\mathcal N}^{\,(\rho^{\prime},\hspace{0.03cm}\rho)}_{\hspace{0.02cm}{\bf k}}(\tau)$ the representation (\ref{eq:6t}). Substituting this representation into the right-hand side of Eq.\,(\ref{eq:9w}) and summing over $\rho^{\prime}$, we derive  
\newpage 
\begin{equation}
\frac{d\hspace{0.03cm}\bigl\langle \mathcal{Q}^{\,a_{1}}_{\hspace{0.03cm}1}(\tau)\hspace{0.02cm}\bigr\rangle}{d\hspace{0.03cm}\tau}
=
\label{eq:9e}
\end{equation}
\begin{align}
	-\,\frac{1}{2}\,\sum_{\rho}\,
	\int\!d\hspace{0.02cm}{\bf k}\,
	{T}^{\hspace{0.03cm}(\rho)}_{\; {\bf k}}(t')
	\biggl(\int{T}^{\,\ast\hspace{0.03cm}(\rho)}_{\; {\bf k}}(t')\hspace{0.03cm}dt'\biggr)
	\Bigl[&{\rm tr}\hspace{0.03cm}
	\bigl(\hspace{0.03cm}T^{\,a^{\prime}_{2}}\hspace{0.03cm}T^{\,a_{1}}\hspace{0.03cm}T^{\,e^{\prime}_{1}}\hspace{0.03cm}T^{\,a^{\prime\prime}_{2}}{\mathcal N}_{\hspace{0.02cm}{\bf k}}\hspace{0.01cm}
	\bigr)
	\hspace{0.01cm}
	\bigl\langle\hspace{0.03cm}\mathcal{Q}^{\hspace{0.03cm}a^{\prime}_{2}}_{2}(\tau)\hspace{0.03cm}\bigr\rangle\,-
	\notag\\
	i\hspace{0.03cm}f^{\hspace{0.03cm}a_{1}\hspace{0.02cm}c^{\prime}\hspace{0.03cm}e_{1}}\,
	&{\rm tr}\hspace{0.03cm}
	\bigl(\hspace{0.03cm}T^{\,c^{\prime}}\hspace{0.03cm}T^{\,a^{\prime\prime}_{2}}\hspace{0.03cm}T^{\,e^{\prime}_{1}}{\mathcal N}_{\hspace{0.02cm}{\bf k}}\hspace{0.01cm}
	\bigr)
	\hspace{0.03cm}\bigl\langle\hspace{0.03cm}\mathcal{Q}^{\hspace{0.03cm}e_{1}}_{1}(\tau)
	\hspace{0.03cm}\bigr\rangle
	\Bigr]
	\hspace{0.01cm}
	\bigl\langle\hspace{0.03cm}\mathcal{Q}^{\hspace{0.03cm}a^{\prime\prime}_{2}}_{2}(\tau)\hspace{0.03cm}\bigr\rangle
	\hspace{0.03cm}\bigl\langle\hspace{0.03cm}\mathcal{Q}^{\hspace{0.03cm}e^{\prime}_{1}}_{1}(\tau)
	\hspace{0.03cm}\bigr\rangle\,-
	\notag
\end{align}
\begin{align}
	\hspace{0.6cm}
	\frac{1}{2}\,\sum_{\rho}\,
	\int\!d\hspace{0.02cm}{\bf k}\,
	{T}^{\,\ast\hspace{0.03cm}(\rho)}_{\; {\bf k}}(t')
	\biggl(\int{T}^{\hspace{0.03cm}(\rho)}_{\;{\bf k}}(t')\hspace{0.03cm}dt'\biggr)
	\Bigl[&{\rm tr}\hspace{0.03cm}\bigl({\mathcal N}_{\hspace{0.02cm}{\bf k}}
	\hspace{0.03cm}T^{\,a^{\prime\prime}_{2}}\hspace{0.03cm}T^{\,e^{\prime}_{1}}\hspace{0.03cm}T^{\,a_{1}}\hspace{0.03cm}T^{\,a^{\prime}_{2}}\hspace{0.01cm}
	\bigr)
	\hspace{0.01cm}
	\bigl\langle\hspace{0.03cm}\mathcal{Q}^{\hspace{0.03cm}a^{\prime}_{2}}_{2}(\tau)\hspace{0.03cm}\bigr\rangle\,+
	\notag\\
	i\hspace{0.03cm}f^{\hspace{0.03cm}a_{1}\hspace{0.02cm}c^{\prime}\hspace{0.03cm}e_{1}}\,
	&{\rm tr}\hspace{0.03cm}\bigl({\mathcal N}_{\hspace{0.02cm}{\bf k}}
	\hspace{0.03cm}T^{\,e^{\prime}_{1}}\hspace{0.03cm}T^{\,a^{\prime\prime}_{2}}\hspace{0.03cm}T^{\,c^{\prime}}\hspace{0.01cm}
	\bigr)
	\hspace{0.03cm}\bigl\langle\hspace{0.03cm}\mathcal{Q}^{\hspace{0.03cm}e_{1}}_{1}(\tau)
	\hspace{0.03cm}\bigr\rangle
	\Bigr]
	\hspace{0.01cm}
	\bigl\langle\hspace{0.03cm}\mathcal{Q}^{\hspace{0.03cm}a^{\prime\prime}_{2}}_{2}(\tau)\hspace{0.03cm}\bigr\rangle
	\hspace{0.03cm}\bigl\langle\hspace{0.03cm}\mathcal{Q}^{\hspace{0.03cm}e^{\prime}_{1}}_{1}(\tau)
	\hspace{0.03cm}\bigr\rangle\,-
	\notag
\end{align}
\begin{align}
	\frac{1}{2}\,\sum_{\rho}\,(-1)^{\rho +1}\!\!
	\int\!d\hspace{0.02cm}{\bf k}\,
	{T}^{\hspace{0.03cm}(\rho)}_{\; {\bf k}}(t')
	\biggl(\int{T}^{\,\ast\hspace{0.03cm}(\rho)}_{\; {\bf k}}(&t')\hspace{0.03cm}dt'\biggr)
	\Bigl[{\rm tr}\hspace{0.03cm}
	\bigl(\hspace{0.03cm}T^{\,a^{\prime}_{2}}\hspace{0.03cm}T^{\,a_{1}}\hspace{0.03cm}T^{\,e^{\prime}_{1}}\hspace{0.03cm}T^{\,a^{\prime\prime}_{2}}{\mathcal W}_{\hspace{0.02cm}{\bf k}}\hspace{0.01cm}
	\bigr)
	\hspace{0.01cm}
	\bigl\langle\hspace{0.03cm}\mathcal{Q}^{\hspace{0.03cm}a^{\prime}_{2}}_{2}(\tau)\hspace{0.03cm}\bigr\rangle\,-
	\notag\\
	i\hspace{0.03cm}f^{\hspace{0.03cm}a_{1}\hspace{0.02cm}c^{\prime}\hspace{0.03cm}e_{1}}\,
	&{\rm tr}\hspace{0.03cm}
	\bigl(\hspace{0.03cm}T^{\,c^{\prime}}\hspace{0.03cm}T^{\,a^{\prime\prime}_{2}}\hspace{0.03cm}T^{\,e^{\prime}_{1}}{\mathcal W}_{\hspace{0.02cm}{\bf k}}\hspace{0.01cm}
	\bigr)
	\hspace{0.03cm}\bigl\langle\hspace{0.03cm}\mathcal{Q}^{\hspace{0.03cm}e_{1}}_{1}(\tau)
	\hspace{0.03cm}\bigr\rangle
	\Bigr]
	\hspace{0.01cm}
	\bigl\langle\hspace{0.03cm}\mathcal{Q}^{\hspace{0.03cm}a^{\prime\prime}_{2}}_{2}(\tau)\hspace{0.03cm}\bigr\rangle
	\hspace{0.03cm}\bigl\langle\hspace{0.03cm}\mathcal{Q}^{\hspace{0.03cm}e^{\prime}_{1}}_{1}(\tau)
	\hspace{0.03cm}\bigr\rangle\,-
	\notag
\end{align}
\begin{align}
	\frac{1}{2}\,\sum_{\rho}\,(-1)^{\rho +1}\!\!
	\int\!d\hspace{0.02cm}{\bf k}\,
	{T}^{\,\ast\hspace{0.03cm}(\rho)}_{\; {\bf k}}(t')
	\biggl(\int{T}^{\hspace{0.03cm}(\rho)}_{\;{\bf k}}(&t')\hspace{0.03cm}dt'\biggr)
	\Bigl[{\rm tr}\hspace{0.03cm}\bigl({\mathcal W}_{\hspace{0.02cm}{\bf k}}
	\hspace{0.03cm}T^{\,a^{\prime\prime}_{2}}\hspace{0.03cm}T^{\,e^{\prime}_{1}}\hspace{0.03cm}T^{\,a_{1}}\hspace{0.03cm}T^{\,a^{\prime}_{2}}\hspace{0.01cm}
	\bigr)
	\hspace{0.01cm}
	\bigl\langle\hspace{0.03cm}\mathcal{Q}^{\hspace{0.03cm}a^{\prime}_{2}}_{2}(\tau)\hspace{0.03cm}\bigr\rangle\,+
	\notag\\
	i\hspace{0.03cm}f^{\hspace{0.03cm}a_{1}\hspace{0.02cm}c^{\prime}\hspace{0.03cm}e_{1}}\,
	&{\rm tr}\hspace{0.03cm}\bigl({\mathcal W}_{\hspace{0.02cm}{\bf k}}
	\hspace{0.03cm}T^{\,e^{\prime}_{1}}\hspace{0.03cm}T^{\,a^{\prime\prime}_{2}}\hspace{0.03cm}T^{\,c^{\prime}}\hspace{0.01cm}
	\bigr)
	\hspace{0.03cm}\bigl\langle\hspace{0.03cm}\mathcal{Q}^{\hspace{0.03cm}e_{1}}_{1}(\tau)
	\hspace{0.03cm}\bigr\rangle
	\Bigr]
	\hspace{0.01cm}
	\bigl\langle\hspace{0.03cm}\mathcal{Q}^{\hspace{0.03cm}a^{\prime\prime}_{2}}_{2}(\tau)\hspace{0.03cm}\bigr\rangle
	\hspace{0.03cm}\bigl\langle\hspace{0.03cm}\mathcal{Q}^{\hspace{0.03cm}e^{\prime}_{1}}_{1}(\tau)
	\hspace{0.03cm}\bigr\rangle\,-
	\notag
\end{align}
\begin{align}
	\sum_{\rho}\,\biggl[\,
	&\int\!d\hspace{0.02cm}{\bf k}\,
	{T}^{\hspace{0.03cm}(\rho)}_{\; {\bf k}}(t')
	\biggl(\int {T}^{\hspace{0.03cm}\ast\hspace{0.03cm}(\rho)}_{\; {\bf k}}(t')\hspace{0.03cm}dt'\biggr)
	-
	\int\!d\hspace{0.02cm}{\bf k}\,
	\hspace{0.03cm}
	{T}^{\,\ast\hspace{0.03cm}(\rho)}_{\; {\bf k}}(t')
	\biggl(\int
	%
	{T}^{\hspace{0.03cm}(\rho)}_{\; {\bf k}}(t')\hspace{0.03cm}dt'\biggr)\biggr]
	\hspace{0.03cm}\times
	\notag\\[1ex]	
	&\bigl(\hspace{0.03cm}T^{\,e_{1}}\hspace{0.03cm}T^{\,a^{\prime}_{2}}\hspace{0.03cm}T^{\,a^{\prime}_{1}}\hspace{0.01cm}
	\bigr)^{\hspace{0.01cm}a^{\phantom{\prime}}_{1}\hspace{0.03cm}a^{\prime\prime}_{2}}\hspace{0.03cm}
	\bigl\langle\hspace{0.03cm}\mathcal{Q}^{\hspace{0.03cm}a^{\prime}_{1}}_{1}(\tau)
	\hspace{0.03cm}\bigr\rangle
	\hspace{0.03cm}\bigl\langle\hspace{0.03cm}\mathcal{Q}^{\hspace{0.03cm}a^{\prime\prime}_{2}}_{2}(\tau)
	\hspace{0.03cm}\bigr\rangle
	\hspace{0.03cm}\bigl\langle\hspace{0.03cm}\mathcal{Q}^{\hspace{0.03cm}a^{\prime}_{2}}_{2}(\tau)
	\hspace{0.03cm}\bigr\rangle
	\hspace{0.01cm}\bigl\langle\hspace{0.03cm}\mathcal{Q}^{\hspace{0.03cm}e_{1}}_{1}(\tau)
	\hspace{0.03cm}\bigr\rangle.
	\notag
\end{align}
Similar equation can be obtained for the second averaged color charge $\bigl\langle\hspace{0.03cm}\mathcal{Q}^{\hspace{0.03cm}a_{2}}_{2}(\tau)
\hspace{0.03cm}\bigr\rangle$. The form of this equation is given in Appendix \ref{appendix_C}, Eq.\,(\ref{ap:C1}). The right-hand side of the resulting equation (\ref{eq:9e}) is real by the construction and has a very complex structure. We are interested in the contributions on the right-hand side of (\ref{eq:9e}) related only to the plasmon bremsstrahlung process. This means that the integrands must contain the Dirac delta function
\[
2\hspace{0.02cm}\pi\hspace{0.03cm}\delta(\omega^{l}_{{\bf k}} - {\bf k}\cdot{\bf v}_{\rho}
-(-1)^{\rho}\, \Delta{\mathbf v}\cdot{\mathbf q}), 
\]
reflecting the energy-momentum conservation law in the bremsstrahlung process, as it takes place in the kinetic equations (\ref{eq:7x}), (\ref{eq:7c}), (\ref{eq:8u}) and (\ref{eq:8i}). Repeating the reasoning of two previous sections
in deriving these kinetic equations, we can represent (\ref{eq:9e}) in the following form:  
\begin{equation}
\frac{d\hspace{0.03cm}\bigl\langle \mathcal{Q}^{\,a_{1}}_{\hspace{0.03cm}1}(\tau)\hspace{0.02cm}\bigr\rangle}{d\hspace{0.03cm}\tau}
=
\label{eq:9r}
\vspace{-0.5cm}
\end{equation}
\begin{align}
-\hspace{0.03cm}&\frac{(2\hspace{0.02cm}\pi)^{3}}{2\hspace{0.03cm}|\Delta{\mathbf v}|}\,\sum_{\rho}\,
\int\!d\hspace{0.02cm}{\bf k}\hspace{0.03cm}d\hspace{0.02cm}{\bf q}\;
\bigl|\hspace{0.03cm}{T}^{\hspace{0.03cm}(\rho)}_{\; {\bf k},\,{\bf q}}\bigr|^{\hspace{0.02cm}2}\,
2\hspace{0.02cm}\pi\hspace{0.03cm}\delta(\omega^{l}_{{\bf k}} - {\bf k}\cdot{\bf v}_{\rho}
-(-1)^{\rho}\, \Delta{\mathbf v}\cdot{\mathbf q})
\,\times
	\notag\\
	&\Bigl[{\rm tr}\hspace{0.03cm}
	\bigl(\hspace{0.03cm}T^{\,a^{\prime}_{2}}\hspace{0.03cm}T^{\,a_{1}}\hspace{0.03cm}T^{\,e^{\prime}_{1}}\hspace{0.03cm}T^{\,a^{\prime\prime}_{2}}{\mathcal N}_{\hspace{0.02cm}{\bf k}}\hspace{0.01cm}
	\bigr)
	\hspace{0.01cm}
	\bigl\langle\hspace{0.03cm}\mathcal{Q}^{\hspace{0.03cm}a^{\prime}_{2}}_{2}(\tau)\hspace{0.03cm}\bigr\rangle
	-	
	i\hspace{0.03cm}f^{\hspace{0.03cm}a_{1}\hspace{0.02cm}c^{\prime}\hspace{0.03cm}e_{1}}\hspace{0.03cm}
	{\rm tr}\hspace{0.03cm}
	\bigl(\hspace{0.03cm}T^{\,c^{\prime}}\hspace{0.03cm}T^{\,a^{\prime\prime}_{2}}\hspace{0.03cm}T^{\,e^{\prime}_{1}}{\mathcal N}_{\hspace{0.02cm}{\bf k}}\hspace{0.01cm}
	\bigr)
	\hspace{0.03cm}\bigl\langle\hspace{0.03cm}\mathcal{Q}^{\hspace{0.03cm}e_{1}}_{1}(\tau)
	\hspace{0.03cm}\bigr\rangle
	\Bigr]
	\hspace{0.01cm}
	\bigl\langle\hspace{0.03cm}\mathcal{Q}^{\hspace{0.03cm}a^{\prime\prime}_{2}}_{2}(\tau)\hspace{0.03cm}\bigr\rangle
	\hspace{0.03cm}\bigl\langle\hspace{0.03cm}\mathcal{Q}^{\hspace{0.03cm}e^{\prime}_{1}}_{1}(\tau)
	\hspace{0.03cm}\bigr\rangle\,-
	\notag
\end{align}
\begin{align}
\hspace{0.4cm}&\frac{(2\hspace{0.02cm}\pi)^{3}}{2\hspace{0.03cm}|\Delta{\mathbf v}|}\,\sum_{\rho}\hspace{0.02cm}(-1)^{\rho + 1}\!\!
	\int\!d\hspace{0.02cm}{\bf k}\hspace{0.03cm}d\hspace{0.02cm}{\bf q}\;
	\bigl|\hspace{0.03cm}{T}^{\hspace{0.03cm}(\rho)}_{\; {\bf k},\,{\bf q}}\bigr|^{\hspace{0.02cm}2}\,
	2\hspace{0.02cm}\pi\hspace{0.03cm}\delta(\omega^{l}_{{\bf k}} - {\bf k}\cdot{\bf v}_{\rho}
	-(-1)^{\rho}\, \Delta{\mathbf v}\cdot{\mathbf q}) 
	\,\times
	\notag\\
	&\Bigl[{\rm tr}\hspace{0.03cm}
	\bigl(\hspace{0.03cm}T^{\,a^{\prime}_{2}}\hspace{0.03cm}T^{\,a_{1}}\hspace{0.03cm}T^{\,e^{\prime}_{1}}\hspace{0.03cm}T^{\,a^{\prime\prime}_{2}}\hspace{0.03cm}{\mathcal W}_{\hspace{0.02cm}{\bf k}}\hspace{0.01cm}
	\bigr)
	\hspace{0.01cm}
	\bigl\langle\hspace{0.03cm}\mathcal{Q}^{\hspace{0.03cm}a^{\prime}_{2}}_{2}(\tau)\hspace{0.03cm}\bigr\rangle
	-
	i\hspace{0.03cm}f^{\hspace{0.03cm}a_{1}\hspace{0.02cm}c^{\prime}\hspace{0.03cm}e_{1}}\hspace{0.03cm}
	{\rm tr}\hspace{0.03cm}
	\bigl(\hspace{0.03cm}T^{\,c^{\prime}}\hspace{0.03cm}T^{\,a^{\prime\prime}_{2}}\hspace{0.03cm}T^{\,e^{\prime}_{1}}\hspace{0.03cm}{\mathcal W}_{\hspace{0.02cm}{\bf k}}\hspace{0.01cm}
	\bigr)
	\hspace{0.03cm}\bigl\langle\hspace{0.03cm}\mathcal{Q}^{\hspace{0.03cm}e_{1}}_{1}(\tau)
	\hspace{0.03cm}\bigr\rangle
	\Bigr]
	\hspace{0.01cm}
	\bigl\langle\hspace{0.03cm}\mathcal{Q}^{\hspace{0.03cm}a^{\prime\prime}_{2}}_{2}(\tau)\hspace{0.03cm}\bigr\rangle
	\hspace{0.03cm}\bigl\langle\hspace{0.03cm}\mathcal{Q}^{\hspace{0.03cm}e^{\prime}_{1}}_{1}(\tau)
	\hspace{0.03cm}\bigr\rangle\,+ 
	\notag\\
	&\ldots\,.
	\notag
\end{align}
The dots on the right side denotes the last contribution to (\ref{eq:9e}), which is independent of the plasmon number densities ${\mathcal N}_{\hspace{0.02cm}{\bf k}}$ and ${\mathcal W}_{\hspace{0.02cm}{\bf k}}$. We analyze this contribution just below. Equation (\ref{eq:9r}), like equation (\ref{eq:8w}), is valid only if the contractions of the type 
\begin{equation}
	\begin{split}
		&{\rm tr}\hspace{0.03cm}
		\bigl(\hspace{0.03cm}T^{\,a^{\prime}_{2}}\hspace{0.03cm}T^{\,a_{1}}\hspace{0.03cm}T^{\,e^{\prime}_{1}}\hspace{0.03cm}T^{\,a^{\prime\prime}_{2}}{\mathcal N}_{\hspace{0.02cm}{\bf k}}\hspace{0.01cm}
		\bigr)
		\hspace{0.01cm}
		\bigl\langle\hspace{0.03cm}\mathcal{Q}^{\hspace{0.03cm}a^{\prime}_{2}}_{2}(\tau)\hspace{0.03cm}\bigr\rangle
		\bigl\langle\hspace{0.03cm}\mathcal{Q}^{\hspace{0.03cm}a^{\prime\prime}_{2}}_{2}(\tau)\hspace{0.03cm}\bigr\rangle
		\hspace{0.03cm}
		\bigl\langle\hspace{0.03cm}\mathcal{Q}^{\hspace{0.03cm}e^{\prime}_{1}}_{1}(\tau)\hspace{0.03cm}\bigr\rangle,\\[1ex]
		i\hspace{0.03cm}&f^{\hspace{0.03cm}a_{1}\hspace{0.02cm}c^{\prime}\hspace{0.03cm}e_{1}}\hspace{0.03cm}
		{\rm tr}\hspace{0.03cm}
		\bigl(\hspace{0.03cm}T^{\,c^{\prime}}\hspace{0.03cm}T^{\,a^{\prime\prime}_{2}}\hspace{0.03cm}T^{\,e^{\prime}_{1}}{\mathcal N}_{\hspace{0.02cm}{\bf k}}\hspace{0.01cm}
		\bigr)
		\hspace{0.03cm}\bigl\langle\hspace{0.03cm}\mathcal{Q}^{\hspace{0.03cm}e_{1}}_{1}(\tau)
		\hspace{0.03cm}\bigr\rangle
		\bigl\langle\hspace{0.03cm}\mathcal{Q}^{\hspace{0.03cm}a^{\prime\prime}_{2}}_{2}(\tau)\hspace{0.03cm}\bigr\rangle
		\hspace{0.03cm}
		\bigl\langle\hspace{0.03cm}\mathcal{Q}^{\hspace{0.03cm}e^{\prime}_{1}}_{1}(\tau)\hspace{0.03cm}\bigr\rangle
	\end{split}
	\label{eq:9t}
\end{equation}	
and so on, are real functions. In the general case, these contractions are complex and therefore the equation (\ref{eq:9r}), as such, does not occur. However, we are only interested in the specific colorless combinations ${\mathfrak q}_{1}(\tau),\,{\mathfrak q}_{2}(\tau),\,{\mathfrak q}_{12}(\tau)$ and $\Lambda^{2}(\tau)$ of the averaged color charges, as were defined by relations (\ref{eq:7w}), for which the imaginary parts of the contractions of the type (\ref{eq:9t}) can be zero.\\
\indent Let us define the equation for the first colorless combination ${\mathfrak q}_{1}(\tau) = 
\hspace{0.03cm}\bigl\langle\hspace{0.03cm}\mathcal{Q}^{\hspace{0.03cm}a_{1}}_{1}
\hspace{0.03cm}\bigr\rangle
\bigl\langle\hspace{0.03cm}\mathcal{Q}^{\hspace{0.03cm}a_{1}}_{1}
\hspace{0.03cm}\bigr\rangle$. The first step is to substitute the color decomposition for the matrix function ${\mathcal N}_{\hspace{0.02cm}{\bf k}}$, Eq.\,(\ref{eq:7q}), into the first contraction in (\ref{eq:9t}). As a result we have 
\begin{equation}
{\rm tr}\hspace{0.03cm}
\bigl(\hspace{0.03cm}T^{\,a^{\prime}_{2}}\hspace{0.03cm}T^{\,a_{1}}\hspace{0.03cm}T^{\,e^{\prime}_{1}}\hspace{0.03cm}T^{\,a^{\prime\prime}_{2}}{\mathcal N}_{\hspace{0.02cm}{\bf k}}\hspace{0.01cm}
\bigr)
=
{N}^{(1)}_{\hspace{0.02cm}{\bf k}}\hspace{0.03cm}
{\rm tr}\hspace{0.03cm}
\bigl(\hspace{0.03cm}T^{\,a^{\prime}_{2}}\hspace{0.03cm}T^{\,a_{1}}\hspace{0.03cm}T^{\,e^{\prime}_{1}}\hspace{0.03cm}T^{\,a^{\prime\prime}_{2}}\hspace{0.01cm}
\bigr)
+
\bigl\langle\hspace{0.03cm}\mathcal{Q}^{\,c}_{1}
\hspace{0.03cm}\bigr\rangle
\hspace{0.03cm}
{N}^{(2)}_{\hspace{0.02cm}{\bf k}}\hspace{0.03cm}
{\rm tr}\hspace{0.03cm}
\bigl(\hspace{0.03cm}T^{\,a^{\prime}_{2}}\hspace{0.03cm}T^{\,a_{1}}\hspace{0.03cm}T^{\,e^{\prime}_{1}}\hspace{0.03cm}T^{\,a^{\prime\prime}_{2}}\hspace{0.03cm}T^{\,c}\hspace{0.01cm}\bigr).
\label{eq:9y}
\end{equation}
The first term on the right-hand side is real, and the second is purely imaginary. We show that the second term in contracting with $\bigl\langle\hspace{0.03cm}\mathcal{Q}^{\hspace{0.03cm}a_{1}}_{1}\hspace{0.03cm}\bigr\rangle$ turns to zero. For this purpose we use the relation for the fifth-order trace (\ref{ap:D10}). By virtue of this relation, the following equality is valid   
\begin{equation}
\begin{split}
&{\rm tr}\hspace{0.03cm}
\bigl(\hspace{0.03cm}T^{\,a^{\prime}_{2}}\hspace{0.03cm}T^{\,a_{1}}\hspace{0.03cm}T^{\,e^{\prime}_{1}}\hspace{0.03cm}T^{\,a^{\prime\prime}_{2}}\hspace{0.03cm}T^{\,c}\hspace{0.01cm}\bigr)
\hspace{0.01cm}
\bigl\langle\hspace{0.03cm}\mathcal{Q}^{\hspace{0.03cm}a^{\prime}_{2}}_{2}(\tau)\hspace{0.03cm}\bigr\rangle
\bigl\langle\hspace{0.03cm}\mathcal{Q}^{\hspace{0.03cm}a^{\prime\prime}_{2}}_{2}(\tau)\hspace{0.03cm}\bigr\rangle
\hspace{0.03cm}
\bigl\langle\hspace{0.03cm}\mathcal{Q}^{\hspace{0.03cm}e^{\prime}_{1}}_{1}(\tau)\hspace{0.03cm}\bigr\rangle
=\\[1ex]
-\hspace{0.03cm} &\frac{i}{2}\,f^{\hspace{0.03cm}e^{\prime}_{1}\hspace{0.02cm}a_{1}\hspace{0.03cm}b}\,
{\rm tr}\hspace{0.03cm}
\bigl(\hspace{0.03cm}T^{\,b}\hspace{0.03cm}T^{\,a^{\prime}_{2}}\hspace{0.03cm}T^{\,c}T^{\,a^{\prime\prime}_{2}}\hspace{0.01cm}\bigr)
\hspace{0.01cm}
\bigl\langle\hspace{0.03cm}\mathcal{Q}^{\hspace{0.03cm}a^{\prime}_{2}}_{2}(\tau)\hspace{0.03cm}\bigr\rangle
\bigl\langle\hspace{0.03cm}\mathcal{Q}^{\hspace{0.03cm}a^{\prime\prime}_{2}}_{2}(\tau)\hspace{0.03cm}\bigr\rangle
\hspace{0.03cm}
\bigl\langle\hspace{0.03cm}\mathcal{Q}^{\hspace{0.03cm}e^{\prime}_{1}}_{1}(\tau)\hspace{0.03cm}\bigr\rangle.
\end{split}
\label{eq:9yyy}
\end{equation}
It is obvious from this expression that when it is contracted with $\bigl\langle\hspace{0.03cm}\mathcal{Q}^{\hspace{0.03cm}a_{1}}_{1}\hspace{0.03cm}\bigr\rangle$, it turns to zero due to the antisymmetry of the structural constants $f^{\hspace{0.03cm}e^{\prime}_{1}\hspace{0.02cm}a_{1}\hspace{0.03cm}b}$. Thus, the imaginary part of the first expression in (\ref{eq:9t}) when it is contracted with $\bigl\langle\hspace{0.03cm}\mathcal{Q}^{\hspace{0.03cm}a_{1}}_{1}\hspace{0.03cm}\bigr\rangle$ vanishes.\\
\indent Now let us analyze the real part in (\ref{eq:9y}), that is, the term with the scalar function ${N}^{(1)}_{\hspace{0.02cm}{\bf k}}$. Using the formula for the fourth-order trace (\ref{ap:D4}), we obtain
\begin{equation}
{N}^{(1)}_{\hspace{0.02cm}{\bf k}}\hspace{0.03cm}
{\rm tr}\hspace{0.03cm}
\bigl(\hspace{0.03cm}T^{\,a^{\prime}_{2}}\hspace{0.03cm}T^{\,a_{1}}\hspace{0.03cm}T^{\,e^{\prime}_{1}}\hspace{0.03cm}T^{\,a^{\prime\prime}_{2}}\hspace{0.01cm}
\bigr)
\hspace{0.01cm}
\bigl\langle\hspace{0.03cm}\mathcal{Q}^{\hspace{0.03cm}a_{1}}_{1}\hspace{0.03cm}
\bigr\rangle
\hspace{0.03cm}
\bigl\langle\hspace{0.03cm}\mathcal{Q}^{\hspace{0.03cm}a^{\prime}_{2}}_{2}(\tau)
\hspace{0.03cm}\bigr\rangle
\bigl\langle\hspace{0.03cm}\mathcal{Q}^{\hspace{0.03cm}a^{\prime\prime}_{2}}_{2}(\tau)\hspace{0.03cm}\bigr\rangle
\hspace{0.03cm}
\bigl\langle\hspace{0.03cm}\mathcal{Q}^{\hspace{0.03cm}e^{\prime}_{1}}_{1}(\tau)\hspace{0.03cm}\bigr\rangle
\label{eq:9u}
\end{equation}
\[
=
{N}^{(1)}_{\hspace{0.02cm}{\bf k}}\hspace{0.03cm}
\biggl\{
{\mathfrak q}_{1}{\mathfrak q}_{2} + {\mathfrak q}^{2}_{12} + \frac{1}{4}\,
N_{c}\,\Omega^{\hspace{0.03cm}e}_{\hspace{0.03cm}11}\hspace{0.03cm}\Omega^{\hspace{0.03cm}e}_{\hspace{0.03cm}22}
\biggr\}, 
\]
where the functions $\Omega^{\hspace{0.03cm}e}_{\hspace{0.03cm}11}$ and $\Omega^{\hspace{0.03cm}e}_{\hspace{0.03cm}22}$ were defined in (\ref{eq:7w}). Next, we will consider the special case -- the $SU(3_{c})$ color group. Then, by virtue of the formula (\ref{ap:E4}), the previous equality takes the form 
\begin{equation}
	{N}^{(1)}_{\hspace{0.02cm}{\bf k}}\hspace{0.03cm}
	{\rm tr}\hspace{0.03cm}
	\bigl(\hspace{0.03cm}T^{\,a^{\prime}_{2}}\hspace{0.03cm}T^{\,a_{1}}\hspace{0.03cm}T^{\,e^{\prime}_{1}}\hspace{0.03cm}T^{\,a^{\prime\prime}_{2}}\hspace{0.01cm}
	\bigr)
	\hspace{0.01cm}
	\bigl\langle\hspace{0.03cm}\mathcal{Q}^{\hspace{0.03cm}a_{1}}_{1}\hspace{0.03cm}
	\bigr\rangle
	\hspace{0.03cm}
	\bigl\langle\hspace{0.03cm}\mathcal{Q}^{\hspace{0.03cm}a^{\prime}_{2}}_{2}(\tau)
	\hspace{0.03cm}\bigr\rangle
	\bigl\langle\hspace{0.03cm}\mathcal{Q}^{\hspace{0.03cm}a^{\prime\prime}_{2}}_{2}(\tau)\hspace{0.03cm}\bigr\rangle
	\hspace{0.03cm}
	\bigl\langle\hspace{0.03cm}\mathcal{Q}^{\hspace{0.03cm}e^{\prime}_{1}}_{1}(\tau)\hspace{0.03cm}\bigr\rangle
	\label{eq:9i}
\end{equation}
\[
=
{N}^{(1)}_{\hspace{0.02cm}{\bf k}}\hspace{0.03cm}
\biggl\{\frac{3}{4}\,{\mathfrak q}_{1}{\mathfrak q}_{2} 
\,+\, 
\frac{3}{2}\,{\mathfrak q}^{2}_{12} \,+\, \frac{1}{2}\,\Lambda^{2} 
\biggr\}. 
\]
\indent Next, we consider the second expression in (\ref{eq:9t}), which, after substitution (\ref{eq:7q}) is rewritten as follows:
\begin{align}
&i\hspace{0.03cm}f^{\hspace{0.03cm}a_{1}\hspace{0.02cm}c^{\prime}\hspace{0.03cm}e_{1}}\hspace{0.03cm}
{\rm tr}\hspace{0.03cm}
\bigl(\hspace{0.03cm}T^{\,c^{\prime}}\hspace{0.03cm}T^{\,a^{\prime\prime}_{2}}
\hspace{0.03cm}T^{\,e^{\prime}_{1}}{\mathcal N}_{\hspace{0.02cm}{\bf k}}\hspace{0.01cm}
\bigr)
\hspace{0.03cm}\bigl\langle\hspace{0.03cm}\mathcal{Q}^{\hspace{0.03cm}e_{1}}_{1}(\tau)
\hspace{0.03cm}\bigr\rangle
\bigl\langle\hspace{0.03cm}\mathcal{Q}^{\hspace{0.03cm}a^{\prime\prime}_{2}}_{2}(\tau)\hspace{0.03cm}\bigr\rangle
\hspace{0.03cm}
\bigl\langle\hspace{0.03cm}\mathcal{Q}^{\hspace{0.03cm}e^{\prime}_{1}}_{1}(\tau)\hspace{0.03cm}\bigr\rangle
=
\label{eq:9ii}\\[1.5ex]
&i\hspace{0.03cm}{N}^{(1)}_{\hspace{0.02cm}{\bf k}}\hspace{0.03cm}f^{\hspace{0.03cm}a_{1}\hspace{0.02cm}c^{\prime}\hspace{0.03cm}e_{1}}\hspace{0.03cm}
{\rm tr}\hspace{0.03cm}
\bigl(\hspace{0.03cm}T^{\,c^{\prime}}\hspace{0.03cm}T^{\,a^{\prime\prime}_{2}}
\hspace{0.03cm}T^{\,e^{\prime}_{1}}\hspace{0.01cm}
\bigr)
\hspace{0.03cm}\bigl\langle\hspace{0.03cm}\mathcal{Q}^{\hspace{0.03cm}e_{1}}_{1}(\tau)
\hspace{0.03cm}\bigr\rangle
\bigl\langle\hspace{0.03cm}\mathcal{Q}^{\hspace{0.03cm}a^{\prime\prime}_{2}}_{2}(\tau)\hspace{0.03cm}\bigr\rangle
\hspace{0.03cm}
\bigl\langle\hspace{0.03cm}\mathcal{Q}^{\hspace{0.03cm}e^{\prime}_{1}}_{1}(\tau)\hspace{0.03cm}\bigr\rangle\,+
\notag\\[1.5ex]
&i\hspace{0.03cm}{N}^{(2)}_{\hspace{0.02cm}{\bf k}}\hspace{0.03cm}	\bigl\langle\hspace{0.03cm}\mathcal{Q}^{\hspace{0.03cm}c}_{1}
\hspace{0.03cm}\bigr\rangle
\hspace{0.03cm}f^{\hspace{0.03cm}a_{1}\hspace{0.02cm}c^{\prime}\hspace{0.03cm}e_{1}}\hspace{0.03cm}
{\rm tr}\hspace{0.03cm}
\bigl(\hspace{0.03cm}T^{\,c^{\prime}}\hspace{0.03cm}T^{\,a^{\prime\prime}_{2}}
\hspace{0.03cm}T^{\,e^{\prime}_{1}}\hspace{0.03cm}T^{\,c}\hspace{0.01cm}
\bigr)
\hspace{0.03cm}\bigl\langle\hspace{0.03cm}\mathcal{Q}^{\hspace{0.03cm}e_{1}}_{1}(\tau)
\hspace{0.03cm}\bigr\rangle
\bigl\langle\hspace{0.03cm}\mathcal{Q}^{\hspace{0.03cm}a^{\prime\prime}_{2}}_{2}(\tau)\hspace{0.03cm}\bigr\rangle
\hspace{0.03cm}
\bigl\langle\hspace{0.03cm}\mathcal{Q}^{\hspace{0.03cm}e^{\prime}_{1}}_{1}(\tau)\hspace{0.03cm}\bigr\rangle.
\notag
\end{align}
The first term on the right-hand side is real, and the second is purely imaginary. Obviously, both the real and imaginary parts vanish when they are contracted with $\bigl\langle\hspace{0.03cm}\mathcal{Q}^{\hspace{0.03cm}a_{1}}_{1}\hspace{0.03cm}\bigr\rangle$.\\
\indent Completely similar reasoning is valid for the contractions in equation (\ref{eq:9r}) containing the color matrix function ${\mathcal W}_{\hspace{0.02cm}{\bf k}}$. Using the color decomposition (\ref{eq:7q}), we get an analogue of the formula (\ref{eq:9i}) with the replacement ${N}^{(1)}_{\hspace{0.02cm}{\bf k}}\rightarrow{W}^{(1)}_{\hspace{0.02cm}{\bf k}}$. As a result, considering all of the above, after contracting the equation (\ref{eq:9r}) with 
$\bigl\langle\hspace{0.03cm}\mathcal{Q}^{\hspace{0.03cm}a_{1}}_{1}\hspace{0.03cm}\bigr\rangle$ we find the desired equation for the first colorless combination ${\mathfrak q}_{1}(\tau)$: 
\begin{align}
	\frac{d\hspace{0.03cm}{\mathfrak q}_{1}(\tau)}{d\hspace{0.03cm}\tau}
	=
	-\hspace{0.03cm}\biggl\{\frac{(2\hspace{0.02cm}\pi)^{3}}{|\Delta{\mathbf v}|}\,\sum_{\rho}\,
	&\int\!d\hspace{0.02cm}{\bf k}\hspace{0.03cm}d\hspace{0.02cm}{\bf q}\;
	\bigl|\hspace{0.03cm}{T}^{\hspace{0.03cm}(\rho)}_{\; {\bf k},\,{\bf q}}\bigr|^{\hspace{0.02cm}2}\hspace{0.03cm}
	{N}^{(1)}_{\hspace{0.02cm}{\bf k}}\hspace{0.03cm}
	2\hspace{0.02cm}\pi\hspace{0.03cm}\delta(\omega^{l}_{{\bf k}} - {\bf k}\cdot{\bf v}_{\rho}
	-(-1)^{\rho}\, \Delta{\mathbf v}\cdot{\mathbf q})\,+
	\label{eq:9o}\\[1ex]
	\frac{(2\hspace{0.02cm}\pi)^{3}}{|\Delta{\mathbf v}|}\,\sum_{\rho}\hspace{0.02cm}(-1)^{\rho + 1}\!\!
	&\int\!d\hspace{0.02cm}{\bf k}\hspace{0.03cm}d\hspace{0.02cm}{\bf q}\;
	\bigl|\hspace{0.03cm}{T}^{\hspace{0.03cm}(\rho)}_{\; {\bf k},\,{\bf q}}\bigr|^{\hspace{0.02cm}2}\hspace{0.03cm}
	{W}^{(1)}_{\hspace{0.02cm}{\bf k}}\hspace{0.03cm}
	2\hspace{0.02cm}\pi\hspace{0.03cm}\delta(\omega^{l}_{{\bf k}} - {\bf k}\cdot{\bf v}_{\rho}
	-(-1)^{\rho}\, \Delta{\mathbf v}\cdot{\mathbf q}) 
	\biggr\}\,\times
	\notag
\end{align}
\[
	\,\biggl(\frac{3}{4}\,{\mathfrak q}_{1}{\mathfrak q}_{2} 
	\,+\, 
	\frac{3}{2}\,{\mathfrak q}^{2}_{12} \,+\, \frac{1}{2}\,\Lambda^{2} 
	\biggr)\,+ 
	\notag\\
	\ldots\,.
\]
The dots on the right-hand side indicates the last contribution to (\ref{eq:9e}) that does not depend on the scalar plasmon number densities ${N}^{(1)}_{\hspace{0.02cm}{\bf k}}$, ${W}^{(1)}_{\hspace{0.02cm}{\bf k}},\ldots,$ and with which we are now dealing.\\
\indent Let us analyze the dependence on the fast time of the contribution on the right-hand side of (\ref{eq:9e}) that does not include the plasmon number density. The obtained expressions (\ref{eq:7a}) and (\ref{eq:7s}) enable us to rewrite the last contribution in (\ref{eq:9e}) in the following form:
\begin{align}
	\sum_{\rho}\,\biggl[\,
	&\int\!d\hspace{0.02cm}{\bf k}\,
	{T}^{\hspace{0.03cm}(\rho)}_{\; {\bf k}}(t')
	\biggl(\int {T}^{\hspace{0.03cm}\ast\hspace{0.03cm}(\rho)}_{\; {\bf k}}(t')\hspace{0.03cm}dt'\biggr)
	-
	\int\!d\hspace{0.02cm}{\bf k}\,
	\hspace{0.03cm}
	{T}^{\,\ast\hspace{0.03cm}(\rho)}_{\; {\bf k}}(t')
	\biggl(\int
	%
	{T}^{\hspace{0.03cm}(\rho)}_{\; {\bf k}}(t')\hspace{0.03cm}dt'\biggr)\biggr]
	\hspace{0.03cm}\times
\label{eq:9yy}\\[1ex]	
	&\bigl(\hspace{0.03cm}T^{\,e_{1}}\hspace{0.03cm}T^{\,a^{\prime}_{2}}\hspace{0.03cm}T^{\,a^{\prime}_{1}}\hspace{0.01cm}
	\bigr)^{\hspace{0.01cm}a^{\phantom{\prime}}_{1}\hspace{0.03cm}a^{\prime\prime}_{2}}\hspace{0.03cm}
	\bigl\langle\hspace{0.03cm}\mathcal{Q}^{\hspace{0.03cm}a^{\prime}_{1}}_{1}(\tau)
	\hspace{0.03cm}\bigr\rangle
	\hspace{0.03cm}\bigl\langle\hspace{0.03cm}\mathcal{Q}^{\hspace{0.03cm}a^{\prime\prime}_{2}}_{2}(\tau)
	\hspace{0.03cm}\bigr\rangle
	\hspace{0.03cm}\bigl\langle\hspace{0.03cm}\mathcal{Q}^{\hspace{0.03cm}a^{\prime}_{2}}_{2}(\tau)
	\hspace{0.03cm}\bigr\rangle
	\hspace{0.01cm}\bigl\langle\hspace{0.03cm}\mathcal{Q}^{\hspace{0.03cm}e_{1}}_{1}(\tau)
	\hspace{0.03cm}\bigr\rangle
	=
	\notag
\end{align}
\hspace{0.04cm}
\[
-\hspace{0.03cm}i\,\sum_{\rho}\,
\!\int\!d\hspace{0.02cm}{\bf k}\,d\hspace{0.02cm}{\bf q}\,d\hspace{0.02cm}{\bf q}'\;
{T}^{\hspace{0.03cm}(\rho)}_{\; {\bf k},\,{\bf q}}\,
{T}^{\,\ast\hspace{0.03cm}(\rho)}_{\; {\bf k},\,{\bf q}'}\,
{\rm e}^{i\hspace{0.03cm}\,(-1)^{\rho}\,\Delta{\mathbf v}\cdot({\mathbf q} - {\mathbf q}')\hspace{0.02cm}t}
\hspace{0.04cm}
{\rm e}^{i\,(-1)^{\rho}\,({\mathbf x}_{0\hspace{0.02cm}1} - {\mathbf x}_{0\hspace{0.02cm}2})\cdot({\bf q} - {\mathbf q}')}\,\times
\]
\[
\biggl(
\frac{1}{\omega^{l}_{{\bf k}} - {\bf k}\cdot{\bf v}_{\rho}
	-(-1)^{\rho}\, \Delta{\mathbf v}\cdot{\mathbf q}' + i0}
\,+\,
\frac{1}{\omega^{l}_{{\bf k}} - {\bf k}\cdot{\bf v}_{\rho}
	-(-1)^{\rho}\, \Delta{\mathbf v}\cdot{\mathbf q} - i0}
\biggr)\times
\]
\hspace{0.06cm}
\[
\bigl(\hspace{0.03cm}T^{\,e_{1}}\hspace{0.03cm}T^{\,a^{\prime}_{2}}\hspace{0.03cm}T^{\,a^{\prime}_{1}}\hspace{0.01cm}
\bigr)^{\hspace{0.01cm}a^{\phantom{\prime}}_{1}\hspace{0.03cm}a^{\prime\prime}_{2}}\hspace{0.03cm}
\bigl\langle\hspace{0.03cm}\mathcal{Q}^{\hspace{0.03cm}a^{\prime}_{1}}_{1}(\tau)
\hspace{0.03cm}\bigr\rangle
\hspace{0.03cm}\bigl\langle\hspace{0.03cm}\mathcal{Q}^{\hspace{0.03cm}a^{\prime\prime}_{2}}_{2}(\tau)
\hspace{0.03cm}\bigr\rangle
\hspace{0.03cm}\bigl\langle\hspace{0.03cm}\mathcal{Q}^{\hspace{0.03cm}a^{\prime}_{2}}_{2}(\tau)
\hspace{0.03cm}\bigr\rangle
\hspace{0.01cm}\bigl\langle\hspace{0.03cm}\mathcal{Q}^{\hspace{0.03cm}e_{1}}_{1}(\tau)
\hspace{0.03cm}\bigr\rangle.
\]
Note that the sign in parentheses on the right-hand side has changed to the opposite of that in the corresponding expression in (\ref{eq:7f}).\\
\indent Next, we proceed as in section \ref{section_7}. The first step is to perform the integration over the fast time $t'$, using the formula (\ref{eq:7j}). Then we introduce into consideration the impact parameter ${\mathbf b}$, according to the expression (\ref{eq:7jj}), and average over it. As a result, we obtain in the integrand of (\ref{eq:9yy}) the $\delta$-function of the following form:
\[
(2\hspace{0.02cm}\pi)^{3}\,\frac{1}{|\Delta{\mathbf v}|}\,
\delta({\mathbf q} - {\mathbf q}^{\prime}),
\]    
which in turn enables us to perform the integration over ${\mathbf q}^{\prime}$. Thus, instead of (\ref{eq:9yy}) we further obtain
\begin{equation}
-\hspace{0.03cm}i\hspace{0.02cm}(2\hspace{0.02cm}\pi)^{3}\,
\frac{1}{|\Delta{\mathbf v}|}\sum_{\rho}\,
\!\int\!d\hspace{0.02cm}{\bf k}\,d\hspace{0.02cm}{\bf q}\;
\Bigl|{T}^{\hspace{0.03cm}(\rho)}_{\; {\bf k},\,{\bf q}}\Bigr|^{2}\,
\times
\label{eq:9uu}
\end{equation}
\[
\biggl(
\frac{1}{\omega^{l}_{{\bf k}} - {\bf k}\cdot{\bf v}_{\rho}
-(-1)^{\rho}\,\Delta{\mathbf v}\cdot{\mathbf q} + i0}
\,+\,
\frac{1}{\omega^{l}_{{\bf k}} - {\bf k}\cdot{\bf v}_{\rho}
	-(-1)^{\rho}\, \Delta{\mathbf v}\cdot{\mathbf q} - i0}
\biggr)\times
\]
\hspace{0.06cm}
\[
\bigl(\hspace{0.03cm}T^{\,e_{1}}\hspace{0.03cm}T^{\,a^{\prime}_{2}}\hspace{0.03cm}T^{\,a^{\prime}_{1}}\hspace{0.01cm}
\bigr)^{\hspace{0.01cm}a^{\phantom{\prime}}_{1}\hspace{0.03cm}a^{\prime\prime}_{2}}\hspace{0.03cm}
\bigl\langle\hspace{0.03cm}\mathcal{Q}^{\hspace{0.03cm}a^{\prime}_{1}}_{1}(\tau)
\hspace{0.03cm}\bigr\rangle
\hspace{0.03cm}\bigl\langle\hspace{0.03cm}\mathcal{Q}^{\hspace{0.03cm}a^{\prime\prime}_{2}}_{2}(\tau)
\hspace{0.03cm}\bigr\rangle
\hspace{0.03cm}\bigl\langle\hspace{0.03cm}\mathcal{Q}^{\hspace{0.03cm}a^{\prime}_{2}}_{2}(\tau)
\hspace{0.03cm}\bigr\rangle
\hspace{0.01cm}\bigl\langle\hspace{0.03cm}\mathcal{Q}^{\hspace{0.03cm}e_{1}}_{1}(\tau)
\hspace{0.03cm}\bigr\rangle
=
\]
\hspace{0.06cm}
\[
-\hspace{0.03cm}2\hspace{0.03cm}i\hspace{0.04cm}\,
\frac{(2\hspace{0.02cm}\pi)^{3}}{|\Delta{\mathbf v}|}\sum_{\rho}\,
\!\int\!d\hspace{0.02cm}{\bf k}\,d\hspace{0.02cm}{\bf q}\;
\Bigl|{T}^{\hspace{0.03cm}(\rho)}_{\; {\bf k},\,{\bf q}}\Bigr|^{2}\,
\frac{\mathcal{P}}{\omega^{l}_{{\bf k}} - {\bf k}\cdot{\bf v}_{\rho}
- (-1)^{\rho}\,\Delta{\mathbf v}\cdot{\mathbf q}}\,
\times
\]
\hspace{0.06cm}
\[
\bigl(\hspace{0.03cm}T^{\,e_{1}}\hspace{0.03cm}T^{\,a^{\prime}_{2}}\hspace{0.03cm}T^{\,a^{\prime}_{1}}\hspace{0.01cm}
\bigr)^{\hspace{0.01cm}a^{\phantom{\prime}}_{1}\hspace{0.03cm}a^{\prime\prime}_{2}}\hspace{0.03cm}
\bigl\langle\hspace{0.03cm}\mathcal{Q}^{\hspace{0.03cm}a^{\prime}_{1}}_{1}(\tau)
\hspace{0.03cm}\bigr\rangle
\hspace{0.03cm}\bigl\langle\hspace{0.03cm}\mathcal{Q}^{\hspace{0.03cm}a^{\prime\prime}_{2}}_{2}(\tau)
\hspace{0.03cm}\bigr\rangle
\hspace{0.03cm}\bigl\langle\hspace{0.03cm}\mathcal{Q}^{\hspace{0.03cm}a^{\prime}_{2}}_{2}(\tau)
\hspace{0.03cm}\bigr\rangle
\hspace{0.01cm}\bigl\langle\hspace{0.03cm}\mathcal{Q}^{\hspace{0.03cm}e_{1}}_{1}(\tau)
\hspace{0.03cm}\bigr\rangle
=
\]
Here, at the last stage, the Sokhotsky formula was used and the symbol $\mathcal{P}$ denotes the principle value. It should be added to the right-hand side of equation (\ref{eq:9r}). However, it is easy to verify that in the case of equation for the colorless combination ${\mathfrak q}_{1}(\tau)$, Eq.\,(\ref{eq:9o}), the contribution (\ref{eq:9uu}) is equal to zero. This immediately follows from a trivial analysis of the contraction
\[
\bigl\langle \mathcal{Q}^{\,a_{1}}_{\hspace{0.03cm}1}(\tau)\hspace{0.02cm}\bigr\rangle
\hspace{0.03cm}
\bigl(\hspace{0.03cm}T^{\,e_{1}}\hspace{0.03cm}T^{\,a^{\prime}_{2}}
\hspace{0.03cm}
T^{\,a^{\prime}_{1}}\hspace{0.01cm}
\bigr)^{\hspace{0.01cm}a^{\phantom{\prime}}_{1}\hspace{0.03cm}
a^{\prime\prime}_{2}}
\hspace{0.03cm}
\bigl\langle\hspace{0.03cm}\mathcal{Q}^{\hspace{0.03cm}a^{\prime}_{1}}_{1}(\tau)
\hspace{0.03cm}\bigr\rangle
\hspace{0.03cm}\bigl\langle\hspace{0.03cm}\mathcal{Q}^{\hspace{0.03cm}
a^{\prime\prime}_{2}}_{2}(\tau)
\hspace{0.03cm}\bigr\rangle
\hspace{0.03cm}\bigl\langle\hspace{0.03cm}\mathcal{Q}^{\hspace{0.03cm}a^{\prime}_{2}}_{2}(\tau)
\hspace{0.03cm}\bigr\rangle
\hspace{0.01cm}\bigl\langle\hspace{0.03cm}\mathcal{Q}^{\hspace{0.03cm}
e_{1}}_{1}(\tau) \hspace{0.03cm}\bigr\rangle =
\]  
\[
\bigl[
i\hspace{0.03cm}f^{\hspace{0.03cm}a_{1}\hspace{0.02cm}e_{1}\hspace{0.03cm}b}\hspace{0.03cm}
\bigl\langle \mathcal{Q}^{\,a_{1}}_{\hspace{0.03cm}1}(\tau)\hspace{0.02cm}\bigr\rangle
\hspace{0.03cm}
\bigl\langle\hspace{0.03cm}\mathcal{Q}^{\hspace{0.03cm}
e_{1}}_{1}(\tau) \hspace{0.03cm}\bigr\rangle
\bigr]
\bigl(\hspace{0.03cm}T^{\,a^{\prime}_{2}}\hspace{0.03cm}
T^{\,a^{\prime}_{1}}\hspace{0.01cm}
\bigr)^{\hspace{0.01cm}b\,a^{\prime\prime}_{2}}
\bigl\langle\hspace{0.03cm}\mathcal{Q}^{\hspace{0.03cm}a^{\prime}_{1}}_{1}(\tau)
\hspace{0.03cm}\bigr\rangle
\hspace{0.03cm}\bigl\langle\hspace{0.03cm}\mathcal{Q}^{\hspace{0.03cm}
	a^{\prime\prime}_{2}}_{2}(\tau)
\hspace{0.03cm}\bigr\rangle
\hspace{0.03cm}\bigl\langle\hspace{0.03cm}\mathcal{Q}^{\hspace{0.03cm}a^{\prime}_{2}}_{2}(\tau)
\hspace{0.03cm}\bigr\rangle
= 0.
\] 
\indent At the end of this section, we will proceed to the derivation of an equation for somewhat more complex colorless combination of the mixed type, namely ${\mathfrak q}_{12} \equiv 
\hspace{0.03cm}\bigl\langle\hspace{0.03cm}\mathcal{Q}^{\hspace{0.03cm}e}_{1}
\hspace{0.03cm}\bigr\rangle
\bigl\langle\hspace{0.03cm}\mathcal{Q}^{\hspace{0.03cm}e}_{2}
\hspace{0.03cm}\bigr\rangle$. It is convenient to represent the derivative of the function ${\mathfrak q}_{12}$ with respect to the slow time $\tau$ in the following form: 
\begin{equation}
\frac{d\hspace{0.03cm}{\mathfrak q}_{12}(\tau)}{d\hspace{0.03cm}\tau}
=
\bigl\langle \mathcal{Q}^{\,a_{1}}_{\hspace{0.03cm}2}(\tau)\hspace{0.02cm}\bigr\rangle
\hspace{0.03cm}
\frac{d\hspace{0.03cm}\bigl\langle \mathcal{Q}^{\,a_{1}}_{\hspace{0.03cm}1}(\tau)\hspace{0.02cm}\bigr\rangle}{d\hspace{0.03cm}\tau}
\,+\,
\bigl\langle \mathcal{Q}^{\,a_{2}}_{\hspace{0.03cm}1}(\tau)\hspace{0.02cm}\bigr\rangle
\hspace{0.03cm}
\frac{d\hspace{0.03cm}\bigl\langle \mathcal{Q}^{\,a_{2}}_{\hspace{0.03cm}2}(\tau)\hspace{0.02cm}\bigr\rangle}{d\hspace{0.03cm}\tau}.
\label{eq:9p}
\end{equation}
Now we are able to analyze only the first term on the right-hand  side. Analysis of the second term, as well as the derivation of equations for the remaining colorless combinations ${\mathfrak q}_{2}$ and $\Lambda^{\!\hspace{0.01cm}2}$ will be performed in the following section.\\
\indent In order to determine $\bigl\langle \mathcal{Q}^{\,a_{1}}_{\hspace{0.03cm}2}(\tau)\hspace{0.02cm}\bigr\rangle
\hspace{0.03cm}
\bigl(d\hspace{0.03cm}\bigl\langle \mathcal{Q}^{\,a_{1}}_{\hspace{0.03cm}1}(\tau)\hspace{0.02cm}\bigr\rangle/{d\hspace{0.03cm}\tau}\bigr)$, we contract our original equation (\ref{eq:9r}) with $\bigl\langle \mathcal{Q}^{\,a_{1}}_{\hspace{0.03cm}2}(\tau)\hspace{0.02cm}\bigr\rangle$. As a first step, take a look at the imaginary part on the right-hand side of (\ref{eq:9y}). According to formulas (\ref{eq:9y}) and (\ref{eq:9yyy}) for this imaginary part, taking into account the expression for the fourth-order trace (\ref{ap:D4}), we have the following chain of equalities
\[
-\hspace{0.03cm}\frac{1}{2}\,{N}^{(2)}_{\hspace{0.02cm}{\bf k}}\hspace{0.03cm}
\bigl\langle\hspace{0.03cm}\mathcal{Q}^{\hspace{0.03cm}c}_{1}
\hspace{0.03cm}\bigr\rangle
\hspace{0.03cm}
f^{\hspace{0.03cm}e^{\prime}_{1}\hspace{0.02cm}a_{1}\hspace{0.03cm}b}\,
{\rm tr}\hspace{0.03cm}
\bigl(\hspace{0.03cm}T^{\,b}\hspace{0.03cm}T^{\,a^{\prime}_{2}}\hspace{0.03cm}T^{\,c}T^{\,a^{\prime\prime}_{2}}\hspace{0.01cm}\bigr)
\hspace{0.01cm}
\bigl\langle \mathcal{Q}^{\,a_{1}}_{\hspace{0.03cm}2}\hspace{0.02cm}\bigr\rangle
\bigl\langle\hspace{0.03cm}\mathcal{Q}^{\hspace{0.03cm}a^{\prime}_{2}}_{2}\hspace{0.03cm}\bigr\rangle
\bigl\langle\hspace{0.03cm}\mathcal{Q}^{\hspace{0.03cm}a^{\prime\prime}_{2}}_{2}\hspace{0.03cm}\bigr\rangle
\hspace{0.03cm}
\bigl\langle\hspace{0.03cm}\mathcal{Q}^{\hspace{0.03cm}e^{\prime}_{1}}_{1}\hspace{0.03cm}\bigr\rangle
=
\]
\[
-\hspace{0.03cm}\frac{1}{2}\,{N}^{(2)}_{\hspace{0.02cm}{\bf k}}\hspace{0.03cm}
\bigl\langle\hspace{0.03cm}\mathcal{Q}^{\hspace{0.03cm}c}_{1}
\hspace{0.03cm}\bigr\rangle
\hspace{0.03cm}\Bigl[
f^{\hspace{0.03cm}e^{\prime}_{1}\hspace{0.02cm}a_{1}\hspace{0.03cm}b}
\bigl\langle\mathcal{Q}^{\,a_{1}}_{\hspace{0.03cm}2}\hspace{0.02cm}\bigr\rangle
\bigl\langle\hspace{0.03cm}\mathcal{Q}^{\hspace{0.03cm}e^{\prime}_{1}}_{1}\hspace{0.03cm}\bigr\rangle\Bigr]
\,
\biggl\{\hspace{0.01cm}
\bigl\langle \mathcal{Q}^{\,b}_{\hspace{0.03cm}2}\hspace{0.02cm}\bigr\rangle
\bigl\langle\hspace{0.03cm}\mathcal{Q}^{\hspace{0.03cm}c}_{2}\hspace{0.03cm}\bigr\rangle
+
\frac{1}{2}\,\Bigl(\bigl\langle \mathcal{Q}^{\,b}_{\hspace{0.03cm}2}\hspace{0.02cm}\bigr\rangle
\bigl\langle\hspace{0.03cm}\mathcal{Q}^{\hspace{0.03cm}c}_{2}\hspace{0.03cm}\bigr\rangle
+
\bigl\langle \mathcal{Q}^{\,a^{\prime}}_{\hspace{0.03cm}2}\hspace{0.02cm}\bigr\rangle
\bigl\langle\hspace{0.03cm}\mathcal{Q}^{\,a^{\prime}}_{\hspace{0.03cm}2}
\hspace{0.03cm}\bigr\rangle\hspace{0.03cm}
\delta^{\hspace{0.03cm}b\hspace{0.03cm}c}\Bigr) 
\,+
\]
\[
\frac{1}{4}\,N_{c}\hspace{0.03cm}\Bigl(f^{\hspace{0.03cm}b\hspace{0.03cm}a^{\prime\prime}_{2}\hspace{0.03cm}e}
\hspace{0.01cm}f^{\hspace{0.03cm}a^{\prime}_{2}\hspace{0.03cm}c\hspace{0.03cm}e}
\bigl\langle\hspace{0.03cm}\mathcal{Q}^{\hspace{0.03cm}a^{\prime\prime}_{2}}_{2}\hspace{0.03cm}\bigr\rangle
\bigl\langle\hspace{0.03cm}\mathcal{Q}^{\hspace{0.03cm}a^{\prime}_{2}}_{2}\hspace{0.03cm}\bigr\rangle
\,+\,
d^{\,b\hspace{0.03cm}a^{\prime\prime}_{2}\hspace{0.03cm}e}
\hspace{0.01cm}d^{\,a^{\prime}_{2}\hspace{0.03cm}c\hspace{0.03cm}e}
\bigl\langle\hspace{0.03cm}\mathcal{Q}^{\hspace{0.03cm}a^{\prime\prime}_{2}}_{2}
\hspace{0.03cm}\bigr\rangle
\bigl\langle\hspace{0.03cm}\mathcal{Q}^{\hspace{0.03cm}a^{\prime}_{2}}_{2}
\hspace{0.03cm}\bigr\rangle\Bigr)\!
\biggr\}
=
\]
\[
-\hspace{0.03cm}\frac{1}{2}\,{N}^{(2)}_{\hspace{0.02cm}{\bf k}}\hspace{0.03cm}
\bigl\langle\hspace{0.03cm}\mathcal{Q}^{\hspace{0.03cm}c}_{1}
\hspace{0.03cm}\bigr\rangle
\hspace{0.03cm}\Lambda^{b}
\biggl\{\hspace{0.03cm}
\frac{3}{2}\,
\bigl\langle \mathcal{Q}^{\,b}_{\hspace{0.03cm}2}\hspace{0.02cm}\bigr\rangle
\bigl\langle\hspace{0.03cm}\mathcal{Q}^{\hspace{0.03cm}c}_{2}\hspace{0.03cm}\bigr\rangle
\,+\,
\frac{1}{2}\,\bigl\langle \mathcal{Q}^{\,a^{\prime}_{2}}_{\hspace{0.03cm}2}\hspace{0.02cm}\bigr\rangle
\bigl\langle\hspace{0.03cm}\mathcal{Q}^{\,a^{\prime}_{2}}_{\hspace{0.03cm}2}
\hspace{0.03cm}\bigr\rangle\hspace{0.03cm}
\delta^{\hspace{0.03cm}b\hspace{0.03cm}c}
\,+
\]
\[
\frac{1}{4}\,N_{c}\hspace{0.03cm}(f^{\hspace{0.03cm}b\hspace{0.03cm}
a^{\prime\prime}_{2}\hspace{0.03cm}e}
\hspace{0.01cm}f^{\hspace{0.03cm}a^{\prime}_{2}\hspace{0.03cm}c\hspace{0.03cm}e}
\bigl\langle\hspace{0.03cm}\mathcal{Q}^{\hspace{0.03cm}a^{\prime\prime}_{2}}_{2}\hspace{0.03cm}\bigr\rangle
\bigl\langle\hspace{0.03cm}\mathcal{Q}^{\hspace{0.03cm}a^{\prime}_{2}}_{2}\hspace{0.03cm}\bigr\rangle
\,+\,
d^{\,b\hspace{0.03cm}a^{\prime\prime}_{2}\hspace{0.03cm}e}
\hspace{0.01cm}d^{\,a^{\prime}_{2}\hspace{0.03cm}c\hspace{0.03cm}e}
\bigl\langle\hspace{0.03cm}\mathcal{Q}^{\hspace{0.03cm}a^{\prime\prime}_{2}}_{2}\hspace{0.03cm}\bigr\rangle
\bigl\langle\hspace{0.03cm}\mathcal{Q}^{\hspace{0.03cm}a^{\prime}_{2}}_{2}\hspace{0.03cm}\bigr\rangle)
\biggr\}
=
\]
\[
-\hspace{0.03cm}\frac{1}{8}\,{N}^{(2)}_{\hspace{0.02cm}{\bf k}}\hspace{0.03cm}N_{c}\hspace{0.03cm}
\biggl(-f^{\hspace{0.03cm}b\hspace{0.03cm}a^{\prime\prime}_{2}
\hspace{0.03cm}e}
\hspace{0.03cm}\Lambda^{b}\hspace{0.03cm}\Lambda^{e}
\,+\,
d^{\,b\hspace{0.03cm}a^{\prime\prime}_{2}\hspace{0.03cm}e}
\hspace{0.03cm}\Lambda^{b}\,\Omega^{\hspace{0.03cm} e}_{12}\biggr)\bigl\langle\hspace{0.03cm}
\mathcal{Q}^{\hspace{0.03cm}a^{\prime\prime}_{2}}_{2}\hspace{0.03cm}\bigr\rangle
=
-\hspace{0.03cm}\frac{1}{8}\,{N}^{(2)}_{\hspace{0.02cm}{\bf k}}\hspace{0.03cm}N_{c}\hspace{0.03cm}
d^{\,b\hspace{0.03cm}a^{\prime\prime}_{2}\hspace{0.03cm}e}\Lambda^{b}\,
\Omega^{\hspace{0.03cm}e}_{12}
\hspace{0.03cm}
\bigl\langle\hspace{0.03cm}\mathcal{Q}^{\hspace{0.03cm}a^{\prime\prime}_{2}}_{2}
\hspace{0.03cm}\bigr\rangle.
\]
Here, for the sake of simplicity, we have suppressed the dependence of the averaged color charges on the slow time $\tau$, taken into account the definition of the functions $\Lambda^{a}$ and $\Omega^{a}_{12}$, Eq.\,(\ref{eq:7w}), and the quite obvious equality 
\[
\Lambda^{a}\bigl\langle\hspace{0.03cm}\mathcal{Q}^{\hspace{0.03cm}a}_{1}
\hspace{0.03cm}\bigr\rangle
=
\Lambda^{a}\bigl\langle\hspace{0.03cm}\mathcal{Q}^{\hspace{0.03cm}a}_{2}
\hspace{0.03cm}\bigr\rangle
= 0.
\]
From the above expression, we see that in the case of an arbitrary group $SU(N_{c})$, this imaginary part is generally not equal to zero. Apparently, it vanishes for the ``trivial'' case of $N_{c}= 2$, when $d^{\,b\hspace{0.03cm}a^{\prime\prime}_{2}\hspace{0.03cm}e} \equiv 0$. However, as was shown in Appendix \ref{appendix_E}, Eq.\,(\ref{ap:E7}), for the case when $N_{c} = 3$, the imaginary part is also zero. Thus, unlike the equation for the colorless combination
 ${\mathfrak q}_{1}(\tau) = 
\hspace{0.03cm}\bigl\langle\hspace{0.03cm}\mathcal{Q}^{\hspace{0.03cm}a_{1}}_{1}
\hspace{0.03cm}\bigr\rangle
\bigl\langle\hspace{0.03cm}\mathcal{Q}^{\hspace{0.03cm}a_{1}}_{1}
\hspace{0.03cm}\bigr\rangle$, where the imaginary part (\ref{eq:9yyy}) trivially reduced to zero for an arbitrary $N_{c}$, for the colorless combination ${\mathfrak q}_{12}(\tau) = 
\hspace{0.03cm}\bigl\langle\hspace{0.03cm}\mathcal{Q}^{\hspace{0.03cm}a_{1}}_{1}
\hspace{0.03cm}\bigr\rangle
\bigl\langle\hspace{0.03cm}\mathcal{Q}^{\hspace{0.03cm}a_{2}}_{2}
\hspace{0.03cm}\bigr\rangle$ we are dealing with a more difficult situation, with specific values $N_{c}$.\\
\indent  Let us now consider the real part for the first expression in (\ref{eq:9t}). In the case under consideration, instead (\ref{eq:9u}) and (\ref{eq:9i}), we now have
\begin{equation}
	{N}^{(1)}_{\hspace{0.02cm}{\bf k}}\hspace{0.03cm}
	{\rm tr}\hspace{0.03cm}
	\bigl(\hspace{0.03cm}T^{\,a^{\prime}_{2}}\hspace{0.03cm}T^{\,a_{1}}\hspace{0.03cm}T^{\,e^{\prime}_{1}}\hspace{0.03cm}T^{\,a^{\prime\prime}_{2}}\hspace{0.01cm}
	\bigr)
	\hspace{0.01cm}
	\bigl\langle\hspace{0.03cm}\mathcal{Q}^{\hspace{0.03cm}a_{1}}_{2}\hspace{0.03cm}
	\bigr\rangle
	\hspace{0.03cm}
	\bigl\langle\hspace{0.03cm}\mathcal{Q}^{\hspace{0.03cm}a^{\prime}_{2}}_{2}(\tau)
	\hspace{0.03cm}\bigr\rangle
	\bigl\langle\hspace{0.03cm}\mathcal{Q}^{\hspace{0.03cm}a^{\prime\prime}_{2}}_{2}(\tau)\hspace{0.03cm}\bigr\rangle
	\hspace{0.03cm}
	\bigl\langle\hspace{0.03cm}\mathcal{Q}^{\hspace{0.03cm}e^{\prime}_{1}}_{1}(\tau)\hspace{0.03cm}\bigr\rangle
	\label{eq:9a}
\end{equation}
\[
=
{N}^{(1)}_{\hspace{0.02cm}{\bf k}}\hspace{0.03cm}
\biggl\{2\hspace{0.03cm}
{\mathfrak q}_{12}\hspace{0.03cm}{\mathfrak q}_{2} \,+\, \frac{1}{4}\,
N_{c}\,\Omega^{\hspace{0.03cm}e}_{\hspace{0.03cm}22}\hspace{0.03cm}
\Omega^{\hspace{0.03cm}e}_{\hspace{0.03cm}12}\biggr\}\bigg|_{N_{c}=3}
=
\frac{9}{4}\,{N}^{(1)}_{\hspace{0.02cm}{\bf k}}\hspace{0.03cm}
{\mathfrak q}_{12}\hspace{0.03cm}{\mathfrak q}_{2}.
\]
Here, in the last step, we used the relation (\ref{ap:E5}), which is valid for $N_{c} = 3$. We will further examine the imaginary part of the second relation in (\ref{eq:9t}). According to the formula (\ref{eq:9ii}), this imaginary part is proportional to 
\[
{N}^{(2)}_{\hspace{0.02cm}{\bf k}}\hspace{0.03cm}\bigl\langle\hspace{0.03cm}\mathcal{Q}^{\hspace{0.03cm}c}_{1}
\hspace{0.03cm}\bigr\rangle
\hspace{0.03cm}f^{\hspace{0.03cm}a_{1}\hspace{0.02cm}c^{\prime}\hspace{0.03cm}e_{1}}\hspace{0.03cm}
{\rm tr}\hspace{0.03cm}
\bigl(\hspace{0.03cm}T^{\,c^{\prime}}\hspace{0.03cm}T^{\,a^{\prime\prime}_{2}}\hspace{0.03cm}T^{\,e^{\prime}_{1}}\hspace{0.03cm}T^{\,c}\hspace{0.01cm}
\bigr)
\hspace{0.03cm}\bigl\langle\hspace{0.03cm}\mathcal{Q}^{\hspace{0.03cm}e_{1}}_{1}(\tau)
\hspace{0.03cm}\bigr\rangle
\bigl\langle\hspace{0.03cm}\mathcal{Q}^{\hspace{0.03cm}a^{\prime\prime}_{2}}_{2}(\tau)\hspace{0.03cm}\bigr\rangle
\hspace{0.03cm}
\bigl\langle\hspace{0.03cm}\mathcal{Q}^{\hspace{0.03cm}e^{\prime}_{1}}_{1}(\tau)
\hspace{0.03cm}\bigr\rangle.
\]     
Contracting the previous expression with $\bigl\langle\hspace{0.03cm}\mathcal{Q}^{\hspace{0.03cm}a_{1}}_{2}\hspace{0.03cm}\bigr\rangle$ and using (\ref{ap:D4}) for the color trace, we find that this imaginary part is equal to
\[
\frac{1}{4}\,N_{c}\,{N}^{(2)}_{\hspace{0.02cm}{\bf k}}\hspace{0.03cm}
d^{\,b_{1}\hspace{0.03cm}c\hspace{0.03cm}e}\Lambda^{c}\,
\Omega^{\hspace{0.03cm}e}_{12}
\hspace{0.03cm}
\bigl\langle\hspace{0.03cm}\mathcal{Q}^{\,b_{1}}_{1}
\hspace{0.03cm}\bigr\rangle.
\] 
As shown in Appendix \ref{appendix_E}, Eq.\,(\ref{ap:E7}), for the case when $N_{c} = 3$, this combination is zero. Thus, the imaginary part for the second relation in (\ref{eq:9t}) does not contribute to the desired equation for ${\mathfrak q}_{12}(\tau)$.\\
\indent Next, for the real part of the second relation in (\ref{eq:9t}), using the third-order trace (\ref{ap:D3}) and contracting with
 $\bigl\langle\hspace{0.03cm}\mathcal{Q}^{\hspace{0.03cm}a_{1}}_{2}\hspace{0.03cm}\bigr\rangle$, we have
\begin{equation}
i\hspace{0.03cm}{N}^{(1)}_{\hspace{0.02cm}{\bf k}}\hspace{0.03cm}f^{\hspace{0.03cm}a_{1}\hspace{0.02cm}c^{\prime}\hspace{0.03cm}e_{1}}\hspace{0.03cm}
{\rm tr}\hspace{0.03cm}
\bigl(\hspace{0.03cm}T^{\,c^{\prime}}\hspace{0.03cm}T^{\,a^{\prime\prime}_{2}}
\hspace{0.03cm}T^{\,e^{\prime}_{1}}\hspace{0.01cm}
\bigr)
\bigl\langle\hspace{0.03cm}\mathcal{Q}^{\hspace{0.03cm}a_{1}}_{2}\hspace{0.03cm}\bigr\rangle
\hspace{0.03cm}\bigl\langle\hspace{0.03cm}\mathcal{Q}^{\hspace{0.03cm}e_{1}}_{1}(\tau)
\hspace{0.03cm}\bigr\rangle
\bigl\langle\hspace{0.03cm}\mathcal{Q}^{\hspace{0.03cm}a^{\prime\prime}_{2}}_{2}(\tau)\hspace{0.03cm}\bigr\rangle
\hspace{0.03cm}
\bigl\langle\hspace{0.03cm}\mathcal{Q}^{\hspace{0.03cm}e^{\prime}_{1}}_{1}(\tau)\hspace{0.03cm}\bigr\rangle
=
\label{eq:9s} 
\end{equation}
\[
\frac{i^{\,2}}{2}\,{N}^{(1)}_{\hspace{0.02cm}{\bf k}}\hspace{0.01cm}N_{c}\hspace{0.03cm}
f^{\hspace{0.03cm}a_{1}\hspace{0.02cm}c^{\prime}\hspace{0.03cm}e_{1}}\hspace{0.03cm}
f^{\hspace{0.03cm}c^{\prime}\hspace{0.02cm}a_{2}^{\prime\prime}\hspace{0.03cm}
e^{\prime}_{1}}\hspace{0.03cm}
\bigl\langle\hspace{0.03cm}\mathcal{Q}^{\hspace{0.03cm}a_{1}}_{2}\hspace{0.03cm}\bigr\rangle
\hspace{0.03cm}\bigl\langle\hspace{0.03cm}\mathcal{Q}^{\hspace{0.03cm}e_{1}}_{1}(\tau)
\hspace{0.03cm}\bigr\rangle
\bigl\langle\hspace{0.03cm}\mathcal{Q}^{\hspace{0.03cm}a^{\prime\prime}_{2}}_{2}(\tau)\hspace{0.03cm}\bigr\rangle
\hspace{0.03cm}
\bigl\langle\hspace{0.03cm}\mathcal{Q}^{\hspace{0.03cm}e^{\prime}_{1}}_{1}(\tau)\hspace{0.03cm}\bigr\rangle
\equiv
\]
\[
\frac{1}{2}\,{N}^{(1)}_{\hspace{0.02cm}{\bf k}}\hspace{0.03cm}N_{c}\hspace{0.03cm}\Lambda^{e}\Lambda^{e}\big|_{N_{c} = 3}
=
\frac{3}{2}\,{N}^{(1)}_{\hspace{0.02cm}{\bf k}}\hspace{0.03cm}\Lambda^{2}.
\]
Completely similar reasonings are valid for the contractions in equation (\ref{eq:9r}) containing the matrix function ${\mathcal W}_{\hspace{0.02cm}{\bf k}}$. Using the color decomposition (\ref{eq:7q}), we obtain the analogues of Eqs.\,(\ref{eq:9a}) and (\ref{eq:9s}) with the only replacement ${N}^{(1)}_{\hspace{0.02cm}{\bf k}}\rightarrow{W}^{(1)}_{\hspace{0.02cm}{\bf k}}$. As a result, taking into account all of the above, after contracting the equation (\ref{eq:9r}) with
$\bigl\langle\hspace{0.03cm}\mathcal{Q}^{\hspace{0.03cm}a_{1}}_{2}\hspace{0.03cm}\bigr\rangle$, we derive the required equation for the first term on the right-hand side of equality (\ref{eq:9p}):
\begin{equation}
\bigl\langle \mathcal{Q}^{\,a_{1}}_{\hspace{0.03cm}2}(\tau)\hspace{0.02cm}\bigr\rangle
\hspace{0.03cm}
\frac{d\hspace{0.03cm}\bigl\langle \mathcal{Q}^{\,a_{1}}_{\hspace{0.03cm}1}(\tau)\hspace{0.02cm}\bigr\rangle}{d\hspace{0.03cm}\tau}
=
\label{eq:9d} 
\end{equation}
\begin{align}
	-\hspace{0.03cm}\biggl\{\frac{(2\hspace{0.02cm}\pi)^{3}}{|\Delta{\mathbf v}|}\,\sum_{\rho}\,
	&\int\!d\hspace{0.02cm}{\bf k}\hspace{0.03cm}d\hspace{0.02cm}{\bf q}\;
	\bigl|\hspace{0.03cm}{T}^{\hspace{0.03cm}(\rho)}_{\; {\bf k},\,{\bf q}}\bigr|^{\hspace{0.02cm}2}\hspace{0.03cm}
	{N}^{(1)}_{\hspace{0.02cm}{\bf k}}\hspace{0.03cm}
	2\hspace{0.02cm}\pi\hspace{0.03cm}\delta(\omega^{l}_{{\bf k}} - {\bf k}\cdot{\bf v}_{\rho}
	-(-1)^{\rho}\, \Delta{\mathbf v}\cdot{\mathbf q})\,+
	\notag\\[1ex]
	\frac{(2\hspace{0.02cm}\pi)^{3}}{|\Delta{\mathbf v}|}\,\sum_{\rho}\hspace{0.02cm}(-1)^{\rho + 1}\!\!
	&\int\!d\hspace{0.02cm}{\bf k}\hspace{0.03cm}d\hspace{0.02cm}{\bf q}\;
	\bigl|\hspace{0.03cm}{T}^{\hspace{0.03cm}(\rho)}_{\; {\bf k},\,{\bf q}}\bigr|^{\hspace{0.02cm}2}\hspace{0.03cm}
	{W}^{(1)}_{\hspace{0.02cm}{\bf k}}\hspace{0.03cm}
	2\hspace{0.02cm}\pi\hspace{0.03cm}\delta(\omega^{l}_{{\bf k}} - {\bf k}\cdot{\bf v}_{\rho}
	-(-1)^{\rho}\, \Delta{\mathbf v}\cdot{\mathbf q}) 
	\biggr\}\,\times
	\notag
\end{align}
\[
\,\biggl(\frac{9}{4}\,{\mathfrak q}_{12}\,{\mathfrak q}_{2} 
\,-\, \frac{3}{2}\,\Lambda^{2} 
\biggr).
\]
The contribution of the term (\ref{eq:9uu}) in this equation vanishes by virtue of  the fact that the color factor here is zero due to the equality 
\[
\bigl[\hspace{0.03cm}f^{\hspace{0.03cm}e_{1}\hspace{0.02cm}a_{1}\hspace{0.03cm}b}\hspace{0.03cm}
\hspace{0.03cm}
\bigl\langle\hspace{0.03cm}\mathcal{Q}^{\hspace{0.03cm}e_{1}}_{1}\hspace{0.03cm}\bigr\rangle
\hspace{0.03cm}\bigl\langle\hspace{0.03cm}\mathcal{Q}^{\hspace{0.03cm}a_{1}}_{2}
\hspace{0.03cm}\bigr\rangle
\bigr]
\hspace{0.03cm}f^{\hspace{0.03cm}a_{2}^{\prime}\hspace{0.02cm}b\hspace{0.03cm}d}
\hspace{0.03cm}\bigl\langle\hspace{0.03cm}\mathcal{Q}^{\hspace{0.03cm}a^{\prime}_{2}}_{2}
\hspace{0.03cm}\bigr\rangle\hspace{0.03cm}
\bigl[\hspace{0.03cm}f^{\hspace{0.03cm}a_{1}^{\prime}\hspace{0.02cm}d\hspace{0.03cm}a^{\prime\prime}}\hspace{0.03cm}
\hspace{0.03cm}
\bigl\langle\hspace{0.03cm}\mathcal{Q}^{\hspace{0.03cm}a_{1}^{\prime}}_{1}\hspace{0.03cm}\bigr\rangle
\hspace{0.03cm}\bigl\langle\hspace{0.03cm}\mathcal{Q}^{\hspace{0.03cm}a_{2}^{\prime\prime}}_{2}
\hspace{0.03cm}\bigr\rangle
\bigr]
=
-\hspace{0.03cm}f^{\hspace{0.03cm}a_{2}^{\prime}\hspace{0.02cm}b\hspace{0.03cm}d}
\hspace{0.03cm}\bigl\langle\hspace{0.03cm}\mathcal{Q}^{\hspace{0.03cm}a^{\prime}_{2}}_{2}
\hspace{0.03cm}\bigr\rangle\hspace{0.03cm}\Lambda^{b}\Lambda^{d}
\equiv 0.
\] 

%
%

\section{\bf Equations for the expected value of color charges ${\mathcal Q}^{\hspace{0.03cm}d}_{\hspace{0.03cm}\alpha}$ (continuation)}
\label{section_10}
\setcounter{equation}{0}

In this section, we proceed with the derivation of equations related to the evolution of the rest of the colorless combinations of two averaged color charges. For the special case of the group $SU(3_{c})$, it remains for us to define the equations for the combinations ${\mathfrak q}_{2} =
\hspace{0.03cm}\bigl\langle\hspace{0.03cm}\mathcal{Q}^{\,e}_{2}
\hspace{0.03cm}\bigr\rangle
\bigl\langle\hspace{0.03cm}\mathcal{Q}^{\,e}_{2}
\hspace{0.03cm}\bigr\rangle,
\,
{\mathfrak q}_{12} = 
\hspace{0.03cm}\bigl\langle\hspace{0.03cm}\mathcal{Q}^{\,e}_{1}
\hspace{0.03cm}\bigr\rangle
\bigl\langle\hspace{0.03cm}\mathcal{Q}^{\,e}_{2}
\hspace{0.03cm}\bigr\rangle$ and $\Lambda^{2} = \Lambda^{e}\Lambda^{e}$. To do this, we need to know the equation for the second averaged color charge
$\bigl\langle\hspace{0.03cm}\mathcal{Q}^{\,a_{2}}_{2}(\tau)\hspace{0.03cm}\bigr\rangle$, in addition to equation (\ref{eq:9r}). The  original equation here is (\ref{ap:C1}) from Appendix \ref{appendix_C}. Using the same line of reasoning that led us from (\ref{eq:9e}) to (\ref{eq:9r}), we find from (\ref{ap:C1})
\begin{equation}
	\frac{d\hspace{0.03cm}\bigl\langle \mathcal{Q}^{\,a_{2}}_{\hspace{0.03cm}2}(\tau)\hspace{0.02cm}\bigr\rangle}{d\hspace{0.03cm}\tau}
	=
	\label{eq:10q}
\end{equation}
\begin{align}
	-\hspace{0.03cm}&\frac{(2\hspace{0.02cm}\pi)^{3}}{2\hspace{0.03cm}|\Delta{\mathbf v}|}\,\sum_{\rho}\,
	\int\!d\hspace{0.02cm}{\bf k}\hspace{0.03cm}d\hspace{0.02cm}{\bf q}\;
	\bigl|\hspace{0.03cm}{T}^{\hspace{0.03cm}(\rho)}_{\; {\bf k},\,{\bf q}}\bigr|^{\hspace{0.02cm}2}\,
	2\hspace{0.02cm}\pi\hspace{0.03cm}\delta(\omega^{l}_{{\bf k}} - {\bf k}\cdot{\bf v}_{\rho}
	-(-1)^{\rho}\, \Delta{\mathbf v}\cdot{\mathbf q})
	\,\times
	\notag\\
&\Bigl[{\rm tr}\hspace{0.03cm}
\bigl(\hspace{0.03cm}T^{\,a^{\prime\prime}_{1}}\hspace{0.03cm}T^{\,a_{2}}\hspace{0.03cm}T^{\,e^{\prime}_{2}}\hspace{0.03cm}T^{\,a^{\prime}_{1}}{\mathcal N}_{\hspace{0.02cm}{\bf k}}\hspace{0.01cm}
\bigr)
\hspace{0.01cm}
\bigl\langle\hspace{0.03cm}\mathcal{Q}^{\hspace{0.03cm}a^{\prime\prime}_{1}}_{1}(\tau)\hspace{0.03cm}\bigr\rangle
-	
	i\hspace{0.03cm}f^{\hspace{0.03cm}a_{2}\hspace{0.02cm}c^{\prime}\hspace{0.03cm}e_{2}}\hspace{0.03cm}
	{\rm tr}\hspace{0.03cm}
	\bigl(\hspace{0.03cm}T^{\,c^{\prime}}\hspace{0.03cm}T^{\,a^{\prime}_{1}}\hspace{0.03cm}T^{\,e^{\prime}_{2}}{\mathcal N}_{\hspace{0.02cm}{\bf k}}\hspace{0.01cm}
	\bigr)
	\hspace{0.03cm}\bigl\langle\hspace{0.03cm}\mathcal{Q}^{\hspace{0.03cm}e_{2}}_{2}(\tau)
	\hspace{0.03cm}\bigr\rangle
	\Bigr]
	\hspace{0.01cm}
	\bigl\langle\hspace{0.03cm}\mathcal{Q}^{\hspace{0.03cm}e^{\prime}_{2}}_{2}(\tau)\hspace{0.03cm}\bigr\rangle
	\hspace{0.03cm}\bigl\langle\hspace{0.03cm}\mathcal{Q}^{\hspace{0.03cm}a^{\prime}_{1}}_{1}(\tau)
	\hspace{0.03cm}\bigr\rangle\,-
	\notag
\end{align}
\begin{align}
	\hspace{0.4cm}&\frac{(2\hspace{0.02cm}\pi)^{3}}{2\hspace{0.03cm}|\Delta{\mathbf v}|}\,\sum_{\rho}\hspace{0.02cm}(-1)^{\rho + 1}\!\!
	\int\!d\hspace{0.02cm}{\bf k}\hspace{0.03cm}d\hspace{0.02cm}{\bf q}\;
	\bigl|\hspace{0.03cm}{T}^{\hspace{0.03cm}(\rho)}_{\; {\bf k},\,{\bf q}}\bigr|^{\hspace{0.02cm}2}\,
	2\hspace{0.02cm}\pi\hspace{0.03cm}\delta(\omega^{l}_{{\bf k}} - {\bf k}\cdot{\bf v}_{\rho}
	-(-1)^{\rho}\, \Delta{\mathbf v}\cdot{\mathbf q}) 
	\,\times
	\notag\\
&\Bigl[{\rm tr}\hspace{0.03cm}
\bigl(\hspace{0.03cm}T^{\,a^{\prime\prime}_{1}}\hspace{0.03cm}T^{\,a_{2}}\hspace{0.03cm}T^{\,e^{\prime}_{2}}\hspace{0.03cm}T^{\,a^{\prime}_{1}}{\mathcal W}_{\hspace{0.02cm}{\bf k}}\hspace{0.01cm}
\bigr)
\hspace{0.01cm}
\bigl\langle\hspace{0.03cm}\mathcal{Q}^{\hspace{0.03cm}a^{\prime\prime}_{1}}_{1}(\tau)\hspace{0.03cm}\bigr\rangle
-	
	i\hspace{0.03cm}f^{\hspace{0.03cm}a_{2}\hspace{0.02cm}c^{\prime}\hspace{0.03cm}e_{2}}\hspace{0.03cm}
	{\rm tr}\hspace{0.03cm}
	\bigl(\hspace{0.03cm}T^{\,c^{\prime}}\hspace{0.03cm}T^{\,a^{\prime}_{1}}\hspace{0.03cm}T^{\,e^{\prime}_{2}}{\mathcal W}_{\hspace{0.02cm}{\bf k}}\hspace{0.01cm}
	\bigr)
	\hspace{0.03cm}\bigl\langle\hspace{0.03cm}\mathcal{Q}^{\hspace{0.03cm}e_{2}}_{2}(\tau)
	\hspace{0.03cm}\bigr\rangle
	\Bigr]
	\hspace{0.01cm}
	\bigl\langle\hspace{0.03cm}\mathcal{Q}^{\hspace{0.03cm}e^{\prime}_{2}}_{2}(\tau)\hspace{0.03cm}\bigr\rangle
	\hspace{0.03cm}\bigl\langle\hspace{0.03cm}\mathcal{Q}^{\hspace{0.03cm}a^{\prime}_{1}}_{1}(\tau)
\hspace{0.03cm}\bigr\rangle\,- 
\notag
\end{align}
\vspace{0.04cm}
\[
-\hspace{0.03cm}2\hspace{0.03cm}i\hspace{0.04cm}\,
\frac{(2\hspace{0.02cm}\pi)^{3}}{|\Delta{\mathbf v}|}\sum_{\rho}\,
\!\int\!d\hspace{0.02cm}{\bf k}\,d\hspace{0.02cm}{\bf q}\;
\Bigl|{T}^{\hspace{0.03cm}(\rho)}_{\; {\bf k},\,{\bf q}}\Bigr|^{2}\,
\frac{\mathcal{P}}{\omega^{l}_{{\bf k}} - {\bf k}\cdot{\bf v}_{\rho}
- (-1)^{\rho}\,\Delta{\mathbf v}\cdot{\mathbf q}}\,
\times
\]
\vspace{0.06cm}
\[
\bigl(\hspace{0.03cm}T^{\,e_{2}}\hspace{0.03cm}T^{\,a_{1}}\hspace{0.03cm}T^{\,a^{\prime}_{2}}\hspace{0.01cm}
\bigr)^{\hspace{0.01cm}a^{\phantom{\prime}}_{2}\hspace{0.03cm}a^{\prime}_{1}}\hspace{0.03cm}
\bigl\langle\hspace{0.03cm}\mathcal{Q}^{\hspace{0.03cm}a^{\prime}_{1}}_{1}(\tau)
\hspace{0.03cm}\bigr\rangle
\hspace{0.03cm}\bigl\langle\hspace{0.03cm}\mathcal{Q}^{\hspace{0.03cm}a^{\prime}_{2}}_{2}(\tau)
\hspace{0.03cm}\bigr\rangle
\hspace{0.03cm}\bigl\langle\hspace{0.03cm}\mathcal{Q}^{\hspace{0.03cm}a_{1}}_{1}(\tau)
\hspace{0.03cm}\bigr\rangle
\hspace{0.01cm}\bigl\langle\hspace{0.03cm}\mathcal{Q}^{\hspace{0.03cm}e_{2}}_{2}(\tau)
\hspace{0.03cm}\bigr\rangle.
\]
As well as  equation (\ref{eq:9r}) for the averaged color charge $\bigl\langle\hspace{0.01cm}\mathcal{Q}^{\,a_{1}}_{\hspace{0.03cm}1}(\tau)
\hspace{0.01cm}\bigr\rangle$, this equation is valid only under the condition that the contractions of the type
\begin{equation}
	\begin{split}
		&{\rm tr}\hspace{0.03cm}
		\bigl(\hspace{0.03cm}T^{\,a^{\prime\prime}_{1}}\hspace{0.03cm}T^{\,a_{2}}\hspace{0.03cm}T^{\,e^{\prime}_{2}}\hspace{0.03cm}T^{\,a^{\prime}_{1}}{\mathcal N}_{\hspace{0.02cm}{\bf k}}\hspace{0.01cm}
		\bigr)
		\hspace{0.01cm}
		\bigl\langle\hspace{0.03cm}\mathcal{Q}^{\hspace{0.03cm}a^{\prime\prime}_{1}}_{1}(\tau)\hspace{0.03cm}\bigr\rangle
		\hspace{0.03cm}
		\bigl\langle\hspace{0.03cm}\mathcal{Q}^{\hspace{0.03cm}e^{\prime}_{2}}_{2}(\tau)\hspace{0.03cm}\bigr\rangle
		\hspace{0.03cm}
		\bigl\langle\hspace{0.03cm}\mathcal{Q}^{\hspace{0.03cm}a^{\prime}_{1}}_{1}(\tau)
		\hspace{0.03cm}\bigr\rangle,\\[1ex]
		i\hspace{0.03cm}&f^{\hspace{0.03cm}a_{2}\hspace{0.02cm}c^{\prime}\hspace{0.03cm}e_{2}}\hspace{0.03cm}
		{\rm tr}\hspace{0.03cm}
		\bigl(\hspace{0.03cm}T^{\,c^{\prime}}\hspace{0.03cm}T^{\,a^{\prime}_{1}}\hspace{0.03cm}T^{\,e^{\prime}_{2}}{\mathcal N}_{\hspace{0.02cm}{\bf k}}\hspace{0.01cm}
		\bigr)
		\hspace{0.03cm}\bigl\langle\hspace{0.03cm}\mathcal{Q}^{\hspace{0.03cm}e_{2}}_{2}(\tau)
		\hspace{0.03cm}\bigr\rangle
		\hspace{0.01cm}
		\bigl\langle\hspace{0.03cm}\mathcal{Q}^{\hspace{0.03cm}e^{\prime}_{2}}_{2}(\tau)\hspace{0.03cm}\bigr\rangle
		\hspace{0.03cm}\bigl\langle\hspace{0.03cm}\mathcal{Q}^{\hspace{0.03cm}a^{\prime}_{1}}_{1}(\tau)
		\hspace{0.03cm}\bigr\rangle
\end{split}
\label{eq:10qq}
\end{equation}	
and so on, are real functions. As we have already discussed in two previous sections, in the general case this circumstance does not occur. But we are interested in specific colorless combinations for which this would be true. The first such combination is ${\mathfrak q}_{2} =
\hspace{0.03cm}\bigl\langle\hspace{0.03cm}\mathcal{Q}^{\,e}_{2}
\hspace{0.03cm}\bigr\rangle
\bigl\langle\hspace{0.03cm}\mathcal{Q}^{\,e}_{2}
\hspace{0.03cm}\bigr\rangle$. In complete analogy with the derivation of Eq.\,(\ref{eq:9o}) for ${\mathfrak q}_{1}$, we obtain for ${\mathfrak q}_{2}$
\begin{align}
	\frac{d\hspace{0.03cm}{\mathfrak q}_{2}(\tau)}{d\hspace{0.03cm}\tau}
	=
	-\hspace{0.03cm}\biggl\{\frac{(2\hspace{0.02cm}\pi)^{3}}{|\Delta{\mathbf v}|}\,\sum_{\rho}\,
	&\int\!d\hspace{0.02cm}{\bf k}\hspace{0.03cm}d\hspace{0.02cm}{\bf q}\;
	\bigl|\hspace{0.03cm}{T}^{\hspace{0.03cm}(\rho)}_{\; {\bf k},\,{\bf q}}\bigr|^{\hspace{0.02cm}2}\hspace{0.03cm}
	{N}^{(1)}_{\hspace{0.02cm}{\bf k}}\hspace{0.03cm}
	2\hspace{0.02cm}\pi\hspace{0.03cm}\delta(\omega^{l}_{{\bf k}} - {\bf k}\cdot{\bf v}_{\rho}
	-(-1)^{\rho}\, \Delta{\mathbf v}\cdot{\mathbf q})\,+
	\label{eq:10w}\\[1ex]
	\frac{(2\hspace{0.02cm}\pi)^{3}}{|\Delta{\mathbf v}|}\,\sum_{\rho}\hspace{0.02cm}(-1)^{\rho + 1}\!\!
	&\int\!d\hspace{0.02cm}{\bf k}\hspace{0.03cm}d\hspace{0.02cm}{\bf q}\;
	\bigl|\hspace{0.03cm}{T}^{\hspace{0.03cm}(\rho)}_{\; {\bf k},\,{\bf q}}\bigr|^{\hspace{0.02cm}2}\hspace{0.03cm}
	{W}^{(1)}_{\hspace{0.02cm}{\bf k}}\hspace{0.03cm}
	2\hspace{0.02cm}\pi\hspace{0.03cm}\delta(\omega^{l}_{{\bf k}} - {\bf k}\cdot{\bf v}_{\rho}
	-(-1)^{\rho}\, \Delta{\mathbf v}\cdot{\mathbf q}) 
	\biggr\}\,\times
	\notag
\end{align}
\[
\,\biggl(\frac{3}{4}\,{\mathfrak q}_{1}{\mathfrak q}_{2} 
\,+\, 
\frac{3}{2}\,{\mathfrak q}^{2}_{12} \,+\, \frac{1}{2}\,\Lambda^{2} 
\biggr).
\] 
Here, the last term in (\ref{eq:10q}) also completely falls out of the dynamics. Note that the right-hand side of (\ref{eq:10w}) exactly coincides with those of the equation for the function ${\mathfrak q}_{1}(\tau)$, Eq.\,(\ref{eq:9o}). This means that there is the following relationship between the two functions ${\mathfrak q}_{1}(\tau)$ and ${\mathfrak q}_{2}(\tau)$:
\begin{equation}
{\mathfrak q}_{2}(\tau) - {\mathfrak q}_{1}(\tau) = C,
\label{eq:10e}
\end{equation}
where $C$ is a certain constant that we assume to be equal to
\[
C = {\mathfrak q}_{2}(0) - {\mathfrak q}_{1}(0)
\equiv
{\mathfrak q}_{2}^{0} - {\mathfrak q}_{1}^{0}.
\] 
\indent Further, the equation (\ref{eq:10q}) allows us to find the second term in the derivative of the function ${\mathfrak q}_{12}(\tau)$, Eq.\,(\ref{eq:9p}). As an analogue to the relation (\ref{eq:9d}), we now have the following expression
\[
	\bigl\langle \mathcal{Q}^{\,a_{2}}_{\hspace{0.03cm}1}(\tau)\hspace{0.02cm}\bigr\rangle
	\hspace{0.03cm}
	\frac{d\hspace{0.03cm}\bigl\langle \mathcal{Q}^{\,a_{2}}_{\hspace{0.03cm}2}(\tau)\hspace{0.02cm}\bigr\rangle}{d\hspace{0.03cm}\tau}
	=
\]
\begin{align}
	-\hspace{0.03cm}\biggl\{&\frac{(2\hspace{0.02cm}\pi)^{3}}{|\Delta{\mathbf v}|}\,\sum_{\rho}\,
	\int\!d\hspace{0.02cm}{\bf k}\hspace{0.03cm}d\hspace{0.02cm}{\bf q}\;
	\bigl|\hspace{0.03cm}{T}^{\hspace{0.03cm}(\rho)}_{\; {\bf k},\,{\bf q}}\bigr|^{\hspace{0.02cm}2}\hspace{0.03cm}
	{N}^{(1)}_{\hspace{0.02cm}{\bf k}}\hspace{0.03cm}
	2\hspace{0.02cm}\pi\hspace{0.03cm}\delta(\omega^{l}_{{\bf k}} - {\bf k}\cdot{\bf v}_{\rho}
	-(-1)^{\rho}\, \Delta{\mathbf v}\cdot{\mathbf q})\,+
	\notag\\[1ex]
	&\frac{(2\hspace{0.02cm}\pi)^{3}}{|\Delta{\mathbf v}|}\,\sum_{\rho}\hspace{0.02cm}(-1)^{\rho + 1}\!\!
	\int\!d\hspace{0.02cm}{\bf k}\hspace{0.03cm}d\hspace{0.02cm}{\bf q}\;
	\bigl|\hspace{0.03cm}{T}^{\hspace{0.03cm}(\rho)}_{\; {\bf k},\,{\bf q}}\bigr|^{\hspace{0.02cm}2}\hspace{0.03cm}
	{W}^{(1)}_{\hspace{0.02cm}{\bf k}}\hspace{0.03cm}
	2\hspace{0.02cm}\pi\hspace{0.03cm}\delta(\omega^{l}_{{\bf k}} - {\bf k}\cdot{\bf v}_{\rho}
	-(-1)^{\rho}\, \Delta{\mathbf v}\cdot{\mathbf q}) 
	\biggr\}\,\times
	\notag
\end{align}
\[
\,\biggl(\frac{9}{4}\,{\mathfrak q}_{12}\,{\mathfrak q}_{1} 
-\, \frac{3}{2}\,\Lambda^{2} 
\biggr).
\]
Here, the contribution of the last term from (\ref{eq:10q}) is also zero. Adding the expression above with (\ref{eq:9d}), we find the desired equation for
the third colorless combination ${\mathfrak q}_{12}(\tau)$:    
\begin{align}
	\frac{d\hspace{0.03cm}{\mathfrak q}_{12}(\tau)}{d\hspace{0.03cm}\tau}
	=
	-\hspace{0.03cm}\biggl\{\frac{(2\hspace{0.02cm}\pi)^{3}}{|\Delta{\mathbf v}|}\,\sum_{\rho}\,
	&\int\!d\hspace{0.02cm}{\bf k}\hspace{0.03cm}d\hspace{0.02cm}{\bf q}\;
	\bigl|\hspace{0.03cm}{T}^{\hspace{0.03cm}(\rho)}_{\; {\bf k},\,{\bf q}}\bigr|^{\hspace{0.02cm}2}\hspace{0.03cm}
	{N}^{(1)}_{\hspace{0.02cm}{\bf k}}\hspace{0.03cm}
	2\hspace{0.02cm}\pi\hspace{0.03cm}\delta(\omega^{l}_{{\bf k}} - {\bf k}\cdot{\bf v}_{\rho}
	-(-1)^{\rho}\, \Delta{\mathbf v}\cdot{\mathbf q})\,+
	\label{eq:10r}\\[1ex]
	\frac{(2\hspace{0.02cm}\pi)^{3}}{|\Delta{\mathbf v}|}\,\sum_{\rho}\hspace{0.02cm}(-1)^{\rho + 1}\!\!
	&\int\!d\hspace{0.02cm}{\bf k}\hspace{0.03cm}d\hspace{0.02cm}{\bf q}\;
	\bigl|\hspace{0.03cm}{T}^{\hspace{0.03cm}(\rho)}_{\; {\bf k},\,{\bf q}}\bigr|^{\hspace{0.02cm}2}\hspace{0.03cm}
	{W}^{(1)}_{\hspace{0.02cm}{\bf k}}\hspace{0.03cm}
	2\hspace{0.02cm}\pi\hspace{0.03cm}\delta(\omega^{l}_{{\bf k}} - {\bf k}\cdot{\bf v}_{\rho}
	-(-1)^{\rho}\, \Delta{\mathbf v}\cdot{\mathbf q}) 
	\biggr\}\,
	\times
	\notag
\end{align}
\[
\,\biggl(\frac{9}{4}\,{\mathfrak q}_{12}\,({\mathfrak q}_{1} + {\mathfrak q}_{2}) - 3\hspace{0.03cm}\Lambda^{2}  
\biggr).
\]
\indent It remains for us to determine the equation for the most complicated colorless function 
\[
\Lambda^{\!2} = \Lambda^{e}\Lambda^{e} = \Lambda^{e} f^{\hspace{0.03cm}e\,a_{1}\hspace{0.03cm}a_{2}\hspace{0.03cm}}
\bigl\langle\hspace{0.03cm}\mathcal{Q}^{\,a_{1}}_{\hspace{0.03cm}1}
(\tau)
\hspace{0.03cm}\bigr\rangle
\bigl\langle\hspace{0.03cm}\mathcal{Q}^{\,a_{2}}_{\hspace{0.03cm}2}
(\tau)
\hspace{0.03cm}\bigr\rangle.
\]
The derivative for this function has the following form: 
\begin{equation}
\frac{d\hspace{0.03cm}\Lambda^{\!\hspace{0.02cm}2} (\tau)}{d\hspace{0.03cm}\tau}
=
2\hspace{0.03cm}\Lambda^{e}(\tau)\,\frac{d\hspace{0.03cm}\Lambda^{e}(\tau)}{d\hspace{0.03cm}\tau}
=
\label{eq:10t} 
\end{equation}
\[
2\hspace{0.03cm}\Lambda^{e}(\tau)\hspace{0.03cm}
f^{\hspace{0.03cm}e\,a_{1}\hspace{0.03cm}a_{2}\hspace{0.03cm}}
\biggl\{\bigl\langle \mathcal{Q}^{\,a_{2}}_{\hspace{0.03cm}2}(\tau)\hspace{0.02cm}\bigr\rangle
\hspace{0.03cm}
\frac{d\hspace{0.03cm}\bigl\langle \mathcal{Q}^{\,a_{1}}_{\hspace{0.03cm}1}(\tau)\hspace{0.02cm}\bigr\rangle}{d\hspace{0.03cm}\tau}
\,+\,
\bigl\langle \mathcal{Q}^{\,a_{1}}_{\hspace{0.03cm}1}(\tau)\hspace{0.02cm}\bigr\rangle
\hspace{0.03cm}
\frac{d\hspace{0.03cm}\bigl\langle \mathcal{Q}^{\,a_{2}}_{\hspace{0.03cm}2}(\tau)\hspace{0.02cm}\bigr\rangle}{d\hspace{0.03cm}\tau}\biggr\}.
\]
First, let us look at the contribution related to the first term in the curly bracket, namely
\begin{equation}
\Lambda^{e}(\tau)\hspace{0.03cm}
f^{\hspace{0.03cm}e\,a_{1}\hspace{0.03cm}a_{2}\hspace{0.03cm}}
\bigl\langle \mathcal{Q}^{\,a_{2}}_{\hspace{0.03cm}2}(\tau)\hspace{0.02cm}\bigr\rangle
\hspace{0.03cm}
\frac{d\hspace{0.03cm}\bigl\langle \mathcal{Q}^{\,a_{1}}_{\hspace{0.03cm}1}(\tau)\hspace{0.02cm}\bigr\rangle}{d\hspace{0.03cm}\tau}.
\label{eq:10y} 
\end{equation}
We use the equation for the color charge $\bigl\langle \mathcal{Q}^{\,a_{1}}_{\hspace{0.03cm}1}(\tau)\hspace{0.02cm}\bigr\rangle$, Eq.\,(\ref{eq:9r}). Let us analyze the real contributions to (\ref{eq:10y}). As usual, the first step is to consider the contraction with the trace in the first term on the right-hand side of (\ref{eq:9y}), which looks like this:
\begin{equation}
{N}^{(1)}_{\hspace{0.02cm}{\bf k}}\hspace{0.03cm}
\Lambda^{e}(\tau)\hspace{0.03cm}
f^{\hspace{0.03cm}e\,a_{1}\hspace{0.03cm}a_{2}\hspace{0.03cm}}
\bigl\langle \mathcal{Q}^{\,a_{2}}_{\hspace{0.03cm}2}(\tau)\hspace{0.02cm}\bigr\rangle
\hspace{0.03cm}
{\rm tr}\hspace{0.03cm}
\bigl(\hspace{0.03cm}T^{\,a^{\prime}_{2}}\hspace{0.03cm}T^{\,a_{1}}\hspace{0.03cm}T^{\,e^{\prime}_{1}}\hspace{0.03cm}T^{\,a^{\prime\prime}_{2}}\hspace{0.01cm}
\bigr)
\hspace{0.01cm}
\bigl\langle\hspace{0.03cm}\mathcal{Q}^{\hspace{0.03cm}a^{\prime}_{2}}_{2}(\tau)\hspace{0.03cm}\bigr\rangle
\bigl\langle\hspace{0.03cm}\mathcal{Q}^{\hspace{0.03cm}a^{\prime\prime}_{2}}_{2}(\tau)\hspace{0.03cm}\bigr\rangle
\hspace{0.03cm}
\bigl\langle\hspace{0.03cm}\mathcal{Q}^{\hspace{0.03cm}e^{\prime}_{1}}_{1}(\tau)\hspace{0.03cm}\bigr\rangle
=
\label{eq:10u} 
\end{equation} 
\[
{N}^{(1)}_{\hspace{0.02cm}{\bf k}}\hspace{0.03cm}
\biggl({\mathfrak q}_{2}\,\Lambda^{2}
+
\frac{1}{4}\,N_{c}
\hspace{0.03cm}
f^{\hspace{0.03cm}e\,a_{1}\hspace{0.03cm}a_{2}\hspace{0.03cm}}
d^{\,a_{1}\hspace{0.03cm}e_{1}^{\prime}\hspace{0.03cm}s}
\Lambda^{e}\,
\Omega^{\hspace{0.03cm}s}_{22}
\hspace{0.03cm}
\bigl\langle\hspace{0.03cm}\mathcal{Q}^{\,e^{\prime}_{1}}_{1}
\hspace{0.03cm}\bigr\rangle
\hspace{0.03cm}
\bigl\langle\hspace{0.03cm}\mathcal{Q}^{\,a_{2}}_{2}
\hspace{0.03cm}\bigr\rangle\biggr).
\] 
Here we have again employed the expression for the fourth-order trace and the definition of the function $\Lambda^{e}$. Let us analyze the last term on the right-hand side of this relation for the special case of the $SU(3_{c})$ color group. For this purpose, we use equality (\ref{ap:D15}), which gives us
\[
d^{\,a_{1}\hspace{0.03cm}e_{1}^{\prime}\hspace{0.03cm}s}
\,\Omega^{\hspace{0.03cm}s}_{22}
\hspace{0.03cm}
\bigl\langle\hspace{0.03cm}\mathcal{Q}^{\,e^{\prime}_{1}}_{1}
\hspace{0.03cm}\bigr\rangle
=
d^{\,a_{1}\hspace{0.03cm}e_{1}^{\prime}\hspace{0.03cm}s}
d^{\,a^{\prime}_{2}\hspace{0.03cm}a_{2}^{\prime\prime}\hspace{0.03cm}s}
\bigl\langle\hspace{0.03cm}\mathcal{Q}^{\,a^{\prime}_{2}}_{2}
\hspace{0.03cm}\bigr\rangle
\hspace{0.03cm}
\bigl\langle\hspace{0.03cm}\mathcal{Q}^{\,a^{\prime\prime}_{2}}_{2}
\hspace{0.03cm}\bigr\rangle
\hspace{0.03cm}
\bigl\langle\hspace{0.03cm}\mathcal{Q}^{\,e^{\prime}_{1}}_{1}
\hspace{0.03cm}\bigr\rangle
=
\] 
\[
\frac{1}{3}\,\bigl[\hspace{0.03cm}2\hspace{0.03cm} f^{\hspace{0.03cm}a_{1}\,a^{\prime}_{2}\hspace{0.03cm}s\hspace{0.03cm}}
\Lambda^{s}\,
\bigl\langle\hspace{0.03cm}\mathcal{Q}^{\,a^{\prime}_{2}}_{2}
\hspace{0.03cm}\bigr\rangle
-
{\mathfrak q}_{2}\hspace{0.03cm}
\bigl\langle\hspace{0.03cm}\mathcal{Q}^{\,a_{1}}_{1}
\hspace{0.03cm}\bigr\rangle
+
2\hspace{0.03cm}{\mathfrak q}_{12}
\bigl\langle\hspace{0.03cm}\mathcal{Q}^{\,a_{1}}_{2}
\hspace{0.03cm}\bigr\rangle\bigr].
\]     
With the use of this relationship, the last term on the right-hand side of (\ref{eq:10u}) for $N_{c} = 3$ takes the form
\[
\frac{3}{4}\,
f^{\hspace{0.03cm}e\,a_{1}\hspace{0.03cm}a_{2}\hspace{0.03cm}}
d^{\,a_{1}\hspace{0.03cm}e_{1}^{\prime}\hspace{0.03cm}s}
\Lambda^{e}\,
\Omega^{\hspace{0.03cm}s}_{22}
\hspace{0.03cm}
\bigl\langle\hspace{0.03cm}\mathcal{Q}^{\,e^{\prime}_{1}}_{1}
\hspace{0.03cm}\bigr\rangle
\hspace{0.03cm}
\bigl\langle\hspace{0.03cm}\mathcal{Q}^{\,a_{2}}_{2}
\hspace{0.03cm}\bigr\rangle
=
-\frac{1}{4}\,{\mathfrak q}_{2}\,\Lambda^{2}
+
\frac{1}{2}\,\bigl(f^{\hspace{0.03cm}e\,a_{1}\hspace{0.03cm}a_{2}\hspace{0.03cm}}
\Lambda^{e}\hspace{0.03cm}
\bigl\langle\hspace{0.03cm}\mathcal{Q}^{\,a_{2}}_{2}
\hspace{0.03cm}\bigr\rangle\bigr)
\bigl(f^{\hspace{0.03cm}e^{\prime}\,a_{1}\hspace{0.03cm}a^{\prime}_{2}\hspace{0.03cm}}
\Lambda^{e^{\prime}}\hspace{0.03cm}
\bigl\langle\hspace{0.03cm}\mathcal{Q}^{\,a^{\prime}_{2}}_{2}
\hspace{0.03cm}\bigr\rangle\bigr),
\] 
and the whole expression (\ref{eq:10u}) is written in its final form
\begin{equation}
	{N}^{(1)}_{\hspace{0.02cm}{\bf k}}\hspace{0.03cm}
	\Lambda^{e}(\tau)\hspace{0.03cm}
	f^{\hspace{0.03cm}e\,a_{1}\hspace{0.03cm}a_{2}\hspace{0.03cm}}
	\bigl\langle \mathcal{Q}^{\,a_{2}}_{\hspace{0.03cm}2}(\tau)\hspace{0.02cm}\bigr\rangle
	\hspace{0.03cm}
	{\rm tr}\hspace{0.03cm}
	\bigl(\hspace{0.03cm}T^{\,a^{\prime}_{2}}\hspace{0.03cm}T^{\,a_{1}}\hspace{0.03cm}T^{\,e^{\prime}_{1}}\hspace{0.03cm}T^{\,a^{\prime\prime}_{2}}\hspace{0.01cm}
	\bigr)
	\hspace{0.01cm}
	\bigl\langle\hspace{0.03cm}\mathcal{Q}^{\hspace{0.03cm}a^{\prime}_{2}}_{2}(\tau)\hspace{0.03cm}\bigr\rangle
	\bigl\langle\hspace{0.03cm}\mathcal{Q}^{\hspace{0.03cm}a^{\prime\prime}_{2}}_{2}(\tau)\hspace{0.03cm}\bigr\rangle
	\hspace{0.03cm}
	\bigl\langle\hspace{0.03cm}\mathcal{Q}^{\hspace{0.03cm}e^{\prime}_{1}}_{1}(\tau)\hspace{0.03cm}\bigr\rangle
	=
\label{eq:10i} 
\end{equation} 
\[
{N}^{(1)}_{\hspace{0.02cm}{\bf k}}\hspace{0.03cm}
\biggl(
\frac{3}{4}\,{\mathfrak q}_{2}\,\Lambda^{2}
+
\frac{1}{2}\,\bigl(f^{\hspace{0.03cm}e\,a_{1}\hspace{0.03cm}a_{2}\hspace{0.03cm}}
\Lambda^{e}\hspace{0.03cm}
\bigl\langle\hspace{0.03cm}\mathcal{Q}^{\,a_{2}}_{2}
\hspace{0.03cm}\bigr\rangle\bigr)
\bigl(f^{\hspace{0.03cm}e^{\prime}\,a_{1}\hspace{0.03cm}a^{\prime}_{2}\hspace{0.03cm}}
\Lambda^{e^{\prime}}
\bigl\langle\hspace{0.03cm}\mathcal{Q}^{\,a^{\prime}_{2}}_{2}
\hspace{0.03cm}\bigr\rangle\bigr)\biggr).
\]
Note that a new colorless structure has appeared here, which is no longer reduced  to the previously introduced colorless ones ${\mathfrak q}_{1},\,{\mathfrak q}_{2},\,{\mathfrak q}_{12}$ and $\Lambda^{2}$, even for the $N_{c} = 3$ case.\\
\indent Let us now consider another real contribution related to the second contraction in (\ref{eq:9t}). Here, we have
\begin{equation}
i\hspace{0.03cm}{N}^{(1)}_{\hspace{0.02cm}{\bf k}}\hspace{0.03cm}
\Lambda^{e}(\tau)\hspace{0.03cm}
f^{\hspace{0.03cm}e\,a_{1}\hspace{0.03cm}a_{2}\hspace{0.03cm}}
\bigl\langle \mathcal{Q}^{\,a_{2}}_{\hspace{0.03cm}2}(\tau)\hspace{0.02cm}\bigr\rangle
\hspace{0.03cm}
f^{\hspace{0.03cm}a_{1}\hspace{0.02cm}c^{\prime}\hspace{0.03cm}e_{1}}\hspace{0.03cm}
{\rm tr}\hspace{0.03cm}
\bigl(\hspace{0.03cm}T^{\,c^{\prime}}\hspace{0.03cm}T^{\,a^{\prime\prime}_{2}}\hspace{0.03cm}T^{\,e^{\prime}_{1}}\hspace{0.01cm}
\bigr)
\hspace{0.03cm}\bigl\langle\hspace{0.03cm}\mathcal{Q}^{\hspace{0.03cm}e_{1}}_{1}(\tau)
\hspace{0.03cm}\bigr\rangle
\bigl\langle\hspace{0.03cm}\mathcal{Q}^{\hspace{0.03cm}a^{\prime\prime}_{2}}_{2}(\tau)\hspace{0.03cm}\bigr\rangle
\hspace{0.03cm}
\bigl\langle\hspace{0.03cm}\mathcal{Q}^{\hspace{0.03cm}e^{\prime}_{1}}_{1}(\tau)\hspace{0.03cm}\bigr\rangle
=
\label{eq:10o} 
\end{equation}  
\[
-\hspace{0.03cm}\frac{1}{2}\,{N}^{(1)}_{\hspace{0.02cm}{\bf k}}\hspace{0.03cm}N_{c}
\hspace{0.03cm}
f^{\hspace{0.03cm}e\,a_{1}\hspace{0.03cm}a_{2}\hspace{0.03cm}}
f^{\hspace{0.03cm}a_{1}\hspace{0.02cm}c^{\prime}\hspace{0.03cm}e_{1}}\hspace{0.03cm}
f^{\hspace{0.03cm}c^{\prime}\hspace{0.02cm}a^{\prime\prime}_{2}\hspace{0.03cm}e^{\prime}_{1}}\hspace{0.03cm}\Lambda^{e}(\tau)\hspace{0.03cm}
\bigl\langle \mathcal{Q}^{\,a_{2}}_{\hspace{0.03cm}2}(\tau)\hspace{0.02cm}\bigr\rangle
\hspace{0.03cm}
\bigl\langle\hspace{0.03cm}\mathcal{Q}^{\hspace{0.03cm}e_{1}}_{1}(\tau)
\hspace{0.03cm}\bigr\rangle
\bigl\langle\hspace{0.03cm}\mathcal{Q}^{\hspace{0.03cm}a^{\prime\prime}_{2}}_{2}(\tau)\hspace{0.03cm}\bigr\rangle
\hspace{0.03cm}
\bigl\langle\hspace{0.03cm}\mathcal{Q}^{\hspace{0.03cm}e^{\prime}_{1}}_{1}(\tau)\hspace{0.03cm}\bigr\rangle
=
\]   
\[
-\,\frac{1}{2}\,{N}^{(1)}_{\hspace{0.02cm}{\bf k}}\hspace{0.03cm}N_{c}
\hspace{0.03cm}
\bigl(f^{\hspace{0.03cm}e\,a_{1}\hspace{0.03cm}a_{2}\hspace{0.03cm}}
\Lambda^{e}\hspace{0.03cm}
\bigl\langle\hspace{0.03cm}\mathcal{Q}^{\,a_{2}}_{2}
\hspace{0.03cm}\bigr\rangle\bigr)
\bigl(f^{\hspace{0.03cm}e^{\prime}\,a_{1}\hspace{0.03cm}e_{1}\hspace{0.03cm}}
\Lambda^{e^{\prime}}
\bigl\langle\hspace{0.03cm}\mathcal{Q}^{\,e_{1}}_{1}
\hspace{0.03cm}\bigr\rangle\bigr),
\]
where we have used the definition of the function $\Lambda^{e^{\prime}}$.\\
\indent Completely similar reasoning is valid for the contractions in equation (\ref{eq:9r}) containing the matrix function ${\mathcal W}_{\hspace{0.02cm}{\bf k}}$. Using the color decomposition (\ref{eq:7q}), we get an analog of the formula (\ref{eq:10i}) and (\ref{eq:10o}) with the replacement ${N}^{(1)}_{\hspace{0.02cm}{\bf k}}\rightarrow{W}^{(1)}_{\hspace{0.02cm}{\bf k}}$.\\
\indent We now need to find similar contributions for the second term in the curly bracket of equality (\ref{eq:10t}):
\begin{equation}
	\Lambda^{e}(\tau)\hspace{0.03cm}
	f^{\hspace{0.03cm}e\,a_{1}\hspace{0.03cm}a_{2}\hspace{0.03cm}}
	\bigl\langle \mathcal{Q}^{\,a_{1}}_{\hspace{0.03cm}1}(\tau)\hspace{0.02cm}\bigr\rangle
	\hspace{0.03cm}
	\frac{d\hspace{0.03cm}\bigl\langle \mathcal{Q}^{\,a_{2}}_{\hspace{0.03cm}2}(\tau)\hspace{0.02cm}\bigr\rangle}{d\hspace{0.03cm}\tau}.
\label{eq:10yy} 
\end{equation}
We will not repeat the calculations performed above, but give the final expression immediately. An analogue of the formula (\ref{eq:10i}) now is  
\begin{equation}
{N}^{(1)}_{\hspace{0.02cm}{\bf k}}\hspace{0.03cm}
\biggl(
\frac{3}{4}\,{\mathfrak q}_{1}\,\Lambda^{2}
-
\frac{1}{2}\,\bigl(f^{\hspace{0.03cm}e\,a_{1}\hspace{0.03cm}a_{2}\hspace{0.03cm}}
\Lambda^{e}\hspace{0.03cm}
\bigl\langle\hspace{0.03cm}\mathcal{Q}^{\,a_{1}}_{1}
\hspace{0.03cm}\bigr\rangle\bigr)
\bigl(f^{\hspace{0.03cm}e^{\prime}\,a_{2}\hspace{0.03cm}a^{\prime}_{1}\hspace{0.03cm}}
\Lambda^{e^{\prime}}
\bigl\langle\hspace{0.03cm}\mathcal{Q}^{\,a^{\prime}_{1}}_{1}
\hspace{0.03cm}\bigr\rangle\bigr)\biggr),
\label{eq:10p} 
\end{equation} 
and an analogue of the formula (\ref{eq:10o}) is
\begin{equation}
-\hspace{0.03cm}\frac{1}{2}\,{N}^{(1)}_{\hspace{0.02cm}{\bf k}}\hspace{0.03cm}N_{c}
\hspace{0.03cm}
\bigl(f^{\hspace{0.03cm}e\,a_{1}\hspace{0.03cm}a_{2}\hspace{0.03cm}}
\Lambda^{e}\hspace{0.03cm}
\bigl\langle\hspace{0.03cm}\mathcal{Q}^{\,a_{1}}_{1}
\hspace{0.03cm}\bigr\rangle\bigr)
\bigl(f^{\hspace{0.03cm}e^{\prime}\,e_{2}\hspace{0.03cm}a_{2}\hspace{0.03cm}}
\Lambda^{e^{\prime}}
\bigl\langle\hspace{0.03cm}\mathcal{Q}^{\,e_{2}}_{2}
\hspace{0.03cm}\bigr\rangle\bigr).
\label{eq:10a} 
\end{equation}  
\indent It remains for us to analyze the contribution that does not depend on the scalar plasmon number densities ${N}^{(1)}_{\hspace{0.02cm}{\bf k}}$, ${W}^{(1)}_{\hspace{0.02cm}{\bf k}},\ldots$ to the derivative (\ref{eq:10t}). We are looking at the term in (\ref{eq:10t}) associated with the derivative $d\hspace{0.03cm}\bigl\langle \mathcal{Q}^{\,a_{1}}_{\hspace{0.03cm}1}(\tau)\hspace{0.02cm}\bigr\rangle/d\hspace{0.03cm}\tau$. Using the color factor for the contribution (\ref{eq:9uu}), we now have the following expression: 
\[
\Lambda^{e}(\tau)\hspace{0.03cm}
\bigl(\hspace{0.03cm}T^{\,e}\hspace{0.03cm}T^{\,e_{1}}\hspace{0.03cm}T^{\,a^{\prime}_{2}}\hspace{0.03cm}T^{\,a^{\prime}_{1}}\hspace{0.01cm}
\bigr)^{\hspace{0.01cm}a^{\phantom{\prime}}_{2}\hspace{0.03cm}a^{\prime\prime}_{2}}\hspace{0.03cm}
\bigl\langle \mathcal{Q}^{\,a_{2}}_{\hspace{0.03cm}2}(\tau)\hspace{0.02cm}\bigr\rangle
\hspace{0.03cm}
\bigl\langle\hspace{0.03cm}\mathcal{Q}^{\hspace{0.03cm}a^{\prime}_{1}}_{1}(\tau)
\hspace{0.03cm}\bigr\rangle
\hspace{0.03cm}\bigl\langle\hspace{0.03cm}\mathcal{Q}^{\hspace{0.03cm}a^{\prime\prime}_{2}}_{2}(\tau)
\hspace{0.03cm}\bigr\rangle
\hspace{0.03cm}\bigl\langle\hspace{0.03cm}\mathcal{Q}^{\hspace{0.03cm}a^{\prime}_{2}}_{2}(\tau)
\hspace{0.03cm}\bigr\rangle
\hspace{0.01cm}\bigl\langle\hspace{0.03cm}\mathcal{Q}^{\hspace{0.03cm}e_{1}}_{1}(\tau)
\hspace{0.03cm}\bigr\rangle
\equiv
\]
\[
\bigl(f^{\hspace{0.03cm}e\,a_{2}\hspace{0.03cm}s\hspace{0.03cm}}
\Lambda^{e}\hspace{0.03cm}
\bigl\langle\hspace{0.03cm}\mathcal{Q}^{\,a_{2}}_{2}
\hspace{0.03cm}\bigr\rangle\bigr)
\bigl(f^{\hspace{0.03cm}e_{1}\hspace{0.01cm}s\hspace{0.03cm}s^{\prime}
\hspace{0.03cm}}\bigl\langle\hspace{0.03cm}\mathcal{Q}^{\,e_{1}}_{1}
\hspace{0.03cm}\bigr\rangle\bigr)
\bigl(f^{\hspace{0.03cm}e^{\prime}\hspace{0.01cm}s^{\prime}\hspace{0.01cm}
a^{\prime}_{2}\hspace{0.03cm}}\Lambda^{e^{\prime}}
\bigl\langle\hspace{0.03cm}\mathcal{Q}^{\,a^{\prime}_{2}}_{2}
\hspace{0.03cm}\bigr\rangle\bigr)
= 0.
\]
Here, we have again used the definition of the function $\Lambda^{e^{\prime}}$. A similar contribution from the derivative $d\hspace{0.03cm}\bigl\langle \mathcal{Q}^{\,a_{2}}_{\hspace{0.03cm}2}(\tau)\hspace{0.02cm}\bigr\rangle/d\hspace{0.03cm}\tau$, as defined in the equation (\ref{eq:10q}) can also be reduced to a similar form and it is equal to
\[
-\bigl(f^{\hspace{0.03cm}e\,a_{1}\hspace{0.03cm}s\hspace{0.03cm}}
\Lambda^{e}\hspace{0.03cm}
\bigl\langle\hspace{0.03cm}\mathcal{Q}^{\,a_{1}}_{1}
\hspace{0.03cm}\bigr\rangle\bigr)
\bigl(f^{\hspace{0.03cm}e_{2}\,s\hspace{0.03cm}s^{\prime}\hspace{0.03cm}}
\bigl\langle\hspace{0.03cm}\mathcal{Q}^{\,e_{2}}_{2}
\hspace{0.03cm}\bigr\rangle\bigr)
\bigl(f^{\hspace{0.03cm}e^{\prime}\hspace{0.01cm}s^{\prime}\hspace{0.01cm}
a^{\prime}_{1}\hspace{0.03cm}}\Lambda^{e^{\prime}}
\bigl\langle\hspace{0.03cm}\mathcal{Q}^{\,a^{\prime}_{1}}_{1}
\hspace{0.03cm}\bigr\rangle\bigr).
\]
Obviously, it is also zero. Thus, the whole contribution without the scalar plasmon number densities ${N}^{(1)}_{\hspace{0.02cm}{\bf k}}$, ${W}^{(1)}_{\hspace{0.02cm}{\bf k}},\ldots$ to equation (\ref{eq:10t}) turns to zero, and the equation itself, with allowance made for the expressions obtained (\ref{eq:10i})\,--\,(\ref{eq:10a}), takes the final form:

{\scalebox{0.97}{
\begin{minipage}{\linewidth}
\begin{align}
	\frac{d\hspace{0.03cm}\Lambda^{\!2} (\tau)}{d\hspace{0.03cm}\tau}
	=
	-\hspace{0.03cm}\biggl\{\frac{(2\hspace{0.02cm}\pi)^{3}}{|\Delta{\mathbf v}|}\,\sum_{\rho}\,
	&\int\!d\hspace{0.02cm}{\bf k}\hspace{0.03cm}d\hspace{0.02cm}{\bf q}\;
	\bigl|\hspace{0.03cm}{T}^{\hspace{0.03cm}(\rho)}_{\; {\bf k},\,{\bf q}}\bigr|^{\hspace{0.02cm}2}\hspace{0.03cm}
	{N}^{(1)}_{\hspace{0.02cm}{\bf k}}\hspace{0.03cm}
	2\hspace{0.02cm}\pi\hspace{0.03cm}\delta(\omega^{l}_{{\bf k}} - {\bf k}\cdot{\bf v}_{\rho}
	-(-1)^{\rho}\, \Delta{\mathbf v}\cdot{\mathbf q})\,+
	\label{eq:10s}\\[1ex]
	\frac{(2\hspace{0.02cm}\pi)^{3}}{|\Delta{\mathbf v}|}\,\sum_{\rho}\hspace{0.02cm}(-1)^{\rho + 1}\!\!
	&\int\!d\hspace{0.02cm}{\bf k}\hspace{0.03cm}d\hspace{0.02cm}{\bf q}\;
	\bigl|\hspace{0.03cm}{T}^{\hspace{0.03cm}(\rho)}_{\; {\bf k},\,{\bf q}}\bigr|^{\hspace{0.02cm}2}\hspace{0.03cm}
	{W}^{(1)}_{\hspace{0.02cm}{\bf k}}\hspace{0.03cm}
	2\hspace{0.02cm}\pi\hspace{0.03cm}\delta(\omega^{l}_{{\bf k}} - {\bf k}\cdot{\bf v}_{\rho}
	-(-1)^{\rho}\, \Delta{\mathbf v}\cdot{\mathbf q}) 
	\biggr\}\,
	\times
	\notag
\end{align}
\end{minipage}
}}
\[
\biggl[\hspace{0.03cm}\frac{3}{4}\,({\mathfrak q}_{1} + {\mathfrak q}_{2})\Lambda^{2} 
\,+\,
\frac{1}{2}\,\bigl(f^{\hspace{0.03cm}e\,a_{2}\hspace{0.03cm}a_{1}\hspace{0.03cm}}
\Lambda^{e}\hspace{0.03cm}
\bigl\langle\hspace{0.03cm}\mathcal{Q}^{\,a_{1}}_{1}
\hspace{0.03cm}\bigr\rangle\bigr)
\bigl(f^{\hspace{0.03cm}e^{\prime}\,a_{2}\hspace{0.03cm}a^{\prime}_{1}\hspace{0.03cm}}
\Lambda^{e^{\prime}}
\bigl\langle\hspace{0.03cm}\mathcal{Q}^{\,a^{\prime}_{1}}_{1}
\hspace{0.03cm}\bigr\rangle
\bigr)
+
\frac{1}{2}\,\bigl(f^{\hspace{0.03cm}e\,a_{1}\hspace{0.03cm}a_{2}\hspace{0.03cm}}
\Lambda^{e}\hspace{0.03cm}
\bigl\langle\hspace{0.03cm}\mathcal{Q}^{\,a_{2}}_{2}
\hspace{0.03cm}\bigr\rangle\bigr)
\bigl(f^{\hspace{0.03cm}e^{\prime}\,a_{1}\hspace{0.03cm}a^{\prime}_{2}\hspace{0.03cm}}
\Lambda^{e^{\prime}}
\bigl\langle\hspace{0.03cm}\mathcal{Q}^{\,a^{\prime}_{2}}_{2}
\hspace{0.03cm}\bigr\rangle\bigr)
\]
\[
+\,
3\hspace{0.03cm}
\bigl(f^{\hspace{0.03cm}e\,a_{1}\hspace{0.03cm}a_{2}\hspace{0.03cm}}
\Lambda^{e}\hspace{0.03cm}
\bigl\langle\hspace{0.03cm}\mathcal{Q}^{\,a_{1}}_{1}
\hspace{0.03cm}\bigr\rangle\bigr)
\bigl(f^{\hspace{0.03cm}e^{\prime}\,e_{2}\hspace{0.03cm}a_{2}\hspace{0.03cm}}
\Lambda^{e^{\prime}}
\bigl\langle\hspace{0.03cm}\mathcal{Q}^{\,e_{2}}_{2}
\hspace{0.03cm}\bigr\rangle\bigr)
\biggr].
\]
\indent The equation (\ref{eq:10s}) like equations (\ref{eq:9o}), (\ref{eq:10w}) and (\ref{eq:10r}) is valid only if the corresponding imaginary parts from the contractions of the type (\ref{eq:9t}) will be equal to zero. For the previous equations, we have shown that the imaginary parts vanish only for specific values $N_{c}= 2$ or $N_{c}=3$. Here we will not provide a proof that the imaginary parts for equation (\ref{eq:10s}), which defines the evolution of $\Lambda^{2}(\tau)$, are also zero for these values due to the cumbersome calculations. Some examples of such calculations are given in the Appendix \ref{appendix_F}.\\
\indent Let us now  give the complete system of differential equations for our colorless structures. We introduce the following notation: 
\begin{equation}
\begin{split}
\mathcal{A}(\tau)
\equiv
-\hspace{0.03cm}\frac{(2\hspace{0.02cm}\pi)^{3}}{|\Delta{\mathbf v}|}\,\sum_{\rho}\,
&\int\!d\hspace{0.02cm}{\bf k}\hspace{0.03cm}d\hspace{0.02cm}{\bf q}\;
\bigl|\hspace{0.03cm}{T}^{\hspace{0.03cm}(\rho)}_{\; {\bf k},\,{\bf q}}\bigr|^{\hspace{0.02cm}2}\hspace{0.03cm}
{N}^{(1)}_{\hspace{0.02cm}{\bf k}}(\tau)\hspace{0.04cm}
2\hspace{0.02cm}\pi\hspace{0.03cm}\delta(\omega^{l}_{{\bf k}} - {\bf k}\cdot{\bf v}_{\rho}
-(-1)^{\rho}\, \Delta{\mathbf v}\cdot{\mathbf q})\,
+
\\[1ex]
\frac{(2\hspace{0.02cm}\pi)^{3}}{|\Delta{\mathbf v}|}\,\sum_{\rho}\hspace{0.02cm}(-1)^{\rho + 1}\!\!
&\int\!d\hspace{0.02cm}{\bf k}\hspace{0.03cm}d\hspace{0.02cm}{\bf q}\;
\bigl|\hspace{0.03cm}{T}^{\hspace{0.03cm}(\rho)}_{\; {\bf k},\,{\bf q}}\bigr|^{\hspace{0.02cm}2}\hspace{0.03cm}
{W}^{(1)}_{\hspace{0.02cm}{\bf k}}(\tau)\hspace{0.04cm}
2\hspace{0.02cm}\pi\hspace{0.03cm}\delta(\omega^{l}_{{\bf k}} - {\bf k}\cdot{\bf v}_{\rho}
-(-1)^{\rho}\, \Delta{\mathbf v}\cdot{\mathbf q}) 
\label{eq:10ss}
\end{split}
\end{equation}
and define the new slow time making use of the formula
\begin{equation}
\tau^{\prime} = \tau^{\prime}(\tau)
=
\int\!\mathcal{A}(\tau)d\tau + C_{1},
\label{eq:10sss}
\end{equation}
where $C_{1}$ is an arbitrary constant of integration. Considering the change of time variable, the relationship between the functions ${\mathfrak q}_{1}$ and ${\mathfrak q}_{2}$, Eq.\,(\ref{eq:10e}), the equations (\ref{eq:9o}), (\ref{eq:10w}), (\ref{eq:10r}) and (\ref{eq:10s}) acquire a more compact form: 
\begin{align}
&\frac{d\hspace{0.03cm}{\mathfrak q}_{1}(\tau^{\prime})}{d\hspace{0.03cm}\tau^{\prime}}
=
\frac{3}{4}\,({\mathfrak q}_{1} + C)\hspace{0.03cm}{\mathfrak q}_{1} 
\,+\, 
\frac{3}{2}\,{\mathfrak q}^{2}_{12} \,+\, \frac{1}{2}\,\Lambda^{2}, 
\notag\\[1ex]
&\frac{d\hspace{0.03cm}{\mathfrak q}_{12}(\tau^{\prime})}{d\hspace{0.03cm}\tau^{\prime}}
=
\frac{9}{4}\,(2\hspace{0.03cm}{\mathfrak q}_{1} + C)\hspace{0.03cm}{\mathfrak q}_{12} - 3\hspace{0.03cm}\Lambda^{2},
\label{eq:10d}\\[1ex]
&\frac{d\hspace{0.03cm}\Lambda^{\!2} (\tau^{\prime})}{d\hspace{0.03cm}\tau^{\prime}}
= 
\frac{3}{4}\,(2\hspace{0.03cm}{\mathfrak q}_{1} + C)\hspace{0.03cm}\Lambda^{2} 
\,+
\notag\\[1ex]
&\hspace{1.95cm}\frac{1}{2}\,\bigl(f^{\hspace{0.03cm}e\,a_{2}\hspace{0.03cm}a_{1}\hspace{0.03cm}}
\Lambda^{e}\hspace{0.03cm}
\bigl\langle\hspace{0.03cm}\mathcal{Q}^{\,a_{1}}_{1}
\hspace{0.03cm}\bigr\rangle\bigr)
\bigl(f^{\hspace{0.03cm}e^{\prime}\,a_{2}\hspace{0.03cm}a^{\prime}_{1}\hspace{0.03cm}}
\Lambda^{e^{\prime}}
\bigl\langle\hspace{0.03cm}\mathcal{Q}^{\,a^{\prime}_{1}}_{1}
\hspace{0.03cm}\bigr\rangle\bigr)
\,+\,
\frac{1}{2}\,\bigl(f^{\hspace{0.03cm}e\,a_{1}\hspace{0.03cm}a_{2}\hspace{0.03cm}}
\Lambda^{e}\hspace{0.03cm}
\bigl\langle\hspace{0.03cm}\mathcal{Q}^{\,a_{2}}_{2}
\hspace{0.03cm}\bigr\rangle\bigr)
\bigl(f^{\hspace{0.03cm}e^{\prime}\,a_{1}\hspace{0.03cm}a^{\prime}_{2}\hspace{0.03cm}}
\Lambda^{e^{\prime}}
\bigl\langle\hspace{0.03cm}\mathcal{Q}^{\,a^{\prime}_{2}}_{2}
\hspace{0.03cm}\bigr\rangle\bigr)\,+
\notag\\[1ex]
&\hspace{1.95cm}3\hspace{0.03cm}
\bigl(f^{\hspace{0.03cm}e\,a_{1}\hspace{0.03cm}a_{2}\hspace{0.03cm}}
\Lambda^{e}\hspace{0.03cm}
\bigl\langle\hspace{0.03cm}\mathcal{Q}^{\,a_{1}}_{1}
\hspace{0.03cm}\bigr\rangle\bigr)
\bigl(f^{\hspace{0.03cm}e^{\prime}\,e_{2}\hspace{0.03cm}a_{2}\hspace{0.03cm}}
\Lambda^{e^{\prime}}
\bigl\langle\hspace{0.03cm}\mathcal{Q}^{\,e_{2}}_{2}
\hspace{0.03cm}\bigr\rangle\bigr).
\notag
\end{align}
We obtained a system of three nonlinear (quadratic) ordinary differential equations of the first order 
describing the evolution of certain colorless structures for the $SU(3_{c})$ group.  As mentioned above, 
this system is not a closed one, since two new colorless structures appeared in the last equation for 
which it is necessary to define their own equations. But more likely, these equations will contain even 
more complex colorless structures. Perhaps this hierarchy of structures would never end and it is necessary 
to simply truncate them at some step to obtain a closed system.      

%
%

\section{\bf A system of the first order linear equations}
\label{section_11}
\setcounter{equation}{0}

To provide some information regarding the behavior of a solution for the system of kinetic equations (\ref{eq:7x}), (\ref{eq:7c}), (\ref{eq:8u}) and (\ref{eq:8i}), we consider the model problem of the evolution of the scalar plasmon number densities $N^{(1)}_{\mathbf k},\ N^{(2)}_{\mathbf k},\,W^{(1)}_{\mathbf k}$ and $W^{(2)}_{\mathbf k}$ in the case when we neglect a change over time of two averaged color charges $\bigl\langle\hspace{0.03cm}\mathcal{Q}^{\hspace{0.03cm}a_{1}}_{1}(\tau)
\hspace{0.03cm}\bigr\rangle$ and
$\bigl\langle\hspace{0.03cm}\mathcal{Q}^{\hspace{0.03cm}a_{2}}_{2}(\tau)
\hspace{0.03cm}\bigr\rangle$.\\
\indent For this purpose, we rewrite the system of kinetic equations (\ref{eq:7x}), (\ref{eq:7c}), (\ref{eq:8u}) and (\ref{eq:8i}) in matrix form. At first we introduce the following notations: 
\begin{align}
A({\mathbf k}) 
&\equiv 
\frac{(2\hspace{0.02cm}\pi)^{3}}{2\hspace{0.03cm}|\Delta{\mathbf v}|}
\sum_{\rho\hspace{0.03cm}=\hspace{0.03cm}1,\hspace{0.03cm}2}
\int\!d\hspace{0.02cm}{\bf q}\;
\bigl|\hspace{0.03cm}{T}^{\hspace{0.03cm}(\rho)}_{\; {\bf k},\,{\bf q}}\bigr|^{\hspace{0.02cm}2}\,
2\hspace{0.02cm}\pi\hspace{0.02cm}\delta(\omega^{l}_{{\bf k}} - {\bf k}\cdot{\bf v}_{\rho}
-(-1)^{\rho}\, \Delta{\mathbf v}\cdot{\mathbf q}), 
\label{eq:11q}\\[1ex]
B({\mathbf k}) 
&\equiv
\frac{(2\hspace{0.02cm}\pi)^{3}}{2\hspace{0.03cm}|\Delta{\mathbf v}|}
\sum_{\rho\hspace{0.03cm}=\hspace{0.03cm}1,\hspace{0.03cm}2}\!
(-1)^{\rho + 1}\!\!
\int\!d\hspace{0.02cm}{\bf q}\;
\bigl|\hspace{0.03cm}{T}^{\hspace{0.03cm}(\rho)}_{\; {\bf k},\,{\bf q}}\bigr|^{\hspace{0.02cm}2}\,
2\hspace{0.02cm}\pi\hspace{0.02cm}\delta(\omega^{l}_{{\bf k}} - {\bf k}\cdot{\bf v}_{\rho}
-(-1)^{\rho}\, \Delta{\mathbf v}\cdot{\mathbf q}), 
\notag\\[1ex]
\kappa_{1}(\tau) 
&\equiv
\frac{3}{2}\,{\mathfrak q}^{\hspace{0.03cm}2}_{12} \,+\, \frac{3}{4}\,{\mathfrak q}_{1}\hspace{0.03cm}{\mathfrak q}_{2}
+ \frac{9}{4}\,{\mathfrak q}_{1}\hspace{0.03cm}{\mathfrak q}_{12} 
\,-\,\Lambda^{2},
\quad\;
\kappa_{2}(\tau)\equiv\frac{3}{2}\,{\mathfrak q}^{\hspace{0.03cm}2}_{12} \,+\, \frac{3}{4}\,{\mathfrak q}_{1}\hspace{0.03cm}{\mathfrak q}_{2}
+ \frac{9}{4}\,{\mathfrak q}_{2}\hspace{0.03cm}{\mathfrak q}_{12} 
\,-\,\Lambda^{2},
\notag
\end{align}
then the system (\ref{eq:7x}), (\ref{eq:7c}), (\ref{eq:8u}) and (\ref{eq:8i}) is rewritten as 
\[
\frac{\partial}{\partial\tau}\!
\left(\!
\begin{array}{c}
N^{(1)}_{\mathbf k}\\[1.8ex]
N^{(2)}_{\mathbf k}\\[1.8ex]
W^{(1)}_{\mathbf k}\\[1.8ex]
W^{(2)}_{\mathbf k}
\end{array}
\!\!\right)	
=\!
\left(\!
\begin{array}{cccc}
0 & \displaystyle\frac{1}{d_{A}}\,\kappa_{1}(\tau)\hspace{0.03cm}A({\mathbf k}) & 0 & \displaystyle\frac{1}{d_{A}}\,\kappa_{2}(\tau)\hspace{0.03cm}B({\mathbf k})	\\[1ex]
\displaystyle\frac{1}{N_{c}}\,\frac{\kappa_{1}(\tau)}{{\mathfrak q}_{1}(\tau)}\,A({\mathbf k}) & 0 & \displaystyle\frac{1}{N_{c}}\,\frac{\kappa_{1}(\tau)}{{\mathfrak q}_{1}(\tau)}\,B({\mathbf k})	& 0 
\\[1ex] 
0 & \displaystyle\frac{1}{d_{A}}\,\kappa_{1}(\tau)\hspace{0.03cm}B({\mathbf k}) & 0 &
\displaystyle\frac{1}{d_{A}}\,\kappa_{2}(\tau)\hspace{0.03cm}A({\mathbf k}) \\[1ex]
\displaystyle\frac{1}{N_{c}}\,\frac{\kappa_{2}(\tau)}{{\mathfrak q}_{2}(\tau)}\,B({\mathbf k}) & 0 & \displaystyle\frac{1}{N_{c}}\,\frac{\kappa_{2}(\tau)}{{\mathfrak q}_{2}(\tau)}\,A({\mathbf k}) & 0
\end{array}
\!\right)\!\!
\left(\!
\begin{array}{c}
	N^{(1)}_{\mathbf k}\\[1.8ex]
	N^{(2)}_{\mathbf k}\\[1.8ex]
	W^{(1)}_{\mathbf k}\\[1.8ex]
	W^{(2)}_{\mathbf k}
\end{array}
\!\!\right)				
\]
\begin{equation}
-\,\frac{1}{2}
\left(
\begin{array}{cccc}
	0 & 0 & 0 & 0	\\[1ex]
	0 & \displaystyle\frac{\partial\ln{\mathfrak q}_{1}(\tau)}{\partial\tau} & 0 & 0 \\[1ex] 
	0 & 0 & 0 & 0 \\[1ex]
	0 & 0 & 0 & \displaystyle\frac{\partial\ln{\mathfrak q}_{2}(\tau)}{\partial\tau}
\end{array}
\right)\!\!
\left(
\begin{array}{c}
	N^{(1)}_{\mathbf k}\\[1.8ex]
	N^{(2)}_{\mathbf k}\\[1.8ex]
	W^{(1)}_{\mathbf k}\\[1.8ex]
	W^{(2)}_{\mathbf k}
\end{array}
\right).
\label{eq:11w}			
\end{equation}
This evolution equation is an extremely complex, since the combinations $\kappa_{1}(\tau),\, \kappa_{2}(\tau)$, ${\mathfrak q}_{1}(\tau)$ and ${\mathfrak q}_{2}(\tau)$ themselves obey nonlinear differential equations that follow from the system (\ref{eq:10d}) and the relation (\ref{eq:10e}). In order to simplify the problem as much as possible, we further assume that the  averaged charges are fixed, i.e. we set
\begin{equation}
\begin{split}
&\kappa_{1}(\tau) = \kappa_{1}(\tau_{0}) \equiv \kappa^{0}_{1},
\quad
\kappa_{2}(\tau) = \kappa_{2}(\tau_{0}) \equiv \kappa^{0}_{2},
\\[1ex]
&{\mathfrak q}_{1}(\tau) = {\mathfrak q}_{1}(\tau_{0}) \equiv {\mathfrak q}^{0}_{1},
\quad\;
{\mathfrak q}_{2}(\tau) = {\mathfrak q}_{2}(\tau_{0}) \equiv {\mathfrak q}^{0}_{2}.
\end{split}
\label{eq:11ww}			
\end{equation}
In this case, the system of equations (\ref{eq:11w}) is noticeably simplified, as it becomes a system of four differential equations with constant coefficients. It is convenient to present this system in block form
\begin{equation}
\frac{\partial}{\partial\tau}\!
\left(
\begin{array}{c}
	{\mathbf N}_{\mathbf k}\\[1ex]
	{\mathbf W}_{\mathbf k}
\end{array}
\right)\!	
=
\left(
\begin{array}{cccc}
	\mathcal{M}_{11} & \mathcal{M}_{12} 	\\[1ex]
	\mathcal{M}_{21} & \mathcal{M}_{22} 
\end{array}
\right)\!\!
\left(
\begin{array}{c}
	{\mathbf N}_{\mathbf k}\\[1ex]
	{\mathbf W}_{\mathbf k}
\end{array}
\right),		
\label{eq:11e}			
\end{equation}
where 
\[
{\mathbf N}_{\mathbf k} \equiv
\left(
\begin{array}{c}
	N^{(1)}_{\mathbf k}\\[1ex]
	N^{(2)}_{\mathbf k}
\end{array}
\right),	
\quad
{\mathbf W}_{\mathbf k} \equiv
\left(
\begin{array}{c}
	W^{(1)}_{\mathbf k}\\[1ex]
	W^{(2)}_{\mathbf k}
\end{array}
\right)	
\]
and
\begin{equation}
\begin{array}{llll}
&\mathcal{M}_{11} \equiv 
\left(
\begin{array}{cc}
	0 & \displaystyle\frac{1}{d_{A}}\,\kappa^{0}_{1}\hspace{0.03cm}A({\mathbf k}) \\[1ex]
	\displaystyle\frac{1}{N_{c}}\,\frac{\kappa^{0}_{1}}{{\mathfrak q}^{0}_{1}}\,A({\mathbf k}) & 0 
\end{array}
\right),
&\mathcal{M}_{12} \equiv
\left(
\begin{array}{cc}
	0 & \displaystyle\frac{1}{d_{A}}\,\kappa^{0}_{2}\hspace{0.03cm}B({\mathbf k})	\\[1ex]
	\displaystyle\frac{1}{N_{c}}\,\frac{\kappa^{0}_{1}}{{\mathfrak q}^{0}_{1}}\,B({\mathbf k})& 0 
\end{array}
\right),\\[6ex]
&\mathcal{M}_{21} \equiv
\left(
\begin{array}{cc}
	0 & \displaystyle\frac{1}{d_{A}}\,\kappa^{0}_{1}\hspace{0.03cm}B({\mathbf k}) 
	 \\[1ex]
	\displaystyle\frac{1}{N_{c}}\,\frac{\kappa^{0}_{2}}{{\mathfrak q}^{0}_{2}}\,B({\mathbf k}) & 0 
\end{array}
\right),
&\mathcal{M}_{22} \equiv
\left(
\begin{array}{cc}
	 0 &
	\displaystyle\frac{1}{d_{A}}\,\kappa^{0}_{2}\hspace{0.03cm}A({\mathbf k}) \\[1ex]
	\displaystyle\frac{1}{N_{c}}\,\frac{\kappa^{0}_{2}}{{\mathfrak q}^{0}_{2}}\,A({\mathbf k}) & 0
\end{array}
\right).
\end{array}
\label{eq:11r}
\end{equation}
To construct the general solution of the homogeneous system (\ref{eq:11e}), it is necessary first to determine the eigenvalues $\lambda_{i},\,i = 1,\ldots  4,$ by solving the characteristic equation 
\begin{equation}
\det
\left(
\begin{array}{cccc}
	\mathcal{M}_{11} - E\hspace{0.01cm}\lambda & \mathcal{M}_{12} 	\\[1ex]
	\mathcal{M}_{21} & \mathcal{M}_{22} - E\hspace{0.01cm}\lambda
\end{array}
\right) = 0,
\label{eq:11t}
\end{equation}
where $E$ is the identity $2\times 2$ matrix. For the sake of definiteness, we consider that the determinant of the matrix $\mathcal{M}_{11} - E\hspace{0.02cm}\lambda$ is nonzero, that is 
\begin{equation}
\det(\mathcal{M}_{11} - E\hspace{0.01cm}\lambda) = \lambda^{2} 
\,-\, 
\displaystyle\frac{1}{d_{A}N_{c}}\,\frac{(\kappa^{0}_{1})^{\hspace{0.02cm}2}}{{\mathfrak q}^{0}_{1}}\,A^{2}({\mathbf k})
\equiv
\label{eq:11yy}
\end{equation}
\[
\biggl[\lambda \,+\, \displaystyle\frac{1}{(d_{A}N_{c})^{1/2}}\,\biggl(\frac{(\kappa^{0}_{1})^{\hspace{0.02cm}2}}{{\mathfrak q}^{0}_{1}}\biggr)^{\!1/2}\!\!A({\mathbf k})\biggr]
\biggl[\lambda \,-\, \displaystyle\frac{1}{(d_{A}N_{c})^{1/2}}\,\biggl(\frac{(\kappa^{0}_{1})^{\hspace{0.02cm}2}}{{\mathfrak q}^{0}_{1}}\biggr)^{\!1/2}\!\!A({\mathbf k})\biggr]
\,\neq\, 0.
\] 
This allow us to evaluate the expression (\ref{eq:11t}) using Schur's formula for the determinant of a block matrix \cite{Schur:1917}. In our case, we have
\begin{equation}
\det
\left(
\begin{array}{cccc}
	\mathcal{M}_{11} - E\hspace{0.01cm}\lambda & \mathcal{M}_{12} 	\\[1ex]
	\mathcal{M}_{21} & \mathcal{M}_{22} - E\hspace{0.01cm}\lambda
\end{array}
\right) =
\label{eq:11y}
\end{equation}
\[
\det(\mathcal{M}_{11} - E\hspace{0.01cm}\lambda)\,
\det\Bigl[(\mathcal{M}_{22} - E\hspace{0.01cm}\lambda)
-
\mathcal{M}_{21}\hspace{0.01cm}
(\mathcal{M}_{11} - E\hspace{0.01cm}\lambda)^{-1}\mathcal{M}_{12}\Bigr]
= 0.
\]
Here, the inverse matrix $(\mathcal{M}_{11} - E\hspace{0.03cm}\lambda)^{-1}$ has the form 
\begin{equation}
(\mathcal{M}_{11} - E\hspace{0.01cm}\lambda)^{-1} = 
\frac{1}{\det(\mathcal{M}_{11} - E\hspace{0.03cm}\lambda)}\,
\left(
\begin{array}{cc}
	-\lambda & -\hspace{0.03cm}\displaystyle\frac{1}{d_{A}}\,\kappa^{0}_{1}\hspace{0.03cm}A({\mathbf k}) 	\\[1ex]
	-\hspace{0.03cm}\displaystyle\frac{1}{N_{c}}\,\frac{\kappa^{0}_{1}}{{\mathfrak q}^{0}_{1}}\,A({\mathbf k}) & -\lambda 
\end{array}
\right).
\label{eq:11u}
\end{equation}
By using the explicit form of the matrices $\mathcal{M}_{21}$ and $\mathcal{M}_{12}$, Eq.\,(\ref{eq:11r}), and of the inverse matrix (\ref{eq:11u}), we are able to determine the product of three matrices on the right-hand side of (\ref{eq:11y}) 
\[
\mathcal{M}_{21}
(\mathcal{M}_{11} - E\hspace{0.03cm}\lambda)^{-1}\mathcal{M}_{12}
=
\]
\[
\frac{1}{\det(\mathcal{M}_{11} - E\hspace{0.01cm}\lambda)}\,
\displaystyle\frac{1}{d_{A}N_{c}}\,B^{2}({\mathbf k})
\left(
\begin{array}{cc}
	-\lambda\,\displaystyle\frac{(\kappa^{0}_{1})^{\hspace{0.02cm}2}}{{\mathfrak q}^{0}_{1}} & -\hspace{0.02cm}\displaystyle\frac{1}{d_{A}}\,\displaystyle\frac{\kappa^{0}_{2}\hspace{0.03cm}(\kappa^{0}_{1})^{\hspace{0.02cm}2}}{{\mathfrak q}^{0}_{1}}A({\mathbf k}) \\[2ex]
	-\hspace{0.02cm}\displaystyle\frac{1}{N_{c}}\,\displaystyle\frac{\kappa^{0}_{2}\hspace{0.03cm}(\kappa^{0}_{1})^{\hspace{0.02cm}2}}{{\mathfrak q}^{0}_{1}{\mathfrak q}^{0}_{2 }}A({\mathbf k}) & -\lambda\,\displaystyle\frac{(\kappa^{0}_{2})^{\hspace{0.02cm}2}}{{\mathfrak q}^{0}_{2}} 
\end{array}
\right).
\]
Let us substitute the last expression into equation (\ref{eq:11y}). After simple algebraic transformations, the characteristic equation (\ref{eq:11y}) can be reduced to the following expression:
\begin{equation}
\det(\mathcal{M}_{11} - E\hspace{0.03cm}\lambda)
\det\Bigl[(\mathcal{M}_{22} - E\hspace{0.01cm}\lambda)
-
\mathcal{M}_{21}\hspace{0.01cm}
(\mathcal{M}_{11} - E\hspace{0.01cm}\lambda)^{-1}\mathcal{M}_{12}\Bigr]
= 
\label{eq:11uu}
\end{equation}
\[
\frac{1}{\det(\mathcal{M}_{11} - E\hspace{0.03cm}\lambda)}\,
\biggl[\lambda^{2}\biggl(\det(\mathcal{M}_{11} - E\hspace{0.01cm}\lambda)
-  
\displaystyle\frac{1}{d_{A}N_{c}}\,\frac{(\kappa^{0}_{1})^{\hspace{0.02cm}2}}{{\mathfrak q}^{0}_{1}}\,B^{2}({\mathbf k})\biggr)
\biggl(\det(\mathcal{M}_{11} - E\hspace{0.01cm}\lambda)
-  
\displaystyle\frac{1}{d_{A}N_{c}}\,\frac{(\kappa^{0}_{2})^{\hspace{0.02cm}2}}{{\mathfrak q}^{0}_{2}}\,B^{2}({\mathbf k})\biggr)
-
\]
\[
\displaystyle\frac{1}{d_{A}N_{c}}\,\frac{(\kappa^{0}_{2})^{\hspace{0.02cm}2}}{{\mathfrak q}^{0}_{2}}\,A^{2}({\mathbf k})
\biggl(\det(\mathcal{M}_{11} - E\hspace{0.01cm}\lambda)
+  
\displaystyle\frac{1}{d_{A}N_{c}}\,\frac{(\kappa^{0}_{1})^{\hspace{0.02cm}2}}{{\mathfrak q}^{0}_{1}}\,B^{2}({\mathbf k})\biggr)^{\!\!2}\,\biggr]
= 0.
\]
Since, in general, determining the roots here is quite difficult, we simplify the task by an additional special assumption that
\[
\frac{(\kappa^{0}_{2})^{\hspace{0.02cm}2}}{{\mathfrak q}^{0}_{2}}
\,=\,
\frac{(\kappa^{0}_{1})^{\hspace{0.02cm}2}}{{\mathfrak q}^{0}_{1}}.
\]
Then the relation (\ref{eq:11uu}) can be factored into a product of two polynomials:
\begin{align}
\biggl(\det(\mathcal{M}_{11} - E\hspace{0.01cm}\lambda)\,
&\biggl[\lambda \,-\, \displaystyle\frac{1}{(d_{A}N_{c})^{1/2}}\,
\biggl(\frac{(\kappa^{0}_{1})^{\hspace{0.02cm}2}}{{\mathfrak q}^{0}_{1}}\biggr)^{\!1/2}\!
A({\mathbf k})\biggr]
\,-\notag\\[1.5ex]
\displaystyle\frac{1}{d_{A}N_{c}}\,\frac{(\kappa^{0}_{1})^{\hspace{0.02cm}2}}{{\mathfrak q}^{0}_{1}}\,B^{2}({\mathbf k})
&\biggl[\lambda \,+\, \displaystyle\frac{1}{(d_{A}N_{c})^{1/2}}\,\biggl(\frac{(\kappa^{0}_{1})^{\hspace{0.02cm}2}}{{\mathfrak q}^{0}_{1}}\biggr)^{\!1/2}\!\!
A({\mathbf k})\biggr]
\biggr)\times
\notag\\[1.5ex]
\biggl(\det(\mathcal{M}_{11} - E\hspace{0.01cm}\lambda)\,
&\biggl[\lambda \,+\, \displaystyle\frac{1}{(d_{A}N_{c})^{1/2}}\,\biggl(\frac{(\kappa^{0}_{1})^{\hspace{0.02cm}2}}{{\mathfrak q}^{0}_{1}}\biggr)^{\!1/2}\!\!A({\mathbf k})\biggr]
\,-\notag\\[1.5ex]
\displaystyle\frac{1}{d_{A}N_{c}}\,\frac{(\kappa^{0}_{1})^{\hspace{0.02cm}2}}{{\mathfrak q}^{0}_{1}}\,B^{2}({\mathbf k})
&\biggl[\lambda \,-\, \displaystyle\frac{1}{(d_{A}N_{c})^{1/2}}\,\biggl(\frac{(\kappa^{0}_{1})^{\hspace{0.02cm}2}}{{\mathfrak q}^{0}_{1}}\biggr)^{\!1/2}\!\!A({\mathbf k})\biggr]
\biggr) = 0.
\notag
\end{align}  
Based on the representation (\ref{eq:11yy}) for the determinant $\det(\mathcal{M}_{11} - E\hspace{0.01cm}\lambda)$, from the preceding expression it is not difficult to find the desired eigenvalues $\lambda_{i}$:
\begin{equation}
\begin{split}
&\lambda_{1}	
=
\displaystyle\frac{1}{(d_{A}N_{c})^{1/2}}\,\biggl(\frac{(\kappa^{0}_{1})^{\hspace{0.02cm}2}}{{\mathfrak q}^{0}_{1}}\biggr)^{\!1/2}\!
\bigl[A({\mathbf k}) + B({\mathbf k})\bigr],
\\[1ex]
&\lambda_{2}	
=
\displaystyle\frac{1}{(d_{A}N_{c})^{1/2}}\,\biggl(\frac{(\kappa^{0}_{1})^{\hspace{0.02cm}2}}{{\mathfrak q}^{0}_{1}}\biggr)^{\!1/2}\!
\bigl[A({\mathbf k}) - B({\mathbf k})\bigr],
\\[1ex]
&\lambda_{3}	
=
-\hspace{0.02cm}\displaystyle\frac{1}{(d_{A}N_{c})^{1/2}}\,\biggl(\frac{(\kappa^{0}_{1})^{\hspace{0.02cm}2}}{{\mathfrak q}^{0}_{1}}\biggr)^{\!1/2}\!
\bigl[A({\mathbf k}) - B({\mathbf k})\bigr],
\\[1ex]
&\lambda_{4}	
=
-\hspace{0.02cm}\displaystyle\frac{1}{(d_{A}N_{c})^{1/2}}\,\biggl(\frac{(\kappa^{0}_{1})^{\hspace{0.02cm}2}}{{\mathfrak q}^{0}_{1}}\biggr)^{\!1/2}\!
\bigl[A({\mathbf k}) + B({\mathbf k})\bigr].
\end{split}
\label{eq:11i}
\end{equation}
Obviously, the following relations hold 
\begin{equation}
\lambda_{3} = - \lambda_{2}, \quad \lambda_{4} = - \lambda_{1}.
\label{eq:11o}
\end{equation}
Further, by virtue of the definition of the functions $A({\mathbf k})$ and $B({\mathbf k})$, Eq.\,(\ref{eq:11q}), we have
\begin{align}
&A({\mathbf k}) + B({\mathbf k}) 
=
\frac{(2\hspace{0.02cm}\pi)^{3}}{|\Delta {\mathbf v}|}\,\!
\int\!d\hspace{0.02cm}{\bf q}\;
\bigl|\hspace{0.03cm}{T}^{\hspace{0.03cm}(1)}_{\; {\bf k},\,{\bf q}}\bigr|^{\hspace{0.02cm}2}\,
2\hspace{0.02cm}\pi\hspace{0.02cm}\delta(\omega^{l}_{{\bf k}} - {\bf k}\cdot{\bf v}_{1} + \Delta{\mathbf v}\cdot{\mathbf q}),
\notag\\[1ex]
&A({\mathbf k}) - B({\mathbf k}) 
=
\frac{(2\hspace{0.02cm}\pi)^{3}}{|\Delta {\mathbf v}|}\,\!
\int\!d\hspace{0.02cm}{\bf q}\;
\bigl|\hspace{0.03cm}{T}^{\hspace{0.03cm}(2)}_{\; {\bf k},\,{\bf q}}\bigr|^{\hspace{0.02cm}2}\,
2\hspace{0.02cm}\pi\hspace{0.02cm}\delta(\omega^{l}_{{\bf k}} - {\bf k}\cdot{\bf v}_{2} - \Delta{\mathbf v}\cdot{\mathbf q}),
\notag
\end{align}
and therefore
\[
\lambda_{1} > 0, \quad \lambda_{2} > 0.
\]
\indent Let us construct the eigenvectors corresponding to the eigenvalues (\ref{eq:11i}). For this purpose, we solve the homogeneous algebraic system 
\begin{equation}
\left(
\begin{array}{cccc}
	\mathcal{M}_{11} - E\hspace{0.01cm}\lambda & \mathcal{M}_{12}\\[1ex]
\mathcal{M}_{21} & \mathcal{M}_{22} - E\hspace{0.01cm}\lambda
\end{array}
\right)
\left(
\begin{array}{c}
	{\mathbf X}\\[1ex]
	{\mathbf Y}
\end{array}
\right) = 0,
\label{eq:11p}
\end{equation}
where ${\mathbf X}$ and ${\mathbf Y}$ are two-component vectors. Let us examine this system in more detail
\begin{align}
(&\mathcal{M}_{11} - E\hspace{0.01cm}\lambda)\hspace{0.03cm}{\mathbf X}
+
\mathcal{M}_{12}\hspace{0.03cm}{\mathbf Y} = 0, 
\notag\\[1ex]
&\mathcal{M}_{21}\hspace{0.03cm}{\mathbf X}
+
(\mathcal{M}_{22} - E\hspace{0.01cm}\lambda)\hspace{0.03cm}{\mathbf Y} = 0.
\notag
\end{align}
From the first equation we define 
\begin{equation}
{\mathbf X} = -\hspace{0.03cm}(\mathcal{M}_{11} - E\hspace{0.01cm}\lambda)^{-1}\mathcal{M}_{12}\hspace{0.03cm}{\mathbf Y}
\label{eq:11a}
\end{equation}
and substitute it into the second equation and as a result,
\[
\Bigl((\mathcal{M}_{22} \,-\,E\hspace{0.01cm}\lambda)
-
\mathcal{M}_{21}\,(\mathcal{M}_{11} - E\hspace{0.01cm}\lambda)^{-1}\mathcal{M}_{12}\Bigr)\hspace{0.03cm}{\mathbf Y} 
= 0.
\]   
Taking into account the specific expressions for the eigenvalues $\lambda = \lambda_{i},\,i=1,\ldots,4$, as were defined by Eq.\,(\ref{eq:11i}), it is not difficult to obtain the explicit form for ${\mathbf Y} = {\mathbf Y}^{(\lambda_{i})}$:
\[
{\mathbf Y}^{(\lambda_{1})}
=
{\mathbf Y}^{(\lambda_{2})}
=
\left(\!
\begin{array}{c}
	1\\[1ex]
	\biggl(\displaystyle\frac{d_{A}}{N_{c}}\biggr)^{\!1/2}\!\!
	\frac{1}{{(\mathfrak q}^{0}_{1})^{1/2}}
\end{array}
\!\right)
Y_{1},
\qquad
{\mathbf Y}^{(\lambda_{3})}
=
{\mathbf Y}^{(\lambda_{4})}
=
\left(\!
\begin{array}{c}
	1\\[1ex]
	-\hspace{0.02cm}\biggl(\displaystyle\frac{d_{A}}{N_{c}}\biggr)^{\!1/2}\!\!
	\frac{1}{{(\mathfrak q}^{0}_{1})^{1/2}}
\end{array}
\!\right)
Y_{1},
\]
where $Y_{1}$ is an arbitrary parameter. Further substituting the obtained expressions ${\mathbf Y}^{(\lambda_{i})}$ into the right-hand side of equation
(\ref{eq:11a}), we find the remaining two-component vectors ${\mathbf X}^{(\lambda_{i})}$: 
\[
{\mathbf X}^{(\lambda_{1})} = {\mathbf Y}^{(\lambda_{1})}, 
\quad
{\mathbf X}^{(\lambda_{2})} = - {\mathbf Y}^{(\lambda_{2})},
\quad
{\mathbf X}^{(\lambda_{3})} = - {\mathbf Y}^{(\lambda_{3})},
\quad
{\mathbf X}^{(\lambda_{4})} = {\mathbf Y}^{(\lambda_{4})}.
\]
The two-component vectors ${\mathbf X}^{(\lambda_{i})}$ and ${\mathbf Y}^{(\lambda_{i})}$ enable us to obtain a set of four linear independent eigenvectors ${\mathbf \Upsilon}^{(\lambda_{i})} = ({\mathbf X}^{(\lambda_{i})}\;{\mathbf Y}^{(\lambda_{i})})^{T}$, describing the structure of solutions to the matrix equation (\ref{eq:11p}). The superscript $T$ denotes the transpose. Setting $Y_{1} = 1$, here we have 
\begin{equation}
\begin{split}	
&{\mathbf \Upsilon}^{(\lambda_{1})}
=
\left(\!
\begin{array}{c}
	1\\[1ex]
	\biggl(\displaystyle\frac{d_{A}}{N_{c}}\biggr)^{\!1/2}\!\!
	\frac{1}{{(\mathfrak q}^{0}_{1})^{1/2}}\\[2ex]
	1\\[1ex]
	\biggl(\displaystyle\frac{d_{A}}{N_{c}}\biggr)^{\!1/2}\!\!
	\frac{1}{{(\mathfrak q}^{0}_{1})^{1/2}}
\end{array}
\!\right),
\qquad\;\;\,
{\mathbf \Upsilon}^{(\lambda_{2})}
=
\left(\!
\begin{array}{c}
	-\hspace{0.03cm}1\\[1ex]
	-\hspace{0.03cm}\biggl(\displaystyle\frac{d_{A}}{N_{c}}\biggr)^{\!1/2}\!\!
	\frac{1}{{(\mathfrak q}^{0}_{1})^{1/2}}\\[2ex]
	1\\[1ex]
	\biggl(\displaystyle\frac{d_{A}}{N_{c}}\biggr)^{\!1/2}\!\!
	\frac{1}{{(\mathfrak q}^{0}_{1})^{1/2}}
\end{array}
\!\right),\\[1ex]
&{\mathbf \Upsilon}^{(\lambda_{3})}
=
\left(\!
\begin{array}{c}
	-\hspace{0.03cm}1\\[1ex]
	\biggl(\displaystyle\frac{d_{A}}{N_{c}}\biggr)^{\!1/2}\!\!
	\frac{1}{{(\mathfrak q}^{0}_{1})^{1/2}}\\[2ex]
	1\\[1ex]
	-\hspace{0.03cm}\biggl(\displaystyle\frac{d_{A}}{N_{c}}\biggr)^{\!1/2}\!\!
	\frac{1}{{(\mathfrak q}^{0}_{1})^{1/2}}
\end{array}
\!\right),
\qquad
{\mathbf \Upsilon}^{(\lambda_{4})}
=
\left(\!
\begin{array}{c}
	1\\[1ex]
	-\hspace{0.03cm}\biggl(\displaystyle\frac{d_{A}}{N_{c}}\biggr)^{\!1/2}\!\!
	\frac{1}{{(\mathfrak q}^{0}_{1})^{1/2}}\\[2ex]
	1\\[1ex]
	-\hspace{0.03cm}\biggl(\displaystyle\frac{d_{A}}{N_{c}}\biggr)^{\!1/2}\!\!
	\frac{1}{{(\mathfrak q}^{0}_{1})^{1/2}}
\end{array}
\!\right). 
\end{split}
\label{eq:11s}
\end{equation} 
The general solution of a system of linear homogeneous ordinary differential equations (\ref{eq:11e}) can be expressed as a linear decomposition of basis solutions derived from the eigenvalues (\ref{eq:11i}) and eigenvectors: (\ref{eq:11s})
\begin{equation}
\left(\!\!
\begin{array}{c}
	{\mathbf N}_{\mathbf k}(\tau)\\[1ex]
	{\mathbf W}_{\mathbf k}(\tau)
\end{array}
\!\!\right)\!	
\,=\,
C_{1}\hspace{0.03cm}{\rm e}^{\lambda_{1}\tau}\,{\mathbf \Upsilon}^{(\lambda_{1})}
\,+\,
C_{2}\hspace{0.03cm} {\rm e}^{\lambda_{2}\tau}\,{\mathbf \Upsilon}^{(\lambda_{2})}
\,+\,
C_{3}\hspace{0.03cm} {\rm e}^{-\lambda_{2}\tau}\,{\mathbf \Upsilon}^{(\lambda_{3})}
\,+\,
C_{4}\hspace{0.03cm} {\rm e}^{-\lambda_{1}\tau}\,{\mathbf \Upsilon}^{(\lambda_{4})}.
\label{eq:11d}
\end{equation}
Here, $C_{i},\, i = 1,\ldots, 4$ is some arbitrary constants and we have used the relation (\ref{eq:11o}) between the eigenvalues $\lambda_{i}$. The constants $C_{i}$ can be expressed in terms of the initial values of the plasmon number densities
\[
{\mathbf N}_{\mathbf k}(0) =
\left(\!\!
\begin{array}{c}
	N^{(1)}_{\mathbf k}(0)\\[1ex]
	N^{(2)}_{\mathbf k}(0)
\end{array}
\!\!\right),	
\quad
{\mathbf W}_{\mathbf k}(0) =
\left(\!\!
\begin{array}{c}
	W^{(1)}_{\mathbf k}(0)\\[1ex]
	W^{(2)}_{\mathbf k}(0)
\end{array}
\!\!\right).	
\]
According to the representation (\ref{eq:11s}), we need to solve the following algebraic system with respect to the constants $C_{i}$
\begin{align}
&C_{1} \,-\, C_{2} \,-\, C_{3} \,+\, C_{4} \,= N^{(1)}_{\mathbf k}(0),
\notag\\[1ex]
&C_{1} \,-\, C_{2} \,+\, C_{3} \,-\, C_{4} \,= N^{(2)}_{\mathbf k}(0)
\biggl(\displaystyle\frac{N_{c}}{d_{A}}\biggr)^{\!1/2}\!\!
{\bigl(\mathfrak q}^{0}_{1}\hspace{0.03cm}\bigr)^{1/2},
\notag\\[1ex]
&C_{1} \,+\, C_{2} \,+\, C_{3} \,+\, C_{4} \,= W^{(1)}_{\mathbf k}(0),
\notag\\[1ex]
&C_{1} \,+\, C_{2} \,-\, C_{3} \,-\, C_{4} \,= W^{(2)}_{\mathbf k}(0)\biggl(\displaystyle\frac{N_{c}}{d_{A}}\biggr)^{\!1/2}\!\!
{\bigl(\mathfrak q}^{0}_{1}\hspace{0.03cm}\bigr)^{1/2}.
\notag
\end{align}
The solution has the following form: 
\begin{align}
&C_{1} = \frac{1}{4}\,\biggl[ N^{(1)}_{\mathbf k}(0) + N^{(2)}_{\mathbf k}(0)
\biggl(\displaystyle\frac{N_{c}}{d_{A}}\biggr)^{\!1/2}\!\!
{\bigl(\mathfrak q}^{0}_{1}\hspace{0.03cm}\bigr)^{1/2} + W^{(1)}_{\mathbf k}(0)
+ 
W^{(2)}_{\mathbf k}(0)\biggl(\displaystyle\frac{N_{c}}{d_{A}}\biggr)^{\!1/2}\!\!
{\bigl(\mathfrak q}^{0}_{1}\hspace{0.03cm}\bigr)^{1/2}\biggr],
\notag\\[1ex]
&C_{2} = -\frac{1}{4}\,\biggl[N^{(1)}_{\mathbf k}(0) + N^{(2)}_{\mathbf k}(0)
\biggl(\displaystyle\frac{N_{c}}{d_{A}}\biggr)^{\!1/2}\!\!
{\bigl(\mathfrak q}^{0}_{1}\hspace{0.03cm}\bigr)^{1/2} - W^{(1)}_{\mathbf k}(0) 
- 
W^{(2)}_{\mathbf k}(0)\biggl(\displaystyle\frac{N_{c}}{d_{A}}\biggr)^{\!1/2}\!\!
{\bigl(\mathfrak q}^{0}_{1}\hspace{0.03cm}\bigr)^{1/2}\biggr],
\notag\\[1ex]
&C_{3} = -\frac{1}{4}\,\biggl[N^{(1)}_{\mathbf k}(0) - N^{(2)}_{\mathbf k}(0)
\biggl(\displaystyle\frac{N_{c}}{d_{A}}\biggr)^{\!1/2}\!\!
{\bigl(\mathfrak q}^{0}_{1}\hspace{0.03cm}\bigr)^{1/2} - W^{(1)}_{\mathbf k}(0) 
+ 
W^{(2)}_{\mathbf k}(0)\biggl(\displaystyle\frac{N_{c}}{d_{A}}\biggr)^{\!1/2}\!\!
{\bigl(\mathfrak q}^{0}_{1}\hspace{0.03cm}\bigr)^{1/2}\biggr],
\notag\\[1ex]
&C_{4} = \frac{1}{4}\,\biggl[ N^{(1)}_{\mathbf k}(0) - N^{(2)}_{\mathbf k}(0)
\biggl(\displaystyle\frac{N_{c}}{d_{A}}\biggr)^{\!1/2}\!\!
{\bigl(\mathfrak q}^{0}_{1}\hspace{0.03cm}\bigr)^{1/2} + W^{(1)}_{\mathbf k}(0) 
- 
W^{(2)}_{\mathbf k}(0)\biggl(\displaystyle\frac{N_{c}}{d_{A}}\biggr)^{\!1/2}\!\!
{\bigl(\mathfrak q}^{0}_{1}\hspace{0.03cm}\bigr)^{1/2}\biggr].
\notag
\end{align}
Substituting these solutions into the general one (\ref{eq:11d}), using the explicit form of eigenvectors (\ref{eq:11s}) and collecting similar terms, we find the required solutions of the original system (\ref{eq:11e}) in a rather compact form. The first pair of solutions has the form
\begin{equation}
\begin{split}
&N^{(1)}_{\mathbf k}(\tau)
\,=\,
\\
&\cosh(\lambda_{+}\tau)\cosh(\lambda_{-}\tau)\hspace{0.03cm}N^{(1)}_{\mathbf k}(0)
\,+\,
\sinh(\lambda_{+}\tau)\cosh(\lambda_{-}\tau)\hspace{0.03cm}N^{(2)}_{\mathbf k}(0)
\biggl(\displaystyle\frac{N_{c}}{d_{A}}\biggr)^{\!1/2}\!\!
{\bigl(\mathfrak q}^{0}_{1}\hspace{0.03cm}\bigr)^{1/2}
\,+\\
&\sinh(\lambda_{+}\tau)\sinh(\lambda_{-}\tau)\hspace{0.03cm}W^{(1)}_{\mathbf k}(0)
\,+\,
\cosh(\lambda_{+}\tau)\sinh(\lambda_{-}\tau)\hspace{0.03cm}W^{(2)}_{\mathbf k}(0)
\biggl(\displaystyle\frac{N_{c}}{d_{A}}\biggr)^{\!1/2}\!\!
{\bigl(\mathfrak q}^{0}_{1}\hspace{0.03cm}\bigr)^{1/2},\\[5ex]
&N^{(2)}_{\mathbf k}(\tau)
\,=\,
\\
&\sinh(\lambda_{+}\tau)\cosh(\lambda_{-}\tau)\hspace{0.03cm}N^{(1)}_{\mathbf k}(0)
\biggl(\displaystyle\frac{d_{A}}{N_{c}}\biggr)^{\!1/2}\!\!
\frac{1}{{\bigl(\mathfrak q}^{0}_{1}\hspace{0.03cm}\bigr)^{1/2}}
\,+\,
\cosh(\lambda_{+}\tau)\cosh(\lambda_{-}\tau)\hspace{0.03cm}N^{(2)}_{\mathbf k}(0)\,+\\
&\cosh(\lambda_{+}\tau)\sinh(\lambda_{-}\tau)\hspace{0.03cm}W^{(1)}_{\mathbf k}(0)
\biggl(\displaystyle\frac{d_{A}}{N_{c}}\biggr)^{\!1/2}\!\!
\frac{1}{{\bigl(\mathfrak q}^{0}_{1}\hspace{0.03cm}\bigr)^{1/2}}
\,+\,
\sinh(\lambda_{+}\tau)\sinh(\lambda_{-}\tau)\hspace{0.03cm}W^{(2)}_{\mathbf k}(0)
\end{split}
\label{eq:11f}
\end{equation}
and the second one is
\begin{equation}
\begin{split}
&W^{(1)}_{\mathbf k}(\tau)
\,=\,
\\
&\sinh(\lambda_{+}\tau)\sinh(\lambda_{-}\tau)\hspace{0.03cm}N^{(1)}_{\mathbf k}(0)
\,+\,
\cosh(\lambda_{+}\tau)\sinh(\lambda_{-}\tau)\hspace{0.03cm}N^{(2)}_{\mathbf k}(0)
\biggl(\displaystyle\frac{N_{c}}{d_{A}}\biggr)^{\!1/2}\!\!
{\bigl(\mathfrak q}^{0}_{1}\hspace{0.03cm}\bigr)^{1/2}
\,+\\[1ex]
&\cosh(\lambda_{+}\tau)\cosh(\lambda_{-}\tau)\hspace{0.03cm}W^{(1)}_{\mathbf k}(0)
\,+\,
\sinh(\lambda_{+}\tau)\cosh(\lambda_{-}\tau)\hspace{0.03cm}W^{(2)}_{\mathbf k}(0)
\biggl(\displaystyle\frac{N_{c}}{d_{A}}\biggr)^{\!1/2}\!\!
{\bigl(\mathfrak q}^{0}_{1}\hspace{0.03cm}\bigr)^{1/2},
\\[5ex]
&W^{(2)}_{\mathbf k}(\tau)
\,=\,
\\
&\cosh(\lambda_{+}\tau)\sinh(\lambda_{-}\tau)\hspace{0.03cm}N^{(1)}_{\mathbf k}(0)
\biggl(\displaystyle\frac{d_{A}}{N_{c}}\biggr)^{\!1/2}\!\!
\frac{1}{{\bigl(\mathfrak q}^{0}_{1}\hspace{0.03cm}\bigr)^{1/2}}
\,+\,
\sinh(\lambda_{+}\tau)\sinh(\lambda_{-}\tau)\hspace{0.03cm}N^{(2)}_{\mathbf k}(0)
\,+
\\[1ex]
&\sinh(\lambda_{+}\tau)\cosh(\lambda_{-}\tau)\hspace{0.03cm}W^{(1)}_{\mathbf k}(0)
\biggl(\displaystyle\frac{d_{A}}{N_{c}}\biggr)^{\!1/2}\!\!
\frac{1}{{\bigl(\mathfrak q}^{0}_{1}\hspace{0.03cm}\bigr)^{1/2}}
\,+\,
\cosh(\lambda_{+}\tau)\cosh(\lambda_{-}\tau)\hspace{0.03cm}W^{(2)}_{\mathbf k}(0),
\end{split}
\label{eq:11g}
\end{equation}
where we have used the hyperbolic sine and cosine sum/difference the formulas and introduced the notations
\[
\lambda_{+} \equiv \frac{1}{2}\, (\lambda_{1} + \lambda_{2})
=
\displaystyle\frac{1}{(d_{A}N_{c})^{1/2}}\,
\biggl(\frac{(\kappa^{0}_{1})^{\hspace{0.02cm}2}}{{\mathfrak q}^{0}_{1}}\biggr)^{\!1/2}\!
A({\mathbf k}),
\]
\[
\lambda_{-} \equiv \frac{1}{2}\, (\lambda_{1} - \lambda_{2})
=
\displaystyle\frac{1}{(d_{A}N_{c})^{1/2}}\,
\biggl(\frac{(\kappa^{0}_{1})^{\hspace{0.02cm}2}}{{\mathfrak q}^{0}_{1}}\biggr)^{\!1/2}\!
B({\mathbf k}).
\]
Here, on the right-hand side we considered the definitions of eigenvalues $\lambda_{1}$ and $\lambda_{2}$, Eq.\,(\ref{eq:11i}). Taking into account the explicit form of the functions $A({\mathbf k})$ and $B({\mathbf k})$, Eq.\,(\ref{eq:11q}), we see that the sum of eigenvalues $\lambda_{+} = \lambda_{+}({\mathbf k})$ is positive for all values of the wave vector ${\mathbf k}$, while the difference $\lambda_{-} = \lambda_{-}({\mathbf k})$ is, generally speaking, indefinite function.\\
\indent As the initial values for the scalar plasmon number densities, it is most natural to choose 
\[
N^{(2)}_{\mathbf k}(0) = W^{(2)}_{\mathbf k}(0) = 0, 
\]
that is, we assume the ``color'' part to be zero at the initial moment of time, and as $N^{(1)}_{\mathbf k}(0)$ and $W^{(1)}_{\mathbf k}(0)$ we choose thermal equilibrium value in the form of the Planck distribution function 
\[
N^{(1)}_{\mathbf k}(0) = W^{(1)}_{\mathbf k}(0)
=
\frac{1}{{\rm e}^{\hspace{0.03cm}\omega^{l}_{\mathbf k}/k_{B}T^{\ast}} - 1}, 
\]
where $T^{\ast}$ is a certain constant, which can be interpreted as a plasmon gas temperature in the statistical equilibrium state. From the solutions obtained (\ref{eq:11f}) and (\ref{eq:11g}), we see that even if there are no ``color'' plasma excitations at the initial moment, they are inevitably generated during bremsstrahlung in scattering of hard color-charged particles on each other.\\
\indent The exact solutions (\ref{eq:11f}) and (\ref{eq:11g}), rather simple in structure, are obtained in the approximation when certain combinations of the averaged color charges in the original system of equations are fixed (\ref{eq:11ww}). Consideration of the evolution of these combinations can drastically changes  the solutions found. For this, it is necessary to solve a system of nonlinear (quadratic) ordinary differential equations (\ref{eq:10d})
with constant coefficients and then return to the originally time $\tau$ through the relations (\ref{eq:10ss}) and (\ref{eq:10sss}). The scalar plasmon number densities $N^{(1)}_{\mathbf k}$ and $W^{(1)}_{\mathbf k}$ enter into the solution in an integrated manner. In order to obtain an analytical solution, here it will also be necessary to make simplifying assumptions. 

%
%

\section{Conclusion}
\label{section_12}
\setcounter{equation}{0}

\begin{figure}[hbtp]
\begin{center}
\begin{tabular*}{0.9\textwidth}{@{}ccccc@{}}
 \hspace{-1cm}	
	\raisebox{-0.35\height}{\resizebox{0.25\textwidth}{!}
		{\includegraphics{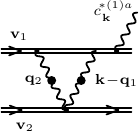}}}
	&$\!{\mathbf +}${ }{ }&
	\raisebox{-0.37\height}{\resizebox{0.35\textwidth}{!}
		{\includegraphics{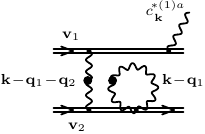}}}
	&$\!{\mathbf +}${ }{ }&
\raisebox{-0.37\height}{\resizebox{0.26\textwidth}{!}
	{\includegraphics{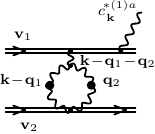}}}
\end{tabular*}
\end{center}
\caption{\small Some soft one-loop corrections to the tree-level plasmon bremsstrahlung process depicted in Fig.\,\ref{fig2-crop}.}
\label{fig5}
\end{figure}

In this study, within the framework of the classical Hamiltonian formalism 
the general theory of calculation of the effective amplitude ${T}^{\hspace{0.03cm}(\rho)\hspace{0.03cm}a\,a_{1}\hspace{0.03cm}a_{2}}_{\; {\bf k},\,{\bf q}}(t)$ describing the plasmon bremsstrahlung in collision of a high-energy color-charged particle with thermal partons in a hot quark-gluon plasma, is developed. Our approach is actually based on three main assumptions. The first of them is associated with the representation of original normal field variable $a^{\hspace{0.02cm}a}_{\hspace{0.02cm}{\bf k}}$ in a ``split'' form, i.e. as a sum of two independent components $a^{\hspace{0.02cm}(1)\hspace{0.03cm}a}_{\hspace{0.02cm}{\bf k}}$ and 
$a^{\hspace{0.02cm}(2)\hspace{0.03cm}a}_{\hspace{0.02cm}{\bf k}}$. Intuitively, it can be understood as the sum of two contributions to the radiation field  from the emission of bremsstrahlung radiation of two independent hard particles. Other main more nontrivial assumption concerns the structure of the free Hamiltonian $H^{(0)}$, Eq.\,(\ref{eq:2p}). This Hamiltonian is not defined in terms of the original field variable $a^{\hspace{0.02cm}a}_{\hspace{0.02cm}{\bf k}}$, as is usually done (see, for example \cite{Markov:2020, Markov:2024}), but in terms of its independent components  $a^{\hspace{0.02cm}(1)\hspace{0.03cm}a}_{\hspace{0.02cm}{\bf k}}$ and 
$a^{\hspace{0.02cm}(2)\hspace{0.03cm}a}_{\hspace{0.02cm}{\bf k}}$. Finally, the third and last assumption is related to two decompositions of the general expression for the matrix plasmon number density ${\mathcal N}^{\,(\alpha,\alpha^{\prime})\hspace{0.03cm}a\hspace{0.03cm}a^{\prime}_{\phantom{1}}\!}_{\hspace{0.02cm}{\bf k}}\!(\tau)$, defined by the definition (\ref{eq:6w}). The decompositions were defined in effective two-particle and color spaces, and represented by the expressions (\ref{eq:6r}) and (\ref{eq:7q}), respectively. The first two of assumptions are fairly straightforward and natural, and the third is not. They turned out to be quite sufficient to build the required classical Hamiltonian formalism for the radiation process in question.\\
\indent In the present work we restricted ourselves to the detailed discussion of only the simplest process of nonlinear interaction of soft purely collective boson excitations in the quark-gluon plasma: the tree-level order plasmon bremsstrahlung by a hard color-charged particle, i.e.
\begin{equation}
	{\rm G}_{1} + {\rm G}_{2} \rightarrow
	{\rm G}^{\hspace{0.02cm}\prime}_{1} + {\rm G}^{\hspace{0.02cm}\prime}_{2} + {\rm g}^{\ast},
	\label{eq:12q}
\end{equation}
where ${\rm g}^{\ast}$ is a plasmon collective excitation and ${\rm G}_{1,2},\, {\rm G}^{\prime}_{1,2}$ are excitations with the characteristic momenta of order of the temperature $T$ and above. For the weakly-excited system corresponding to the level of thermal fluctuations, the given radiation process is dominant. The approach developed allows us to consider more complicated radiation processes, for example, bremsstrahlung of two plasmons 
\[
{\rm G}_{1} + {\rm G}_{2} \rightarrow
{\rm G}^{\hspace{0.02cm}\prime}_{1} + {\rm G}^{\hspace{0.02cm}\prime}_{2} 
+ {\rm g}^{\ast}_{1} + {\rm g}^{\ast}_{2}.
\]
The corresponding effective Hamiltonian of this radiation process has the following form  
\begin{align}
{\mathcal H}^{(6)} 
=
&\sum_{\rho_{1},\hspace{0.03cm}\rho_{2}}
\int\!d\hspace{0.02cm}{\bf k}_{1}\hspace{0.03cm}d\hspace{0.02cm}{\bf k}_{2}\,
{T}^{\hspace{0.03cm}(\rho_{1}\rho_{2})\hspace{0.03cm}a_{1}\,a_{2}\hspace{0.03cm}a_{3}\hspace{0.03cm}a_{4}}_{\; {\bf k}_{1},\,{\bf k}_{2}}\,
c^{\hspace{0.02cm}(\rho_{1})\hspace{0.02cm}a_{1}}_{\hspace{0.02cm}{\bf k}_{1}}\hspace{0.02cm}
c^{\hspace{0.02cm}(\rho_{2})\hspace{0.02cm}a_{2}}_{\hspace{0.02cm}{\bf k}_{2}}\hspace{0.02cm}
{\mathcal Q}^{\hspace{0.03cm}a_{3}}_{\hspace{0.03cm}1}\hspace{0.02cm}
{\mathcal Q}^{\hspace{0.03cm}a_{4}}_{\hspace{0.03cm}2}
\,+
\notag\\[1ex]
&\sum_{\rho_{1},\hspace{0.03cm}\rho_{2}}
\int\!d\hspace{0.02cm}{\bf k}_{1}\hspace{0.03cm}d\hspace{0.02cm}{\bf k}_{2}\,
{T}^{\hspace{0.03cm}\ast\hspace{0.03cm}(\rho_{1}\rho_{2})\hspace{0.03cm}a_{1}\,a_{2}\hspace{0.03cm}a_{3}\hspace{0.03cm}a_{4}}_{\; {\bf k}_{1},\,{\bf k}_{2}}\,
c^{\ast\ \!\!(\rho_{1})\hspace{0.02cm}a_{1}}_{\hspace{0.02cm}{\bf k}_{1}}\hspace{0.02cm}
c^{\ast\ \!\!(\rho_{2})\hspace{0.02cm}a_{2}}_{\hspace{0.02cm}{\bf k}_{2}}\hspace{0.02cm}
{\mathcal Q}^{\hspace{0.03cm}a_{3}}_{\hspace{0.03cm}1}\hspace{0.02cm}
{\mathcal Q}^{\hspace{0.03cm}a_{4}}_{\hspace{0.03cm}2}.
\notag
\end{align}
The practical computation of the effective amplitude ${T}^{\hspace{0.03cm}(\rho_{1}\rho_{2})\hspace{0.03cm}a_{1}\,a_{2}\hspace{0.03cm}a_{3}\hspace{0.03cm}a_{4}}_{\; {\bf k}_{1},\,{\bf k}_{2}}$ and the corresponding kinetic equations becomes considerably more cumbersome and, as a consequence, ineffective. Other approaches that are not directly related to the Hamiltonian formalism are more appropriate here (see for example \cite{Markov:2005qe}).\\
\indent However, even for the simplest bremsstrahlung radiation process (\ref{eq:12q}) the story does not end here. In fact, the most general form of the effective Hamiltonian describing one-plasmon bremsstrahlung in scattering of an high-energy color-charged particle by a hard thermal color-charged one, instead of (\ref{eq:4y}), is 
\[
{\mathcal H}
=
	\sum_{\rho}
	\int\!d\hspace{0.02cm}{\bf k}\,
	{T}^{\hspace{0.03cm}(\rho)\hspace{0.03cm}a}_{\; {\bf k}}(\mathcal{Q}_{1}, \mathcal{Q}_{2})\,
	c^{\hspace{0.02cm}(\rho)\hspace{0.02cm}a}_{\hspace{0.02cm}{\bf k}}
	\,+\,
	\sum_{\rho}
	\int\!d\hspace{0.02cm}{\bf k}\,
	{T}^{\hspace{0.03cm}\ast\hspace{0.03cm}(\rho)\hspace{0.03cm}a}_{\; {\bf k}}(\mathcal{Q}_{1}, \mathcal{Q}_{2})\,
	c^{\ast\ \!\!(\rho)\hspace{0.02cm}a}_{\hspace{0.02cm}{\bf k}},
\]
where the effective amplitude ${T}^{\hspace{0.03cm}(\rho)\hspace{0.03cm}a}_{\; {\bf k}}(\mathcal{Q}_{1}, \mathcal{Q}_{2})$ is an arbitrary function of the color vectors $\mathcal{Q}_{1} = (\mathcal{Q}^{\,a_{1}}_{1})$ and $\mathcal{Q}_{2} = (\mathcal{Q}^{\,a_{2}}_{2})$. We can represent this amplitude by a series expansion in the color charges $\mathcal{Q}^{\,a_{1}}_{1}$ and $\mathcal{Q}^{\,a_{2}}_{2}$ of two energetic particles
\begin{equation}
{T}^{\hspace{0.03cm}(\rho)\hspace{0.03cm}a}_{\; {\bf k}}(\mathcal{Q}_{1}, \mathcal{Q}_{2})
=
{T}^{\hspace{0.03cm}(\rho)\hspace{0.03cm}a\,a_{1}\hspace{0.03cm}a_{2}}_{\; {\bf k}}\,{\mathcal Q}^{\hspace{0.03cm}a_{1}}_{\hspace{0.03cm}1}\hspace{0.02cm}{\mathcal Q}^{\hspace{0.03cm}a_{2}}_{\hspace{0.03cm}2}
\,+
\label{eq:12w}
\end{equation}
\[
\frac{1}{2}\,\Bigl(
{T}^{\hspace{0.03cm}(\rho)\hspace{0.03cm}a\,a^{\phantom{\prime}}_{1}
\hspace{0.03cm}a^{\phantom{\prime}}_{2}
\hspace{0.03cm}a^{\prime}_{2}}_{\; {\bf k}}\,
{\mathcal Q}^{\hspace{0.03cm}a^{\phantom{\prime}}_{1}}_{\hspace{0.03cm}1}
\hspace{0.02cm}{\mathcal Q}^{\hspace{0.03cm}a^{\phantom{\prime}}_{2}}_{\hspace{0.03cm}2}
\hspace{0.02cm}{\mathcal Q}^{\hspace{0.03cm}a^{\prime}_{2}}_{\hspace{0.03cm}2}
+
{T}^{\hspace{0.03cm}(\rho)\hspace{0.03cm}a\,a^{\phantom{\prime}}_{1}\hspace{0.03cm}a^{\prime}_{1}\hspace{0.03cm}a^{\phantom{\prime}}_{2}}_{\; {\bf k}}\,
{\mathcal Q}^{\hspace{0.03cm}a^{\phantom{\prime}}_{1}}_{\hspace{0.03cm}1}
\hspace{0.02cm}{\mathcal Q}^{\hspace{0.03cm}a^{\prime}_{1}}_{\hspace{0.03cm}1}
\hspace{0.02cm}{\mathcal Q}^{\hspace{0.03cm}a^{\phantom{\prime}}_{2}}_{\hspace{0.03cm}2}
\Bigr)
+\hspace{0.03cm} \ldots\,.
\]
Thus, our expression for the effective amplitude (\ref{eq:4_1u})\,--\,(\ref{eq:4uu}) represents only the first term of this expansion. It is implicitly assumed that the remaining higher-order terms of the expansion are suppressed by the coupling constant. However, what is their physical meaning or interpretation? At least the next contribution in the expansion (\ref{eq:12w}) can be calculated explicitly within the framework of our approach. An example of calculating the next-to-leading effective amplitude can be found in \cite{Markov:2005qe}. The diagrammatic interpretation of certain terms in ${T}^{\hspace{0.03cm}(\rho)\hspace{0.03cm}a\,a^{\phantom{\prime}}_{1}
\hspace{0.03cm}a^{\phantom{\prime}}_{2}	\hspace{0.03cm}a^{\prime}_{2}}_{\; {\bf k}}$ is presented in Fig.\,\ref{fig5}.
From the view of these diagrams, we can conclude that what is being discussed here is soft ``one-loop'' corrections to the tree-level plasmon bremsstrahlung radiation. These one-loop corrections are suppressed by a power of $g^{2}$ compared with the tree approximation (\ref{eq:4_1u})\,--\,(\ref{eq:4uu}). The next-to-leading order effective amplitude ${T}^{\hspace{0.03cm}(\rho)\hspace{0.03cm}a\,a^{\phantom{\prime}}_{1}
	\hspace{0.03cm}a^{\phantom{\prime}}_{2}	\hspace{0.03cm}a^{\prime}_{2}}_{\; {\bf k}}$ in (\ref{eq:12w}) must have the following structure, instead of (\ref{eq:4_1u}):
\[
{T}^{\hspace{0.03cm}(\rho)\hspace{0.03cm}a\,a^{\phantom{\prime}}_{1}
	\hspace{0.03cm}a^{\phantom{\prime}}_{2}	\hspace{0.03cm}a^{\prime}_{2}}_{\; {\bf k}}
=
\int\!d\hspace{0.02cm}{\bf q}_{1}\hspace{0.03cm}d\hspace{0.02cm}{\bf q}_{2}\,
{T}^{\hspace{0.03cm}(\rho)\hspace{0.03cm}a\,a^{\phantom{\prime}}_{1}
\hspace{0.03cm}a^{\phantom{\prime}}_{2}	\hspace{0.03cm}a^{\prime}_{2}}_{\; {\bf k},\,{\bf q}_{1},\,{\bf q}_{2}},
\]
i.e., here we have two momentum transfer ${\bf q}_{1}$ and ${\bf q}_{2}$. We assume that the higher-order terms in the expansion (\ref{eq:12w}) are a kind of fluctuation deviation from the bremsstrahlung radiation in the tree-level approximation. In particular, the first diagram in Fig.\,\ref{fig5} represents the one-loop correction to the propagator of   hard particle with the label 1, defined by the expression (\ref{eq:4aa}).\\    
%
%
\indent A generalization of the results obtained in the present study to the fermionic sector of hard and soft excitations in the quark-gluon plasma is of significant theoretical and practical interest. It is clear that to describe the color degrees of freedom of both the hard color-charged particles with half-integer spin and the soft normal fermionic field variables within the framework of classical Hamiltonian formalism, we need to use functions taking values in Grassmann algebra. In constructing a complete Hamiltonian wave theory of QGP including bosonic and fermionic, as well as hard and soft degrees of freedom it will be necessary to define a generalized (nonlinear) system of dynamical equations of the Wong type describing the evolution of both the ordinary (commutative) classical color charge $\mathcal{Q}^{\hspace{0.02cm}a}$ of a hard test particle with integer spin and the Grassmann-valued color charges ${\uptheta}^{\,\ast\ \!\!i}$ and ${\uptheta}^{\phantom{\ast}\!\!i},\,i = 1,\ldots,N_{c}$, of other hard test particle with half-integer spin in external stochastic bosonic and fermionic fields. The Grassmann color charges belong to the {\it fundamental} representation of the $SU(N_{c})$ color group. Here, it will also be necessary to generalize the corresponding canonical transformations, which include simultaneously the boson and fermion degrees of freedom of the collective excitations of the quark-gluon plasma, and the degrees of freedom associated with the commutative and anticommutative color charges of hard particles.\\ 
\indent It should be noted, however, that the consideration of radiative processes that change the statistics of soft and hard modes seems to be quite complicated and involved. Here, we need to create an appropriate apparatus that adequately addresses this problem (see, for example, \cite{Markov:2006wp, Markov:2008wz}). It can be assumed that the structure of the effective Hamiltonian describing bremsstrahlung process of purely collective soft fermion excitations (plasminos) will be noticeably more complicated than the effective Hamiltonian for soft boson excitations obtained in the present work, Eq.\,(\ref{eq:4y}). Such an effective Hamiltonian should most likely have the following form:
\begin{equation}
\begin{split}
{\mathcal H}^{(5)}
=
&\sum_{\rho}
\int\!d\hspace{0.02cm}{\bf p}\,
{T}^{\hspace{0.03cm}(\rho)\hspace{0.03cm}i\,i_{1}\hspace{0.03cm}
a_{2}}_{\; {\bf p}}\,
b^{\hspace{0.02cm}(\rho)\hspace{0.02cm}i}_{\hspace{0.02cm}{\bf p}}\hspace{0.02cm}
{\uptheta}^{\,\ast\ \!\!i_{1}}_{\hspace{0.03cm}1}
\hspace{0.02cm}
{\mathcal Q}^{\hspace{0.03cm}a_{2}}_{\hspace{0.03cm}2}
\,+\,
\sum_{\rho}
\int\!d\hspace{0.02cm}{\bf p}\,
{T}^{\hspace{0.03cm}\ast\hspace{0.03cm}(\rho)\hspace{0.03cm}i\,i_{1}
\hspace{0.03cm}a_{2}}_{\; {\bf p}}\,
b^{\ast\ \!\!(\rho)\hspace{0.02cm}i}_{\hspace{0.02cm}{\bf p}}
\hspace{0.02cm}
{\uptheta}^{\phantom{\ast}\!\!i_{1}}_{\hspace{0.03cm}1}
\hspace{0.02cm}
{\mathcal Q}^{\hspace{0.03cm}a_{2}}_{\hspace{0.03cm}2}
\,+\\[1ex]
&\sum_{\rho}
\int\!d\hspace{0.02cm}{\bf p}\,
{T}^{\hspace{0.03cm}(\rho)\hspace{0.03cm}i\,i_{2}\hspace{0.03cm}
a_{1}}_{\; {\bf p}}\,
b^{\hspace{0.02cm}(\rho)\hspace{0.02cm}i}_{\hspace{0.02cm}{\bf p}}
\hspace{0.02cm}
{\uptheta}^{\,\ast\ \!\!i_{2}}_{\hspace{0.03cm}2}
\hspace{0.02cm}
{\mathcal Q}^{\hspace{0.03cm}a_{1}}_{\hspace{0.03cm}1}
\,+\,
\sum_{\rho}
\int\!d\hspace{0.02cm}{\bf p}\,
{T}^{\hspace{0.03cm}\ast\hspace{0.03cm}(\rho)\hspace{0.03cm}i\,i_{2}
\hspace{0.03cm}a_{1}}_{\; {\bf p}}\,
b^{\ast\ \!\!(\rho)\hspace{0.02cm}i}_{\hspace{0.02cm}{\bf p}}
\hspace{0.02cm}
{\uptheta}^{\phantom{\ast}\!\!i_{2}}_{\hspace{0.03cm}2}
\hspace{0.02cm}
{\mathcal Q}^{\hspace{0.03cm}a_{1}}_{\hspace{0.03cm}1}.
\end{split} 
\label{eq:12e}
\end{equation}
Here, the Grassmann-valued functions $b^{\hspace{0.02cm}(\rho)\hspace{0.02cm}i}_{\hspace{0.03cm}{\bf p}},\,\rho = 1,2,$ are the independent components of the total normal fermion field variable $b^{\hspace{0.02cm}i}_{\hspace{0.02cm}{\bf p}}$, i.e.
\[
b^{\hspace{0.02cm}i}_{\hspace{0.02cm}{\bf p}} =
b^{\hspace{0.02cm}(1)\hspace{0.02cm}i}_{\hspace{0.02cm}{\bf p}}
+
b^{\hspace{0.02cm}(2)\hspace{0.02cm}i}_{\hspace{0.02cm}{\bf p}},
\]
which in turn is amplitude for the plasmino mode excitations as it follows from the decomposition of the collective quark-antiquark field into plane waves 
\[
{\Psi}^{i}_{\alpha}(x) 
\!=\! 
\sum_{\;\lambda\hspace{0.02cm} =\hspace{0.02cm} \pm\hspace{0.02cm} 1}\hspace{0.01cm} 
\int\!\!\frac{d\hspace{0.02cm}{\bf p}}{(2\pi)^{3}}\left(\!\frac{Z_{-}({\bf p})}
{2}\right)^{\!\!1/2}\!\left[\hspace{0.03cm}b_{\bf p}^{\,i}(\lambda)
\hspace{0.03cm}u^{(-)}_{\alpha}(\hat{\bf p}, \lambda)\,
e^{-i\hspace{0.03cm}\omega^{-}_{{\bf p}}\hspace{0.02cm}t\hspace{0.03cm} +
\hspace{0.03cm} i\hspace{0.03cm}{\bf p}\hspace{0.02cm}\cdot\hspace{0.02cm} 
{\bf x}}
\!+
d_{\bf p}^{\,\ast\hspace{0.03cm}i}(\lambda)\hspace{0.03cm}v^{(-)}_{\alpha}
(\hat{\bf p}, \lambda)\,
e^{\hspace{0.03cm}i\hspace{0.03cm}\omega^{-}_{{\bf p}}
\hspace{0.02cm}t\hspace{0.03cm} -\hspace{0.03cm} i\hspace{0.03cm}
{\bf p}\hspace{0.02cm}\cdot\hspace{0.02cm}{\bf x}}\hspace{0.03cm}\right]\!,
\]
where $\alpha = 1, 2, 3, 4$, the spinors $u^{(s)}_{\alpha}(\hat{\bf p}, \lambda)$ and $v^{(s)}_{\alpha}(\hat{\bf p}, \lambda)$ denote solutions of the free massless Dirac equation and $\omega^{-}_{{\bf p}}$ is the dispersion relation of plasmino mode. The Hamiltonian (\ref{eq:12e}) describes radiative process with the change of statistics that occurs when two color-charged particles with integer and half-integer spins collide with each other.\\
\indent One more remark should be made regarding one-plasmon bremsstrahlung.
To the effective Hamiltonian (\ref{eq:4y}), we need to add another Hamiltonian that is associated with the hard Fermi excitations of the quark-gluon plasma, namely  
\[
{\mathcal H}^{(5)}
=
\sum_{\rho}
\int\!d\hspace{0.02cm}{\bf k}\,
{T}^{\hspace{0.03cm}(\rho)\hspace{0.03cm}a\,i_{1}\hspace{0.03cm}i_{2}}_{\; {\bf k}}\,
c^{\hspace{0.02cm}(\rho)\hspace{0.02cm}a}_{\hspace{0.02cm}{\bf k}}\hspace{0.02cm}
\hspace{0.02cm}
{\uptheta}^{\,\ast\ \!\!i_{1}}_{\hspace{0.03cm}1}
\hspace{0.02cm}
{\uptheta}^{\phantom{\ast}\!\!i_{2}}_{\hspace{0.03cm}2}
\,+\,
\sum_{\rho}
\int\!d\hspace{0.02cm}{\bf k}\,
{T}^{\hspace{0.03cm}\ast\hspace{0.03cm}(\rho)\hspace{0.03cm}a\,i_{1}
\hspace{0.03cm}i_{2}}_{\; {\bf k}}\,
c^{\ast\ \!\!(\rho)\hspace{0.02cm}a}_{\hspace{0.02cm}{\bf k}}
\hspace{0.02cm}
{\uptheta}^{\,\ast\ \!\!i_{2}}_{\hspace{0.03cm}2}
\hspace{0.02cm}
{\uptheta}^{\phantom{\ast}\!\!i_{1}}_{\hspace{0.03cm}1}.
\]
It will allow to take into account all relevant contributions to the single plasmon bremsstrahlung process.\\
\indent As an application of the effective Hamilton theory of plasmon (and plasmino) bremsstrahlung produced by a hard color-charged particle propagating in the hot quark-gluon plasma, we intend to consider the application to the problem of energy losses of the fast particle. This task will be the subject of our next work and be carried out within the framework of the basic ideas of the paper \cite{Markov:2025}, i.e. with the help of the construction of corresponding classical scattering matrix \cite{Zakharov:1982, Shulman:1985, Zakharov:1988}. However, the direct transference of ideas in \cite{Markov:2025} meets with certain difficulties because the approach developed in the present study requires some ``adjustment and adaptation'' to the multiple-time-scale perturbation method, which we have used extensively throughout the paper \cite{Markov:2025}.


\section*{\bf Acknowledgment}

The research was funded by the Ministry of Education and Science of the Russian Federation within the framework of the project ``Development of analytical  and numerical methods of description in problems of mathematical physics, continuum mechanics, quantum field theory and nuclear physics'' (no. of state registration: 126021217175-3).

\begin{appendices}
\numberwithin{equation}{section}


\numberwithin{equation}{section}
\section{Effective three-gluon vertex} 
\numberwithin{equation}{section}
\label{appendix_A}

In this Appendix, we have provided the explicit form of the vertex functions and gluon propagator in the hard thermal loop (HTL) approximation \cite{Blaizot:2002, Ghiglieri:2020, Braaten:1990}.\\
\indent Effective three-gluon vertex
\begin{equation}
\,^{\ast} \Gamma^{\hspace{0.03cm}\mu\hspace{0.02cm} \nu  \rho}(k, k_{1}, k_{2}) \equiv
\Gamma^{\hspace{0.03cm}\mu\hspace{0.02cm} \nu  \rho}(k, k_{1}, k_{2}) +
\delta\hspace{0.025cm} \Gamma^{\hspace{0.03cm}\mu\hspace{0.02cm} \nu  \rho}(k, k_{1}, k_{2})
\label{ap:A1}
\end{equation}
is the sum of bare three-gluon vertex
\begin{equation}
	\Gamma^{\hspace{0.03cm}\mu\hspace{0.02cm}\nu\hspace{0.02cm}\rho}(k, k_{1}, k_{2}) =
	g^{\hspace{0.03cm}\mu\hspace{0.02cm}\nu} (k - k_{1})^{\rho} + g^{\hspace{0.03cm}\nu\hspace{0.02cm}\rho} (k_{1} - k_{2})^{\mu} +
	g^{\hspace{0.03cm}\mu\hspace{0.02cm}\rho} (k_{2} - k)^{\nu}
	\label{ap:A2}
\end{equation}
and the corresponding HTL correction
\begin{equation}
	\delta\hspace{0.025cm} \Gamma^{\hspace{0.03cm}\mu\hspace{0.02cm} \nu  \rho}(k, k_{1}, k_{2}) =
	3\hspace{0.035cm}\omega^{\hspace{0.02cm} 2}_{\rm pl}\!\int\!\frac{d\hspace{0.035cm}\Omega}{4 \pi} \,
	\frac{v^{\hspace{0.03cm}\mu}\hspace{0.02cm} v^{\hspace{0.03cm}\nu} v^{\hspace{0.03cm}\rho}}{v\cdot k + i\hspace{0.025cm}\epsilon} \,
	\Biggl(\frac{\omega_{2}}{v\cdot k_{2} - i\epsilon} -
	\frac{\omega_1}{v\cdot k_{1} - i\epsilon}\Biggr),
	\quad \epsilon\rightarrow +\hspace{0.02cm}0,
	\label{ap:A3}
\end{equation}
where $v^{\hspace{0.03cm}\mu} = (1,{\bf {\bf v}})$, $k^{\hspace{0.03cm}\mu} = (\omega, {\bf k})$ is a gluon four-momentum with $k  + k_{1} + k_{2} = 0$, $d\hspace{0.035cm}\Omega$ is a differential solid angle and $\omega_{\rm pl}^2 = g^2(2N_c + N_f)T^2/18$ is the quark-gluon plasma frequency squared. We present below useful properties of the three-gluon HTL-resummed vertex function for complex conjugation and permutation of momenta:
\[
	\left(\!\,^{\ast}\Gamma_{\mu\hspace{0.02cm} \mu_{1} \mu_{2}}(-k_{1} - k_{2}, k_{1}, k_{2})\right)^{\ast} =
	-\!\,^{\ast}\Gamma_{\mu\hspace{0.02cm} \mu_{1} \mu_{2}}(k_{1} + k_{2}, -k_{1}, -k_{2}) 
	= \!\,^{\ast}\Gamma_{\mu\hspace{0.02cm} \mu_{2}\mu_{1}}(k_{1} + k_{2}, -k_{2}, -k_{1}).
\]


\section{Kinetic equation for the function ${\mathcal W}_{\hspace{0.02cm}{\bf k}}(\tau)$}
\numberwithin{equation}{section}
\label{appendix_B}

Here, we present the kinetic equation for the second color matrix function ${\mathcal W}_{\hspace{0.02cm}{\bf k}}(\tau) = \bigl({\mathcal W}^{\hspace{0.03cm}a\hspace{0.03cm}a^{\prime}_{\phantom{1}}\!}_{\hspace{0.02cm}{\bf k}}(\tau)\bigr)$. The kinetic equation follows from the general one (\ref{eq:6e}) by substituting into it the representation (\ref{eq:6t}). Then we must contract the left- and right-hand sides of Eq.\, (\ref{eq:6e}) with the Pauli matrix $({\sigma}_{3})^{\alpha^{\prime}\hspace{0.01cm}\alpha}$ considering that ${\rm tr}(\sigma_{3})^{2} = 2$. As a result we have
\begin{equation}
	\frac{\partial\hspace{0.04cm}{\mathcal W}^{\hspace{0.03cm}a\hspace{0.03cm}a^{\prime}_{\phantom{1}}\!}_{\hspace{0.02cm}{\bf k}}\!(\tau)}{\partial\hspace{0.03cm}\tau}
	=
	\label{ap:B1}
\end{equation}
\begin{align}
	-\,&\frac{1}{2}\,
	\sum_{\rho}\,{T}^{\hspace{0.03cm}(\rho)}_{\; {\bf k}}(t')
	\,\biggl(\int{T}^{\,\ast\hspace{0.03cm}(\rho)}_{\; {\bf k}}(t')\hspace{0.03cm}dt'\biggr)
	\Bigl\{
	\bigl(\hspace{0.03cm}T^{\,a_{2}}\hspace{0.03cm}T^{\,e_{1}}\hspace{0.03cm}T^{\,a^{\prime}_{2}}{\mathcal W}_{\hspace{0.02cm}{\bf k}}\hspace{0.01cm}
	\bigr)^{\hspace{0.01cm}a\hspace{0.03cm}a^{\prime}_{\phantom{1}}\!}\!
	\hspace{0.01cm}\bigl\langle\hspace{0.03cm}\mathcal{Q}^{\hspace{0.03cm}a^{\prime}_{2}}_{2}
	\hspace{0.03cm}\bigr\rangle
	\hspace{0.03cm}\bigl\langle\hspace{0.03cm}\mathcal{Q}^{\hspace{0.03cm}e_{1}}_{1}
	\hspace{0.03cm}\bigr\rangle
	\hspace{0.01cm}\bigl\langle\hspace{0.03cm}\mathcal{Q}^{\hspace{0.03cm}a_{2}}_{2}
	\hspace{0.03cm}\bigr\rangle\,+
	\notag\\[1ex]
	&\hspace{8cm}
	\bigl(\hspace{0.03cm}T^{\,a_{1}}\hspace{0.03cm}T^{\,e_{2}}\hspace{0.03cm}T^{\,a^{\prime}_{1}}{\mathcal W}_{\hspace{0.02cm}{\bf k}}\hspace{0.01cm}
	\bigr)^{\hspace{0.01cm}a\hspace{0.03cm}a^{\prime}_{\phantom{1}}\!}\!
	\hspace{0.01cm}\bigl\langle\hspace{0.03cm}\mathcal{Q}^{\hspace{0.03cm}a^{\prime}_{1}}_{1}
	\hspace{0.03cm}\bigr\rangle
	\hspace{0.03cm}\bigl\langle\hspace{0.03cm}\mathcal{Q}^{\hspace{0.03cm}e_{2}}_{2}
	\hspace{0.03cm}\bigr\rangle
	\hspace{0.01cm}\bigl\langle\hspace{0.03cm}\mathcal{Q}^{\hspace{0.03cm}a_{1}}_{1}
	\hspace{0.03cm}\bigr\rangle	
	\Bigr\}\,-
	\notag\\[1ex]
	&\frac{1}{2}\,\sum_{\rho}\,{T}^{\,\ast\hspace{0.03cm}({\rho})}_{\; {\bf k}}(t')
	\,
	\biggl(\int{T}^{\hspace{0.03cm}(\rho)}_{\; {\bf k}}(t')\hspace{0.03cm}dt'\biggr)
	\Bigl\{
	\bigl(\hspace{0.03cm}{\mathcal W}_{\hspace{0.02cm}{\bf k}}
	\hspace{0.03cm}
	T^{\,a^{\prime}_{2}}\hspace{0.03cm}T^{\,e_{1}}\hspace{0.03cm}
	T^{\,a_{2}}\hspace{0.01cm}
	\bigr)^{\hspace{0.01cm}a\hspace{0.03cm}a^{\prime}_{\phantom{1}}\!}
	\hspace{0.01cm}
	\bigl\langle\hspace{0.03cm}\mathcal{Q}^{\hspace{0.03cm}a^{\prime}_{2}}_{2}
	\hspace{0.03cm}\bigr\rangle
	\hspace{0.03cm}\bigl\langle\hspace{0.03cm}\mathcal{Q}^{\hspace{0.03cm}e_{1}}_{1}
	\hspace{0.03cm}\bigr\rangle
	\hspace{0.01cm}\bigl\langle\hspace{0.03cm}\mathcal{Q}^{\hspace{0.03cm}a_{2}}_{2}
	\hspace{0.03cm}\bigr\rangle\,+
	\notag\\[1ex]
	&\hspace{8cm}
	\bigl(\hspace{0.03cm}{\mathcal W}_{\hspace{0.02cm}{\bf k}}
	\hspace{0.03cm}
	T^{\,a^{\prime}_{1}}\hspace{0.03cm}T^{\,e_{2}}\hspace{0.03cm}
	T^{\,a_{1}}\hspace{0.01cm}
	\bigr)^{\hspace{0.01cm}a\hspace{0.03cm}a^{\prime}_{\phantom{1}}\!}
	\hspace{0.01cm}
	\bigl\langle\hspace{0.03cm}\mathcal{Q}^{\hspace{0.03cm}a^{\prime}_{1}}_{1}
	\hspace{0.03cm}\bigr\rangle
	\hspace{0.03cm}\bigl\langle\hspace{0.03cm}\mathcal{Q}^{\hspace{0.03cm}e_{2}}_{2}
	\hspace{0.03cm}\bigr\rangle
	\hspace{0.01cm}\bigl\langle\hspace{0.03cm}\mathcal{Q}^{\hspace{0.03cm}a_{1}}_{1}
	\hspace{0.03cm}\bigr\rangle
	\Bigr\}\,-
	\notag\\[1ex]
	&\frac{1}{2}\,\sum_{\rho}\hspace{0.02cm}(-1)^{\rho + 1}\,{T}^{\hspace{0.03cm}(\rho)}_{\; {\bf k}}(t')
	\,\biggl(\int{T}^{\,\ast\hspace{0.03cm}(\rho)}_{\; {\bf k}}(t')\hspace{0.03cm}dt'\biggr)
	\Bigl\{
	\bigl(\hspace{0.03cm}T^{\,a_{2}}\hspace{0.03cm}T^{\,e_{1}}\hspace{0.03cm}T^{\,a^{\prime}_{2}}{\mathcal N}_{\hspace{0.02cm}{\bf k}}\hspace{0.01cm}
	\bigr)^{\hspace{0.01cm}a\hspace{0.03cm}a^{\prime}_{\phantom{1}}\!}\!
	\hspace{0.01cm}\bigl\langle\hspace{0.03cm}\mathcal{Q}^{\hspace{0.03cm}a^{\prime}_{2}}_{2}
	\hspace{0.03cm}\bigr\rangle\hspace{0.03cm}\bigl\langle\hspace{0.03cm}
	\mathcal{Q}^{\hspace{0.03cm}e_{1}}_{1}
	\hspace{0.03cm}\bigr\rangle	\hspace{0.01cm}\bigl\langle\hspace{0.03cm}
	\mathcal{Q}^{\hspace{0.03cm}a_{2}}_{2}\hspace{0.03cm}\bigr\rangle\,
	+
	\notag\\[1ex]
	&\hspace{8cm}
	\bigl(\hspace{0.03cm}T^{\,a_{1}}\hspace{0.03cm}T^{\,e_{2}}\hspace{0.03cm}T^{\,a^{\prime}_{1}}{\mathcal N}_{\hspace{0.02cm}{\bf k}}\hspace{0.01cm}
	\bigr)^{\hspace{0.01cm}a\hspace{0.03cm}a^{\prime}_{\phantom{1}}\!}\!
	\hspace{0.01cm}\bigl\langle\hspace{0.03cm}\mathcal{Q}^{\hspace{0.03cm}a^{\prime}_{1}}_{1}
	\hspace{0.03cm}\bigr\rangle
	\hspace{0.03cm}\bigl\langle\hspace{0.03cm}\mathcal{Q}^{\hspace{0.03cm}e_{2}}_{2}
	\hspace{0.03cm}\bigr\rangle
	\hspace{0.01cm}\bigl\langle\hspace{0.03cm}\mathcal{Q}^{\hspace{0.03cm}a_{1}}_{1}
	\hspace{0.03cm}\bigr\rangle	
	\Bigr\}\,-
	\notag\\[1ex]
	&\frac{1}{2}\,\sum_{\rho}\hspace{0.02cm}(-1)^{\rho + 1}\,{T}^{\,\ast\hspace{0.03cm}({\rho})}_{\; {\bf k}}(t')
	\,
	\biggl(\int{T}^{\hspace{0.03cm}(\rho)}_{\; {\bf k}}(t')\hspace{0.03cm}dt'\biggr)
	\Bigl\{
	\bigl(\hspace{0.03cm}{\mathcal N}_{\hspace{0.02cm}{\bf k}}
	\hspace{0.03cm}
	T^{\,a^{\prime}_{2}}\hspace{0.03cm}T^{\,e_{1}}\hspace{0.03cm}
	T^{\,a_{2}}\hspace{0.01cm}
	\bigr)^{\hspace{0.01cm}a\hspace{0.03cm}a^{\prime}_{\phantom{1}}\!}
	\hspace{0.01cm}
	\bigl\langle\hspace{0.03cm}\mathcal{Q}^{\hspace{0.03cm}a^{\prime}_{2}}_{2}
	\hspace{0.03cm}\bigr\rangle\hspace{0.03cm}\bigl\langle\hspace{0.03cm}
	\mathcal{Q}^{\hspace{0.03cm}e_{1}}_{1}	\hspace{0.03cm}\bigr\rangle
	\hspace{0.01cm}\bigl\langle\hspace{0.03cm}
	\mathcal{Q}^{\hspace{0.03cm}a_{2}}_{2}\hspace{0.03cm}\bigr\rangle\,
	+
	\notag\\[1ex]
	&\hspace{8cm}
	\bigl(\hspace{0.03cm}{\mathcal N}_{\hspace{0.02cm}{\bf k}}
	\hspace{0.03cm}
	T^{\,a^{\prime}_{1}}\hspace{0.03cm}T^{\,e_{2}}\hspace{0.03cm}
	T^{\,a_{1}}\hspace{0.01cm}
	\bigr)^{\hspace{0.01cm}a\hspace{0.03cm}a^{\prime}_{\phantom{1}}\!}
	\hspace{0.01cm}
	\bigl\langle\hspace{0.03cm}\mathcal{Q}^{\hspace{0.03cm}a^{\prime}_{1}}_{1}
	\hspace{0.03cm}\bigr\rangle
	\hspace{0.03cm}\bigl\langle\hspace{0.03cm}\mathcal{Q}^{\hspace{0.03cm}e_{2}}_{2}
	\hspace{0.03cm}\bigr\rangle
	\hspace{0.01cm}\bigl\langle\hspace{0.03cm}\mathcal{Q}^{\hspace{0.03cm}a_{1}}_{1}
	\hspace{0.03cm}\bigr\rangle
	\Bigr\}.
	\notag
\end{align}
This equation closes the kinetic equation (\ref{eq:6y}) for the first color matrix function ${\mathcal N}_{\hspace{0.02cm}{\bf k}} = ({\mathcal N}^{\,a\hspace{0.03cm}a^{\prime}_{\phantom{1}}\!}_{\hspace{0.02cm}{\bf k}})$ in (\ref{eq:6t}).


\numberwithin{equation}{section}
\section{Equation\,for\,the\,mean value of color charge~${\mathcal Q}^{\hspace{0.03cm}a_{2}}_{\hspace{0.03cm}2}$}
\numberwithin{equation}{section}
\label{appendix_C}

This Appendix provides a general form of the evolution equation for the average value of the second color charge ${\mathcal Q}^{\hspace{0.03cm}a_{2}}_{\hspace{0.03cm}2}$. The derivation of this equation coincides exactly with the one of a similar equation for the average value of the first color charge ${\mathcal Q}^{\hspace{0.03cm}a_{1}}_{\hspace{0.03cm}1}$, given in section \ref{section_9}. It has the following form: 
\begin{equation}
\frac{d\hspace{0.03cm}\bigl\langle \mathcal{Q}^{\,a_{2}}_{\hspace{0.03cm}2}(\tau)\hspace{0.02cm}\bigr\rangle}{d\hspace{0.03cm}\tau}
=
\label{ap:C1}
\end{equation}
\begin{align}
	-\,\frac{1}{2}\,\sum_{\rho}\,
	\int\!d\hspace{0.02cm}{\bf k}\,
	{T}^{\hspace{0.03cm}(\rho)}_{\; {\bf k}}(t')
	\biggl(\int{T}^{\,\ast\hspace{0.03cm}(\rho)}_{\; {\bf k}}(t')\hspace{0.03cm}dt'\biggr)
	\Bigl[&{\rm tr}\hspace{0.03cm}
	\bigl(\hspace{0.03cm}T^{\,a_{1}}\hspace{0.03cm}T^{\,a_{2}}\hspace{0.03cm}T^{\,e^{\prime}_{2}}\hspace{0.03cm}T^{\,a^{\prime}_{1}}{\mathcal N}_{\hspace{0.02cm}{\bf k}}\hspace{0.01cm}
	\bigr)
	\hspace{0.01cm}
	\bigl\langle\hspace{0.03cm}\mathcal{Q}^{\hspace{0.03cm}a_{1}}_{1}(\tau)\hspace{0.03cm}\bigr\rangle\,-
	\notag\\
	i\hspace{0.03cm}f^{\hspace{0.03cm}a_{2}\hspace{0.02cm}c^{\prime}\hspace{0.03cm}e_{2}}\,
	&{\rm tr}\hspace{0.03cm}
	\bigl(\hspace{0.03cm}T^{\,c^{\prime}}\hspace{0.03cm}T^{\,a^{\prime}_{1}}\hspace{0.03cm}T^{\,e^{\prime}_{2}}{\mathcal N}_{\hspace{0.02cm}{\bf k}}\hspace{0.01cm}
	\bigr)
	\hspace{0.03cm}\bigl\langle\hspace{0.03cm}\mathcal{Q}^{\hspace{0.03cm}e_{2}}_{2}(\tau)
	\hspace{0.03cm}\bigr\rangle
	\Bigr]
	\hspace{0.01cm}
	\bigl\langle\hspace{0.03cm}\mathcal{Q}^{\hspace{0.03cm}e^{\prime}_{2}}_{2}(\tau)\hspace{0.03cm}\bigr\rangle
	\hspace{0.03cm}\bigl\langle\hspace{0.03cm}\mathcal{Q}^{\hspace{0.03cm}a^{\prime}_{1}}_{1}(\tau)
	\hspace{0.03cm}\bigr\rangle\,-
	\notag
\end{align}
\begin{align}
	\hspace{0.6cm}
	\frac{1}{2}\,\sum_{\rho}\,
	\int\!d\hspace{0.02cm}{\bf k}\,
	{T}^{\,\ast\hspace{0.03cm}(\rho)}_{\; {\bf k}}(t')
	\biggl(\int{T}^{\hspace{0.03cm}(\rho)}_{\;{\bf k}}(t')\hspace{0.03cm}dt'\biggr)
	\Bigl[&{\rm tr}\hspace{0.03cm}\bigl({\mathcal N}_{\hspace{0.02cm}{\bf k}}
	\hspace{0.03cm}T^{\,a^{\prime}_{1}}\hspace{0.03cm}T^{\,e^{\prime}_{2}}\hspace{0.03cm}T^{\,a_{2}}\hspace{0.03cm}T^{\,a^{\prime}_{1}}\hspace{0.01cm}
	\bigr)
	\hspace{0.01cm}
	\bigl\langle\hspace{0.03cm}\mathcal{Q}^{\hspace{0.03cm}a_{1}}_{1}(\tau)\hspace{0.03cm}\bigr\rangle\,+
	\notag\\
	i\hspace{0.03cm}f^{\hspace{0.03cm}a_{2}\hspace{0.02cm}c^{\prime}\hspace{0.03cm}e_{2}}\,
	&{\rm tr}\hspace{0.03cm}\bigl({\mathcal N}_{\hspace{0.02cm}{\bf k}}
	\hspace{0.03cm}T^{\,e^{\prime}_{2}}\hspace{0.03cm}T^{\,a^{\prime}_{1}}\hspace{0.03cm}T^{\,c^{\prime}}\hspace{0.01cm}
	\bigr)
	\hspace{0.03cm}\bigl\langle\hspace{0.03cm}\mathcal{Q}^{\hspace{0.03cm}e_{2}}_{2}(\tau)
	\hspace{0.03cm}\bigr\rangle
	\Bigr]
	\hspace{0.01cm}
	\bigl\langle\hspace{0.03cm}\mathcal{Q}^{\hspace{0.03cm}e^{\prime}_{2}}_{2}(\tau)\hspace{0.03cm}\bigr\rangle
	\hspace{0.03cm}\bigl\langle\hspace{0.03cm}\mathcal{Q}^{\hspace{0.03cm}a^{\prime}_{1}}_{1}(\tau)
	\hspace{0.03cm}\bigr\rangle\,-
	\notag
\end{align}
\begin{align}
	\frac{1}{2}\,\sum_{\rho}\,(-1)^{\rho +1}\!\!
	\int\!d\hspace{0.02cm}{\bf k}\,
	{T}^{\hspace{0.03cm}(\rho)}_{\; {\bf k}}(t')
	\biggl(\int{T}^{\,\ast\hspace{0.03cm}(\rho)}_{\; {\bf k}}(&t')\hspace{0.03cm}dt'\biggr)
	\Bigl[{\rm tr}\hspace{0.03cm}
	\bigl(\hspace{0.03cm}T^{\,a_{1}}\hspace{0.03cm}T^{\,a_{2}}\hspace{0.03cm}T^{\,e^{\prime}_{2}}\hspace{0.03cm}T^{\,a^{\prime}_{1}}\hspace{0.03cm}{\mathcal W}_{\hspace{0.02cm}{\bf k}}\hspace{0.01cm}
	\bigr)
	\hspace{0.01cm}
	\bigl\langle\hspace{0.03cm}\mathcal{Q}^{\hspace{0.03cm}a_{1}}_{1}(\tau)\hspace{0.03cm}\bigr\rangle\,-
	\notag\\
	i\hspace{0.03cm}f^{\hspace{0.03cm}a_{2}\hspace{0.02cm}c^{\prime}\hspace{0.03cm}e_{2}}\,
	&{\rm tr}\hspace{0.03cm}
	\bigl(\hspace{0.03cm}T^{\,c^{\prime}}\hspace{0.03cm}T^{\,a^{\prime}_{1}}\hspace{0.03cm}T^{\,e^{\prime}_{2}}\hspace{0.03cm}{\mathcal W}_{\hspace{0.02cm}{\bf k}}\hspace{0.01cm}
	\bigr)
	\hspace{0.03cm}\bigl\langle\hspace{0.03cm}\mathcal{Q}^{\hspace{0.03cm}e_{2}}_{2}(\tau)
	\hspace{0.03cm}\bigr\rangle
	\Bigr]
	\hspace{0.01cm}
	\bigl\langle\hspace{0.03cm}\mathcal{Q}^{\hspace{0.03cm}e^{\prime}_{2}}_{2}(\tau)\hspace{0.03cm}\bigr\rangle
	\hspace{0.03cm}\bigl\langle\hspace{0.03cm}\mathcal{Q}^{\hspace{0.03cm}a^{\prime}_{1}}_{1}(\tau)
	\hspace{0.03cm}\bigr\rangle\,-
	\notag
\end{align}
\begin{align}
	\frac{1}{2}\,\sum_{\rho}\,(-1)^{\rho +1}\!\!
	\int\!d\hspace{0.02cm}{\bf k}\,
	{T}^{\,\ast\hspace{0.03cm}(\rho)}_{\; {\bf k}}(t')
	\biggl(\int{T}^{\hspace{0.03cm}(\rho)}_{\;{\bf k}}(&t')\hspace{0.03cm}dt'\biggr)
	\Bigl[{\rm tr}\hspace{0.03cm}\bigl(\hspace{0.03cm}{\mathcal W}_{\hspace{0.02cm}{\bf k}}
	\hspace{0.03cm}T^{\,a^{\prime}_{1}}\hspace{0.03cm}T^{\,e^{\prime}_{2}}\hspace{0.03cm}T^{\,a_{2}}\hspace{0.03cm}T^{\,a^{\prime}_{1}}\hspace{0.01cm}
	\bigr)
	\hspace{0.01cm}
	\bigl\langle\hspace{0.03cm}\mathcal{Q}^{\hspace{0.03cm}a_{1}}_{1}(\tau)\hspace{0.03cm}\bigr\rangle\,+
	\notag\\
	i\hspace{0.03cm}f^{\hspace{0.03cm}a_{2}\hspace{0.02cm}c^{\prime}\hspace{0.03cm}e_{2}}\,
	&{\rm tr}\hspace{0.03cm}\bigl(\hspace{0.03cm}{\mathcal W}_{\hspace{0.02cm}{\bf k}}
	\hspace{0.03cm}T^{\,e^{\prime}_{2}}\hspace{0.03cm}T^{\,a_{1}}\hspace{0.03cm}T^{\,c^{\prime}}\hspace{0.01cm}
	\bigr)
	\hspace{0.03cm}\bigl\langle\hspace{0.03cm}\mathcal{Q}^{\hspace{0.03cm}e_{2}}_{2}(\tau)
	\hspace{0.03cm}\bigr\rangle
	\Bigr]
	\hspace{0.01cm}
	\bigl\langle\hspace{0.03cm}\mathcal{Q}^{\hspace{0.03cm}e^{\prime}_{2}}_{2}(\tau)\hspace{0.03cm}\bigr\rangle
	\hspace{0.03cm}\bigl\langle\hspace{0.03cm}\mathcal{Q}^{\hspace{0.03cm}a^{\prime}_{1}}_{1}(\tau)
	\hspace{0.03cm}\bigr\rangle\,-
	\notag
\end{align}
\begin{align}
	\sum_{\rho}\,\biggl[\,
	&\int\!d\hspace{0.02cm}{\bf k}\,
	{\rm e}^{-i\hspace{0.03cm}(\omega^{l}_{{\bf k}}\hspace{0.03cm} -\hspace{0.03cm}{\bf v}_{\rho}\hspace{0.02cm}\cdot\hspace{0.02cm}{\bf k})\hspace{0.02cm}t'}
	\hspace{0.03cm}
	{T}^{\hspace{0.03cm}(\rho)}_{\; {\bf k}}(t')
	\biggl(\int{\rm e}^{i\hspace{0.03cm}(\omega^{l}_{{\bf k}}\hspace{0.03cm} -\hspace{0.03cm}{\bf v}_{\rho}\hspace{0.02cm}\cdot\hspace{0.02cm}{\bf k})\hspace{0.02cm}t'}
	\hspace{0.03cm}
	{T}^{\hspace{0.03cm}\ast\hspace{0.03cm}(\rho)}_{\; {\bf k}}(t')\hspace{0.03cm}dt'\biggr)\,-
	\notag\\[1ex]
	&\int\!d\hspace{0.02cm}{\bf k}\,
	{\rm e}^{i\hspace{0.03cm}(\omega^{l}_{{\bf k}}\hspace{0.03cm} -\hspace{0.03cm}{\bf v}_{\rho}\hspace{0.02cm}\cdot\hspace{0.02cm}{\bf k})\hspace{0.02cm}t'}
	\hspace{0.03cm}
	{T}^{\,\ast\hspace{0.03cm}(\rho)}_{\; {\bf k}}(t')
	\biggl(\int{\rm e}^{-i\hspace{0.03cm}(\omega^{l}_{{\bf k}}\hspace{0.03cm} -\hspace{0.03cm}{\bf v}_{\rho}\hspace{0.02cm}\cdot\hspace{0.02cm}{\bf k})\hspace{0.02cm}t'}
	\hspace{0.03cm}
	{T}^{\hspace{0.03cm}(\rho)}_{\; {\bf k}}(t')\hspace{0.03cm}dt'\biggr)\biggr]
	\hspace{0.03cm}\times
	\notag	
\end{align}
\[
\bigl(\hspace{0.03cm}T^{\,e_{2}}\hspace{0.03cm}T^{\,a^{\phantom{\prime}}_{1}}\hspace{0.03cm}T^{\,a^{\prime}_{2}}\hspace{0.01cm}
\bigr)^{\hspace{0.01cm}a^{\phantom{\prime}}_{2}\hspace{0.03cm}a^{\prime}_{1}}\hspace{0.03cm}
\bigl\langle\hspace{0.03cm}\mathcal{Q}^{\hspace{0.03cm}a^{\prime}_{1}}_{1}(\tau)
\hspace{0.03cm}\bigr\rangle
\hspace{0.03cm}\bigl\langle\hspace{0.03cm}\mathcal{Q}^{\hspace{0.03cm}a^{\prime}_{2}}_{2}(\tau)
\hspace{0.03cm}\bigr\rangle
\hspace{0.03cm}\bigl\langle\hspace{0.03cm}\mathcal{Q}^{\hspace{0.03cm}a^{\phantom{\prime}}_{1}}_{1}(\tau)
\hspace{0.03cm}\bigr\rangle
\hspace{0.01cm}\bigl\langle\hspace{0.03cm}\mathcal{Q}^{\hspace{0.03cm}e_{2}}_{2}(\tau)
\hspace{0.03cm}\bigr\rangle.
\]


\section{Traces for generators in the adjoint repre\-sentation}
\numberwithin{equation}{section}
\label{appendix_D}

In this Appendix, we have given the traces for generators in the adjoint representation, which we use throughout the text of this paper. A comprehensive list of the various traces, relations and identities for the color matrices in the adjoint representation can be found in \cite{Kaplan:1967, Macfarlane:1968, Azcarraga:1998, Fadin:2005, Nikolaev:2005, Haber:2021}. The original definition of matrices $T^{\,a}$ is
\[
\bigl(T^{\,a}\bigr)^{\hspace{0.01cm}b\hspace{0.03cm}c} \equiv -\hspace{0.02cm}i\hspace{0.03cm}
f^{\hspace{0.03cm}a\hspace{0.02cm}b\hspace{0.03cm}c},
\]
where $f^{\hspace{0.03cm}a\hspace{0.02cm}b\hspace{0.03cm}c}$ are the totally antisymmetric structure constants for the $\mathfrak{su}(N_{c})$ Lie algebra. These matrices are traceless, i.e.
\[
{\rm tr}\,T^{\,a} = 0
\]
and satisfy the following commutation relation
\[
\bigl[\hspace{0.02cm}T^{\,a},T^{\,b}\hspace{0.02cm}\bigr] = i\hspace{0.02cm}f^{\hspace{0.03cm}a\hspace{0.02cm}b\hspace{0.03cm}c}
\,T^{\,c}.
\]
The traces of a product of two and three generators are
\begin{align}
&{\rm tr}\hspace{0.03cm}
\bigl(T^{\,a}\hspace{0.03cm}T^{\,b}\bigr) = N_{c}\hspace{0.04cm}\delta^{\hspace{0.03cm}a\hspace{0.03cm}b},
\notag\\[1ex]
&{\rm tr}\hspace{0.03cm}
\bigl(T^{\,a}\hspace{0.03cm}T^{\,b}\hspace{0.03cm}T^{\,c}\bigr) = \displaystyle\frac{i}{2}\,N_{c}\hspace{0.03cm}
f^{\hspace{0.03cm}a\hspace{0.02cm}b\hspace{0.03cm}c}.
\label{ap:D3}
\end{align}
We will also need the trace of four generators 
\begin{align}
&{\rm tr}\hspace{0.03cm}\bigl(T^{\,a}\hspace{0.02cm} T^{\,b}\hspace{0.02cm} T^{\,c}\hspace{0.02cm} 
T^{\,d}\hspace{0.03cm}\bigr)
=
\delta^{\hspace{0.02cm}a\hspace{0.02cm}d}\hspace{0.03cm}\delta^{\hspace{0.02cm}b\hspace{0.03cm}c}
+
\frac{1}{2}\,\bigl(\hspace{0.02cm}
\delta^{\hspace{0.02cm}a\hspace{0.02cm}b}\hspace{0.03cm}\delta^{\hspace{0.02cm}c\hspace{0.03cm}d}
+
\delta^{\hspace{0.02cm}a\hspace{0.02cm}c}\hspace{0.03cm}
\delta^{\hspace{0.02cm}b\hspace{0.03cm}d}\hspace{0.02cm}\bigr)
+
\frac{1}{4}\,N_{c}\hspace{0.02cm}\bigl(\hspace{0.02cm}
f^{\hspace{0.03cm}a\hspace{0.03cm}d\hspace{0.03cm}e}
\hspace{0.01cm}f^{\hspace{0.03cm}b\hspace{0.03cm}c\hspace{0.03cm}e}
\!+
d^{\hspace{0.04cm}a\hspace{0.03cm}d\hspace{0.03cm}e}
\hspace{0.02cm}d^{\hspace{0.04cm}b\hspace{0.03cm}c\hspace{0.03cm}e}
\hspace{0.02cm}\bigr),
\label{ap:D4}
\end{align}
where $d^{\hspace{0.04cm}a\hspace{0.03cm}b\hspace{0.03cm}c}$ are the totally symmetric structure constants for the $\mathfrak{su}(N_{c})$ Lie algebra. 
If one uses the relation  	
\begin{equation}
f^{\hspace{0.03cm}a\hspace{0.03cm}b\hspace{0.03cm}e}
\hspace{0.01cm}f^{\hspace{0.03cm}c\hspace{0.03cm}d\hspace{0.03cm}e}
=
\frac{2}{N_{c}}\,\bigl(\hspace{0.02cm}
\delta^{\hspace{0.02cm}a\hspace{0.02cm}c}\hspace{0.03cm}
\delta^{\hspace{0.02cm}b\hspace{0.03cm}d}
-
\delta^{\hspace{0.02cm}a\hspace{0.02cm}d}\hspace{0.03cm}
\delta^{\hspace{0.02cm}b\hspace{0.03cm}c}\hspace{0.02cm}\bigr)
+
\bigl(\hspace{0.02cm}d^{\hspace{0.03cm}a\hspace{0.03cm}
c\hspace{0.03cm}e}
\hspace{0.01cm}d^{\hspace{0.03cm}b\hspace{0.03cm}d\hspace{0.03cm}e}
\!-
d^{\hspace{0.04cm}b\hspace{0.03cm}c\hspace{0.03cm}e}
\hspace{0.03cm}d^{\hspace{0.04cm}a\hspace{0.03cm}d\hspace{0.03cm}e}
\hspace{0.02cm}\bigr),
\label{ap:D6}
\end{equation}
then the trace (\ref{ap:D4}) can also be presented in a slightly different form
\begin{equation}
{\rm tr}\hspace{0.03cm}\bigl(T^{\,a}\hspace{0.02cm} T^{\,b}\hspace{0.02cm} T^{\,c}\hspace{0.02cm} 
T^{\,d}\hspace{0.03cm}\bigr)
=
\delta^{\hspace{0.02cm}a\hspace{0.02cm}b}\hspace{0.03cm}
\delta^{\hspace{0.02cm}c\hspace{0.03cm}d}
+		\delta^{\hspace{0.02cm}a\hspace{0.02cm}d}\hspace{0.03cm}
\delta^{\hspace{0.02cm}b\hspace{0.03cm}c}
+
\frac{1}{4}\,N_{c}\hspace{0.02cm}\bigl(\hspace{0.02cm}
d^{\hspace{0.03cm}a\hspace{0.03cm}b\hspace{0.03cm}e}
\hspace{0.01cm}d^{\hspace{0.03cm}c\hspace{0.03cm}d\hspace{0.03cm}e}
\!+
d^{\hspace{0.04cm}a\hspace{0.03cm}d\hspace{0.03cm}e}
\hspace{0.03cm}d^{\hspace{0.04cm}b\hspace{0.03cm}c\hspace{0.03cm}e}
-
d^{\hspace{0.04cm}a\hspace{0.03cm}c\hspace{0.03cm}e}
\hspace{0.03cm}d^{\hspace{0.04cm}b\hspace{0.03cm}d\hspace{0.03cm}e}
\hspace{0.02cm}\bigr).
\label{ap:D7}
\end{equation}
The trace (\ref{ap:D7}) is written in such a form, which shows its symmetry under permutation of the indices $a$ and $c$ (and, correspondingly, $b$ and $d$\hspace{0.03cm}), i.e.
\begin{equation}
{\rm tr}\hspace{0.03cm}\bigl(T^{\,a}\hspace{0.02cm} T^{\,b}\hspace{0.02cm} T^{\,c}\hspace{0.02cm} 
T^{\,d}\hspace{0.03cm}\bigr)
=
{\rm tr}\hspace{0.03cm}\bigl(T^{\,c}\hspace{0.02cm} T^{\,b}\hspace{0.02cm} T^{\,a}\hspace{0.02cm}T^{\,d}\hspace{0.03cm}\bigr)
=
{\rm tr}\hspace{0.03cm}\bigl(T^{\,a}\hspace{0.02cm} T^{\,d}\hspace{0.02cm} T^{\,c}\hspace{0.02cm} 
T^{\,b}\hspace{0.03cm}\bigr).
\label{ap:D8}
\end{equation}
\indent The trace of five generators $T^{\,a}$ can be represented as a linear combination of the traces of four generators \cite{Ritbergen:1999}\hspace{0.03cm}\footnote{\hspace{0.03cm}In the cited paper in formula (45) for the trace of five generators in one of the terms on the right-hand side, two indices are incorrectly placed.}
\begin{equation}
{\rm tr}\hspace{0.03cm}\bigl(T^{\,a_{1}}\hspace{0.02cm} T^{\,a_{2}}\hspace{0.02cm} T^{\,a_{3}}\hspace{0.02cm} 
T^{\,a_{4}}\hspace{0.03cm}T^{\,a_{5}}\hspace{0.02cm}\bigr)
\label{ap:D9}
\end{equation}
\[
\begin{split}
=
-\hspace{0.03cm}\frac{i}{2}\,\Bigl\{ 
&f^{\hspace{0.03cm}a_{4}\hspace{0.02cm}a_{3}\hspace{0.02cm} b}\hspace{0.05cm}
{\rm tr}\hspace{0.03cm}\bigl(T^{\,a_{1}}\hspace{0.02cm} T^{\,a_{2}}\hspace{0.02cm} T^{\,b}\hspace{0.02cm} 
T^{\,a_{5}}\hspace{0.03cm}\bigr)
+
 f^{\hspace{0.03cm}a_{5}\hspace{0.02cm}a_{3}\hspace{0.02cm} b}\hspace{0.05cm}
{\rm tr}\hspace{0.03cm}\bigl(T^{\,a_{1}}\hspace{0.02cm} T^{\,a_{2}}\hspace{0.02cm} 
T^{\,a_{4}}\hspace{0.02cm}T^{\,b}\hspace{0.03cm}\bigr)\\[0.7ex]
+\,
&f^{\hspace{0.03cm}a_{5}\hspace{0.02cm}a_{4}\hspace{0.02cm} b}\hspace{0.05cm}
{\rm tr}\hspace{0.03cm}\bigl(T^{\,a_{1}}\hspace{0.02cm}T^{\,a_{2}}
\hspace{0.02cm} T^{\,b}\hspace{0.02cm} 
T^{\,a_{3}}\hspace{0.03cm}\bigr)
+
f^{\hspace{0.03cm}a_{2}\hspace{0.02cm}a_{1}\hspace{0.02cm} b}\hspace{0.05cm}
{\rm tr}\hspace{0.03cm}\bigl(T^{\,b}\hspace{0.02cm} T^{\,a_{5}}\hspace{0.02cm} T^{\,a_{4}}\hspace{0.02cm} 
T^{\,a_{3}}\hspace{0.03cm}\bigr)\Bigr\}.
\end{split}
\]
This expression is a consequence of the sign reversal property of the permutation of matrices $T^{\,a}$ under the trace sign in inverse order 
\[
{\rm tr}\hspace{0.03cm}\bigl(T^{\,a_{1}}\hspace{0.02cm} T^{\,a_{2}}\hspace{0.02cm} T^{\,a_{3}}\hspace{0.02cm} 
T^{\,a_{4}}\hspace{0.03cm}T^{\,a_{5}}\hspace{0.02cm}\bigr)
=
-{\rm tr}\hspace{0.03cm}\bigl(T^{\,a_{5}}\hspace{0.02cm} T^{\,a_{4}}\hspace{0.02cm} T^{\,a_{3}}\hspace{0.02cm} 
T^{\,a_{2}}\hspace{0.03cm}T^{\,a_{1}}\hspace{0.02cm}\bigr),
\]
which, in turn, is a trivial consequence of the identity 
\[
{\rm tr}\hspace{0.03cm}\bigl(T^{\,a_{1}}\hspace{0.02cm} T^{\,a_{2}}\hspace{0.02cm} T^{\,a_{3}}\hspace{0.02cm} 
T^{\,a_{4}}\hspace{0.03cm}T^{\,a_{5}}\hspace{0.02cm}\bigr)
=
-\hspace{0.03cm}2\hspace{0.03cm}
{\rm tr}\hspace{0.03cm}\bigl(t^{\,b}\hspace{0.02cm}\bigl[\hspace{0.03cm}
t^{\,a_{1}},\bigl[\hspace{0.02cm} t^{\,a_{2}},\bigl[\hspace{0.02cm} t^{\,a_{3}},\bigl[\hspace{0.02cm}t^{\,a_{4}},\bigl[\hspace{0.03cm}t^{\,a_{5}},t^{b}\bigr]\bigr]\bigr]\bigr]\bigr]\hspace{0.02cm}\bigr),
\]
where $t^{\,a}$ are the $N^{\hspace{0.03cm}2}_{c} - 1$ generators in the defining representation of the $\mathfrak{su}(N_{c})$ Lie algebra.\\ 
\indent Based on the representation (\ref{ap:D9}) and considering the symmetry property (\ref{ap:D8}), it is easy to see that the following useful relation exists for the fifth-order trace under permutations of indices $a_{3}$ and $a_{5}$:
\begin{equation}	
{\rm tr}\hspace{0.03cm}\bigl(T^{\,a_{1}}\hspace{0.02cm} T^{\,a_{2}}\hspace{0.02cm} T^{\,a_{3}}\hspace{0.02cm} 
T^{\,a_{4}}\hspace{0.03cm}T^{\,a_{5}}\hspace{0.02cm}\bigr)
+
{\rm tr}\hspace{0.03cm}\bigl(T^{\,a_{1}}\hspace{0.02cm} T^{\,a_{2}}\hspace{0.02cm} T^{\,a_{5}}\hspace{0.02cm} 
T^{\,a_{4}}\hspace{0.03cm}T^{\,a_{3}}\hspace{0.02cm}\bigr)
=
-\hspace{0.03cm}if^{\hspace{0.03cm}a_{2}\hspace{0.02cm}a_{1}
\hspace{0.01cm}b}\hspace{0.04cm}
{\rm tr}\hspace{0.03cm}\bigl(T^{\,b}\hspace{0.03cm} T^{\,a_{5}}\hspace{0.02cm} T^{\,a_{4}}\hspace{0.02cm} 
T^{\,a_{3}}\hspace{0.03cm}\bigr).
\label{ap:D10}
\end{equation}	
We use this relation in section \ref{section_9} and in Appendix \ref{appendix_F} to calculate imaginary parts that appear when determining the equations for the certain colorless combinations constructed from the averaged color charges $\bigl\langle\hspace{0.03cm}\mathcal{Q}^{\hspace{0.03cm}a_{1}}_{1}
\hspace{0.03cm}\bigr\rangle$ and $\bigl\langle\hspace{0.03cm}\mathcal{Q}^{\hspace{0.03cm}a_{2}}_{2}
\hspace{0.03cm}\bigr\rangle$ (see the next Appendix).\\ 
\indent There are also two additional identities for the special case of $N_{c} = 3$ \cite{Macfarlane:1968, Haber:2021}. The first and main of them has the following form:
\begin{equation}
d^{\,a\hspace{0.02cm}b\hspace{0.03cm}e}\hspace{0.02cm} d^{\,c\hspace{0.02cm}d\hspace{0.03cm}e}
+
d^{\,a\hspace{0.02cm}c\hspace{0.03cm}e}\hspace{0.02cm} d^{\,b\hspace{0.02cm}d\hspace{0.03cm}e}
+
d^{\,a\hspace{0.02cm}d\hspace{0.03cm}e}\hspace{0.02cm} d^{\,b\hspace{0.02cm}c\hspace{0.03cm}e}
= 
\frac{1}{3}\,\bigl(\hspace{0.02cm}
\delta^{\hspace{0.02cm}a\hspace{0.02cm}b}\delta^{\hspace{0.02cm}
c\hspace{0.02cm} d} 
+
\delta^{\hspace{0.02cm}a\hspace{0.02cm}c}\delta^{\hspace{0.02cm}
b\hspace{0.02cm}d}
+
\delta^{\hspace{0.02cm}a\hspace{0.02cm}d}\delta^{\hspace{0.02cm}
b\hspace{0.02cm} c}\hspace{0.02cm}\bigr).
\label{ap:D13}
\end{equation}
This relation can also be presented in a slightly different form 
\[
\bigl\{\hspace{0.02cm}D^{\,a},D^{\,b}\hspace{0.02cm}\bigr\}^{c\hspace{0.02cm} d}
= 
-\hspace{0.03cm}d^{\,a\hspace{0.02cm}b\hspace{0.03cm}e}\hspace{0.02cm} \bigl(D^{\,e}\bigr)^{cd}
+
\frac{1}{3}\,\bigl(
\delta^{\hspace{0.02cm}a\hspace{0.03cm}b}
\delta^{\hspace{0.02cm}c\hspace{0.03cm}d} 
+
\delta^{\hspace{0.02cm}a\hspace{0.03cm}d}
\delta^{\hspace{0.02cm}b\hspace{0.03cm}c}
+
\delta^{\hspace{0.02cm}a\hspace{0.03cm}c}
\delta^{\hspace{0.02cm}b\hspace{0.03cm}d}\bigr),
\]
where $\bigl(D^{\,a}\bigr)^{\hspace{0.01cm}b\hspace{0.03cm}c} \equiv 
d^{\hspace{0.03cm}a\hspace{0.02cm}b\hspace{0.03cm}c}$. A useful consequence of it is the identity involving antisymmetric structural constants
\begin{equation}
\bigl\{\hspace{0.02cm}T^{\,a},T^{\,b}\hspace{0.02cm}\bigr\}^{c\hspace{0.02cm} d}
= 
3\hspace{0.03cm}d^{\,a\hspace{0.02cm}b\hspace{0.03cm}e}\hspace{0.02cm} \bigl(D^{\,e}\bigr)^{cd}
+
\delta^{\hspace{0.02cm}a\hspace{0.03cm}b}
\delta^{\hspace{0.02cm}c\hspace{0.03cm}d} 
-
\delta^{\hspace{0.02cm}a\hspace{0.03cm}d}
\delta^{\hspace{0.02cm}b\hspace{0.03cm}c}
-
\delta^{\hspace{0.02cm}a\hspace{0.03cm}c}
\delta^{\hspace{0.02cm}b\hspace{0.03cm}d}.
\label{ap:D15}
\end{equation}


\numberwithin{equation}{section}
\section{Some relations with the averaged color charges for the special case of $N_{c} = 3$}
\numberwithin{equation}{section}
\label{appendix_E}

In the section \ref{section_7}, we introduced the following combinations involving two averaged color charges:
\begin{equation}
	{\mathfrak q}_{1} \equiv 
	\hspace{0.03cm}\bigl\langle\hspace{0.03cm}\mathcal{Q}^{\hspace{0.03cm}e}_{1}
	\hspace{0.03cm}\bigr\rangle
	\bigl\langle\hspace{0.03cm}\mathcal{Q}^{\hspace{0.03cm}e}_{1}
	\hspace{0.03cm}\bigr\rangle,
	\quad
	{\mathfrak q}_{2} \equiv 
	\hspace{0.03cm}\bigl\langle\hspace{0.03cm}\mathcal{Q}^{\hspace{0.03cm}e}_{2}
	\hspace{0.03cm}\bigr\rangle
	\bigl\langle\hspace{0.03cm}\mathcal{Q}^{\hspace{0.03cm}e}_{2}
	\hspace{0.03cm}\bigr\rangle,
	\quad
	{\mathfrak q}_{12} \equiv 
	\hspace{0.03cm}\bigl\langle\hspace{0.03cm}\mathcal{Q}^{\hspace{0.03cm}e}_{1}
	\hspace{0.03cm}\bigr\rangle
	\bigl\langle\hspace{0.03cm}\mathcal{Q}^{\hspace{0.03cm}e}_{2}
	\hspace{0.03cm}\bigr\rangle,
	\vspace{-0.3cm}
\label{ap:E1}
\end{equation}
\begin{align}
	&\Lambda^{c} =  \Lambda^{c}(\tau)
	\equiv 
	f^{\hspace{0.03cm}c\,b_{1}\hspace{0.03cm}b_{2}\hspace{0.03cm}}
	\bigl\langle\hspace{0.03cm}\mathcal{Q}^{\,b_{1}}_{\hspace{0.03cm}1}
	(\tau)
	\hspace{0.03cm}\bigr\rangle
	\bigl\langle\hspace{0.03cm}\mathcal{Q}^{\,b_{2}}_{\hspace{0.03cm}2}
	(\tau)
	\hspace{0.03cm}\bigr\rangle,
	\notag\\[1ex]
	&\Omega^{\hspace{0.03cm}c}_{11} = \Omega^{\hspace{0.03cm}c}_{11}(\tau)
	\equiv 
	d^{\,c\,b_{1}\hspace{0.03cm}b^{\prime}_{1}\hspace{0.03cm}}
	\bigl\langle\hspace{0.03cm}\mathcal{Q}^{\,b_{1}}_{\hspace{0.03cm}1}
	(\tau)
	\hspace{0.03cm}\bigr\rangle
	\bigl\langle\hspace{0.03cm}\mathcal{Q}^{\,b^{\prime}_{1}}_{\hspace{0.03cm}1}
	(\tau)
	\hspace{0.03cm}\bigr\rangle,
	\notag\\[1ex]
	&\Omega^{\hspace{0.03cm}c}_{22} = \Omega^{\hspace{0.03cm}c}_{22}(\tau)
	\equiv 
	d^{\,c\,b_{2}\hspace{0.03cm}b^{\prime}_{2}\hspace{0.03cm}}
	\bigl\langle\hspace{0.03cm}\mathcal{Q}^{\,b_{2}}_{\hspace{0.03cm}2}
	(\tau)
	\hspace{0.03cm}\bigr\rangle
	\bigl\langle\hspace{0.03cm}\mathcal{Q}^{\,b^{\prime}_{2}}_{\hspace{0.03cm}2}
	(\tau)
	\hspace{0.03cm}\bigr\rangle,
	\notag\\[1ex]
	&\Omega^{\hspace{0.03cm}c}_{12} = \Omega^{\hspace{0.03cm}c}_{12}(\tau)
	\equiv 
	d^{\,c\,b_{1}\hspace{0.03cm}b^{\prime}_{2}\hspace{0.03cm}}
	\bigl\langle\hspace{0.03cm}\mathcal{Q}^{\,b_{1}}_{\hspace{0.03cm}1}
	(\tau)
	\hspace{0.03cm}\bigr\rangle
	\bigl\langle\hspace{0.03cm}\mathcal{Q}^{\,b^{\prime}_{2}}_{\hspace{0.03cm}2}
	(\tau)
	\hspace{0.03cm}\bigr\rangle.
	\notag
\end{align}
These functions are included in the kinetic equations (\ref{eq:7x}), (\ref{eq:7c}), (\ref{eq:8u}) and (\ref{eq:8i}) in the form of colorless combinations of the type
$$
{\mathfrak q}_{1},\;\;
{\mathfrak q}_{2},\;\;
{\mathfrak q}_{12},\quad
\Lambda^{2} \equiv \Lambda^{e}\Lambda^{e},\quad
\Omega^{\hspace{0.03cm}2}_{12}
\equiv \Omega^{\hspace{0.03cm}e}_{12}\hspace{0.03cm}\Omega^{\hspace{0.03cm}e}_{12}
\quad \mbox{and}\quad \Omega^{\hspace{0.03cm}e}_{11}\hspace{0.03cm}\Omega^{\hspace{0.03cm}e}_{22}.
$$
In this Appendix, for a particular and physically most important case when $N_{c} = 3$, we derive formulas that allow us to get rid of at least the last two colorless combinations associated with the symmetric structural constants.\\
\indent As a first step, we take a look at the relation (\ref{ap:D6}), valid for an arbitrary $N_{c}$, and contract it with $\bigl\langle\hspace{0.03cm}\mathcal{Q}^{\hspace{0.03cm}a}_{1}
\hspace{0.03cm}\bigr\rangle
\bigl\langle\hspace{0.03cm}\mathcal{Q}^{\hspace{0.03cm}b}_{2}
\hspace{0.03cm}\bigr\rangle
\hspace{0.03cm}
\bigl\langle\hspace{0.03cm}\mathcal{Q}^{\hspace{0.03cm}c}_{1}
\hspace{0.03cm}\bigr\rangle
\bigl\langle\hspace{0.03cm}\mathcal{Q}^{\hspace{0.03cm}d}_{2}
\hspace{0.03cm}\bigr\rangle$.
Making use of the definitions (\ref{ap:E1}), as a result we have
\begin{equation} 
\Lambda^{2} = \frac{2}{N_{c}}\,\bigl({\mathfrak q}_{1}\hspace{0.03cm}
{\mathfrak q}_{2} - {\mathfrak q}^{2}_{12}\bigr)
\,+\,
\bigl(\hspace{0.03cm}\Omega^{\hspace{0.03cm}e}_{11}\hspace{0.03cm}\Omega^{\hspace{0.03cm}e}_{22} -	
\Omega^{\hspace{0.03cm}2}_{12}\bigr).
\label{ap:E2}
\end{equation}	
For the $SU(3_{c})$ color group in the previous relation, we can eliminate  either the colorless combination 
$\Omega^{\hspace{0.03cm}e}_{11}\hspace{0.03cm}\Omega^{\hspace{0.03cm}e}_{22}$ or the colorless one $\Omega^{\hspace{0.03cm}2}_{12}$. For this purpose, we use the equality (\ref{ap:D13}). We contract (\ref{ap:D13}) again with the same combination of the averaged color charges, namely with 
$\bigl\langle\hspace{0.03cm}\mathcal{Q}^{\hspace{0.03cm}a}_{1}
\hspace{0.03cm}\bigr\rangle
\bigl\langle\hspace{0.03cm}\mathcal{Q}^{\hspace{0.03cm}b}_{2}
\hspace{0.03cm}\bigr\rangle
\hspace{0.03cm}
\bigl\langle\hspace{0.03cm}\mathcal{Q}^{\hspace{0.03cm}c}_{1}
\hspace{0.03cm}\bigr\rangle
\bigl\langle\hspace{0.03cm}\mathcal{Q}^{\hspace{0.03cm}d}_{2}
\hspace{0.03cm}\bigr\rangle$, what gives us this time the relation of the form
$$
2\hspace{0.04cm}\Omega^{\hspace{0.03cm}2}_{12}
+\,
\Omega^{\hspace{0.03cm}e}_{11}\hspace{0.03cm}\Omega^{\hspace{0.03cm}e}_{22}
=
\frac{1}{3}\,\bigl(2\hspace{0.03cm}{\mathfrak q}^{2}_{12} 
+\,
{\mathfrak q}_{1}\hspace{0.03cm}{\mathfrak q}_{2}\bigr).
$$ 
It enables to exclude in (\ref{ap:E2}), for the particular case of $N_{c} = 3$, either  $\Omega^{\hspace{0.03cm}e}_{11}\hspace{0.03cm}\Omega^{\hspace{0.03cm}e}_{22}$ or $\Omega^{\hspace{0.03cm}2}_{12}$. This gives us the following two relations:
\begin{equation}
\Lambda^{2} = {\mathfrak q}_{1}\hspace{0.03cm}{\mathfrak q}_{2} 
\,-\,
3\hspace{0.04cm}\Omega^{\hspace{0.03cm}2}_{12}
\label{ap:E3}
\end{equation}
and
\begin{equation}
\Lambda^{2} = \frac{1}{2}\,{\mathfrak q}_{1}\hspace{0.03cm}
{\mathfrak q}_{2} - {\mathfrak q}^{2}_{12}
\,+\,
\frac{3}{2}\,\Omega^{\hspace{0.03cm}e}_{11}\hspace{0.03cm}\Omega^{\hspace{0.03cm}e}_{22}. 
\label{ap:E4}
\end{equation}
Note that in the first equality, the contribution proportional to ${\mathfrak q}^{2}_{12}$ exactly reduced. Let us give two more useful relations that follow from (\ref{ap:D13}) when contracting it with  $\bigl\langle\hspace{0.03cm}\mathcal{Q}^{\hspace{0.03cm}a}_{1}
\hspace{0.03cm}\bigr\rangle
\bigl\langle\hspace{0.03cm}\mathcal{Q}^{\hspace{0.03cm}b}_{1}
\hspace{0.03cm}\bigr\rangle
\hspace{0.03cm}
\bigl\langle\hspace{0.03cm}\mathcal{Q}^{\hspace{0.03cm}c}_{1}
\hspace{0.03cm}\bigr\rangle
\bigl\langle\hspace{0.03cm}\mathcal{Q}^{\hspace{0.03cm}d}_{2}
\hspace{0.03cm}\bigr\rangle$ 
or
$\bigl\langle\hspace{0.03cm}\mathcal{Q}^{\hspace{0.03cm}a}_{2}
\hspace{0.03cm}\bigr\rangle
\bigl\langle\hspace{0.03cm}\mathcal{Q}^{\hspace{0.03cm}b}_{2}
\hspace{0.03cm}\bigr\rangle
\hspace{0.03cm}
\bigl\langle\hspace{0.03cm}\mathcal{Q}^{\hspace{0.03cm}c}_{2}
\hspace{0.03cm}\bigr\rangle
\bigl\langle\hspace{0.03cm}\mathcal{Q}^{\hspace{0.03cm}d}_{1}
\hspace{0.03cm}\bigr\rangle$. Here, we get, correspondingly
\begin{equation}
\Omega^{\hspace{0.03cm}e}_{11}\hspace{0.03cm}\Omega^{\hspace{0.03cm}e}_{12}	
=
\frac{1}{3}\,{\mathfrak q}_{1}\hspace{0.03cm}{\mathfrak q}_{12}
\quad\mbox{or}\quad
\Omega^{\hspace{0.03cm}e}_{22}\hspace{0.03cm}\Omega^{\hspace{0.03cm}e}_{12}	
=
\frac{1}{3}\,{\mathfrak q}_{2}\hspace{0.03cm}{\mathfrak q}_{12}.
\label{ap:E5}
\end{equation}
\indent Finally, we consider another type of the contractions that arise in sections \ref{section_9} and \ref{section_10}, namely,
\begin{equation}
d^{\,b_{1}\hspace{0.03cm}c\hspace{0.03cm}e}\Lambda^{c}\,
\Omega^{\hspace{0.03cm}e}_{12}
\hspace{0.03cm}
\bigl\langle\hspace{0.03cm}\mathcal{Q}^{\,b_{1}}_{1}
\hspace{0.03cm}\bigr\rangle
\quad
\mbox{and}
\quad
d^{\,b_{2}\hspace{0.03cm}c\hspace{0.03cm}e}\Lambda^{c}\,
\Omega^{\hspace{0.03cm}e}_{12}
\hspace{0.03cm}
\bigl\langle\hspace{0.03cm}\mathcal{Q}^{\,b_{2}}_{2}
\hspace{0.03cm}\bigr\rangle.
\label{ap:E6}
\end{equation}
For the sake of specificity, let us examine the first of these. Here, we have
\[
d^{\,b_{1}\hspace{0.03cm}c\hspace{0.03cm}e}\Lambda^{c}\,
\Omega^{\hspace{0.03cm}e}_{12}
\hspace{0.03cm}
\bigl\langle\hspace{0.03cm}\mathcal{Q}^{\,b_{1}}_{1}
\hspace{0.03cm}\bigr\rangle
=
\Lambda^{c}\hspace{0.03cm}d^{\,b_{1}\hspace{0.03cm}c\hspace{0.03cm}e}\,
d^{\,e\hspace{0.03cm}a_{1}\hspace{0.03cm}a_{2}}
\hspace{0.03cm}
\bigl\langle\hspace{0.03cm}\mathcal{Q}^{\,a_{1}}_{1}
\hspace{0.03cm}\bigr\rangle
\hspace{0.03cm}
\bigl\langle\hspace{0.03cm}\mathcal{Q}^{\,a_{2}}_{2}
\hspace{0.03cm}\bigr\rangle
\hspace{0.03cm}
\bigl\langle\hspace{0.03cm}\mathcal{Q}^{\,b_{1}}_{1}
\hspace{0.03cm}\bigr\rangle
\equiv
\]
\[
\Lambda^{c}\hspace{0.03cm}
\biggl[
\frac{1}{3}\,\bigl\{T^{\,a_{1}},T^{\,a_{2}}
\bigr\}^{b_{1}\hspace{0.03cm}c}
-
\frac{1}{3}\,\bigl(
\delta^{\hspace{0.02cm}a_{1}\hspace{0.03cm}a_{2}}
\delta^{\hspace{0.02cm}b_{1}\hspace{0.03cm}c} 
-
\delta^{\hspace{0.02cm}a_{1}\hspace{0.03cm}c}
\delta^{\hspace{0.02cm}a_{2}\hspace{0.03cm}b_{1}}
-
\delta^{\hspace{0.02cm}a_{1}\hspace{0.03cm}b_{1}}
\delta^{\hspace{0.02cm}a_{2}\hspace{0.03cm}c}\hspace{0.03cm}\bigr)
\biggr]
\hspace{0.03cm}
\bigl\langle\hspace{0.03cm}\mathcal{Q}^{\,a_{1}}_{1}
\hspace{0.03cm}\bigr\rangle
\hspace{0.03cm}
\bigl\langle\hspace{0.03cm}\mathcal{Q}^{\,a_{2}}_{2}
\hspace{0.03cm}\bigr\rangle
\hspace{0.03cm}
\bigl\langle\hspace{0.03cm}\mathcal{Q}^{\,b_{1}}_{1}
\hspace{0.03cm}\bigr\rangle.
\]
In the last step the relation (\ref{ap:D15}) was used. Obviously, all terms with the Kronecker delta symbol bringing zero contributions by virtue of 
\[
\Lambda^{c}\bigl\langle\hspace{0.03cm}\mathcal{Q}^{\,c}_{1}
\hspace{0.03cm}\bigr\rangle
=
\Lambda^{c}\bigl\langle\hspace{0.03cm}\mathcal{Q}^{\,c}_{2}
\hspace{0.03cm}\bigr\rangle
= 0,
\] 
and thus the last expression goes into
\[
-\frac{1}{3}\,\Lambda^{c}\hspace{0.03cm}\bigl[
f^{\hspace{0.03cm}a_{1}\,b_{1}\hspace{0.03cm}e\hspace{0.03cm}}
f^{\hspace{0.03cm}a_{2}\,e\hspace{0.03cm}c\hspace{0.03cm}}
+
f^{\hspace{0.03cm}a_{2}\,b_{1}\hspace{0.03cm}e\hspace{0.03cm}}
f^{\hspace{0.03cm}a_{1}\,e\hspace{0.03cm}c\hspace{0.03cm}}
\bigr]
\hspace{0.03cm}
\bigl\langle\hspace{0.03cm}\mathcal{Q}^{\,a_{1}}_{1}
\hspace{0.03cm}\bigr\rangle
\hspace{0.03cm}
\bigl\langle\hspace{0.03cm}\mathcal{Q}^{\,a_{2}}_{2}
\hspace{0.03cm}\bigr\rangle
\hspace{0.03cm}
\bigl\langle\hspace{0.03cm}\mathcal{Q}^{\,b_{1}}_{1}
\hspace{0.03cm}\bigr\rangle
\equiv
\frac{1}{3}\,f^{\hspace{0.03cm}a_{1}\,e\hspace{0.03cm}c\hspace{0.03cm}}
\Lambda^{e}\Lambda^{c}\hspace{0.03cm}
\bigl\langle\hspace{0.03cm}\mathcal{Q}^{\,a_{1}}_{1}
\hspace{0.03cm}\bigr\rangle
= 0.
\]
Thus, for the contractions (\ref{ap:E6}) in the case of $N_{c} = 3$, we have the following equalities:
\begin{equation}
	d^{\;b_{1}\hspace{0.03cm}c\hspace{0.03cm}e}\Lambda^{c}\,
	\Omega^{\hspace{0.03cm}e}_{12}
	\hspace{0.03cm}
	\bigl\langle\hspace{0.03cm}\mathcal{Q}^{\,b_{1}}_{1}
	\hspace{0.03cm}\bigr\rangle = 0
	\quad
	\mbox{and}
	\quad
	d^{\;b_{2}\hspace{0.03cm}c\hspace{0.03cm}e}\Lambda^{c}\,
	\Omega^{\hspace{0.03cm}e}_{12}
	\hspace{0.03cm}
	\bigl\langle\hspace{0.03cm}\mathcal{Q}^{\,b_{2}}_{2}
	\hspace{0.03cm}\bigr\rangle = 0.
	\label{ap:E7}
\end{equation}


\numberwithin{equation}{section}
\section{Imaginary contributions to the equation for~$\Lambda^{2}$}
\numberwithin{equation}{section}
\label{appendix_F}

In this Appendix, we give a couple of specific examples of calculating the imaginary contributions to the evolution equation for the colorless combination 
$\Lambda^{2}$, Eq.\,(\ref{eq:10t}), and prove that these contributions vanish for $N_{c} = 3$. Consider first the imaginary contributions to the expression (\ref{eq:10y}). By virtue of the first contraction in (\ref{eq:9t}), here we have the following chain of relations for the purely imaginary part:
\begin{align}
&{N}^{(2)}_{\hspace{0.02cm}{\bf k}}\hspace{0.03cm}
\Lambda^{e}\hspace{0.03cm}
f^{\hspace{0.03cm}e\,a_{1}\hspace{0.03cm}a_{2}\hspace{0.03cm}}
\bigl\langle \mathcal{Q}^{\,a_{2}}_{\hspace{0.03cm}2}\hspace{0.02cm}\bigr\rangle
\hspace{0.03cm}
{\rm tr}\hspace{0.03cm}
\bigl(\hspace{0.03cm}T^{\,a^{\prime}_{2}}\hspace{0.03cm}T^{\,a_{1}}\hspace{0.03cm}T^{\,e^{\prime}_{1}}\hspace{0.03cm}T^{\,a^{\prime\prime}_{2}}\hspace{0.03cm}
T^{\,c_{1}}\hspace{0.01cm}\bigr)
\hspace{0.02cm}
\bigl\langle\hspace{0.03cm}\mathcal{Q}^{\,c_{1}}_{1}
\hspace{0.03cm}\bigr\rangle
\hspace{0.03cm}
\bigl\langle\hspace{0.03cm}\mathcal{Q}^{\hspace{0.03cm}a^{\prime}_{2}}_{2}
\hspace{0.03cm}\bigr\rangle
\bigl\langle\hspace{0.03cm}\mathcal{Q}^{\hspace{0.03cm}a^{\prime\prime}_{2}}_{2}\hspace{0.03cm}\bigr\rangle
\hspace{0.03cm}
\bigl\langle\hspace{0.03cm}\mathcal{Q}^{\hspace{0.03cm}e^{\prime}_{1}}_{1}\hspace{0.03cm}\bigr\rangle
=
\notag\\[1ex]
-\hspace{0.03cm}\frac{i}{2}\,
&{N}^{(2)}_{\hspace{0.02cm}{\bf k}}\hspace{0.03cm}
\Lambda^{e}\hspace{0.03cm}
f^{\hspace{0.03cm}e\,a_{1}\hspace{0.03cm}a_{2}\hspace{0.03cm}}
f^{\hspace{0.03cm}e^{\prime}_{1}\hspace{0.02cm}a_{1}\hspace{0.03cm}b}\,
{\rm tr}\hspace{0.03cm}
\bigl(\hspace{0.03cm}T^{\,b}\hspace{0.03cm}T^{\,a^{\prime}_{2}}\hspace{0.03cm}
T^{\,c_{1}}T^{\,a^{\prime\prime}_{2}}\hspace{0.01cm}\bigr)
\hspace{0.03cm}
\bigl\langle \mathcal{Q}^{\,a_{2}}_{\hspace{0.03cm}2}\hspace{0.02cm}\bigr\rangle
\hspace{0.03cm}
\bigl\langle\hspace{0.03cm}\mathcal{Q}^{\,c_{1}}_{1}
\hspace{0.03cm}\bigr\rangle
\hspace{0.03cm}
\bigl\langle\hspace{0.03cm}\mathcal{Q}^{\hspace{0.03cm}a^{\prime}_{2}}_{2}\hspace{0.03cm}\bigr\rangle
\bigl\langle\hspace{0.03cm}\mathcal{Q}^{\hspace{0.03cm}a^{\prime\prime}_{2}}_{2}\hspace{0.03cm}\bigr\rangle
\hspace{0.03cm}
\bigl\langle\hspace{0.03cm}\mathcal{Q}^{\hspace{0.03cm}e^{\prime}_{1}}_{1}\hspace{0.03cm}\bigr\rangle
=
\label{ap:F1}
\end{align}
\[
-\hspace{0.03cm}\frac{i}{2}\,{N}^{(2)}_{\hspace{0.02cm}{\bf k}}\hspace{0.03cm}
\biggl[-\hspace{0.03cm}\frac{3}{2}\,\Lambda^{e}\hspace{0.03cm}\Lambda^{a_{1}}
\!f^{\hspace{0.03cm}e\,a_{1}\hspace{0.03cm}a_{2}\hspace{0.03cm}} 
\bigl\langle\hspace{0.03cm}
\mathcal{Q}^{\,a_{2}}_{\hspace{0.03cm}2}\hspace{0.02cm}\bigr\rangle
\hspace{0.03cm}{\mathfrak q}_{12}
\,-\,
\frac{1}{4}\,N_{c}\hspace{0.03cm}\Lambda^{e}\hspace{0.03cm}\Lambda^{d}
\hspace{0.03cm}
f^{\hspace{0.03cm}e\,a_{1}\hspace{0.03cm}a_{2}\hspace{0.03cm}}
f^{\hspace{0.03cm}e^{\prime}_{1}\hspace{0.02cm}a_{1}\hspace{0.03cm}b}\,
f^{\hspace{0.03cm}b\,a^{\prime\prime}_{2}\hspace{0.03cm}d\hspace{0.03cm}}
\hspace{0.02cm}
\bigl\langle\hspace{0.03cm}\mathcal{Q}^{\,a_{2}}_{2}
\hspace{0.03cm}\bigr\rangle
\hspace{0.03cm}
\bigl\langle\hspace{0.03cm}\mathcal{Q}^{\hspace{0.03cm}a^{\prime\prime}_{2}}_{2}\hspace{0.03cm}\bigr\rangle
\hspace{0.03cm}
\bigl\langle\hspace{0.03cm}\mathcal{Q}^{\hspace{0.03cm}e^{\prime}_{1}}_{1}\hspace{0.03cm}\bigr\rangle\,
+
\]
\[
\frac{1}{4}\,N_{c}\hspace{0.03cm}\Lambda^{e}\hspace{0.03cm}\Omega^{d}_{12}
\hspace{0.03cm}
f^{\hspace{0.03cm}e\,a_{1}\hspace{0.03cm}a_{2}\hspace{0.03cm}}
f^{\hspace{0.03cm}e^{\prime}_{1}\hspace{0.02cm}a_{1}\hspace{0.03cm}b}\,
d^{\,b\,a^{\prime\prime}_{2}\hspace{0.03cm}d\hspace{0.03cm}}
\hspace{0.02cm}
\bigl\langle\hspace{0.03cm}\mathcal{Q}^{\,a_{2}}_{2}
\hspace{0.03cm}\bigr\rangle
\hspace{0.03cm}
\bigl\langle\hspace{0.03cm}\mathcal{Q}^{\hspace{0.03cm}a^{\prime\prime}_{2}}_{2}\hspace{0.03cm}\bigr\rangle
\hspace{0.03cm}
\bigl\langle\hspace{0.03cm}\mathcal{Q}^{\hspace{0.03cm}e^{\prime}_{1}}_{1}\hspace{0.03cm}\bigr\rangle
\biggr],
\]
where we have used the symmetry relation (\ref{ap:D10}) connecting the fifth-order trace with the fourth-order trace, an expression for a fourth-order trace (\ref{ap:D4}), and the corresponding definitions for color combinations, Eq.\,(\ref{ap:E1}). Obviously, the first term in square brackets on the most right-hand side of (\ref{ap:F1}) goes to zero. The second term can be represented as follows:
\begin{equation}
\Lambda^{e}\hspace{0.03cm}\Lambda^{d}
\hspace{0.03cm}
f^{\hspace{0.03cm}e\,a_{1}\hspace{0.03cm}a_{2}\hspace{0.03cm}}
f^{\hspace{0.03cm}e^{\prime}_{1}\hspace{0.02cm}a_{1}\hspace{0.03cm}b}\,
f^{\hspace{0.03cm}b\,a^{\prime\prime}_{2}\hspace{0.03cm}d\hspace{0.03cm}}
\hspace{0.02cm}
\bigl\langle\hspace{0.03cm}\mathcal{Q}^{\,a_{2}}_{2}
\hspace{0.03cm}\bigr\rangle
\hspace{0.03cm}
\bigl\langle\hspace{0.03cm}\mathcal{Q}^{\hspace{0.03cm}a^{\prime\prime}_{2}}_{2}\hspace{0.03cm}\bigr\rangle
\hspace{0.03cm}
\bigl\langle\hspace{0.03cm}\mathcal{Q}^{\hspace{0.03cm}e^{\prime}_{1}}_{1}\hspace{0.03cm}\bigr\rangle
=
\label{ap:F2}
\end{equation}
\[
i\hspace{0.03cm}
\bigl(\hspace{0.03cm}T^{\,a_{2}}\hspace{0.03cm}T^{\,e^{\prime}_{1}}\hspace{0.03cm}T^{\,a^{\prime\prime}_{2}}\hspace{0.03cm}\bigr)^{e\hspace{0.03cm}d}
\Lambda^{e}\Lambda^{d}
\hspace{0.02cm}
\bigl\langle\hspace{0.03cm}\mathcal{Q}^{\,a_{2}}_{2}
\hspace{0.03cm}\bigr\rangle
\hspace{0.03cm}
\bigl\langle\hspace{0.03cm}\mathcal{Q}^{\hspace{0.03cm}a^{\prime\prime}_{2}}_{2}\hspace{0.03cm}\bigr\rangle
\hspace{0.03cm}
\bigl\langle\hspace{0.03cm}\mathcal{Q}^{\hspace{0.03cm}e^{\prime}_{1}}_{1}\hspace{0.03cm}\bigr\rangle
\equiv
i\hspace{0.03cm}(-1)^{3}\hspace{0.03cm}
\bigl(\hspace{0.03cm}T^{\,a^{\prime\prime}_{2}}\hspace{0.03cm}T^{\,e^{\prime}_{1}}\hspace{0.03cm}T^{\,a_{2}}\hspace{0.03cm}\bigr)^{d\hspace{0.03cm}e}
\Lambda^{e}\Lambda^{d}
\hspace{0.02cm}
\bigl\langle\hspace{0.03cm}\mathcal{Q}^{\,a_{2}}_{2}
\hspace{0.03cm}\bigr\rangle
\hspace{0.03cm}
\bigl\langle\hspace{0.03cm}\mathcal{Q}^{\hspace{0.03cm}a^{\prime\prime}_{2}}_{2}\hspace{0.03cm}\bigr\rangle
\hspace{0.03cm}
\bigl\langle\hspace{0.03cm}\mathcal{Q}^{\hspace{0.03cm}e^{\prime}_{1}}_{1}\hspace{0.03cm}\bigr\rangle.
\]
Here, in the last step, we rearranged the matrices in parentheses in reverse order, which led to the appearance of the factor $(-1)^{3}$ and the permutation of the indices $d$ and $e$. Due to the fact that the common multiplier on the right-hand side of (\ref{ap:F2}) is symmetric in the indices $d$ and $e$, as well as in the indices $a_{2}$ and $a^{\prime\prime}_{2}$, this expression also turns to zero.\\
\indent Further, by virtue of the second contraction in (\ref{eq:9t}), making use of the expression for the trace (\ref{ap:D4}), here we have the following basis relation for the imaginary part:
\begin{equation}
i\hspace{0.03cm}{N}^{(2)}_{\hspace{0.02cm}{\bf k}}\hspace{0.03cm}
\Lambda^{e}
f^{\hspace{0.03cm}e\,a_{1}\hspace{0.03cm}a_{2}\hspace{0.03cm}}
f^{\hspace{0.03cm}a_{1}\hspace{0.02cm}c^{\prime}\hspace{0.03cm}e_{1}}\hspace{0.03cm}
{\rm tr}\hspace{0.03cm}
\bigl(\hspace{0.03cm}T^{\,c^{\prime}}\hspace{0.03cm}T^{\,e^{\prime}_{1}}
\hspace{0.03cm}T^{\,a^{\prime\prime}_{2}}\hspace{0.03cm}T^{\,c_{1}}
\hspace{0.01cm}
\bigr)
\hspace{0.03cm}
\bigl\langle\hspace{0.03cm}
\mathcal{Q}^{\,a_{2}}_{\hspace{0.03cm}2}\hspace{0.02cm}\bigr\rangle
\hspace{0.03cm}
\bigl\langle\hspace{0.03cm}\mathcal{Q}^{\hspace{0.03cm}e_{1}}_{1}
\hspace{0.03cm}\bigl\rangle
\hspace{0.03cm}
\bigl\langle\hspace{0.03cm}\mathcal{Q}^{\hspace{0.03cm}a^{\prime\prime}_{2}}_{2}\hspace{0.03cm}\bigr\rangle
\hspace{0.03cm}
\bigl\langle\hspace{0.03cm}\mathcal{Q}^{\hspace{0.03cm}e^{\prime}_{1}}_{1}
\hspace{0.03cm}\bigr\rangle
\hspace{0.03cm}
\bigl\langle\hspace{0.03cm}\mathcal{Q}^{\hspace{0.03cm}c_{1}}_{1}\hspace{0.03cm}\bigl\rangle\,
=
\label{ap:F3}
\end{equation}
\[
i\hspace{0.03cm}{N}^{(2)}_{\hspace{0.02cm}{\bf k}}\hspace{0.03cm}
\biggl[-\hspace{0.03cm}\frac{1}{2}\,\Lambda^{e}\hspace{0.03cm}\Lambda^{a_{1}}
\!f^{\hspace{0.03cm}e\,a_{1}\hspace{0.03cm}a_{2}\hspace{0.03cm}} 
\bigl\langle\hspace{0.03cm}
\mathcal{Q}^{\,a_{2}}_{\hspace{0.03cm}2}\hspace{0.02cm}\bigr\rangle
\hspace{0.03cm}{\mathfrak q}_{1}
\,+\,
\frac{1}{4}\,N_{c}\hspace{0.03cm}\Lambda^{e}\hspace{0.03cm}\Lambda^{e^{\prime}}
\hspace{0.03cm}
f^{\hspace{0.03cm}e\,a_{1}\hspace{0.03cm}a_{2}\hspace{0.03cm}}
f^{\hspace{0.03cm}a_{1}\hspace{0.02cm}c^{\prime}\hspace{0.03cm}e_{1}}
f^{\hspace{0.03cm}c^{\prime}\,c_{1}\hspace{0.03cm}e^{\prime}\hspace{0.03cm}}
\hspace{0.02cm}
\bigl\langle\hspace{0.03cm}\mathcal{Q}^{\,a_{2}}_{2}
\hspace{0.03cm}\bigr\rangle
\hspace{0.03cm}
\bigl\langle\hspace{0.03cm}\mathcal{Q}^{\hspace{0.03cm}e_{1}}_{1}\hspace{0.03cm}\bigr\rangle
\hspace{0.03cm}
\bigl\langle\hspace{0.03cm}\mathcal{Q}^{\hspace{0.03cm}c_{1}}_{1}\hspace{0.03cm}\bigr\rangle\,
+
\]
\[
\frac{1}{4}\,N_{c}\hspace{0.03cm}\Lambda^{e}\hspace{0.03cm}
\Omega^{e^{\prime}}_{12}
\hspace{0.03cm}
f^{\hspace{0.03cm}e\,a_{1}\hspace{0.03cm}a_{2}\hspace{0.03cm}}
f^{\hspace{0.03cm}a_{1}\hspace{0.02cm}c^{\prime}\hspace{0.03cm}e_{1}}
d^{\,c^{\prime}\,c_{1}\hspace{0.03cm}e^{\prime}\hspace{0.03cm}}
\hspace{0.02cm}
\bigl\langle\hspace{0.03cm}\mathcal{Q}^{\,a_{2}}_{2}
\hspace{0.03cm}\bigr\rangle
\hspace{0.03cm}
\bigl\langle\hspace{0.03cm}\mathcal{Q}^{\hspace{0.03cm}e_{1}}_{1}\hspace{0.03cm}\bigr\rangle
\hspace{0.03cm}
\bigl\langle\hspace{0.03cm}\mathcal{Q}^{\hspace{0.03cm}c_{1}}_{1}\hspace{0.03cm}\bigr\rangle
\biggr].
\]
The first term in square brackets is zero. The last term in (\ref{ap:F3}) we will discuss just below when consider the imaginary contributions from contraction (\ref{eq:10yy}). It cancel with the corresponding contribution of this contraction.\\
\indent Let us now examine the imaginary contributions from the second contraction (\ref{eq:10yy}). We analyze the first contraction in (\ref{eq:10qq}) in complete analogy with the expression (\ref{ap:F1}). Here, we have
\begin{align}
	&{N}^{(2)}_{\hspace{0.02cm}{\bf k}}\hspace{0.03cm}
	\Lambda^{e}\hspace{0.03cm}
	f^{\hspace{0.03cm}e\,a_{1}\hspace{0.03cm}a_{2}\hspace{0.03cm}}
	\bigl\langle \mathcal{Q}^{\,a_{1}}_{\hspace{0.03cm}1}\hspace{0.02cm}\bigr\rangle
	\hspace{0.03cm}
	{\rm tr}\hspace{0.03cm}
	\bigl(\hspace{0.03cm}T^{\,a^{\prime\prime}_{1}}\hspace{0.03cm}T^{\,a_{2}}\hspace{0.03cm}T^{\,e^{\prime}_{2}}\hspace{0.03cm}T^{\,a^{\prime}_{1}}\hspace{0.03cm}
	T^{\,c_{1}}\hspace{0.01cm}\bigr)
	\hspace{0.02cm}
	\bigl\langle\hspace{0.03cm}\mathcal{Q}^{\,c_{1}}_{1}
	\hspace{0.03cm}\bigr\rangle
	\hspace{0.03cm}
	\bigl\langle\hspace{0.03cm}\mathcal{Q}^{\hspace{0.03cm}a^{\prime\prime}_{1}}_{1}
	\hspace{0.03cm}\bigr\rangle
	\bigl\langle\hspace{0.03cm}\mathcal{Q}^{\hspace{0.03cm}e^{\prime}_{2}}_{2}\hspace{0.03cm}\bigr\rangle
	\hspace{0.03cm}
	\bigl\langle\hspace{0.03cm}\mathcal{Q}^{\hspace{0.03cm}a^{\prime}_{1}}_{1}\hspace{0.03cm}\bigr\rangle
	=
	\notag\\[1ex]
	-\hspace{0.03cm}\frac{i}{2}\,
	&{N}^{(2)}_{\hspace{0.02cm}{\bf k}}\hspace{0.03cm}
	\Lambda^{e}
	f^{\hspace{0.03cm}e\,a_{1}\hspace{0.03cm}a_{2}\hspace{0.03cm}}
	f^{\hspace{0.03cm}e^{\prime}_{2}\hspace{0.02cm}a_{2}\hspace{0.03cm}b}\,
	{\rm tr}\hspace{0.03cm}
	\bigl(\hspace{0.03cm}T^{\,b}\hspace{0.03cm}T^{\,a^{\prime\prime}_{1}}\hspace{0.03cm}
	T^{\,c_{1}}T^{\,a^{\prime}_{1}}\hspace{0.01cm}\bigr)
	\hspace{0.03cm}
	\bigl\langle \mathcal{Q}^{\,a_{1}}_{\hspace{0.03cm}1}\hspace{0.02cm}\bigr\rangle
	\hspace{0.03cm}
	\bigl\langle\hspace{0.03cm}\mathcal{Q}^{\,c_{1}}_{1}
	\hspace{0.03cm}\bigr\rangle
	\hspace{0.03cm}
	\bigl\langle\hspace{0.03cm}\mathcal{Q}^{\hspace{0.03cm}a^{\prime\prime}_{1}}_{1}\hspace{0.03cm}\bigr\rangle
	\bigl\langle\hspace{0.03cm}\mathcal{Q}^{\hspace{0.03cm}e^{\prime}_{2}}_{2}\hspace{0.03cm}\bigr\rangle
	\hspace{0.03cm}
	\bigl\langle\hspace{0.03cm}\mathcal{Q}^{\hspace{0.03cm}a^{\prime}_{1}}_{1}\hspace{0.03cm}\bigr\rangle
	=
	\label{ap:F4}
\end{align}
\[
-\hspace{0.03cm}\frac{i}{2}\,{N}^{(2)}_{\hspace{0.02cm}{\bf k}}
\hspace{0.03cm}
\biggl[2\hspace{0.03cm}\Lambda^{e}\hspace{0.03cm}\Lambda^{a_{2}}
f^{\hspace{0.03cm}e\,a_{1}\hspace{0.03cm}a_{2}\hspace{0.03cm}} 
\bigl\langle\hspace{0.03cm}
\mathcal{Q}^{\,a_{1}}_{\hspace{0.03cm}1}\hspace{0.02cm}\bigr\rangle
\hspace{0.03cm}{\mathfrak q}_{1}
\,+\,
\frac{1}{4}\,N_{c}\hspace{0.04cm}\Lambda^{e}\hspace{0.04cm}\Omega^{\,d}_{11}
\hspace{0.03cm}
f^{\hspace{0.03cm}e\,a_{1}\hspace{0.03cm}a_{2}\hspace{0.03cm}}
f^{\hspace{0.03cm}e^{\prime}_{2}\hspace{0.02cm}a_{2}\hspace{0.03cm}b}\,
d^{\;b\,a^{\prime}_{1}\hspace{0.01cm}d\hspace{0.03cm}}
\hspace{0.02cm}
\bigl\langle\hspace{0.03cm}\mathcal{Q}^{\,a_{1}}_{1}
\hspace{0.03cm}\bigr\rangle
\hspace{0.03cm}
\bigl\langle\hspace{0.03cm}\mathcal{Q}^{\hspace{0.03cm}e^{\prime}_{2}}_{2}\hspace{0.03cm}\bigr\rangle
\hspace{0.03cm}
\bigl\langle\hspace{0.03cm}\mathcal{Q}^{\hspace{0.03cm}a^{\prime}_{1}}_{1}\hspace{0.03cm}\bigr\rangle\,
\biggr].
\]
The first term in square brackets is obviously zero. The second term for an arbitrary value of $N_{c}$, is nonzero. We consider the special case, when $N_{c} = 3$. Then by virtue of the definition of the function $\Omega^{\,d}_{11}$, Eq.\,(\ref{ap:E1}), and the relation (\ref{ap:D13}) we  have 
\[
d^{\;b\,a^{\prime}_{1}d\hspace{0.04cm}}
\hspace{0.02cm}\Omega^{\,d}_{11}
\hspace{0.03cm}
\bigl\langle\hspace{0.03cm}\mathcal{Q}^{\hspace{0.03cm}a^{\prime}_{1}}_{1}\hspace{0.03cm}\bigr\rangle
=
\frac{1}{3}\,
\bigl\langle\hspace{0.03cm}\mathcal{Q}^{\,b_{1}}_{1}\hspace{0.03cm}\bigr\rangle
\hspace{0.03cm}{\mathfrak q}_{1}.
\]
Taking into account this equality and the definition of the function $\Lambda^{a}$, for  $N_{c} = 3$, instead of (\ref{ap:F4}) we obtain finally
\[
-\hspace{0.03cm}\frac{i}{2}\,{N}^{(2)}_{\hspace{0.02cm}{\bf k}}
\hspace{0.03cm}
\biggl(\frac{1}{4}\,\Lambda^{e}\hspace{0.03cm}
\Lambda^{a_{2}}
f^{\hspace{0.03cm}e\,a_{1}\hspace{0.03cm}a_{2}\hspace{0.03cm}}
\hspace{0.02cm}
\bigl\langle\hspace{0.03cm}\mathcal{Q}^{\,a_{1}}_{1}
\hspace{0.03cm}\bigr\rangle\hspace{0.03cm}{\mathfrak q}_{1}\!
\biggr)
\equiv 0.
\]
Thus, for the special case of the $SU(3_{c})$ color group, this imaginary contribution goes to zero.\\
\indent It remains for us to consider the imaginary contribution associated with the second contraction in (\ref{eq:10qq}). Using the formula for the fourth-order trace (\ref{ap:D4}), here we have the following original expression for the imaginary part: 
\begin{equation}
i\hspace{0.03cm}{N}^{(2)}_{\hspace{0.02cm}{\bf k}}\hspace{0.03cm}
\Lambda^{e}
f^{\hspace{0.03cm}e\,a_{1}\hspace{0.03cm}a_{2}\hspace{0.03cm}}
f^{\hspace{0.03cm}a_{2}\hspace{0.02cm}c^{\prime}\hspace{0.03cm}e_{2}}
\hspace{0.03cm}
{\rm tr}\hspace{0.03cm}
\bigl(\hspace{0.03cm}T^{\,c^{\prime}}\hspace{0.03cm}T^{\,a^{\prime}_{1}}\hspace{0.03cm}T^{\,e^{\prime}_{2}}\hspace{0.03cm}T^{\,c_{1}}\hspace{0.01cm}
\bigr)
\hspace{0.03cm}
\bigl\langle\hspace{0.03cm}
\mathcal{Q}^{\,a_{1}}_{\hspace{0.03cm}1}\hspace{0.02cm}\bigr\rangle
\hspace{0.03cm}
\bigl\langle\hspace{0.03cm}
\mathcal{Q}^{\,c_{1}}_{\hspace{0.03cm}1}\hspace{0.02cm}\bigr\rangle
\hspace{0.03cm}
\bigl\langle\hspace{0.03cm}\mathcal{Q}^{\hspace{0.03cm}e_{2}}_{2}
\hspace{0.03cm}\bigr\rangle
\hspace{0.03cm}
\bigl\langle\hspace{0.03cm}\mathcal{Q}^{\hspace{0.03cm}e^{\prime}_{2}}_{2}
\hspace{0.03cm}\bigr\rangle
\hspace{0.03cm}
\bigl\langle\hspace{0.03cm}\mathcal{Q}^{\hspace{0.03cm}a^{\prime}_{1}}_{1}
\hspace{0.03cm}\bigr\rangle\,
=
\label{ap:F5}
\end{equation}
\[
i\hspace{0.03cm}{N}^{(2)}_{\hspace{0.02cm}{\bf k}}\hspace{0.03cm}
\biggl[\hspace{0.03cm}\frac{3}{2}\,\Lambda^{e}\hspace{0.03cm}\Lambda^{a_{2}}
f^{\hspace{0.03cm}e\,a_{1}\hspace{0.03cm}a_{2}\hspace{0.03cm}} 
\bigl\langle\hspace{0.03cm}
\mathcal{Q}^{\,a_{1}}_{\hspace{0.03cm}1}\hspace{0.02cm}\bigr\rangle
\hspace{0.03cm}{\mathfrak q}_{12}
\,+\,
\frac{1}{4}\,N_{c}\hspace{0.03cm}\Lambda^{e}\hspace{0.03cm}\Lambda^{d}
	\hspace{0.01cm}
	f^{\hspace{0.03cm}e\,a_{1}\hspace{0.03cm}a_{2}\hspace{0.03cm}}
	f^{\hspace{0.03cm}a_{2}\hspace{0.02cm}c^{\prime}\hspace{0.03cm}e_{2}}
	f^{\hspace{0.03cm}c^{\prime}\,c_{1}\hspace{0.03cm}d\hspace{0.03cm}}
	\hspace{0.02cm}
	\bigl\langle\hspace{0.03cm}\mathcal{Q}^{\,a_{1}}_{1}
	\hspace{0.03cm}\bigr\rangle
	\hspace{0.03cm}
	\bigl\langle\hspace{0.03cm}\mathcal{Q}^{\hspace{0.03cm}e_{2}}_{2}
	\hspace{0.03cm}\bigr\rangle
	\hspace{0.03cm}
	\bigl\langle\hspace{0.03cm}\mathcal{Q}^{\hspace{0.03cm}c_{1}}_{1}
	\hspace{0.03cm}\bigr\rangle\,
+
\]
\[
\frac{1}{4}\,N_{c}\hspace{0.03cm}\Lambda^{e}\hspace{0.04cm}
\Omega^{\,d}_{12}
\hspace{0.04cm}
f^{\hspace{0.03cm}e\,a_{1}\hspace{0.03cm}a_{2}\hspace{0.03cm}}
f^{\hspace{0.03cm}a_{2}\hspace{0.02cm}c^{\prime}\hspace{0.03cm}e_{2}}
d^{\,c^{\prime}\,c_{1}\hspace{0.03cm}d\hspace{0.03cm}}
\bigl\langle\hspace{0.03cm}\mathcal{Q}^{\,a_{1}}_{1}
\hspace{0.03cm}\bigr\rangle
\hspace{0.03cm}
\bigl\langle\hspace{0.03cm}\mathcal{Q}^{\hspace{0.03cm}c_{1}}_{1}\hspace{0.03cm}
\bigr\rangle
\hspace{0.03cm}
\bigl\langle\hspace{0.03cm}\mathcal{Q}^{\hspace{0.03cm}e_{2}}_{2}\hspace{0.03cm}
\bigr\rangle
\biggr].
\]
As usual, the first term in square brackets turns to zero. The second term is zero due to the same reasoning that led us to the identity (\ref{ap:F2}). Finally, we add the last remaining term to the last remaining one in square brackets in (\ref{ap:F3}), changing, where it is necessary, the color indices of summation:
\[
\frac{i}{4}\hspace{0.03cm}{N}^{(2)}_{\hspace{0.02cm}{\bf k}}\hspace{0.03cm}
N_{c}\hspace{0.03cm}\Lambda^{e}\hspace{0.04cm}\Omega^{\,d}_{12} 
\bigl(\hspace{0.03cm}
f^{\hspace{0.03cm}e\,a_{2}\hspace{0.03cm}e_{2}\hspace{0.03cm}}
f^{\hspace{0.03cm}a_{2}\hspace{0.02cm}c^{\prime}\hspace{0.03cm}a_{1}}
+
f^{\hspace{0.03cm}e\,a_{1}\hspace{0.03cm}a_{2}\hspace{0.03cm}}
f^{\hspace{0.03cm}a_{2}\hspace{0.02cm}c^{\prime}\hspace{0.03cm}e_{2}}
\bigr)
d^{\,c^{\prime}\,c_{1}\hspace{0.03cm}d\hspace{0.03cm}}
\bigl\langle\hspace{0.03cm}\mathcal{Q}^{\,a_{1}}_{1}
\hspace{0.03cm}\bigr\rangle
\hspace{0.03cm}
\bigl\langle\hspace{0.03cm}\mathcal{Q}^{\hspace{0.03cm}c_{1}}_{1}\hspace{0.03cm}
\bigr\rangle
\hspace{0.03cm}
\bigl\langle\hspace{0.03cm}\mathcal{Q}^{\hspace{0.03cm}e_{2}}_{2}\hspace{0.03cm}
\bigr\rangle\,
=
\]
\[
\frac{i}{4}\hspace{0.03cm}{N}^{(2)}_{\hspace{0.02cm}{\bf k}}\hspace{0.03cm}
N_{c}\hspace{0.03cm}\Lambda^{e}\hspace{0.04cm}\Omega^{\,d}_{12} 
\hspace{0.03cm}\bigl(\hspace{0.03cm}
f^{\hspace{0.03cm}e\,c^{\prime}\hspace{0.03cm}s\hspace{0.03cm}}
f^{\hspace{0.03cm}s\hspace{0.02cm}a_{1}\hspace{0.03cm}e_{2}}
\bigr)
d^{\,c^{\prime}\,c_{1}\hspace{0.03cm}d\hspace{0.03cm}}
\bigl\langle\hspace{0.03cm}\mathcal{Q}^{\,a_{1}}_{1}
\hspace{0.03cm}\bigr\rangle
\hspace{0.03cm}
\bigl\langle\hspace{0.03cm}\mathcal{Q}^{\hspace{0.03cm}c_{1}}_{1}\hspace{0.03cm}
\bigr\rangle
\hspace{0.03cm}
\bigl\langle\hspace{0.03cm}\mathcal{Q}^{\hspace{0.03cm}e_{2}}_{2}\hspace{0.03cm}
\bigr\rangle
\equiv
\frac{i}{4}\hspace{0.03cm}{N}^{(2)}_{\hspace{0.02cm}{\bf k}}\hspace{0.03cm}
N_{c}\hspace{0.03cm}\Omega^{\,d}_{12}\hspace{0.03cm}
\bigl(\Lambda^{e}\Lambda^{s}f^{\hspace{0.03cm}e\,c^{\prime}\hspace{0.03cm}s\hspace{0.03cm}}\bigr)
\hspace{0.03cm}
d^{\,c^{\prime}\,c_{1}\hspace{0.03cm}d\hspace{0.03cm}}
\bigl\langle\hspace{0.03cm}\mathcal{Q}^{\hspace{0.03cm}c_{1}}_{1}\hspace{0.03cm}
\bigr\rangle.
\]
Here, in the last step, we have used the definition for the function $\Lambda^{s}$. Obviously, this relation is zero. Thus, the sum of the last two contributions in (\ref{ap:F3}) and (\ref{ap:F5}) turns to zero for an arbitrary value of $N_{c}$. Thus, we have explicitly shown that most of imaginary contributions of the contractions (\ref{eq:9t}) and (\ref{eq:10qq}) involving the matrix function $\mathcal{N}_{\bf k}$, vanish for $N_{c} = 3$. Unfortunately, we are left with two terms (the last term in (\ref{ap:F1}) and the second term in (\ref{ap:F3})) for which we have not been able to prove that they are equal to (or not equal to) zero, even for the special case of $N_{c} = 3$.\\
\indent All of the above reasoning also apply to the imaginary contributions from similar contractions of the type (\ref{eq:9t}) and (\ref{eq:10w}) only with another matrix function  $\mathcal{W}_{\bf k}$.
\end{appendices}
\newpage


\bibliographystyle{unsrt}
\bibliography{C:/OldDisk/D/Users/Markov/PAPERS/MarkovBibliography.bib}

\end{document}